\documentclass[aps, prl, twocolumn, nofootinbib, superscriptaddress]{revtex4-2}
\PassOptionsToPackage{dvipsnames}{xcolor}
\usepackage{amsmath}
\usepackage{mathrsfs}
\usepackage{amssymb}
\usepackage{amsfonts}
\usepackage{booktabs, array}
\usepackage[T1]{fontenc}

\usepackage[caption=false]{subfig}
\usepackage{tikz}
\usepackage{graphicx} 
\usepackage{multirow}
\usepackage{makecell}
\usepackage{tabularx, makecell, adjustbox}
\usepackage{longtable}

\def\apj{{\ Astrophys. J.}}
\def\mnras{{\ Mon. Not. Roy. Astron. Soc.}}
\def\aap{{\ Astronomy \& Astrophysics}}
\def\prd{{\ Phys. Rev. D}}

\def\araa{{\ Annual Review of Astronomy \& Astrophysics}}

\def\apjl{{\ Astroph. J. Lett.}}
\def\apjs{{\ Astroph. J. S.} }
\def\icarus{{\ Icarus}}

\usepackage{xcolor}
\definecolor{linkcolor}{rgb}{0.0,0.3,0.5}
\usepackage[
    hypertexnames=false,
    unicode,
    colorlinks=true,
    linkcolor=linkcolor,
    citecolor=linkcolor,
    filecolor=linkcolor,
    urlcolor=linkcolor,
    pdfusetitle
]{hyperref}
\usepackage{orcidlink}

\newcolumntype{L}[1]{>{\raggedright\arraybackslash}p{#1}}
\newcolumntype{C}[1]{>{\centering\arraybackslash}p{#1}}
\newcolumntype{R}[1]{>{\raggedleft\arraybackslash}p{#1}}

\graphicspath{./Plots}
\begin{document}
    \title{Probing  Formation Channels of Extreme Mass-Ratio Inspirals }
    \author{Houyi Sun}
\affiliation{Department of Astronomy, Tsinghua University, Beijing 100084, China}
\author{Ya-Ping Li}
\affiliation{Shanghai Astronomical Observatory, Chinese Academy of Sciences, Shanghai 200030, China}
\author{Zhen Pan\orcidlink{0000-0001-9608-009X}}
\affiliation{Tsung-Dao Lee Institute, Shanghai Jiao Tong University, Shanghai, 1 Lisuo Road, 201210, China}
\affiliation{School of Physics \& Astronomy, Shanghai Jiao Tong University, Shanghai, 800 Dongchuan Road, 200240, China}
    \author{Huan Yang \orcidlink{0000-0002-9965-3030}}
    \email{hyangdoa@tsinghua.edu.cn} \affiliation{Department of Astronomy, Tsinghua University, Beijing 100084, China}
    \date{August 2025}
    \begin{abstract}
        The population study of stellar-mass black hole (sBH) binaries with ground-based gravitational wave detection has achieved tremendous success in recent years. Future observation of extreme mass-ratio inspirals will similarly require proper population analysis that identify the formation channels, measuring the branch ratio(s) and characterizing major properties within each major channel. In this work, we propose that the measurement of eccentricity, inclination, and component mass provides critical information to distinguish different formation channels and probe detailed formation mechanisms. Focusing on the dry and wet extreme mass-ratio inspirals, we establish the theoretical expectation of these observables in each formation channel. We also discuss how their distributions can be used to probe  lifetime and turbulence level of active galactic nuclei disks, accretion patterns of supermassive black holes and population properties of sBHs within nuclear star clusters. 
        
    \end{abstract}
    \maketitle

    \section{Introduction}
    \label{sec:introduction}

The study of gravitational wave source populations generally provides a demographic description of the source distribution for each formation channel and offers critical insights into key aspects of individual formation mechanisms. One successful example is the measurement of binary black hole (BBH) merger events \cite{Abbott2021_GWTC2Pop, Abbott2023PopulationGWTC3,LIGOScientific:2025pvj} , which utilizes information about distances, component masses, and spins to begin inferring the source redshift distributions, branching ratios of various formation mechanisms, delay times, etc. With the rapidly increasing number of detection events from LIGO-Virgo-KAGRA observations, many important questions related to the formation of BBHs are expected to see significant progress in the near future \cite{Mapelli2021BBHformation,Mandel:2021smh,Gerosa:2021mno,Grobner:2020drr,Antonini:2016gqe}.

\begin{table*}[htbp]
\centering
\caption{Comparison between Wet EMRIs and Dry EMRIs at $r_{\rm p} = 10 M_\bullet$}
\label{tab:all}
\renewcommand\arraystretch{1.4} 
\begin{tabular}{@{}>{\bfseries}l C{0.4\textwidth} C{0.4\textwidth}@{}}
\toprule\toprule
 & \textbf{Wet EMRI} & \textbf{Dry EMRI} \\
\midrule
Eccentricity & 
\makecell[{{C{0.4\textwidth}}}]{Multi-body dynamics: $\lesssim 0.01$, Fig.~\ref{fig:dis}\\[2pt]
Turbulence: $\simeq 10^{-4}$, Fig.~\ref{fig:ecc_sca}} & 
$\gtrsim 0.01$, Fig.~\ref{fig:dry_e} \\
\midrule
Inclination &
\makecell[{{C{0.4\textwidth}}}]{Coherent accretion: aligned orbit\\[2pt]
Chaotic accretion: tilted distribution, Fig.~\ref{fig:inc_1e6}} & 
Tilted distribution, Fig.~\ref{fig:R_theta} \\
\midrule
sBH Mass & 
\makecell[{{C{0.4\textwidth}}}]{Capture: larger 35$M_\odot$ peak\\[2pt]
Accretion:  component mass increases\\[2pt] 
Mergers: new peaks around 20$M_\odot$ and 45$M_\odot$\\ Fig.~\ref{fig:mass_toal}} & 
Larger $35M_\odot$ peak, Fig.~\ref{fig:mass_dry}  \\
\bottomrule
\end{tabular}
\end{table*}

For space-borne gravitational wave detectors \cite{Babak2017_LISA_EMRI, Fan2020TianQinEMRI}, extreme mass-ratio inspirals (EMRIs) are among the primary extragalactic sources. Identifying and characterizing their formation channels are key scientific goals of these missions. Several major formation channels have been proposed. First, within the nuclear star cluster surrounding a massive black hole (MBH), stellar-mass black holes (sBHs) can be gravitationally scattered into extremely low-angular-momentum orbits. In such cases, gravitational wave emission during pericenter passages may efficiently damp these highly eccentric orbits before two-body scattering disrupts their trajectory toward the so-called ``loss cone'' \cite{Cutler1994SchwarzschildRR}. These ``dry EMRIs'' may contribute $\mathcal{O}(1)-\mathcal{O}(10^3)$ events to space-borne detectors, with main uncertainties arising from 
 the mass function of MBHs, the composition and distribution of nuclear star clusters \cite{Babak2017_LISA_EMRI}.  
Second, in a small fraction ($\mathcal{O}(1\%)-\mathcal{O}(10\%)$), according to observations \cite{MillerAGNfrac} of galaxies that host active galactic nuclei (AGNs), sBHs may be captured through interactions with the accretion disks and subsequently migrate toward the central MBHs. It turns out that disk-driven migration is a more efficient mechanism for generating EMRIs than the ``loss-cone'' mechanism, such that the overall number of ``wet EMRIs'' may be comparable to or even greater than that of ``dry EMRIs'' \cite{Levin2007,Pan202101,Pan:2021oob,Derdzinski:2022ltb}, despite the small fraction of galaxies hosting AGNs. The main uncertainties in the rate predictions arise from the modeling of nuclear star clusters, the mass function of MBHs, and the structure of accretion disks. Third, additional formation channels for EMRIs have been proposed, including the tidal disruption of sBH binaries near MBHs (the ``Hills'' mechanism \cite{Hills1988TidalDisruption}) and the Kozai-Lidov mechanism in systems with supermassive black hole (SMBH) binaries \cite{Naoz2022EKL_EMRI}. Determining the rate of the Hills channel requires a quantitative population study of the number and distribution of sBH binaries within nuclear star clusters, as well as the probability of binary tidal disruption and subsequent EMRI formation. Previous simulations show that Hills EMRIs are actually quite similar to dry EMRIs in the sensitivity band of space-borne gravitational wave missions \cite{Raveh2021}.
On the other hand, the EMRI formation rate in systems with SMBH binaries remains under debate \cite{Mazzolari2022MBHB_EMRI, BodeWegg2014EMRIinMBHB}. Therefore, in this work, we focus on the dry and wet EMRI channels, leaving the discussion of additional formation pathways to future studies when more quantitative predictions become available.

To diagnose EMRI formation channels, a primary objective is to identify the key observables and their distributions for each channel. For example, the spin magnitudes and orientations of sBHs provide crucial information for distinguishing field binaries from those formed in dynamical environments, particularly in the context of ground-based detectors \cite{Rodriguez2016SpinFormation}. For EMRIs, we will show that while dry EMRIs naturally retain eccentricities of $\geq 0.01$ in the Laser Interferometer Space Antenna (LISA) band, wet EMRIs are not perfectly {\it circular} either. Two possible mechanisms for exciting eccentricity in wet EMRIs are likely to operate: turbulence in the AGN disk and multi-body resonance effects within the disk. As a result, the eccentricities of wet EMRIs may be pumped to $\mathcal{O}(10^{-4})$-$\mathcal{O}(10^{-2})$ in the LISA band, depending on the level of turbulence and whether multiple sBHs migrate together. It should be noted that the eccentricity measurement precision for LISA, Taiji, and TianQin is expected to reach $\mathcal{O}(10^{-5})$ \cite{Babak2017_LISA_EMRI}. Such high precision may not only help distinguish  different formation channels, but also enable the characterization of key channel properties, such as the relevant surface density and average turbulence strength in AGN disks.

Recent studies have explored the possibility that accretion disks may directly introduce observable phase shifts into EMRI waveforms, which potentially enables the direct detection of disk effects from wet EMRIs \cite{Kocsis2011DiskEMRI, Derdzinski2020GasDiskEMRI}. This necessarily requires a rather high disk surface density ($\gtrsim 10^7 \ \mathrm{g \ cm^{-2}}$) in order to produce sufficiently strong effects \cite{Kocsis2011DiskEMRI}. This requirement cannot be satisfied by the canonical $\alpha$-disk scenario, but may be fulfilled by one of its variants, the so-called ``$\beta$-disk'' model \cite{Sakimoto1981MagneticViscosity}. The plausibility of this scenario remains to be tested by further numerical simulations and future observations. However, population studies involving eccentricity and inclination angle do not require a direct disk imprint on the waveform. Instead, they infer disk effects during the earlier stages of wet EMRIs by measuring orbital parameters. This type of {\rm indirect} measurement offers a unique opportunity to probe the formation and early dynamical evolution of EMRIs. 

In addition to eccentricity, the inclination angle relative to the spin axis of the central MBH is also a key observable in population studies. Our results show that dry EMRIs tend to favor prograde orbits, with the event rate varying by more than an order of magnitude across different inclination angles. The detailed distribution of the inclination angle is derived and can be directly compared with future observational data. For wet EMRIs, the inclination angle distribution depends on whether the MBH’s spin is more chaotic or coherent \cite{Dotti2013BHspinOrientation}, as well as on how the disk lifetime compares to the Bardeen-Petterson timescale \cite{Natara1998_Alignment}. 

We further investigate the mass distribution of sBHs associated with different formation channels, each characterized by distinct signatures. In particular, for BBHs detected by the LIGO-Virgo Collaboration, the component mass distribution shows prominent peaks near $10\ M_\odot$ and $35\ M_\odot$. We adopt a phenomenological pairing function that characterizes the efficiency of binary BH formation from isolated sBHs, and use it to infer the underlying mass function of isolated sBHs. For dry EMRIs, the mass function of their sBHs exhibits more prominent peaks around $35\,M_\odot$ compared to the mass function of isolated sBHs, due to the more efficient gravitational scattering of more massive sBHs and the effect of mass segregation. For wet EMRIs, the mass function is further shaped by additional processes, including the capture of sBHs by the AGN disk, mergers of sBHs in the AGN disk, and mass growth due to gas accretion. Multiple new features on the mass function may appear, as summarized in Table \ref{tab:all}, along with the main predictions of the distributions of the observables considered in the wet and dry formation channels.  Other observables, such as the spins of secondary black holes, have also been proposed to probe different formation channels \cite{Cui:2025bgu}. 
However,  the {\it in situ} distribution of isolated sBHs remains unknown, and the accretion processes of these objects within AGN disks still require systematic characterization through numerical simulations. Moreover, secondary spin measurement precision degrades dramatically for low-eccentricity systems (such as wet EMRIs) \cite{Cui:2025bgu}. As a result, we do not perform quantitative analysis in spin distributions in this work.

This paper is organized as follows. In Section~\ref{sec:ecc}, we discuss the eccentricity distribution of EMRIs, covering dry EMRIs in Section~\ref{sec:drye} and wet EMRIs in Section~\ref{sec:wete}. For the latter, we examine two mechanisms that can excite eccentricity: multi-body resonances (Section~\ref{sec:mulres}) and turbulence (Section~\ref{sec:turb}). Section~\ref{sec:inc} focuses on the distribution of inclination angles, with dry and wet EMRIs discussed in Sections~\ref{sec:iotadry} and \ref{sec:iotawet}, respectively. In Section~\ref{sec:lvk}, we analyze the mass distribution, with corresponding discussions for dry and wet EMRIs in Sections~\ref{sec:massdry} and \ref{sec:masswet}. Finally, Section~\ref{sec:conc} summarizes our conclusions and outlines the prospects for future studies of EMRI populations. If no special instructions are given, in this paper we use the $G = c = 1$ units, and we adopt the notation $M_{\bullet}$ for the mass of the MBH. Table~\ref{tab:symbols} (in Appendix.~\ref{app:phy_q}) provides a summary of the physical quantities employed in this study.

\section{Eccentricity}\label{sec:ecc}

In this section, we investigate the expected eccentricity distributions associated with the dry and wet EMRI formation channels. In particular, since EMRI waveforms typically consist of a large number of cycles ($\gtrsim 10^4$), even small deviations in the intrinsic parameters of the system can accumulate and result in substantial phase differences at the end of the waveform. This implies that the measurement precision of many system parameters for EMRIs is significantly higher than that for comparable-mass binary systems. For example, as noted in \cite{Babak2017_LISA_EMRI}, the eccentricity of a typical EMRI event can be measured with precision of the order of $10^{-5}$. This exceptional accuracy not only enables a clear distinction between EMRIs originating from different formation channels but also opens the possibility of probing properties of the EMRI environment during the early stages of its evolution, as will be discussed later in this section.

\subsection{Dry EMRI}\label{sec:drye}

The dry EMRI channel, originating from two-body gravitational scatterings in nuclear star clusters, was the first major EMRI formation mechanism proposed for detection by space-based gravitational wave observatories. Several studies have been conducted on the formation rates and eccentricity distributions of EMRI \cite{Hopman2005_OrbitalStatistics,Amaro-Seoane:2010dzj,Amaro-Seoane:2012lgq,Bar-Or2016,Babak2017_LISA_EMRI}. For completeness, in this section we briefly outline the main procedure for computing the distribution of dry EMRIs using the Fokker-Planck (FP) method, with particular emphasis on the eccentricity distribution. A more detailed discussion of the formalism can be found in \cite{Pan202101}.

\begin{figure}[h]
    \centering
    \includegraphics[width=1.0\linewidth]{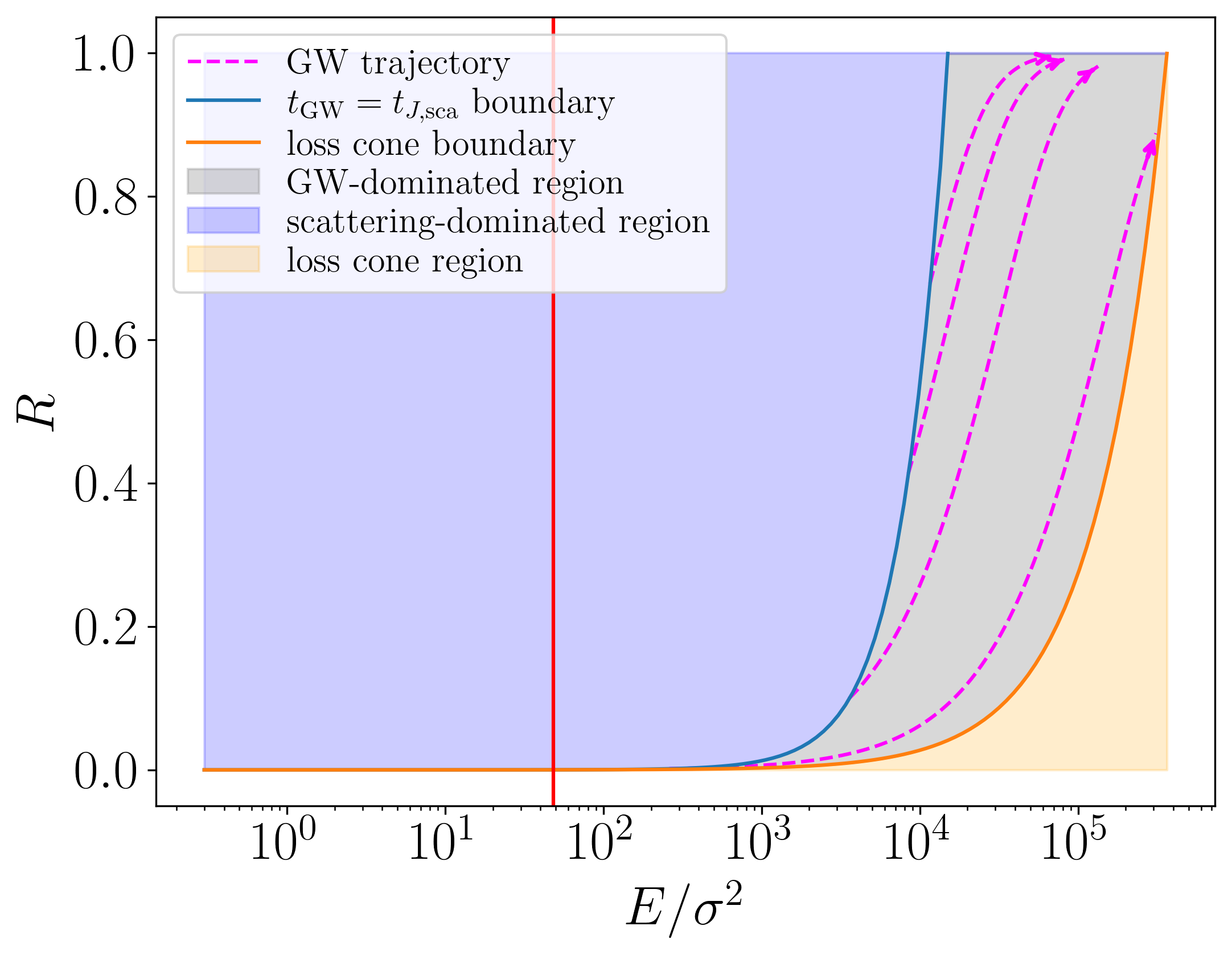}
    \caption{The boundary in phase space in FP equation. The gray shaded region indicates where gravitational-wave emission dominates, while the purple shaded region is dominated by gravitational scattering. We also plot the trajectory of sBH in the GW-dominated region because of gravitational radiation. The vertical red solid line show the critical Energy value $E_{\rm cr}$, where loss cone boundary and $t_{\rm GW}=t_{\rm J}$ boundary intersect. On the loss cone boundary, when the energy is larger than $E_{\rm cr}$, the flux through the boundary corresponds to EMRI flux.}
    \label{fig:boundary}
\end{figure}

Within the FP formalism, the phase-space distribution of stars and sBHs in a nuclear star cluster is modeled under the assumption of spherical symmetry. Each object’s orbit is characterized by its energy $E$ and angular momentum $J$. For convenience, the angular momentum is often re-parameterized using the dimensionless variable
$R = J^2/J_c^2(E)$,
where $J_c(E)$ is the angular momentum of a circular orbit with energy $E$. This choice simplifies the treatment of diffusion in angular momentum space, as $R \in [0, 1]$.

There is a distinct region in the phase space $(E, R)$, bounded by the energy-dependent angular momentum $J_{\rm lc}(E)$, as shown by the orange area in Fig.~\ref{fig:boundary}. Objects such as stellar-mass black holes that initially occupy this region will plunge into the MBH within a single orbital period. This region is known as the ``loss cone'' \cite{Cutler1994SchwarzschildRR}, and its boundary, $J_{\rm lc}(E)$, referred to as the ``loss-cone boundary'' (i.e., the last stable orbit), is represented by the orange line in Fig.~\ref{fig:boundary}.

Since dry EMRIs typically have eccentricities close to unity, it is a good approximation to define the loss cone boundary using $J_{\rm lc} = 4M_{\bullet}$ \cite{Cutler1994SchwarzschildRR}, assuming a non-spinning (Schwarzschild) MBH. This expression becomes exact in the limit of extreme eccentricity. Note that in this figure we rescale the specific energy using $\sigma$, defined as the stellar velocity dispersion following \cite{Pan202101, Gultekin2009MSigmaML}.

\begin{figure*}[t]
\centering
\vspace{0.6cm}
\includegraphics[width=0.45\linewidth]{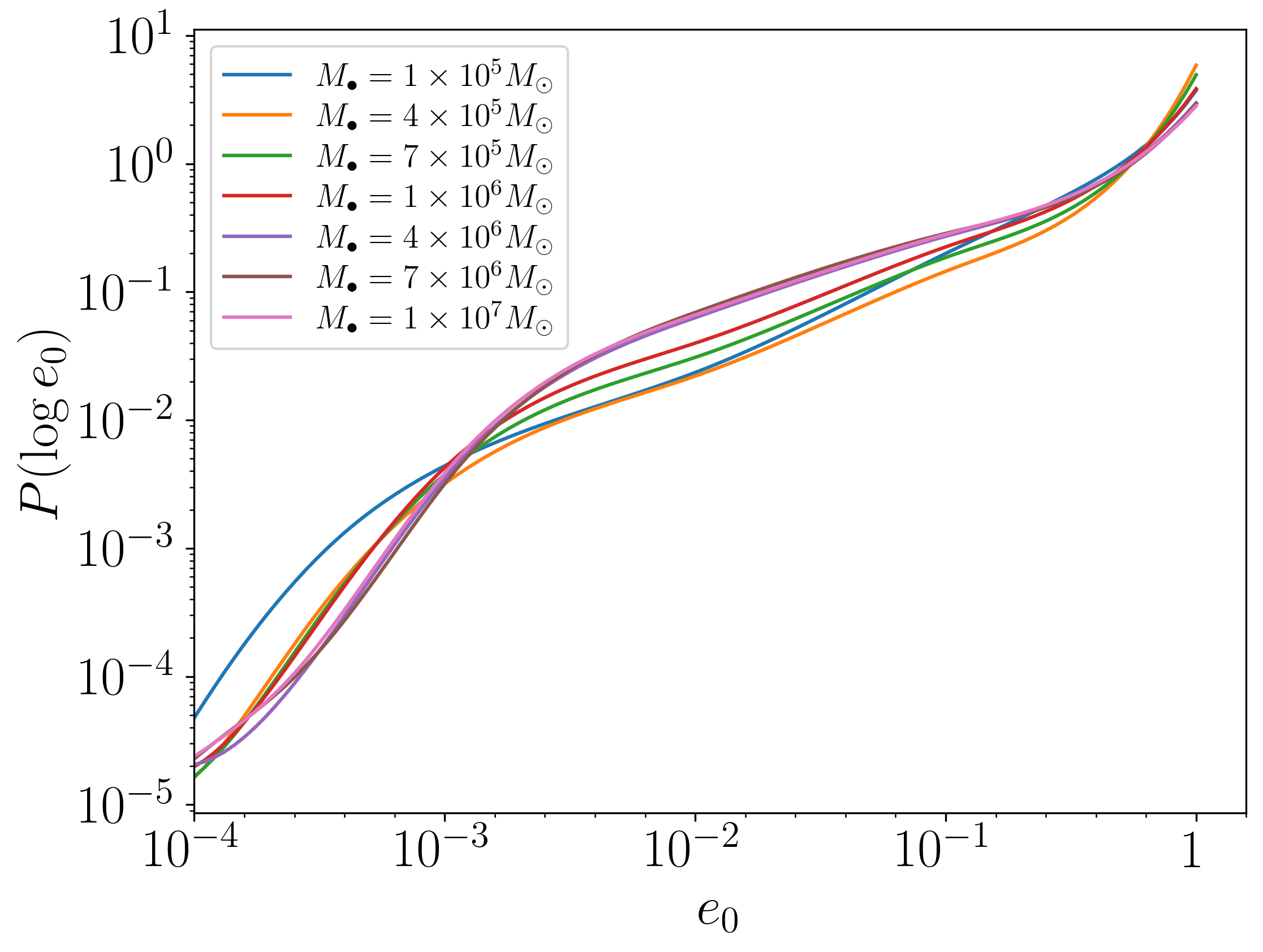}
\includegraphics[width=0.45\linewidth]{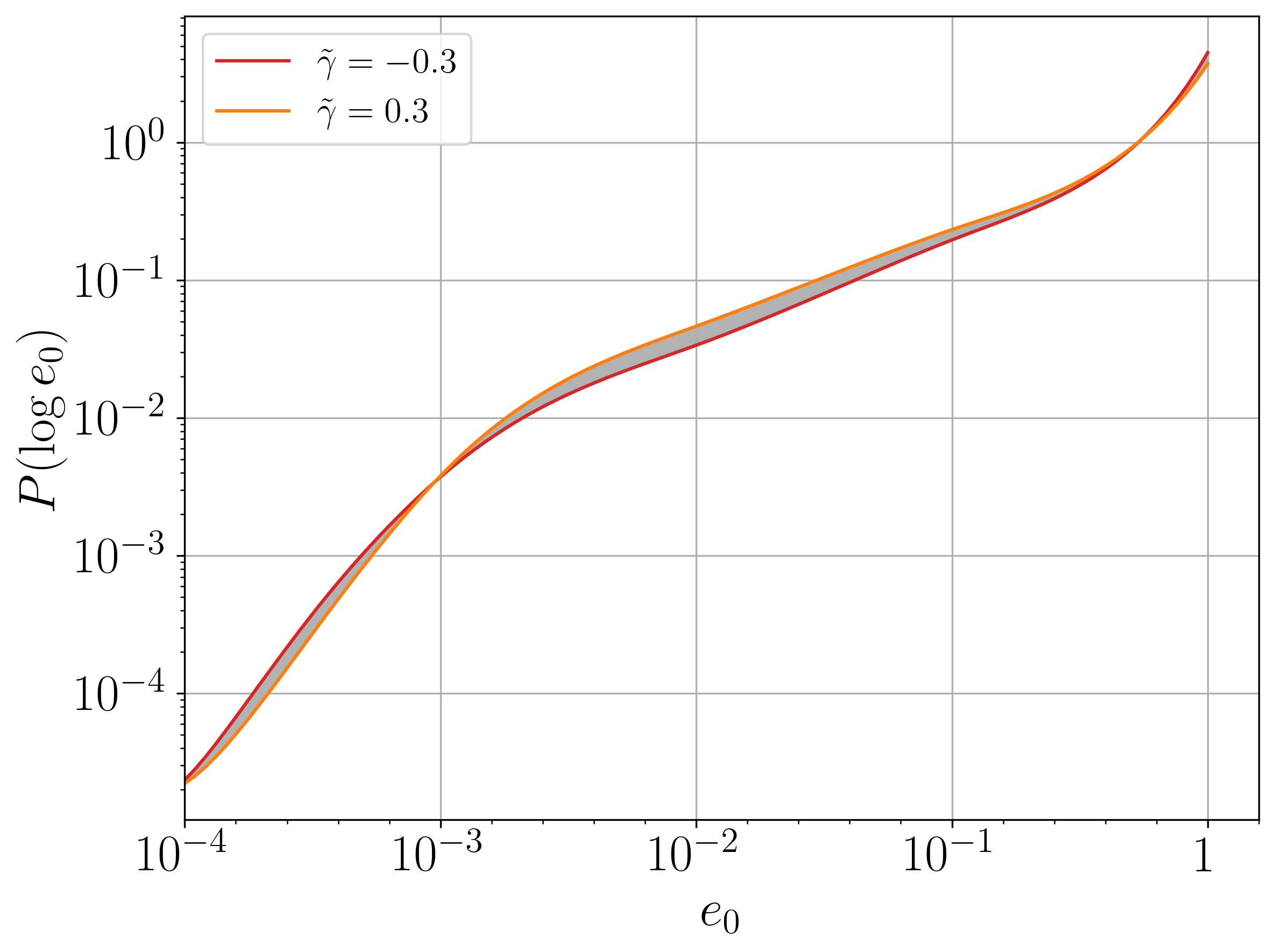}
\vspace{0.3cm}
\caption{Dry EMRI eccentricity distribution at $r_p\approx10M_{\bullet}$. The left panel shows the eccentricity distribution for MBHs of different masses, and right panel shows the overall eccentricity distribution for different MBH mass functions. }
\label{fig:dry_e}
\end{figure*}

Assume that $f$ denotes the number density in position-velocity space, and $N$ represents the distribution function in the $(E, R)$ phase space.  
The two are connected through the following relation \cite{Cohn1978_StellarDistBH, Cohn1979}:

\begin{align}
    N(E,R)dEdR = \int_{r_{p}}^{r_{a}} d^3r d^3v \ f(E,R)
\end{align}
in which $r_p(E,R)$ and $r_a(E,R)$ are the orbital pericenter and apocenter, respectively \cite{Pan202101}. This number density relation can be translated to
\begin{align}
    N(E,R) &= 4\pi^2P(E,R)J^2_c(E)f(E,R) \notag \\
    :&= C(E,R)f(E,R),
\end{align}
where $P(E,R)$ is the orbital period and $C(E,R)$ is often referred to as the weight function.
The FP equation that describes the time-dependent evolution of the distribution is 
\begin{align}
    C(E,R)\frac{\partial f}{\partial t} =-\frac{\partial F_E}{\partial E} - \frac{\partial F_R}{\partial R}\,.
\end{align}
Here, $F_{E,R}$ is the phase-space flux along the $E/R$ direction. They can be obtained from
\begin{align}
    -F_E &= D_{EE}\frac{\partial f}{\partial E} + D_{ER}\frac{\partial f}{\partial R} + D_E f \,,\notag \\
    -F_R &=  D_{RR} \frac{\partial f}{\partial R} + D_{ER}\frac{\partial f}{\partial E} + D_R f\,.
\end{align}
The diffusion coefficients $\{D_{EE}, D_{ER}, D_{RR}\}$ and the advection coefficients $\{D_E, D_R\}$ can be found in \cite{Pan202101,Pan202112}.

There are two key timescales that govern the evolution of a nuclear star cluster: the gravitational scattering timescale, $t_{J, \rm sca}$, which characterizes the rate at which an object’s angular momentum changes due to interactions with other objects in the cluster; and the gravitational wave radiation timescale, $t_{\rm GW}$, which describes the orbital evolution of a two-body system composed of the object and the central MBH, driven by gravitational wave emission. Based on \cite{Pan202112, Hopman2005_OrbitalStatistics, Zwick2020, Zwick2021a, VazquezAceves2022}, these two timescales are
\begin{align}
    t_{\rm GW} &= \frac{5a^4}{256M_{\bullet}^2m}\frac{(1-e^2)^{\frac{7}{2}}}{1+\frac{73}{24}e^2+\frac{37}{96}e^4}8^{1-\sqrt{1-e}}\exp\left\{\frac{5M_{\bullet}}{a(1-e)}\right\}\,,  \notag \\
    t_{J, \rm sca} &\approx \frac{J^2}{J_c^2(E)}t_E(E,R)=\frac{J^2}{J_c^2(E)}\frac{E^2}{2D_{EE}(E,R=1)}\,,
\end{align}
where $t_{\rm GW}$ above is modified Peter's formula for a highly eccentric binary system with semi-major axis $a$, eccentricity $e$, and mass components $M_{\bullet}$ and $m$.

In the $(E, R)$ phase space, it is instructive to define a line where the gravitational wave inspiral timescale equals the angular momentum relaxation timescale, i.e., $t_{\rm GW} = t_{J, \rm sca}$.
This condition is shown as the blue line in Fig.~\ref{fig:boundary}. It also intersects the loss-cone boundary, and the corresponding energy at the intersection is denoted by $E_{\rm cr}$. It is evident that in the region between the loss cone boundary and the blue line, the gravitational-wave timescale $t_{\rm GW}$ is shorter than the angular momentum diffusion timescale $t_{J, \rm sca}$, implying that gravitational-wave emission dominates the secular orbital evolution. This region is commonly referred to as the ``EMRI regime'', where a highly eccentric EMRI initially forms and subsequently evolves toward the loss cone along the dashed trajectories shown in Fig.~\ref{fig:boundary}. To compute the EMRI formation rate, one may evaluate the integrated flux along either the blue line or at the loss-cone boundary. In a quasi-steady state, both approaches yield the same result due to flux conservation:

\begin{align}
    \Gamma_\text{EMRI} = \int_{E_\text{cr}}^{E_\text{max}} dE \left ( -F_R + F_E \frac{dR}{dE}\right ),
\end{align}
where $E_{\rm max}$ corresponds to the point where the loss-cone boundary intersects the boundary $R=1$, i.e. the circular orbit with angular momentum $J_c(E)=J_{\rm lc}=4M_{\bullet}$.

For a typical dry EMRI, the eccentricity is initially close to unity. During repeated pericenter passages, the system emits substantial gravitational radiation, resulting in the loss of both energy and angular momentum. Consequently, the eccentricity decreases rapidly from cycle to cycle. Because the initial semi-major axis is large compared to the size of the central MBH (and thus the system is only weakly bound), the semi-major axis $a$ is more strongly affected by gravitational radiation than the pericenter distance $r_p$. As a result, in the early stages of evolution, $a$ can decrease by orders of magnitude while $r_p$ experiences only modest fractional changes. Once the eccentricity drops below approximately 0.9, the waveform enters the valid parameter range of state-of-the-art waveform models, such as \texttt{fastemriwaveforms} (FEW) \cite{Katz2021FastEMRIWaveforms}.

The eccentricity distribution of dry EMRIs can be derived from the phase-space distribution function in $(E, R)$, obtained by solving the FP equation. At each point in the phase space characterized by $(E, R)$, the corresponding pericenter distance $r_p(E, R)$ and apocenter distance $r_a(E, R)$ can be calculated. The semimajor axis a and eccentricity e are then given by
$a = (r_a + r_p)/2, \quad e = (r_a - r_p)/(r_a + r_p)$. Within the EMRI regime, as the orbital parameters $(e, a)$ evolve under gravitational wave-driven dynamics \cite{Peters1964a},  we define the reference eccentricity of each trajectory as $e_0 := e |_{r_p = 10M_{\bullet}}$. The eccentricity distribution is then obtained by evaluating the EMRI rate -measured at the boundary where $t_{\rm GW} = t_{J, \rm sca}$ - for different trajectories such as 

\begin{align}
    P(\log e_0)=\Gamma_{\rm EMRI}^{-1}\left (-F_R+F_E\frac{dR}{dE} \right )\frac{dE}{d\log e_0}\,.
\end{align}

For an MBH with mass $M_{\bullet}$, the resulting time-averaged distribution of $e_0$ is shown in the left panel of Fig.~\ref{fig:dry_e}. We find that the eccentricities of dry EMRIs fall predominantly within the range $10^{-2}$ to $1$, representing more than $95\%$ of the total population. This is consistent with the results in \cite{Hopman2005_OrbitalStatistics}. 

To assess the overall statistical distribution of the dry EMRI eccentricity across the population of MBHs in the universe, we incorporate the MBH mass function. Specifically, we assume a parametric form for the comoving number density of MBHs as a function of the MBH mass given by 
\begin{align}
\frac{dn}{d\log M_{\bullet}} =\tilde{C} \left(\frac{M_{\bullet}}{10^6M_\odot}\right)^{\tilde{\gamma}} \text{Mpc}^{-3},
\end{align}

where $\tilde{\gamma}$ is a power law index that characterizes the slope of the MBH mass function and $\tilde{C}$ is a constant that does not affect the eccentricity distribution. 
We consider the range of index from $\tilde\gamma=-0.3 $ to $\tilde\gamma = 0.3$ \cite{Babak2017_LISA_EMRI} to reflect plausible uncertainties in the MBH population models inferred from observations and simulations.

The mass-weighted average eccentricity distribution for a given mass function is
\begin{align}
\langle P(e_0)\rangle_{M_{\bullet}} = \frac{\int dM_{\bullet} \ \Gamma_{\rm EMRI}(M_{\bullet})P(e_0) \ \frac{dn}{dM_{\bullet}}}{\int dM_{\bullet} \Gamma_{\rm EMRI}(M_{\bullet}) \ \frac{dn}{dM_{\bullet}}},
\end{align}
in which $\Gamma_{\rm EMRI}(M_{\bullet})$
calculated from the FP equation, and we have only considered MBHs in the mass range $\in (10^5,10^7)M_\odot$, relevant for spaceborne GW detection. The overall distribution is shown in the right panel of Fig.~\ref{fig:dry_e}. In general, the change in the power law index $\tilde\gamma$ only slightly shifts the distribution around $e_0 \sim 10^{-2}$.

We also compare the eccentricity distribution computed here with a recent study by \cite{Mancieri2025_EccentricityEMRI}. Although there are a few differences in the setup, the results are broadly consistent. More detailed comparison can be found in Appendix.~\ref{app:compare}.

\subsection{Wet EMRI}\label{sec:wete}

\begin{figure*}
\centering
\includegraphics[width=0.45\textwidth,clip=true]{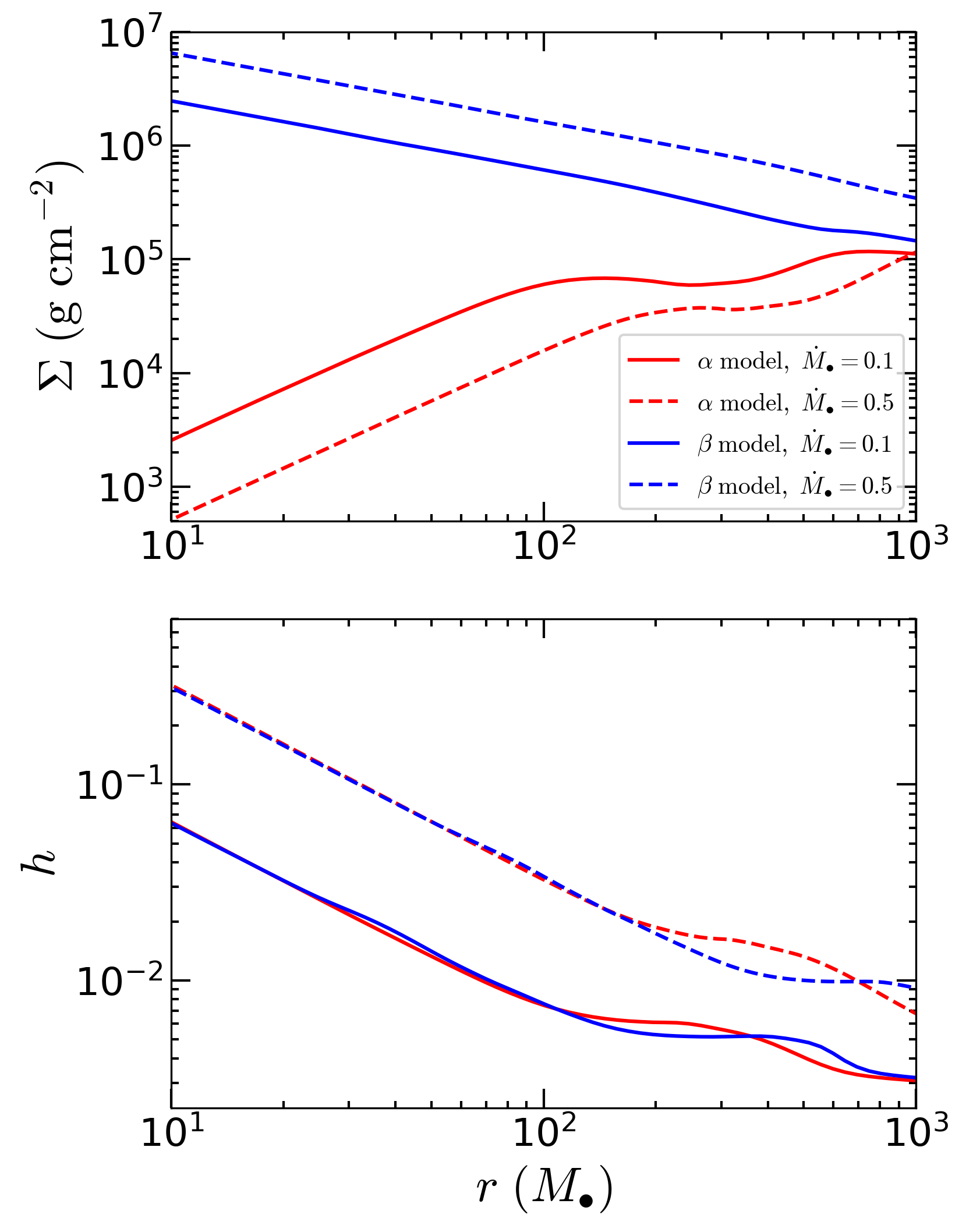}
\includegraphics[width=0.45\textwidth,clip=true]{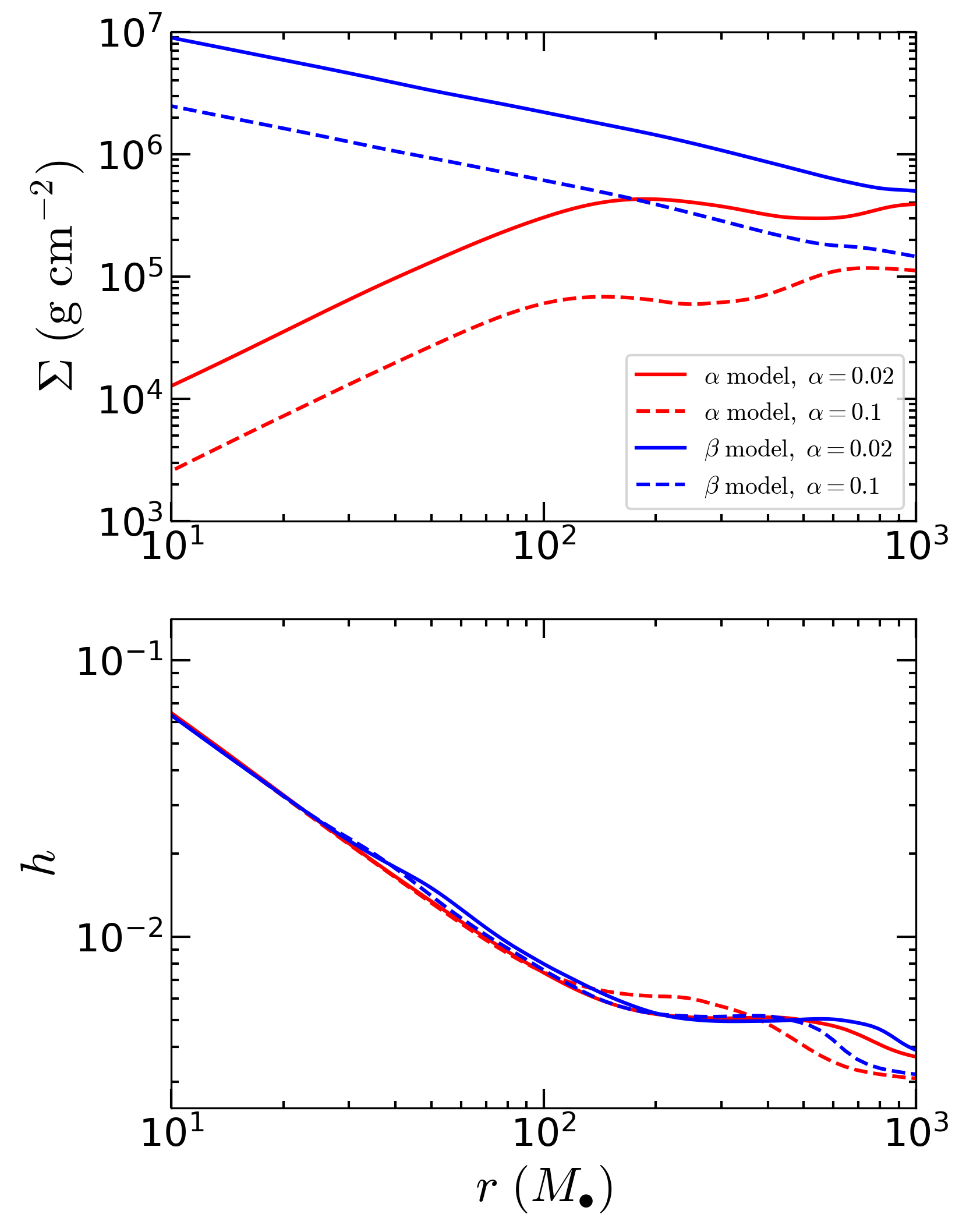}
\caption{The disk surface density $\Sigma$ (upper panels) and aspect ratio $h$ (lower panels) for an $\alpha$ (red lines) and $\beta$ (blue lines) model of the AGN disk. The left panels show the disk profiles with different disk accretion rate $\dot{M}_{\bullet}$ with $\alpha=0.1$, and the right panels show the dependence on the disk viscosity $\alpha$ with $\dot{M}_{\bullet}=0.1\dot{M}_{\rm Edd}$. 
}
 \label{fig:disk_model}
\end{figure*}

Wet EMRIs, on the other hand, may exhibit diminishing eccentricities according to standard disk migration theory. However, we will show that two novel mechanisms can potentially excite the eccentricities of wet EMRIs: one arising from multibody interactions and resonances within the accretion disk (see Sec.~\ref{sec:mulres}), and the other from turbulent fluctuations in the disk (see Sec.~\ref{sec:turb}). Accurate eccentricity measurements offer a valuable opportunity to probe the evolutionary history and environmental conditions of wet EMRIs.

We consider the orbital evolution of EMRI using parametrized AGN disk models. For an AGN disk, the disk profiles within $100\ M_{\bullet}$ can be parameterized as \cite{Kocsis2011DiskEMRI}

\begin{align}
h \simeq 0.7\dot{M}_{\bullet,0.1} \frac{M_{\bullet}}{r}
\end{align}
and 
\begin{align}
\Sigma(r) =\Sigma_0 \left (\frac{r}{100M_{\bullet}} \right )^n\,.
\end{align}
For the $\alpha$ model with $\alpha=0.1$ and $\dot{M}_{0.1,\bullet}=\dot{M}_{\bullet}/0.1\dot{M}_{\rm Edd}$ where $\dot{M}_{\rm Edd}$ is the Eddington accretion rate for the MBH, we have $n=1.5,\Sigma_0=7\times10^{4}\ {\rm g\ cm^{-2}}$. For the $\beta$ model with the same model parameters, we have $n\simeq-0.6,\Sigma_0=6\times10^{5}\ {\rm g\ cm^{-2}}$. The disk profiles with different $\alpha$ and $\dot{M}_{\bullet}$ for $\alpha$ and $\beta$ models are shown in Fig.~\ref{fig:disk_model}. The real disk profiles may deviate from both $\alpha$- and $\beta$-disk prescriptions, so it is instructive to consider the parametrized form of the disk, especially for the discussion in Sec.~\ref{sec:mulres} and Sec.~\ref{sec:turb}.

\subsubsection{Multi-body resonance}\label{sec:mulres}


\begin{figure*}[t]
\centering
\vspace{0.6cm}
\includegraphics[width=0.45\linewidth]{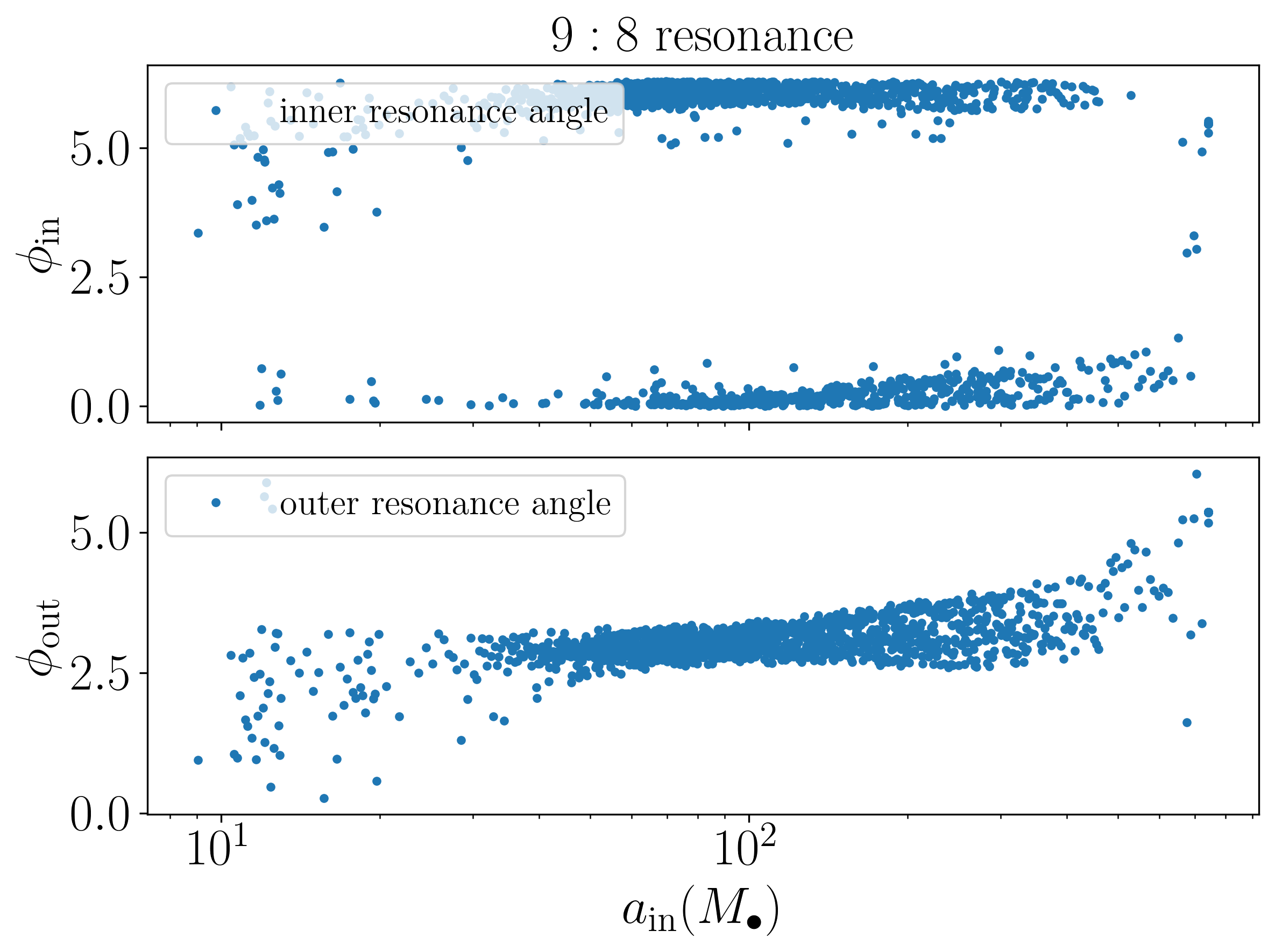}
 \includegraphics[width=0.45\linewidth]{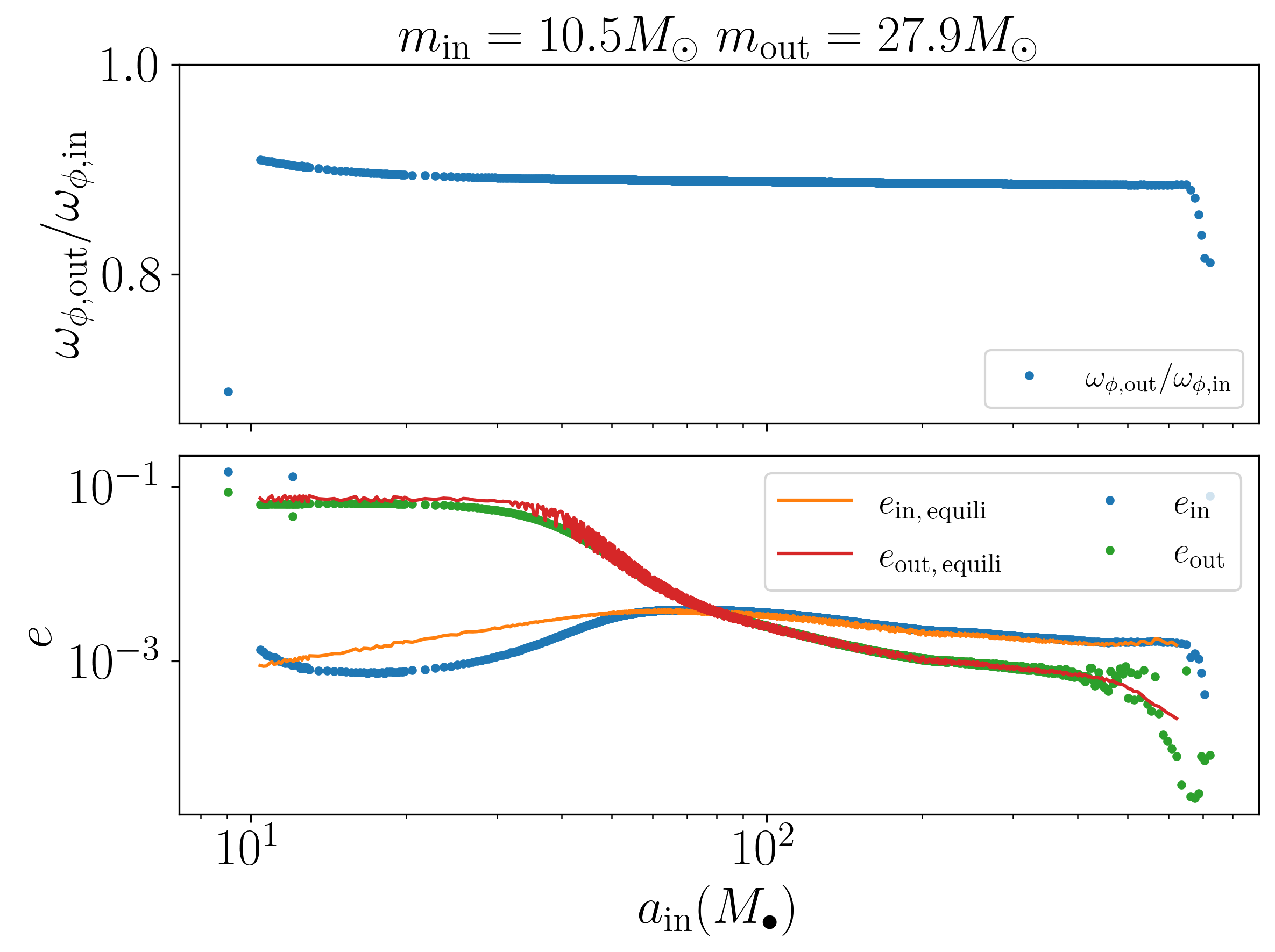}
\vspace{0.3cm}
\caption{A pair of sBHs is locked into both inner and outer $(8,1)$ resonances, as shown by the resonant angles in the left panel. The ratio between $\omega_{\phi,{\rm in}}$ and $\omega_{\phi,{\rm out}}$ is approximately $9:8$. In the right panel, the evolution of (azimuthal) period ratios and the eccentricity are shown, with $e_{\rm equi}$ denoting the
equilibrium value obtained from analytical approximations discussed in Appendix.~\ref{app:equi}. At very small semi-major axes ($\sim 10M_{\bullet}$), the continuing orbital shrinkage driven by the sustained resonance brings the two sBHs into close encounter, triggering a scattering event. This interaction excites the eccentricity of the inner object and ultimately disrupts the resonance.}
\label{fig:res}
\end{figure*}

An {\it isolated} sBH embedded in a thin accretion disk generically excites density waves that carry away energy and angular momentum, analogous to the behavior of planets in protoplanetary disks. The corresponding damping timescales for angular momentum ($t_J$)
and eccentricity ($t_{e}$) are 

\begin{align}\label{eq:tje}
    t_{J}&=-\frac{J}{\dot{J}}\sim \frac{M_{\bullet}}{m}\frac{M_{\bullet}}{\Sigma r^2}\frac{h^2}{\Omega}\,, \notag \\
    t_e&=-\frac{e}{\dot{e}}\sim \frac{M_{\bullet}}{m}\frac{M_{\bullet}}{\Sigma r^2}\frac{h^4}{\Omega},
\end{align}
where $m$ is the sBH mass, $\Sigma$ is the disk surface density, $h$ is the disk aspect ratio, and $\Omega$ is Keplerian angular velocity. All these quantities related to the disk are measured at the semi-major axis $a$ of the sBH orbit. For a thin disk the damping timescale of eccentricity is much smaller than that of the energy and angular momentum (by a factor $h^2\ll1$), so the eccentricity would quickly decrease below $10^{-5}$, as can be seen from hydrodynamical simulations \cite{Cresswell2007EccIncEvol,Li2019}. However, it is instructive to examine the underlying assumption of this argument, e.g., an isolated sBH, as multibody interactions may significantly modify the dynamics of sBHs.

In fact, since migration within the AGN is dominated by disk interactions at large radii and gravitational wave back reaction at small radii, the effective lifetime (measured by $a/\dot{a}$) is maximized at a few hundred gravitational radii, according to the analysis in \cite{Pan202101} (see also Fig.~\ref{fig:adot_nbody} below). The refined population study (following a similar formalism) in \cite{Bell2016} predicts that there are on average one or more sBH staying at a few hundred gravitational radii. This number is larger for shorter disk lifetimes, as the wet EMRI rate generally decreases over time due to decreasing supply of sBHs \cite{Pan202101,Pan202112}. Therefore, it is natural to expect that a fraction of wet EMRIs are accompanied by close companions. We will need to jointly evolve the multibody system to determine its long-term dynamics. In this work, for the sake of simplicity, we focus on three-body systems with two sBHs initially placed at two different radii within a thin accretion disk. The motion is assumed to be planar.


To numerically evolve the three-body system, we adopt the N-body code REBOUND \cite{ReinLiu2012} with the following modifications. First, since the later stages of the EMRI trajectory enter the strong-gravity regime, we include the 1PN, 2PN, and 2.5PN Post-Newtonian (PN) corrections to the equations of motion of the sBHs \cite{Pati2002_PNII}:

\begin{align}
    \vec{a}_\text{PN} = &\frac{M}{r^2}\left[\frac{\vec{r}}{r}\left(\text{A}_\text{1PN} + \text{A}_\text{2PN}\right)+v_r\vec{v}\left(\text{B}_\text{1PN}+\text{B}_\text{2PN}\right)\right] \notag \\
    &+\frac{8}{5} \eta \frac{M}{r^2}\frac{M}{r}\left[v_r\frac{\vec{r}}{r}\text{A}_\text{2.5PN} - \vec{v} \ \text{B}_\text{2.5PN}\right],
\end{align}
where $M=M_{\bullet}+m$ is the total mass of center MBH and sBH, $\eta=M_{\bullet} m/(M_{\bullet}+m)^2$ is symmetrical mass ratio, $v_{r}$ is the radial velocity of the sBH, $\vec{r}$ and $\vec{v}$ are the position and velocity vector of the sBH.  A and B are the PN coefficients that are presented  in Appendix.~\ref{app:pn}. Notice that we have not included PN corrections to the gravitational force between the two sBHs, because this force at Newtonian order is already $m/M_{\bullet}$ times smaller than the force between an sBH and the MBH.

Second, we also include the disk force in the equation of motion. Assuming the characteristic timescales $t_{\rm J}, t_{\rm e}$, the effective acceleration term can be written as 
\begin{align}
   \vec{a}_\text{mig}=-\frac{\vec{v}}{t_J}-2\frac{v_r}{t_e}\frac{\vec{r}}{r}\, .
\end{align}
We apply the approximation formula in Eq.~(\ref{eq:tje}) to implement such disk-force component. 

In the following analysis, we adopt a central MBH mass of $M_{\bullet} = 4\times 10^6 M_\odot$ and use $\alpha$-disk parameters with $\alpha = 0.1$ and $\dot{M}_\bullet = 0.1 \dot{M}_{\rm Edd}$ \cite{Pan202101}.


\begin{figure*}[t]
\centering
\vspace{0.6cm}
\includegraphics[width=0.45\linewidth]{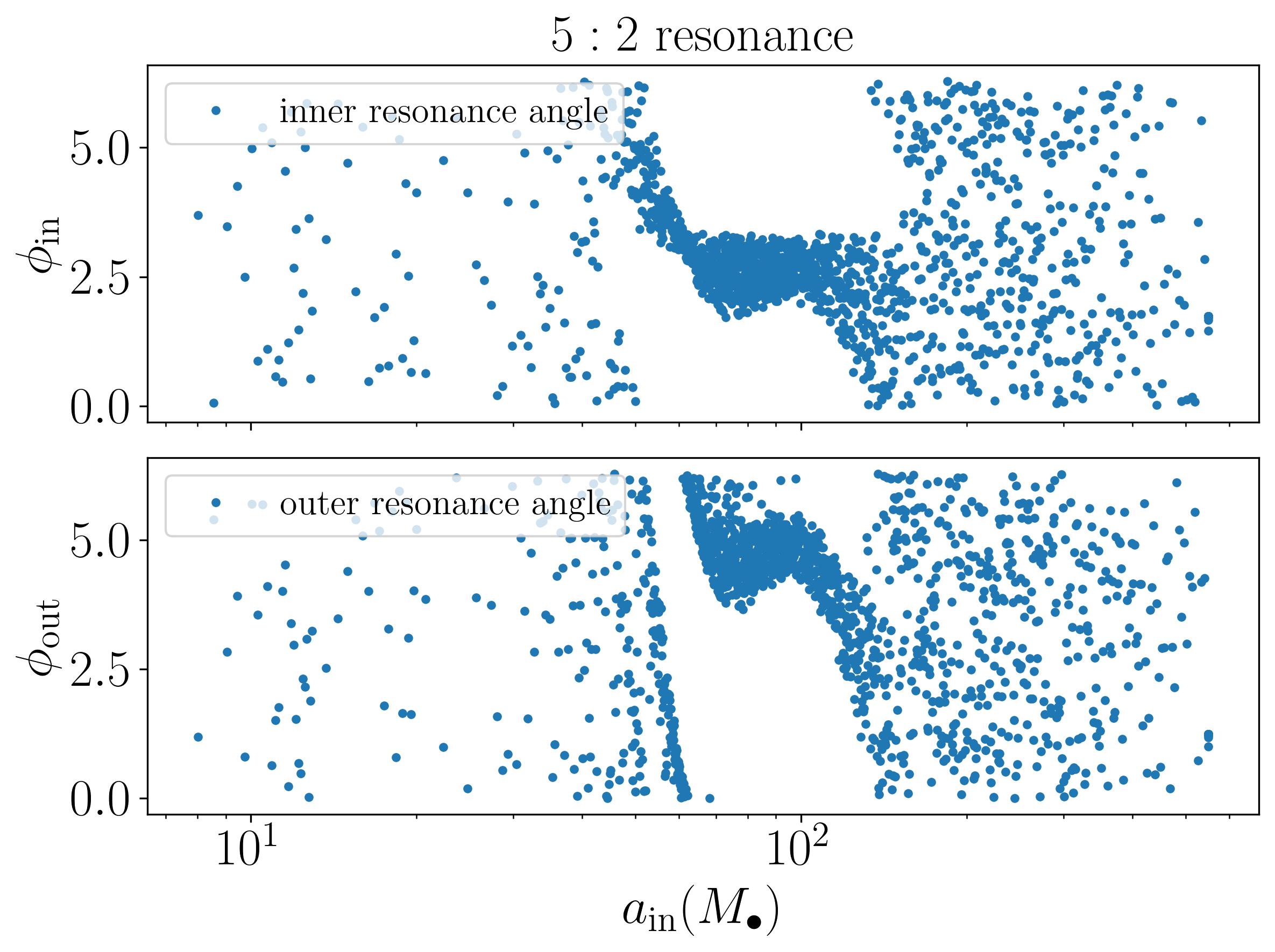}
 \includegraphics[width=0.45\linewidth]{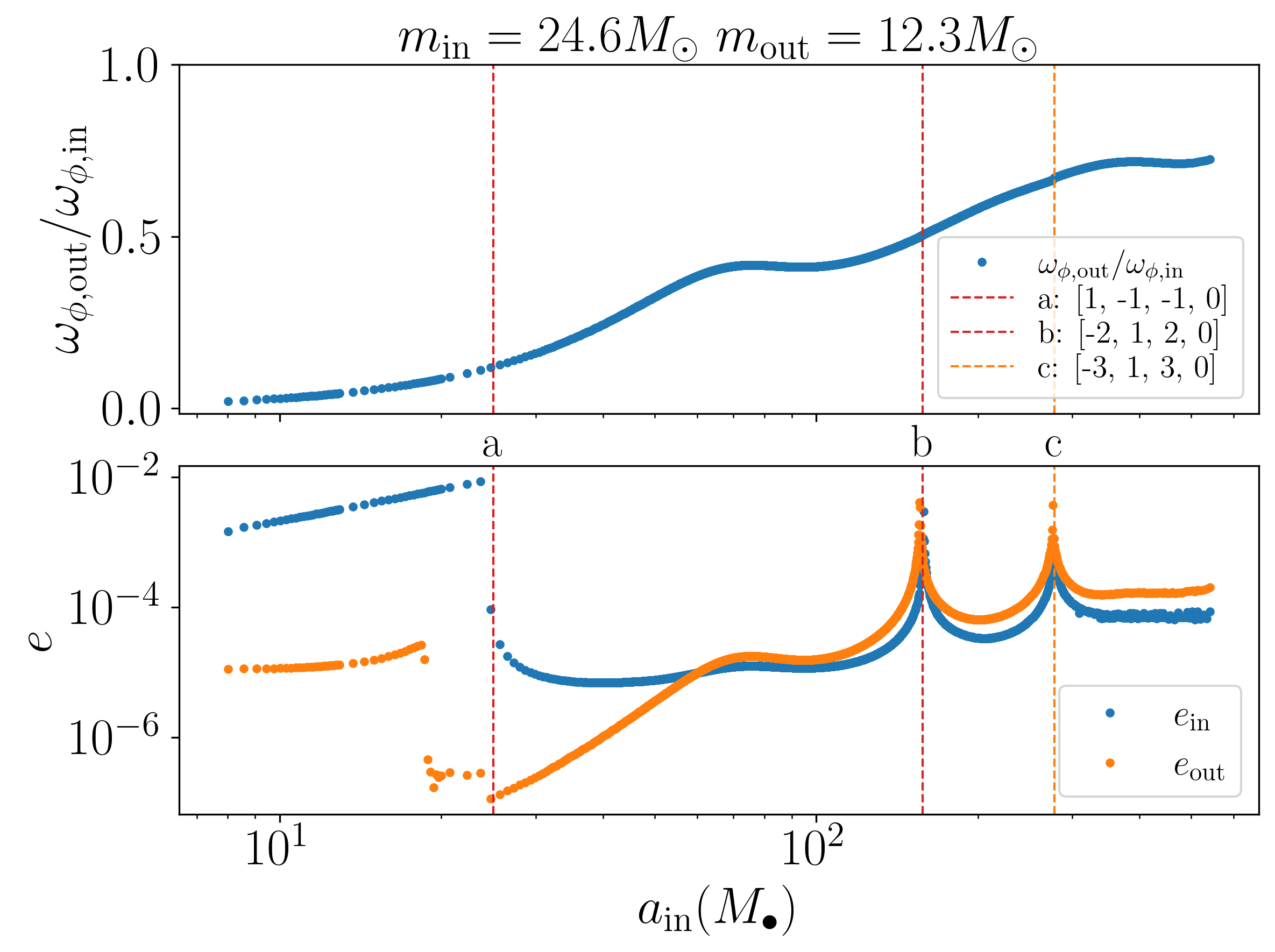}
\vspace{0.3cm}
\caption{A sample evolution with $m_{\rm in} =24.6 M_\odot$ and $m_{\rm out} = 12.3 M_\odot$. The left panel shows the resonance angles of the inner and outer objects, while the right panel shows the period ratio and eccentricity evolution with semi-major axis of inner object. The system is out of resonance for most of the time except briefly at $r_{\rm in} \sim 100 M_{\bullet}$. We mark the locations where the eccentricities are excited by three transient resonances $a,b,c$. In the right upper panel, we also show the resonance coefficients defined in Eq.~(\ref{eq:k_int}). }
\label{fig:large_in}
\end{figure*}

\paragraph{\bf Dynamics.}  Depending on initial conditions of the system, the multibody dynamics can be classified into resonance, non-resonant migration, and scattering regimes. In principle, dynamical formation of sBH binaries can also be classified as the scattering case, as it typically involves close gravitational encounters. However, as simulations \cite{Mishra2024CircumSingleMHD,LiJ2023,Wang2025} suggest that it often requires the modeling of circum-single disks, we shall not consider this scenario in this study.

Resonance locking generally happens if the outer object migrates faster (e.g., being the more massive object) than the inner object, so that the relative motion is convergent. Similarly to mean-motion resonances in planetary dynamics \cite{Cresswell2006MultiProtoplanets}, we can define the resonant angles for inner and outer resonances as  

\begin{align}
    \phi_{\text{in}} &= (p+q)\lambda_{\text{out}}-p\lambda_{\text{in}}-q\varpi_\text{in}\,, \notag \\
    \phi_{\text{out}}&=(p+q)\lambda_{\text{out}}-p\lambda_{\text{in}}-q\varpi_\text{out}\,,
\end{align}
where the $\lambda$ is mean longitude and $\varpi$ is longitude of pericenter of true physical orbit. A system is trapped within a $(p + q) : p$ inner/outer mean-motion resonance if $\phi_{\rm in}/\phi_{\rm out}$ is bound around a fixed value. In this case, notice that the time-averaged derivative of $\lambda$ corresponds to $\omega_\phi$ and the time-averaged derivative of $\lambda-\varpi$ corresponds to $\omega_r$, where $\omega_\phi$ and $\omega_r$ denote the angular frequency in the $\phi$ and $r$ direction, respectively. These two frequencies are the same in Newtonian theory, but start to differ from each other when PN corrections are included, because of the PN pericenter precession effect. The bounded value of $\phi_{\rm in}/\phi_{\rm out}$ also implies that
\begin{align}
(p+q)\omega_{\phi, {\rm out}}-(p+q)\omega_{\phi, {\rm in}}+q \omega_{r,{\rm in}}=0\,,
\end{align}
for the inner resonance and 
\begin{align}
p\omega_{\phi, {\rm out}}-p\omega_{\phi, {\rm in}}+q \omega_{r,{\rm out}}=0\,,
\end{align}
for the outer resonance. In the following discussion, we will use $(p,q)_{\rm in}$ and $(p,q)_{\rm out}$ to denote these resonances.

\begin{figure*}[t]
\centering
\vspace{0.6cm}
\includegraphics[width=0.45\linewidth]{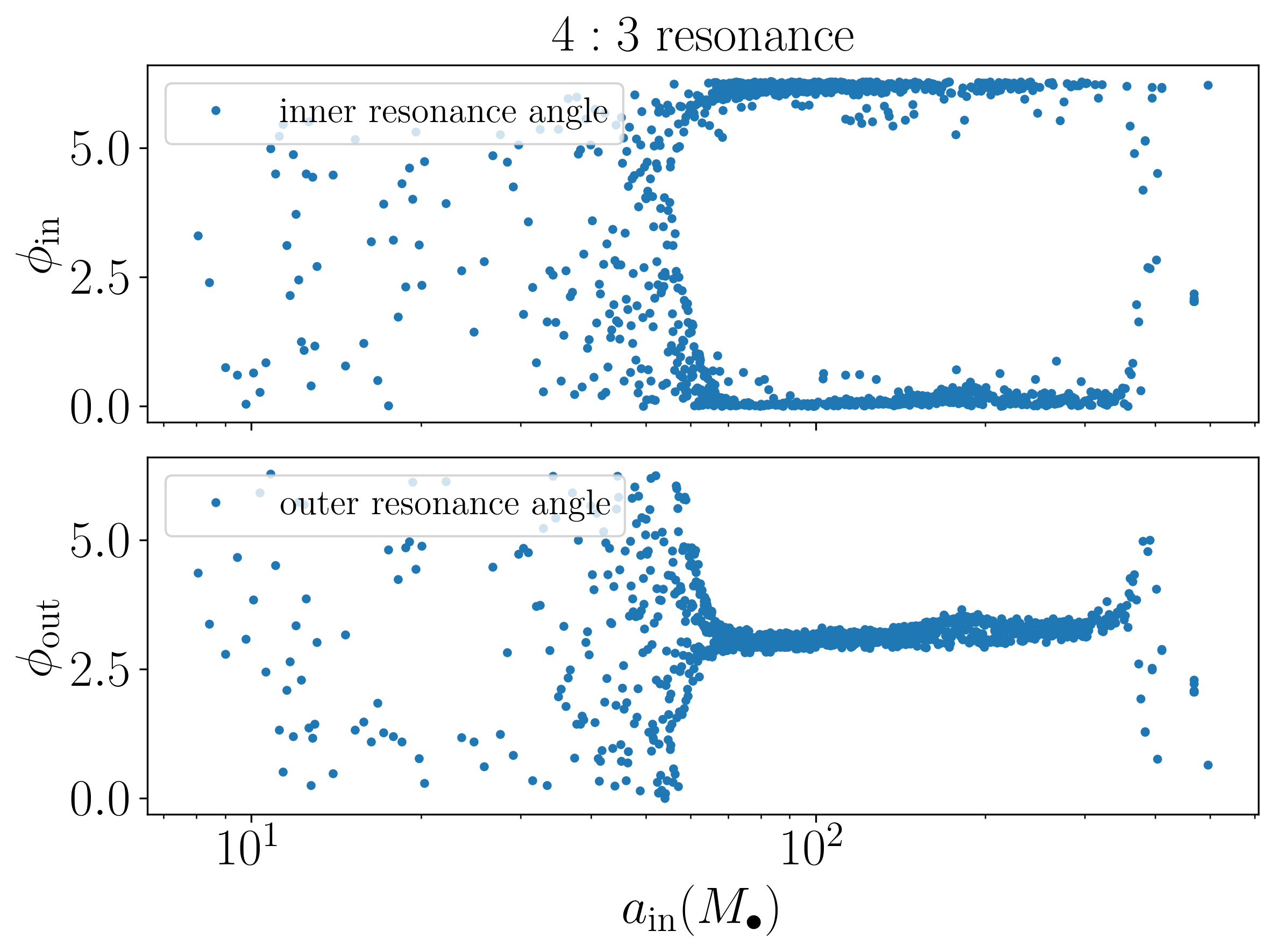}
 \includegraphics[width=0.45\linewidth]{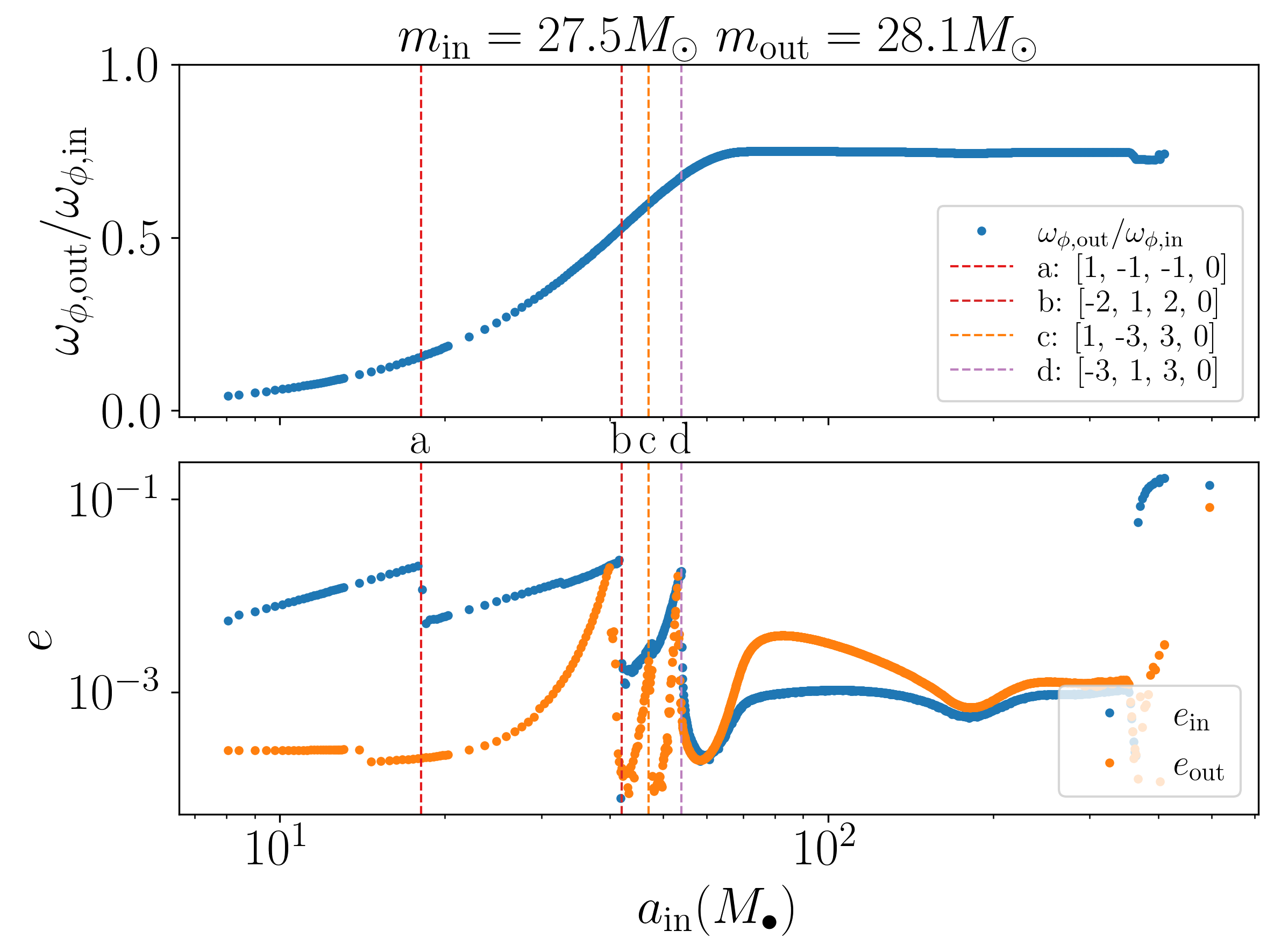}
\vspace{0.3cm}
\caption{A sample evolution with $m_{\rm in} =27.5 M_\odot$ and $m_{\rm out} = 28.1 M_\odot$. The system is rapidly locked into a $4:3$ resonance from the initial condition, 
which is broken at  $r_{\rm in} \approx 80 M_{\bullet}$. In right panels, we mark the locations where the eccentricities are excited from four transient resonance $a, b, c, d$. The resonance coefficients defined in Eq.~(\ref{eq:k_int}) are also shown in the right upper panel.}
\label{fig:43}
\end{figure*}

In reality, because of the dissipation due to disk density waves and gravitational wave emission, the system gradually loses energy and angular momentum, causing sBHs to migrate towards the MBH. If the migration timescale is longer than the liberation timescale of the particular resonance, the resonance state tends to be maintained. An example evolution is shown in Fig.~\ref{fig:res}, for which a pair of sBHs are trapped in both $(8,1)_{\rm in}$ and $(8,1)_{\rm out}$ resonances (as indicated by the resonant angles $\phi_{\rm in, out}$) and jointly migrate inward. The angular period ratio remains approximately constant.

The evolution of eccentricities of the inner and outer sBHs also shows distinctive features.  They are defined as
\begin{align}
    e=\frac{r_a-r_p}{r_a+r_p}
\end{align}
computed with the pericenter $r_p$ and apocenter $r_a$ of an orbit. 
Unlike isolated sBH migrations, we observe significant eccentricities after the initial stage without resonance locking. As shown by the analytical theory in Appendix.~\ref{app:equi}, the ``equilibrium'' eccentricity at any time depends on the particular resonance trapped in, and also on the decay time scales $t_{J}, t_{e}$. The migration dynamics competes with the resonance dynamics between the two sBHs, as the migration timescale becomes smaller compared with the libration timescale, the eccentricity increases in time. This behavior of  eccentricity evolution was previously seen in dynamical theories developed for planet migrations \cite{Cresswell2006MultiProtoplanets}, and is still applicable to the sBH migration in AGN disks \cite{Yang2019RelativisticMMR}. When the sBH migrates inward further, the PN pericenter precession starts to modify the resonance dynamics. As we include the PN precession term in the analytical theory, we observe that the inner ``equilibrium eccentricity'' indeed starts to decrease within a certain radius, which matches the numerical evolution of $e_{\rm in}$ in Fig.~\ref{fig:res}.  The difference between analytical theory and numerical evolution between $e_{\rm in}$ and $e_{\rm in, equi}$ ($e_{\rm out}$ and $e_{\rm out, equi}$ show remarkable agreement) may arise from the simplified treatment in analytical theory, as only the 1PN precession term is included therein. However, this is the first time that a PN effect on the eccentricity of mean-motion resonances is identified, leading to eccentricity $\sim 10^{-3}$ as the inner sBH reaches $10M_{\bullet}$ for this particular system.

\begin{figure*}[t]
\centering
\vspace{0.6cm}
\includegraphics[width=0.45\linewidth]{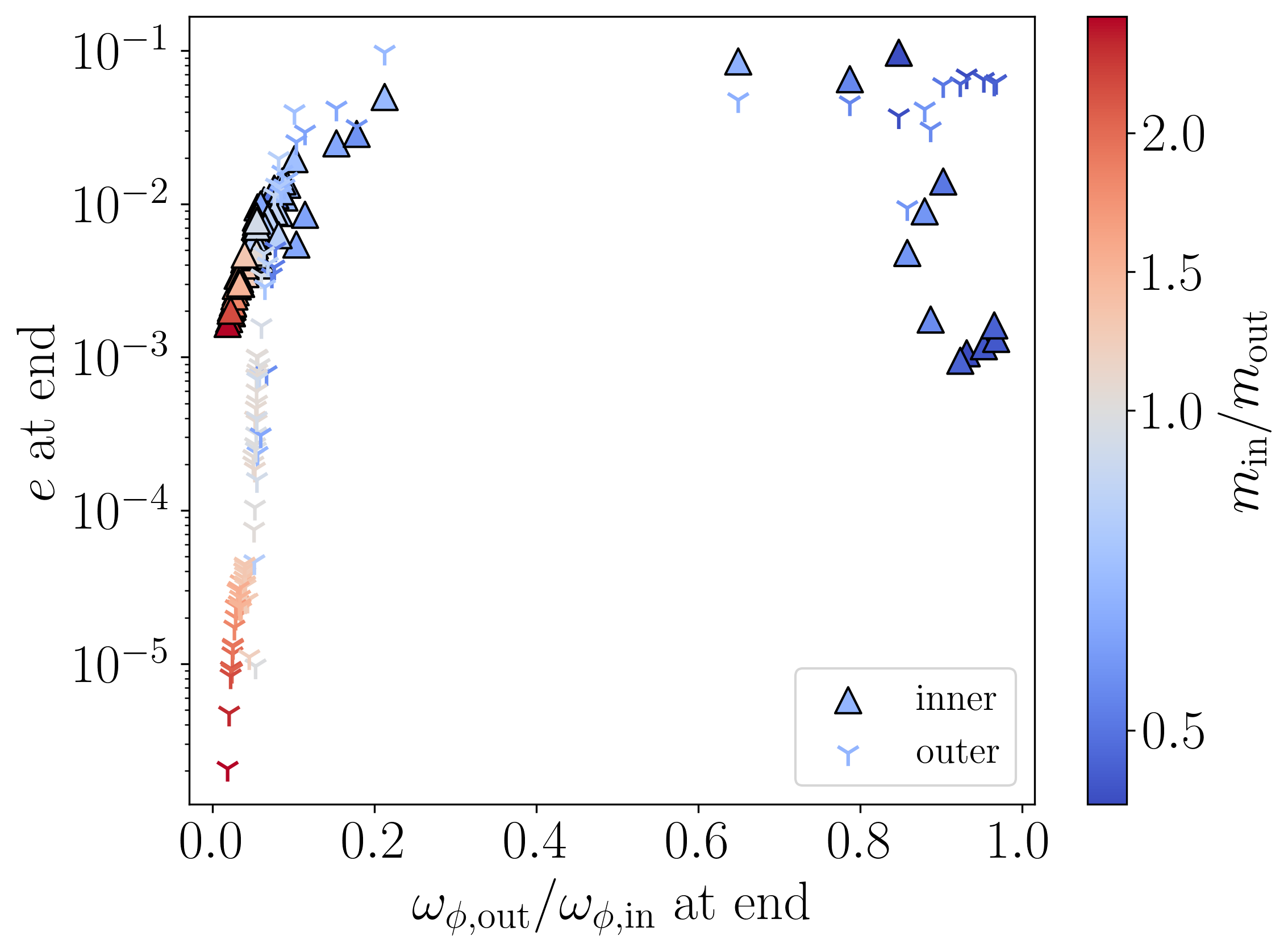}
 \includegraphics[width=0.45\linewidth]{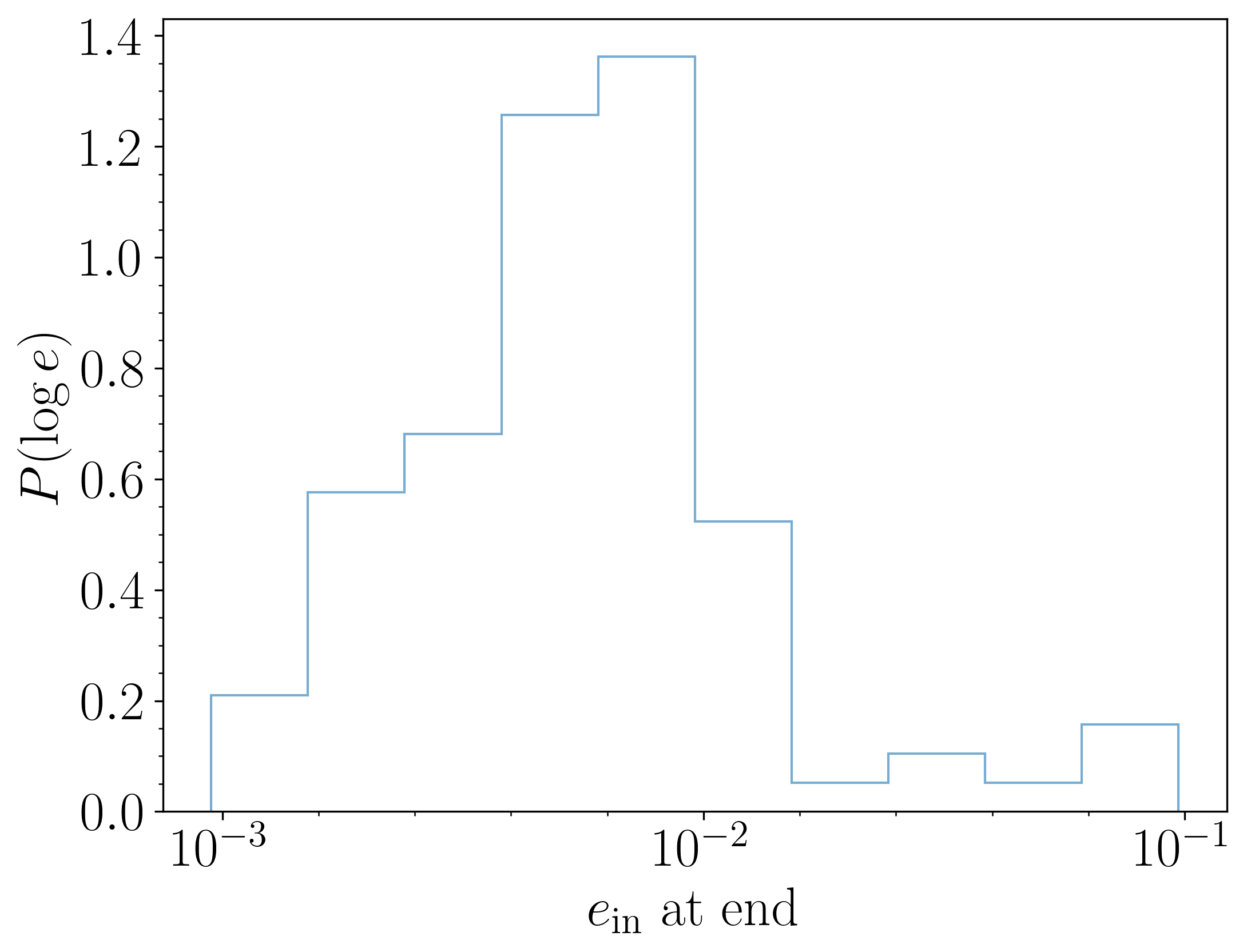}
\vspace{0.3cm}
\caption{The two dimensional scatter plot between the eccentricity and period ratios is shown in the left panel for an $\alpha$-disk model with $\dot{M}_{\bullet}=0.1\dot{M}_{\rm Edd}$, $\alpha=0.1$, and the histogram of eccentricities for the inner object at $r=10 M_{\bullet}$ is shown in the right panel. Different symbols in the left panel correspond to the inner and outer objects of the pair.}
\label{fig:dis}
\end{figure*}

\begin{figure}
\centering
\includegraphics[width=1.0\linewidth]{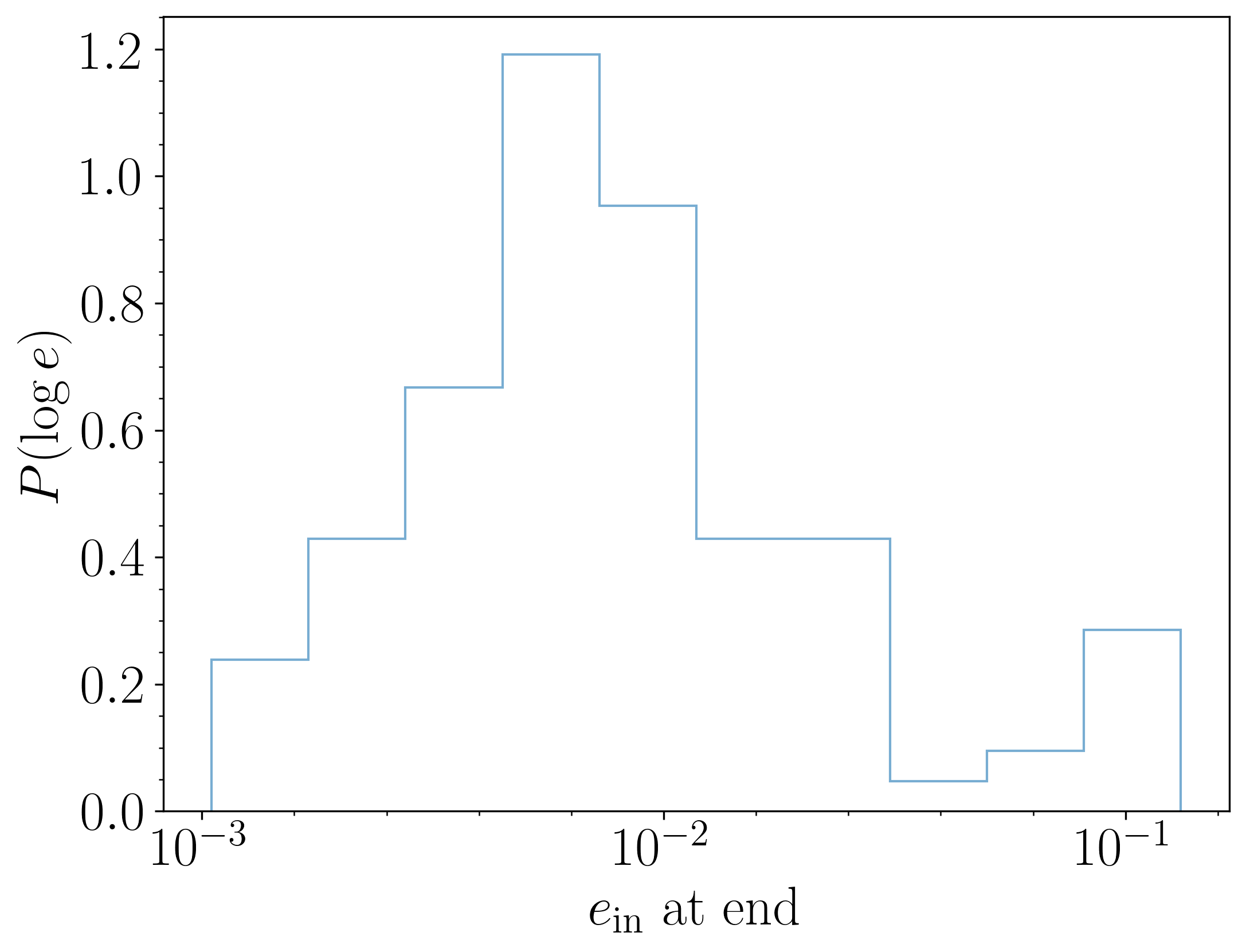}
\caption{Similar to the right panel of Fig.~\ref{fig:dis} but for the $\beta$-disk model. The eccentricity distribution is similar to $\alpha$-disk model in Fig.~\ref{fig:dis}.}
\label{fig:beta_dis}
\end{figure}

When the two sBHs are not locked in resonance, their period ratios no longer remain constant, and eccentricities decrease over time because of dissipation from the disk and gravitational wave emission. The mutual gravitational force between sBHs in general does not contribute secular effects at the leading order, except during the tidal resonances with \cite{Yang2019RelativisticMMR}
\begin{align}
k_1\omega_{\phi,\text{in}}+k_2\omega_{r,\text{in}}+k_3\omega_{\phi,\text{out}}+k_4\omega_{r,\text{out}}=0
\label{eq:k_int}
\end{align}
where $(k_1, k_2, k_3, k_4)$ are integers. Notice that in the most general cases all six frequencies of $\omega_r, \omega_\theta, \omega_\phi$ are allowed in the above relation, if the restriction on planar motion is removed. The conserved quantities of motion may undergo finite changes within any individual tidal resonance, so that one observes eccentricity excitations across tidal resonances.

It is instructive to show a few such examples. In Fig.~\ref{fig:large_in} a pair of sBHs migrate out of resonance except briefly locked into the $5:2$ resonance around $r_{\rm in} \sim 100M_{\bullet}$. During the non-resonant phase, the eccentricities decrease over time, but significant eccentricity excitations are also observed across at least three transient tidal resonances. The inner sBH's eccentricity is around $10^{-3}$ at $r=10 M_{\bullet}$. 

In Fig.~\ref{fig:43} another pair of sBHs is shown, exhibiting a clear separation between the mean-motion resonance phase and non-resonant phase. During the $4:3$ resonance phase we observe behavior similar to that in Fig.~\ref{fig:res}. During the non-resonant phase we observe both the eccentricity damping when the system is away from major tidal resonances and eccentricity pumping across tidal resonances. The final eccentricity of the inner object at $r=10M_{\bullet}$ is around $0.01$.

In addition to the resonant and non-resonant regimes, a scattering regime arises when the two sBHs are in close proximity. As illustrated in Fig.~\ref{fig:res}, at very small semi-major axes the sustained resonance further reduces their separation, leading to strong mutual scattering. This interaction excites the eccentricity of the inner sBH to relatively high values, up to $\sim 0.1$, and breaks their sustained resonance.

\paragraph{\bf Population.} To study the rich phenomena of multibody dynamics, we perform a statistical simulation of 100 examples. The central MBH has mass $M_{\bullet}=4\times10^6M_\odot$, the surface density of disk $\Sigma(r)$ and aspect ratio $h(r)$ are chosen as the same setup as the $\alpha$-disk in \cite{Pan202101} with $\dot{M}_{\bullet}=0.1\dot{M}_{\rm Edd}$, $\alpha=0.1$ as shown in Fig.~\ref{fig:disk_model}. The component mass of sBHs are randomly chosen  in the range $\{ m_{\text{in}}, m_{\text{out}} \} \in (10,30) M_\odot$, the initial semi-major axis $(a_{\text{in}}, a_{\text{out}})$ in the range $(400, 1000)M_{\bullet}$, the initial eccentricities $(e_{\text{in}}, e_{\text{out}})$ in the range $(0.0, 0.2)$, the initial true anomaly $(f_{\text{in}}, f_{\text{out}})$ in the range $(0, 2\pi)$. The initial longitudes of pericenter are set to zero, and the ascending nodes are also zero due to planar motion. The simulation is terminated when the inner sBH migrates to $r=8M_{\bullet}$. In order to quantity the eccentricity distribution, we measure the benchmark eccentricity for each system when the inner object migrates to $r=10M_{\bullet}$.

Under this set of initial conditions, there are cases where the two sBHs start in close proximity and undergo strong gravitational scattering, resulting in one of them being ejected and becoming unbound. Such scenarios are excluded from the population study, as our focus is on bound states of two sBHs.

The configurations of the systems in their final state are shown in Fig.~\ref{fig:dis}. First, we observe a gap in the period ratios for the sBH pair. 
The right branch corresponds to systems that remain in mean motion resonances for most of the evolution time, with period ratios closely matching the resonance values.  Since locking into and maintaining resonance generally requires faster intrinsic migration of the outer object, we find that in the right branch, the outer objects are typically more massive than the inner ones. The left branch, on the other hand, is contributed by pairs that have never been locked into a mean-motion resonance or pairs that are locked into a resonance briefly and escape the resonance midway, so that the period ratios continuously change over time as the inner object migrates faster than the outer object. For systems with large $m_{\rm in}/m_{\rm out}$, the influence of tidal resonances on the outer object is small (though not necessarily on the inner object), allowing the outer eccentricity to be damped to very small values by the end of the evolution. The gap between the left and right branches results from the relatively short time systems spend in these period ratios during the evolution.

The complete sample of eccentricities for the inner object at $r=10M_{\bullet}$ is summarized in the histogram shown in the right panel of Fig.~\ref{fig:dis}. We find that resonance effects generally lead to $e_{\rm in} \sim (10^{-3}, 10^{-2})$ when the inner object enters the sensitivity band of space-borne detectors.

To examine the influence of the disk model on the final eccentricity, we also consider a $\beta$-disk with $\alpha = 0.1$ and $\dot{M}_{\bullet} = 0.1 \dot{M}_{\rm Edd}$, using the same initial conditions for the 100 realizations of $\alpha$-disk model. The result, shown in Fig.~\ref{fig:beta_dis}, exhibits a distribution similar to that of the $\alpha$-disk in the right panel of Fig.~\ref{fig:dis}, suggesting that the disk model has only a minor influence compared to the PN effect.

\subsubsection{Turbulence}\label{sec:turb}

AGN disks are expected to exhibit ubiquitous strong turbulence driven by magneto-rotational instability \citep{Balbus1991}. The turbulent eddies associated with such disks exert stochastic forces on embedded sBHs, thereby exciting their orbital eccentricities. In order to quantify the eccentricity evolution of sBHs embedded in AGN disks, we first perform a series of hydrodynamical simulations tracking the evolution of a single sBH in a turbulent disk. Based on these hydrodynamical simulations, we can extract the scaling relations for the turbulence forces acting on the sBH. By incorporating these prescribed scaling relations into an \textit{N}-body code, we can then investigate the long-term evolution of the sBH and infer the eccentricity when the EMRI enters the sensitivity band of detectors, for example $r=10\ M_{\bullet}$. The inferred eccentricity of the EMRI typically depends on the properties of the turbulent disks.  Conversely, these eccentricities can be used to constrain key disk properties, such as the turbulence level, surface density, and scale height, etc.

\paragraph{\textbf{ Numerical Setup for Hydrodynamical Simulations.}}
\label{sec:setup}

To explore EMRI dynamics within an AGN disk, 
we use the hydrodynamic grid-based code \texttt{FARGO3D} \citep{FARGO3D} in our simulations. 

For the disk model, 
we adopt a locally isothermal disk model with a constant aspect ratio $h \equiv H/r= h_0$. 
We also set the initial disk gas 
surface density profile to be $\Sigma = \Sigma_0 (r/r_0)^{-1/2} \left[M_{\bullet}/r_{0}^{2}\right]$. 
Here $\Sigma_0$ is a normalization constant relevant to disk mass, $r_0$ is the code unit length and corresponds to the initial inner planet's orbital radius,
$M_{\bullet}$ is the mass of the central MBH and the code unit for mass,
and $\Omega=\sqrt{G M_{\bullet}/r^3}$ is the Keplerian angular frequency.  Note that even though the slopes for the surface density and aspect ratio are not exactly the same as the $\alpha$- or $\beta$-disk models presented in Fig.~\ref{fig:disk_model}, they should not significantly influence our hydrodynamical simulation results. 
Given the low disk masses considered in our simulations, we neglect the self-gravity of the disk.  
The indirect term associated with the motion of the central object is included,
as the central MBH is fixed at the origin of the reference frame.

We do not include the classical $\alpha$ viscosity \citep{shakura1973} in our disk model. 
Instead, we incorporate a phenomenological turbulence prescription into the accretion disk \citep{Ogihara2007,BaruteauLin2010,Pierens2011, ChenLin2023,Wu2024chaotic,Chen2025} based on previous MHD simulations
\citep{Laughlin2004} to study the orbital evolution in turbulent disks. 
Specifically, we add a fluctuating potential $\Phi_{\rm turb}$ to the momentum equation consisting of 50 stochastic modes at each time step:
\begin{equation}
    \Phi_{\rm turb }(r, \phi, t)=\gamma r^{2} \Omega^{2} \sum_{k=1}^{50} \Lambda_{k}(n_{k}, r, \varphi, t),
    \label{eq:phiturb}
\end{equation}
where $\gamma$ is a dimensionless characteristic amplitude of turbulence. Each mode, denoted as $\Lambda_{k}$, 
is given by
\begin{equation}\label{eq:Lambda}
    \Lambda_{k} = \xi_k e^{-(r-r_k)^2/\sigma_k^2} \cos(n_{k}\phi -\phi_k-\Omega_k \tilde{t}_k) \sin (\pi \tilde{t}_k/\Delta t_k),
\end{equation}
which is associated with
a wavenumber $n_{k}$ drawn from a logarithmically uniform distribution between $n_{k}=1$ and the maximum value $n_{\rm max}$, corresponding to the azimuthal grid scale.
The initial radial position $r_k$ and azimuthal angle $\phi_k$ of each mode are sampled from a uniform distribution. 
The radial extent of each mode is defined as $\sigma_k = \pi r_k /4n_{k}$. 
The modes are activated at time $t_{0,k}$ 
and persist for a duration $\Delta t_k = 0.2\pi r_k / n_{k} c_s$, 
where $c_s$ denotes the local sound speed. 
$\Omega_k$ represents the Keplerian frequency at $r_k$, $\tilde{t}_k = t - t_{0,k}$, 
and $\xi_k$ is a dimensionless constant sampled from a Gaussian distribution with unit width. 
Following \citet{BaruteauLin2010}, 
the parameter choices for this turbulence driver
emulate a Kolmogorov cascade power spectrum, maintaining a Scale Law of $n_{k}^{-5/3}$. 

Based on the velocity fluctuations induced by this turbulence driver, an effective time-averaged Reynolds stress parameter $\langle \alpha_{\rm R} \rangle$ (hereafter referred to as the dimensionless viscosity parameter $\alpha$)  in the vicinity of the sBH  can be calibrated as \citep{BaruteauLin2010}
\begin{equation}\label{eq:alpha_R}
    \alpha \simeq 35 (\gamma/h_0)^2.
\end{equation}  
This corresponds to an effective kinematic viscosity of $\nu=\alpha H^{2}\Omega$, which is consistent with previous calibrations \citep{ChenLin2023,Wu2024chaotic,Chen2025}.

Our simulations are performed in a 2D $\left(r,\phi\right)$ coordinate system, 
with a computational domain ranging radially from 0.4 $r_0$ to 3.2 $r_0$ and azimuthally from 0 to $2\pi$. 
The domain is resolved by 512 logarithmic-spaced grid cells in the radial direction and 1536 uniformly spaced cells in the azimuthal direction. 
We apply wave-damping radial boundary conditions \citep{deValBorro2006} to both the inner edge ($[0.4-0.5]r_{0}$) and the outer ($[2.8-3.2]r_{0}$) edge to avoid wave reflection. 
Tests with different numerical resolutions have been performed to confirm the convergence of our simulation results.
To avoid numerical artifacts caused by damping of turbulence at the boundaries,
we limit the turbulent region to a radial range between 0.5 $r_0$ and 2.8 $r_0$, instead of extending turbulence across the entire disk. 

In all simulations presented in this work, 
the sBH is initially located at an orbital radius of $r = 1.0~r_0$ with a mass ratio of sBH to MBH of $q\equiv m/M_{\bullet}$. 
We fully account for the orbital evolution of the sBH because of the gravitational interaction with the disk. 
We will explore the orbital evolution of the sBH with different turbulence strengths $\gamma$, mass ratios $q$, disk surface densities $\Sigma_{0}$, and different disk aspect ratios $h_{0}$.

\paragraph{\textbf{Orbital Perturbation with Turbulent Forces.}}

Before presenting the results of the hydrodynamical simulation, we first discuss how turbulent disk forces induce orbital perturbations.
Considering the sBH perturbed by the (turbulent) fluctuating gravitational force from the disk, the equation of motion is given by \citep{Li2025}
\begin{align}
&\ddot{r} - r \dot{\phi}^2 +\frac{M_{\bullet}}{r^2} = f_r\,, \nonumber \\
& \ddot{\phi}r^2+2 \dot{\phi} \dot{r} r =t_\phi\,,
\end{align}
with $f_r$ being the specific force in the radial direction and $t_\phi$ the specific torque in the azimuthal direction. We consider the case where sBH is weakly perturbed from a circular orbit.

\begin{align}
r=r_0 +\delta r,\quad \phi = \Omega_{0} t+ \delta \phi,
\end{align}
with $\Omega_{0}^{2} = M_{\bullet}/r^3_0$. The equation of motion can be simplified as

\begin{align}
&\ddot{r} -3\Omega_{0}^{2} \delta r -2 r_0 \Omega_{0} \delta \dot{\phi} =f_r,\nonumber \\
& \delta \ddot{\phi}r^2_0+2 \Omega_{0} r_0 \delta \dot{r} =t_\phi\,.
\end{align}

In the frequency domain, we have the replacement rules $\partial_t \rightarrow - i \omega, \partial^2_t \rightarrow -\omega^2$, so the equation of motion further becomes
\begin{align}\label{eq:eom1}
&-\omega^2 \delta r-3 \Omega_{0}^{2} \delta r+2 i r_0 \Omega_{0} \omega \delta \phi = f_r(\omega),\nonumber \\
& -\omega^2 \delta \phi r^2_0-2 i \Omega_{0}\omega r_0 \delta r =t_\phi(\omega).
\end{align}

The solution is given by
\begin{align}
\delta r & = -\frac{\omega r_0 f_r + 2 i \Omega_{0} t_\phi}{\omega r_0(\omega^2-\Omega_{0}^{2})},\nonumber \\
\delta \phi & = -\frac{\omega^2 t_\phi - 2 i r_0 \omega \Omega_{0} f_r +3 \Omega_{0}^{2} t_\phi}{\omega^2 r^2_0 (\omega^2-\Omega_{0}^{2})}.
\end{align}
The induced eccentricity, which is described by the magnitude of radial fluctuations, can be estimated as
\begin{align}
\langle e^2 \rangle  & = \int d \omega \frac{S_{\delta r}(\omega)}{r^2_0} \nonumber \\
& =\int d \omega \frac{S_{f_r}}{r^2_0(\omega^2-\Omega_{0}^{2})^2}+ \int d \omega \frac{4 \Omega_{0}^{2} S_{t_\phi}}{\omega^2 r^4_0(\omega^2-\Omega_{0}^{2})^2}.
\end{align}

This equation tends to yield divergent results if $S_{f_r}(\Omega_{0})$ or $S_{t_\phi}(\Omega_{0})$ is nonzero. To regularize the equilibrium eccentricity, we recognize that the co-orbital Lindblad resonances and density waves in the presence of turbulence tend to damp the eccentricity,
say at a rate $\gamma_e =1/\tau_e \ll \omega$. Therefore, we effectively modify the first equation of Eq.~(\ref{eq:eom1}) by 
\begin{align}
-\omega^2 \delta r-i \omega \gamma_e \delta r-3 \Omega_{0}^{2} \delta r+2 i r_0 \Omega_{0} \omega \delta \phi = f_r(\omega),
\end{align}
so that 
\begin{align}
\langle e^2 \rangle  & = \int d \omega \frac{S_{\delta r}(\omega)}{r^2_0} \nonumber \\
& =\int d \omega \frac{S_{f_r}}{r^2_0[(\omega^2-\Omega_{0}^{2})^2+\gamma_e^2 \omega^2]} \nonumber \\
&+ \int d \omega \frac{4 \Omega_{0}^{2} S_{t_\phi}}{\omega^2 r^4_0[(\omega^2-\Omega_{0}^{2})^2+\gamma_e^2 \omega^2]}.
\label{eq:e_smooth}
\end{align}

The pole or resonance part of this integral is given by (assuming $\gamma_e \ll \Omega_{0}$)
\begin{align}
\langle e^2 \rangle_r =\frac{\pi S_{f_r}(\Omega_{0})}{2 r^2_0 \Omega_{0}^{2} \gamma_e}+\frac{2 \pi S_{t_\phi}(\Omega_{0})}{\Omega_{0}^{2} r^4_0 \gamma_e}\,.
\label{eq:e_pol}
\end{align}
The equilibrium eccentricity can be estimated by combining the smooth component of Eq.~(\ref{eq:e_smooth}) and the pole component  (Eq.~\ref{eq:e_pol}) of the power spectrum. 
In other words, the underlying eccentricity damping rate, $\gamma_e$, can be constrained based on the simulated eccentricity (or the orbital perturbation $\delta r/r$) of the embedded object and the turbulent force acting on the embedded object.
In most cases, we find that the radial forces dominate over the azimuthal forces; thus it is the first term that mainly contributes to the eccentricity evolution of the EMRI.

\begin{figure}
\centering
\includegraphics[width=0.45\textwidth,clip=true]{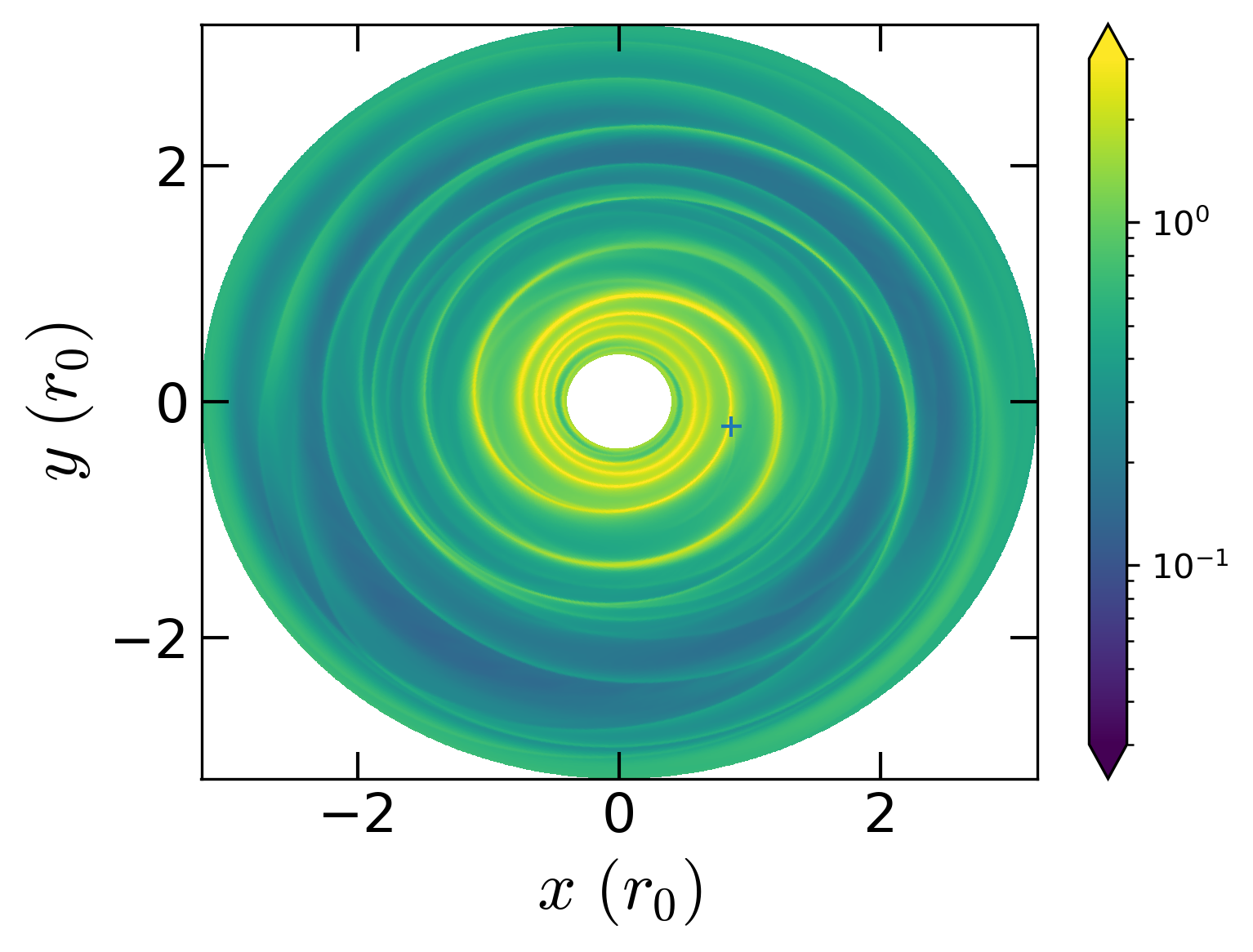}
\caption{A typical snapshot of gas surface density (in unit of $\Sigma_{0}$) for a turbulent disk at 4000 $P_{0}$. $P_{0}$ is the orbital period at its initial location $r_{0}$. The disk aspect ratio is $h=0.03$, turbulence strength $\gamma=1.6\times10^{-3}$ (or equivalently $\alpha=0.1$), mass ratio $q=5\times10^{-6}$. The `+' symbol indicate the position of the embedded object.}
 \label{fig:sigma_fid}
\end{figure}

\begin{figure}
\centering
\includegraphics[width=0.45\textwidth,clip=true]{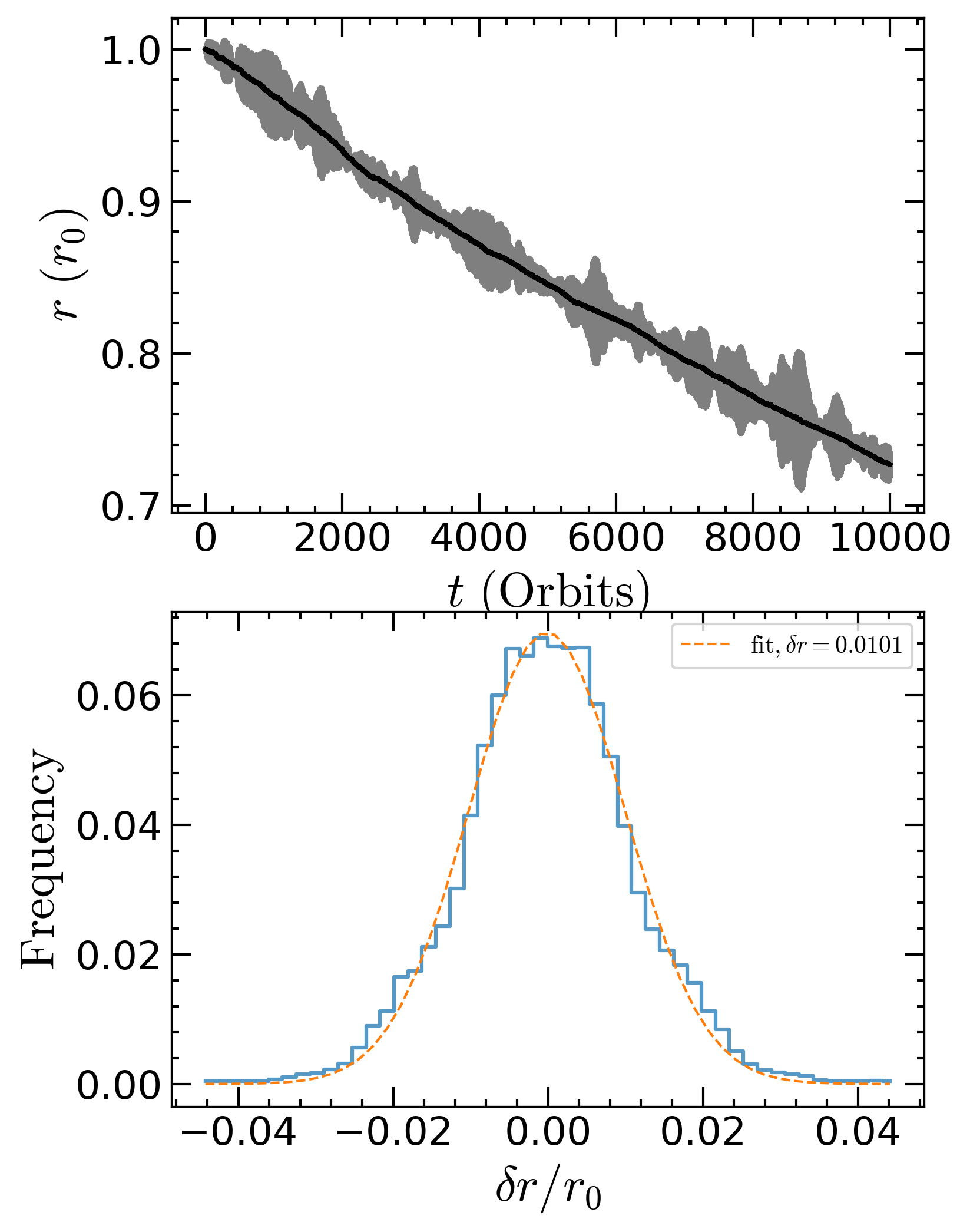}
\caption{Upper panel: The time evolution of the orbital radius (shaded region) and semi-major axis (black line) of the EMRI for a fiducial run. Lower panel: the histogram of the radial oscillation $\delta r$ for EMRI, which can be fitted with a Gaussian profile of a standard deviation $\delta r\simeq0.01\ r_{0}$. Here, we have adopted a mass ratio of $q=5\times10^{-6}$, a turbulence strength of $\gamma=1.6\times 10^{-3}$, a disk aspect ratio $h_{0}=0.03$ and a surface density $\Sigma_{0}=10^{-4}\ M_{\bullet}/r_{0}^{2}$ at $r_{0}$. The turbulence strength and the disk aspect ratio correspond to the viscosity parameter $\alpha=0.1$.}
 \label{fig:dr_fid}
\end{figure}

\paragraph{\textbf{Hydrodynamical Evolution of EMRI.}}

We first present the simulation results for an sBH embedded in a disk with a typical model parameters.  The mass ratio of the sBH to the MBH is set to  $q=m/M_{\bullet}=5\times10^{-6}$, where the mass of the sBH is  $m=20\ M_{\odot}$ and the mass of the MBH is $M_{\bullet}=4\times10^{6}\ M_{\odot}$. The other disk parameters are $\Sigma_{0}=10^{-4}\ M_{\bullet}/r_{0}^{2}$ and $h_{0}=0.03$ at $r_{0}$. We consider a turbulent disk with a turbulence strength of $\gamma=1.6\times10^{-3}$, which corresponds to the viscosity parameter $\alpha=0.1$. 

\begin{figure*}[t]
\centering
\includegraphics[width=0.9\textwidth,clip=true]{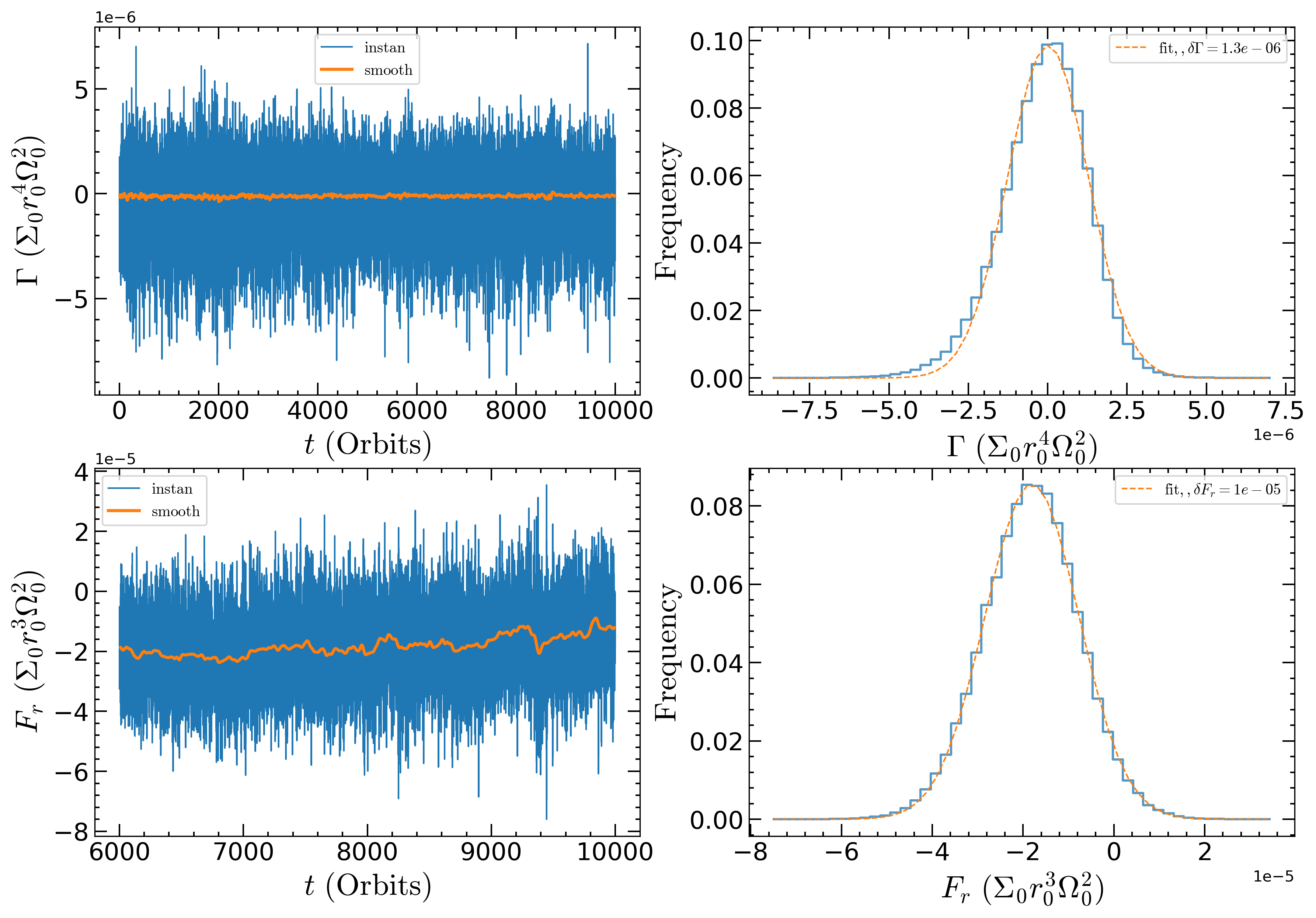}
\caption{The evolution of the turbulent forces for the model shown in Fig.~\ref{fig:dr_fid}. The upper left panel show the instantaneous torque acting on the EMRI with the its timing average indicated by the orange line. The upper right panel shows the histogram of the turbulent torque. Lower panel: similar to the upper panel but for the radial force.}
 \label{fig:force_fid}
\end{figure*}

A snapshot of the disk surface density at 4000 orbits (measured in units of the orbital period $P_{0}$ at $r_{0}$) is shown in Fig.~\ref{fig:sigma_fid}. Strong density fluctuations are present throughout the disk as a result of the injection of external turbulence. We have confirmed that the density power spectrum follows a power-law scaling consistent with $n_{k}^{-5/3}$. The low mass ratio results in a weak density wave around the embedded object. 

Under this circumstance, the inward migration rate is approximately 2-3 times faster than the classical type I migration rate as shown in the upper panel of Fig.~\ref{fig:dr_fid}, possibly due to turbulent density structures. During inward migration of the EMRI, an important feature is that the eccentricity can be excited as shown in Fig.~\ref{fig:dr_fid}. As shown in the lower panel of Fig.~\ref{fig:dr_fid}, the standard deviation of the radial oscillation from the mean radial motion $\delta r/r$, which is a measurement of the mean eccentricity of the EMRI $e$, is $\sim0.01$. We have confirmed that such a radial oscillation (or the orbital eccentricity) does not evolve with time significantly after a few thousand orbits, indicating a convergence in time. The eccentricity excitation arises from the turbulent density structures, which generate the stochastic radial and azimuthal acceleration of the EMRI, as shown in Fig.~\ref{fig:force_fid}. Based on the radial and azimuthal acceleration profile for the EMRI, we can obtain an effective eccentricity damping rate of the turbulent disk $\gamma_{e}\simeq10^{-4}\Omega_{0}$ for this disk model using Eqs.~(\ref{eq:e_smooth}) and (\ref{eq:e_pol}). This damping rate for this particular model is on the same order of magnitude as the type I damping rate of the orbital eccentricity in the subthermal regime $\sim qh_{0}^{-4} \Sigma_{0}r_{0}^{2}/M_{\bullet}\Omega_{0}\simeq 5\times10^{-4}\Omega_{0}$ \citep[e.g.,][]{Papaloizou2000,Li2019,Ida2020}. 
However, as we will show later, the parameter dependence of this damping rate differs from that of the classical Type I eccentricity damping rate.

\begin{figure*}
\centering
\includegraphics[width=0.9\textwidth,clip=true]{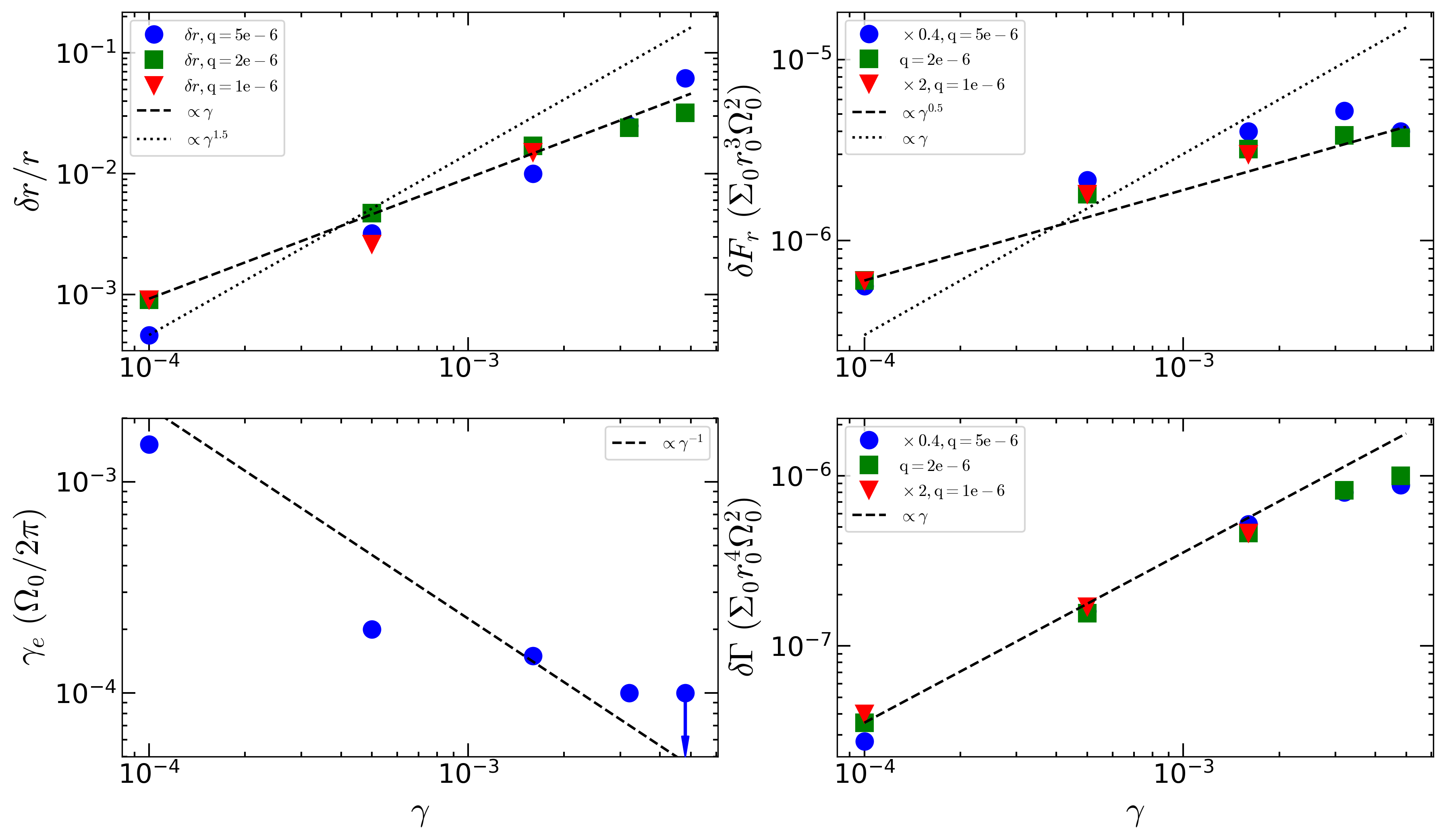}
\caption{The scaling relation of $\delta r/r$ (upper left panel), radial force $F_{r}$ (upper right panel), eccentricity damping rate $\gamma_{e}$ (lower left), and torque $\Gamma$ (lower right panel)  as a function of the disk turbulence strength $\gamma$. Different symbols indicate different mass ratio for the EMRI. Here, we have fixed the disk aspect ratio of $h_{0}=0.03$, and a surface density $\Sigma=10^{-4}\ M_{\bullet}/r_{0}^{2}$.}
 \label{fig:scale_gamma}
\end{figure*}

\begin{figure*}[t]
\centering
\includegraphics[width=0.9\textwidth,clip=true]{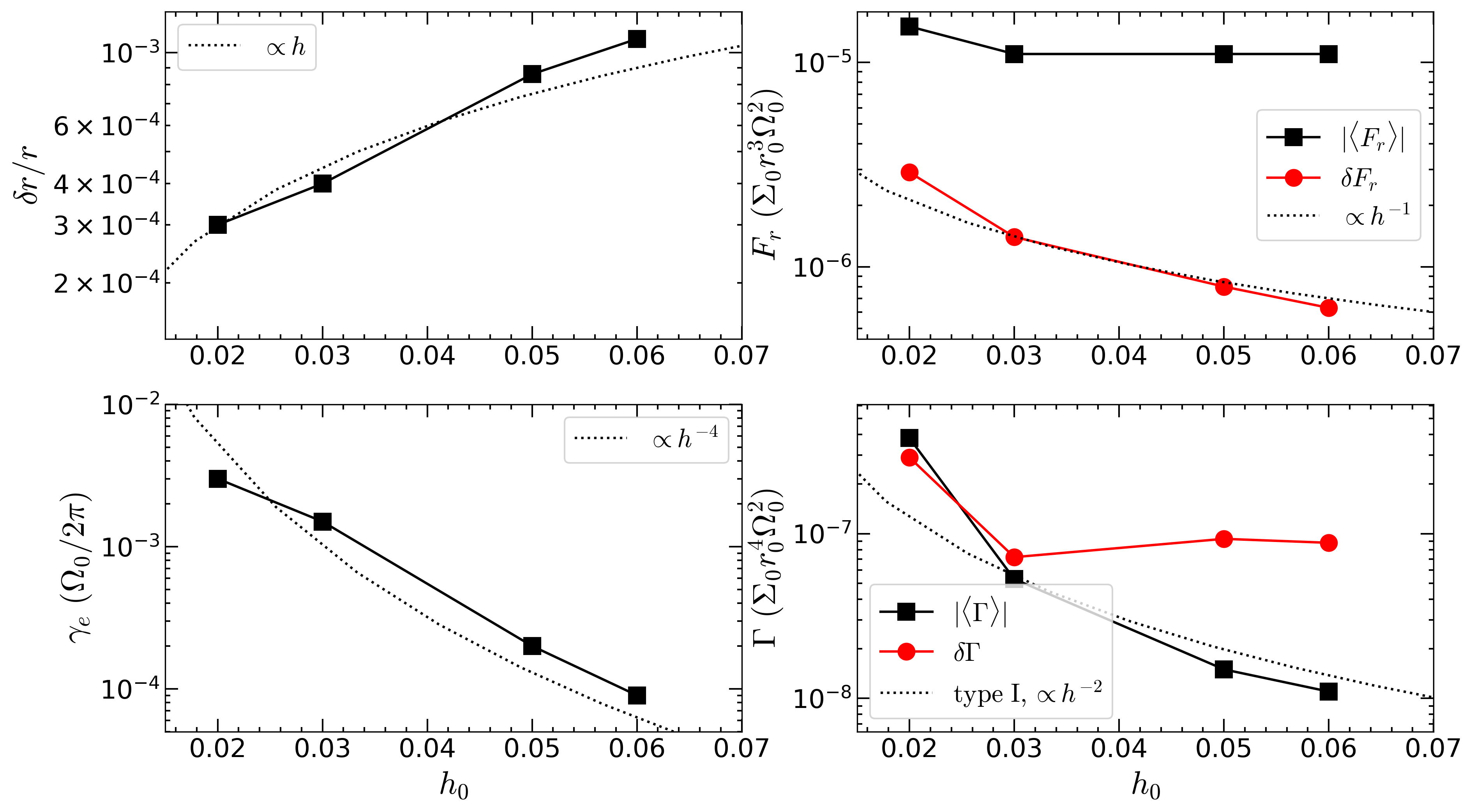}
\caption{The scaling relation of $\delta r/r$ (upper left), radial force $F_{r}$ (upper right), eccentricity damping rate $\gamma_{e}$ (lower left), and torque $\Gamma$  (lower right) as a function of the disk aspect ratio $h_{0}$. Here, we have fixed the turbulence strength at $\gamma=10^{-4}$, a surface density $\Sigma=10^{-4}M_{\bullet}/r_{0}^{2}$, and mass ratio of $q=5\times10^{-6}$.}
 \label{fig:scale_h}
\end{figure*}

\begin{figure}
\centering
\includegraphics[width=0.45\textwidth,clip=true]{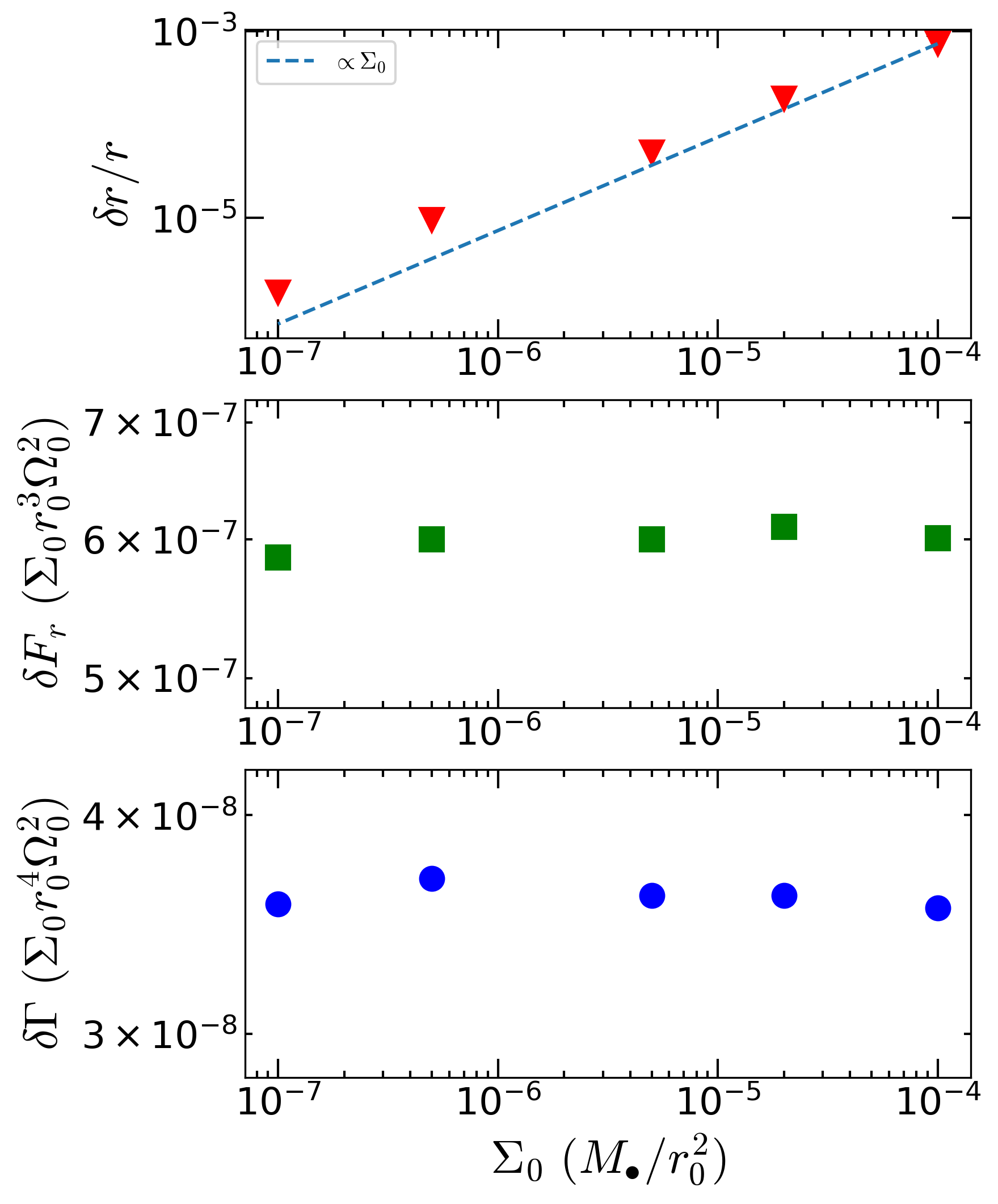}
\caption{The scaling relation of $\delta r/r$ (upper panel), radial force $\delta F_{r}$ (middle panel) and torque $\delta \Gamma$ (lower panel)  as a function of the disk surface density $\Sigma_{0}$. The dashed line in the upper panel corresponds to the linear relation $\propto \Sigma_{0}$. Here, we have fixed the disk aspect ratio of $h_{0}=0.03$, a mass ratio of $q=2\times10^{-6}$, and a turbulence strength of $\gamma=10^{-4}$.}
 \label{fig:scale_sigma}
\end{figure}

\paragraph{\textbf{Scaling Relation for Turbulent Disks.}} \label{sec:turb_scal}

The evolution of EMRI eccentricity can depend on various disk parameters, such as the turbulence strength $\gamma$, the disk aspect ratio $h_0$, the disk surface density $\Sigma_{0}$ and the EMRI mass ratio $q$. To comprehensively characterize the turbulence-induced eccentricity under general conditions, we conduct an extensive parameter survey across a range of model parameters.

The dependence of $\delta r$, the turbulence forces ($\delta F_{r}$ and $\delta \Gamma$), and the damping rate $\gamma_{e}$ on the disk turbulence strength $\gamma$ and the mass ratio $q$ are shown in Fig.~\ref{fig:scale_gamma}.
We have fixed the disk aspect ratio of $h_{0}=0.03$, and a surface density $\Sigma_{0}=10^{-4}\ M_{\bullet}/r_{0}^{2}$. The pow-law scaling for different quantities is shown as a dashed or dotted line in each panel. It can be seen that the relations $\delta F_{r}\propto q\gamma$, $\delta\Gamma\propto q\gamma$, $\gamma_{e}\propto \gamma^{-1}$ 
can roughly match the parameter dependence. The linear dependence of the turbulence force acting on the sBH results in a constant specific force ($f_r\equiv \delta F_r/q$, and $\tau_{\phi}=\delta\Gamma/q$) for the sBH. This behavior can be attributed to the fact that the turbulent potential is independent of the mass ratio $q$. The linear dependence of $\gamma$ is directly related to the scaling of the turbulence potential of Eq.~(\ref{eq:phiturb}). The scaling of the eccentricity damping rate $\gamma_{e}$ in the turbulence strength suggests a suppressed eccentricity damping for the stronger turbulence, which is probably due to the weaker coherent structures and/or Lindblad resonances in strongly turbulent environments.
For the parameter explored here, we have $f_{r}\gg \tau_{\phi}/r_{0}$. Therefore, the radial fluctuation follows  $\delta r\propto (f_{r}^{2}/\gamma_{e})^{1/2} \propto \gamma^{1.5} q^{0}$, which is roughly consistent with the scaling results shown in the upper left panel of Fig.~\ref{fig:scale_gamma}. 
The deviation from linear scaling, especially the flattening of the scaling, at large $\gamma$ may arise from strong turbulence-induced disturbances in the disk, leading to a nonlinear response.  
To account for this deviation in the $\delta F_r-\gamma$ relation, we also tested an alternative scaling, $f_r \propto \gamma^{1/2}$, which appears to better capture the flattening of the trend at higher turbulence levels. The resulting scaling for $\delta r$, $\delta r \propto \gamma$, is shown as the dashed line in the upper left panel of Fig.~\ref{fig:scale_gamma}. 
In the following, we will adopt both scalings in long-term N-body simulations to predict the orbital evolution of the EMRI in the vicinity of the MBH.

Now we further explore the scaling relation with respect to the disk aspect ratio. Here we have fixed a disk surface density $\Sigma_{0}=10^{-4}\ M_{\bullet}/r_{0}^{2}$, a turbulence strength $\gamma=10^{-4}$, and the EMRI mass ratio $q=5\times10^{-6}$. The simulation results for different $h_{0}$ are shown in Fig.~\ref{fig:scale_h}.
We can see that $\delta F_{r}\propto h_{0}^{-1}$. There is a weak dependence of $\delta \Gamma$ on $h_{0}$, so we treat $\delta \Gamma$ as a constant with respect to $h_{0}$. For the damping rate $\gamma_{e}$, we identify a scaling relation $\gamma_{e}\propto h_{0}^{-4}$, which coincidentally resembles the classical type I scaling \citep[e.g.,][]{Papaloizou2000,Li2019,Ida2020}. 
Given these scalings, the radial fluctuation follows $\delta r$ as $\delta r\propto (f_{r}^2/\gamma_{e})^{1/2}\propto h_{0}$, which is consistent with the simulation results shown in the upper left panel of Fig.~\ref{fig:scale_h}. 

We also present the scaling of the time-averaged background force magnitude $|\langle F_{r}\rangle|$ and $|\langle \Gamma\rangle|$ in the right panels of Fig.~\ref{fig:scale_h}. The ``quasi-steady'' (time-averaged) component of the radial force is comparable to or even larger than the stochastic one and shows almost no dependence on the disk aspect ratio $h_{0}$. In contrast, the ``quasi-steady'' component of the torque is generally weaker than the stochastic component and exhibits a power-law decay with $h_{0}^{-2}$, consistent with the scaling of the type I migration torque \citep{Tanaka2002,Paardekooper2023}.

Finally, we perform a set of additional simulations with different disk masses by varying $\Sigma_{0}$ from $10^{-7}M_{\bullet}/r_{0}^{2}$ to $10^{-4}M_{\bullet}/r_{0}^{2}$. The other model parameters are chosen as the turbulence strength $\gamma=10^{-4}$, a disk aspect ratio of $h_{0}=0.03$, and a mass ratio for EMRI $q=2\times10^{-6}$. The simulation results are shown in Fig.~\ref{fig:scale_sigma}. We find that all the quantities $\delta F_{r}$, $\delta \Gamma$, and $\delta r$ are linearly proportional to $\Sigma_{0}$. This  implies that the damping rate $\gamma_e$ should be independent of $\Sigma_{0}$ over the three orders of magnitude explored, which is remarkably different from the classical type I scaling ($\propto \Sigma_{0}$) for the orbital eccentricity. The distinct scaling behavior of the turbulence-induced $\gamma_e$ highlights the need for further dedicated studies to uncover its physical origin.  

Now we can summarize the full scaling of stochastic forces. For the stochastic radial force,
\begin{eqnarray}
    \delta F_{r}&\simeq& 3\times10^{-6}\Sigma_{0}r_{0}^{3}\Omega_{0}^{2}\left(\frac{h_{0}}{0.03}\right)^{-1}\left(\frac{\gamma}{1.6\times10^{-3}}\right)^{s} \nonumber \\
    &\times& \left(\frac{q}{5\times10^{-6}}\right),
     \label{eq:fr_sca}
\end{eqnarray}
where $s=0.5$ or $1.0$ based on the scaling of $\delta F_{r}$ on the turbulence strength $\gamma$ shown in Fig.~\ref{fig:scale_gamma}. For the stochastic torque,
\begin{equation}
    \delta \Gamma \simeq 5\times10^{-7}\Sigma_{0}r_{0}^{4}\Omega_{0}^{2}\left(\frac{\gamma}{1.6\times10^{-3}}\right)\left(\frac{q}{5\times10^{-6}}\right),
     \label{eq:torque_sca}
\end{equation}
and the damping rate is
\begin{equation}
    \gamma_{e} \simeq 10^{-4}\Omega_{0}/(2\pi)\left(\frac{\gamma}{1.6\times10^{-3}}\right)^{-1}\left(\frac{h_{0}}{0.03}\right)^{-4}.
     \label{eq:gammae_sca}
\end{equation}

When the stochastic radial force dominates (e.g., for smaller $h_{0}$)  the orbital eccentricity evolution, we have the equilibrium orbital perturbation, namely eccentricity as $\delta r/r \propto \Sigma_{0}h_{0}\gamma^{s+0.5}r^{1.25}$. Therefore, we can obtain

\begin{eqnarray}
     \frac{\delta r}{r} &\simeq& 3\times 10^{-5} \left(\frac{\Sigma_{0}}{6\times10^{5}\ {\rm g\ cm^{-2}}}\right) \left(\frac{\alpha}{0.1}\right)^{s/2+0.25}  \nonumber \\
     && \left(\frac{h_{0}}{0.03}\right)^{s+1.5}\left(\frac{r}{100\ M_{\bullet}}\right)^{1.25}.
     \label{eq:dr_sca1}
\end{eqnarray}
Otherwise, for the $\tau_{\phi}$ dominating case (e.g., for larger $h_{0}$), $\delta r/r \propto \Sigma_{0}h_{0}^{2}\gamma^{1.5} r^{1.25}$, we have

\begin{eqnarray}
     \frac{\delta r}{r} &\simeq& 5\times 10^{-6} \left(\frac{\Sigma_{0}}{6\times10^{5}\ {\rm g\ cm^{-2}}}\right) \left(\frac{\alpha}{0.1}\right)^{0.75}  \nonumber \\
     && \left(\frac{h_{0}}{0.03}\right)^{3.5}\left(\frac{r}{100\ M_{\bullet}}\right)^{1.25}.
     \label{eq:dr_sca2}
\end{eqnarray}
Note that these scalings depend solely on local disk conditions, neither on global radial profiles nor specific AGN disk models.

\begin{figure}
\centering
\includegraphics[width=0.45\textwidth,clip=true]{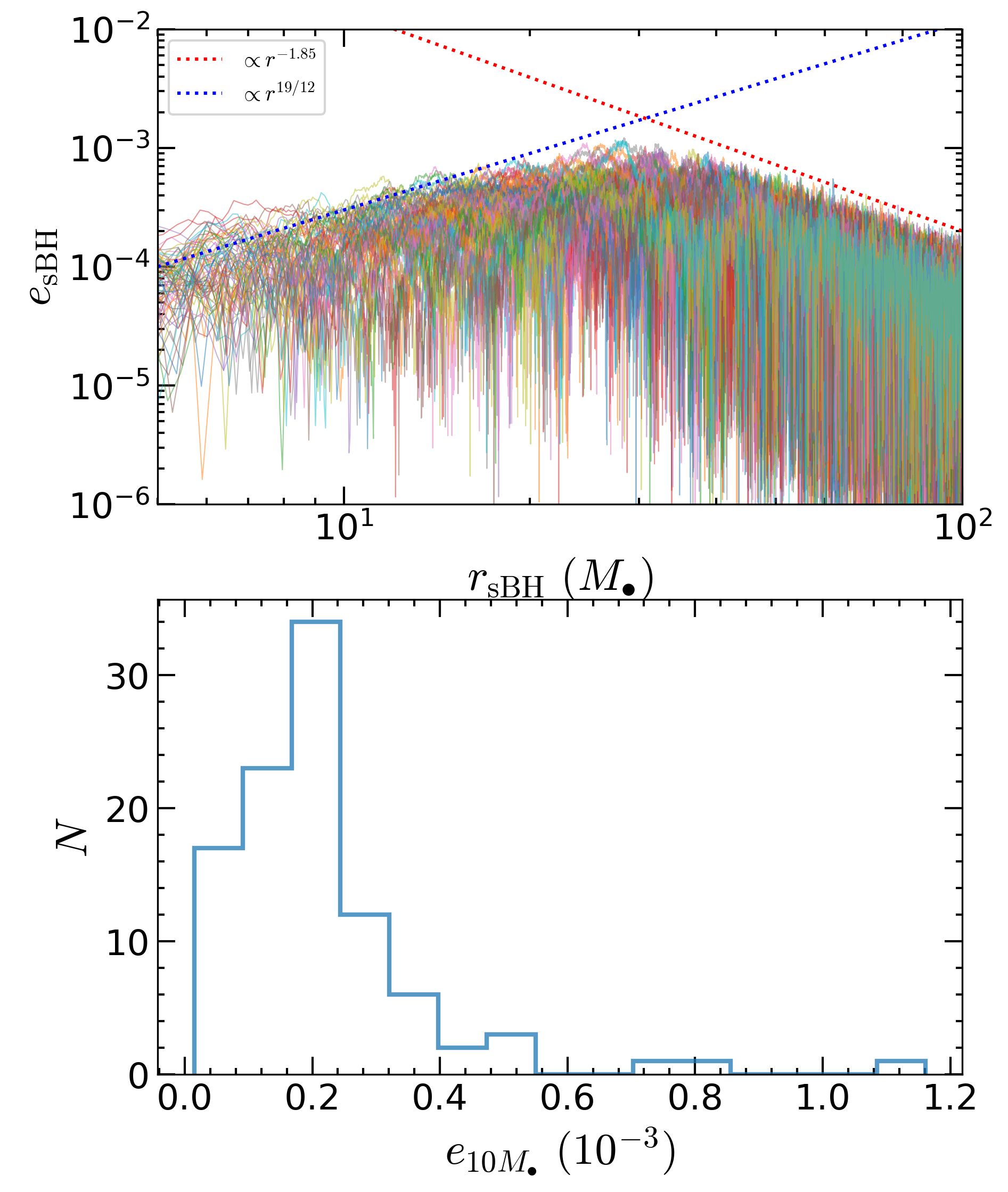}
\caption{Upper panel: the eccentricity evolution for a $\beta$-disk model with 100 realizations. We adopt the model parameters with $\alpha=0.1$, $\dot{M}_{\bullet}=0.5\dot{M}_{\rm Edd}$, $s=1.0$, $M_{\bullet}=4\times10^{6}M_{\odot}$ and $q=5\times10^{-6}$. Different lines in the upper panel correspond to different Monte Carlo realizations, and the dashed line shows the scaling of $r^{-1.85}$ expected from the turbulence-driven equilibrium eccentricity evolution. The eccentricity decays in the inner disk region is due to the gravitational wave radiation which follows the scaling of $r^{19/12}$. Lower panel: the histogram of the eccentricity at $10\ M_{\bullet}$ for different realizations.}
 \label{fig:ecc_nbody}
\end{figure}

The above discussion assumes quasi-equilibrium in the evolution of eccentricities at each radius $r$, driven by the stochastic force.
As the sBH inspirals towards the MBH, the gravitational wave radiation becomes increasingly significant. Consequently, there exists a critical radius $r_{\rm peak}$, within which the eccentricity damping timescale $1/\gamma_{\rm e,gw}$ by gravitational wave radiation becomes shorter than the turbulent disk equilibrium timescale $1/\gamma_{e}$ (also see Fig.~\ref{fig:ecc_nbody}). At radii smaller than this critical value (that is, $r\lesssim r_{\rm peak}$), the eccentricity evolution is predominantly governed by gravitational wave radiation, following a radius scaling of $r^{19/12}$ \citep{Peters1964a}.

\paragraph{\textbf{Longterm Evolution and Implications.}}\label{sec:turb_nboby}

We implement the stochastic radial $f_{r}$ and torque $\tau_{\phi}$  with the damping rate $\gamma_{e}$ following the above scaling relation into \textit{N}-body code Rebound \citep{ReinLiu2012}.
We use the WHFast integrator \cite{ReinTamayo2015} in Rebound to perform the long-term 
 calculation. Migration and turbulence forces are implemented in Reboundx \citep{Tamayo2020} to study the long-term evolution of EMRI. The ``quasi-steady'' component of the turbulence torque drives the inward migration of the sBH, whereas the stochastic components of the radial and azimuthal forces can excite the eccentricity of the EMRI.
The eccentricity and semi-major axis damping due to gravitational wave radiation are also included in the \textit{N}-body calculation.

As an illustrative example, we first consider a $\beta$-disk model, in which the disk surface density is higher in the inner region ($r\lesssim 100\ M_{\bullet}$) compared to a $\alpha$-disk model with the same disk parameter (refer to Fig.~\ref{fig:disk_model}). As a result, the turbulence-driven eccentricity in the inner disk is higher. We choose the model parameter as $\alpha=0.1$, $\dot{M}_{\bullet}=0.5\dot{M}_{\rm Edd}$, $M_{\bullet}=4\times10^{6}M_{\odot}$ and $q=5\times10^{-6}$.
The disk profiles are shown in Fig.~\ref{fig:disk_model}.
The scaling index $s$ for the radial turbulence force is set to $s=1.0$.  
Since the turbulence force is stochastic in nature, we simulate 100 Monte Carlo realizations with the same model parameters to explore the statistics properties of the EMRI.

The results for these \textit{N}-body simulations are shown in Fig.~\ref{fig:ecc_nbody}. As the EMRI migrates inward, its orbital eccentricity increases as shown in the upper panel of Fig.~\ref{fig:ecc_nbody}, although there are some fluctuations due to the implementation of stochastic forces. The increase in eccentricity is attributed to the radial dependence of the turbulence force and damping rate, such that the equilibrium eccentricity varies with orbital radius. Based on the scaling shown in the previous Section, we find $e_{\rm sBH}\propto r^{-1.85}$, shown as the dashed line in the upper panel of Fig.~\ref{fig:ecc_nbody}. This radial dependence is in good agreement with our \textit{N}-body simulation results. Within a radius of $r\sim30\ M_{\bullet}$, the gravitational wave radiation becomes dominant and the eccentricity of the sBH decays with radius following $r^{19/12}$, as expected.

The lower panel of Fig.~\ref{fig:ecc_nbody} shows the statistical distribution of the orbital eccentricity during the last inspiral stage of EMRI, for example, at $10\ M_{\bullet}$. We can see that the predicated EMRI eccentricity at $r=10\ M_{\bullet}$ is around $e_{\rm sBH}\simeq2\times10^{-4}$, which is detectable for future LISA/Taiji/TianQin observations \cite{Babak2017_LISA_EMRI,Fan2020TianQinEMRI}.

We have also explored the case where the power law index for the radial turbulence force is $s=0.5$. In this case, the radial dependence of the eccentricity is $e_{\rm sBH}\propto r^{-1.35}$ based on the scaling of the radial turbulence force. Due to the much steeper radial dependence of the azimuthal turbulence forces, they can also contribute comparably to the orbital eccentricity around $10\ M_{\bullet}$. As a result, the equilibrium eccentricity reaches a level comparable to $10^{-4}$ at $r\simeq 10\ M_{\bullet}$.

For an $\alpha$-disk model with the same model parameters, the disk scale height remains almost the same; however, the disk surface density scales as $\Sigma\simeq 10^{5} (r/100\ M_{\bullet})^{1.5}\ {\rm g\ cm^{-2}}$ in the inner region of the AGN disk ($r\lesssim100 M_{\bullet}$). This results in a significantly lower disk density at $10\ M_{\bullet}$, resulting in a much lower eccentricity of the EMRI $e_{\rm sBH}\lesssim  10^{-7}$ induced by disk turbulence. Such a low eccentricity lies below the detection limit of the LISA/Taiji/TianQin and could be undetectable.

\begin{figure}
\centering
\includegraphics[width=0.5\textwidth,clip=true]{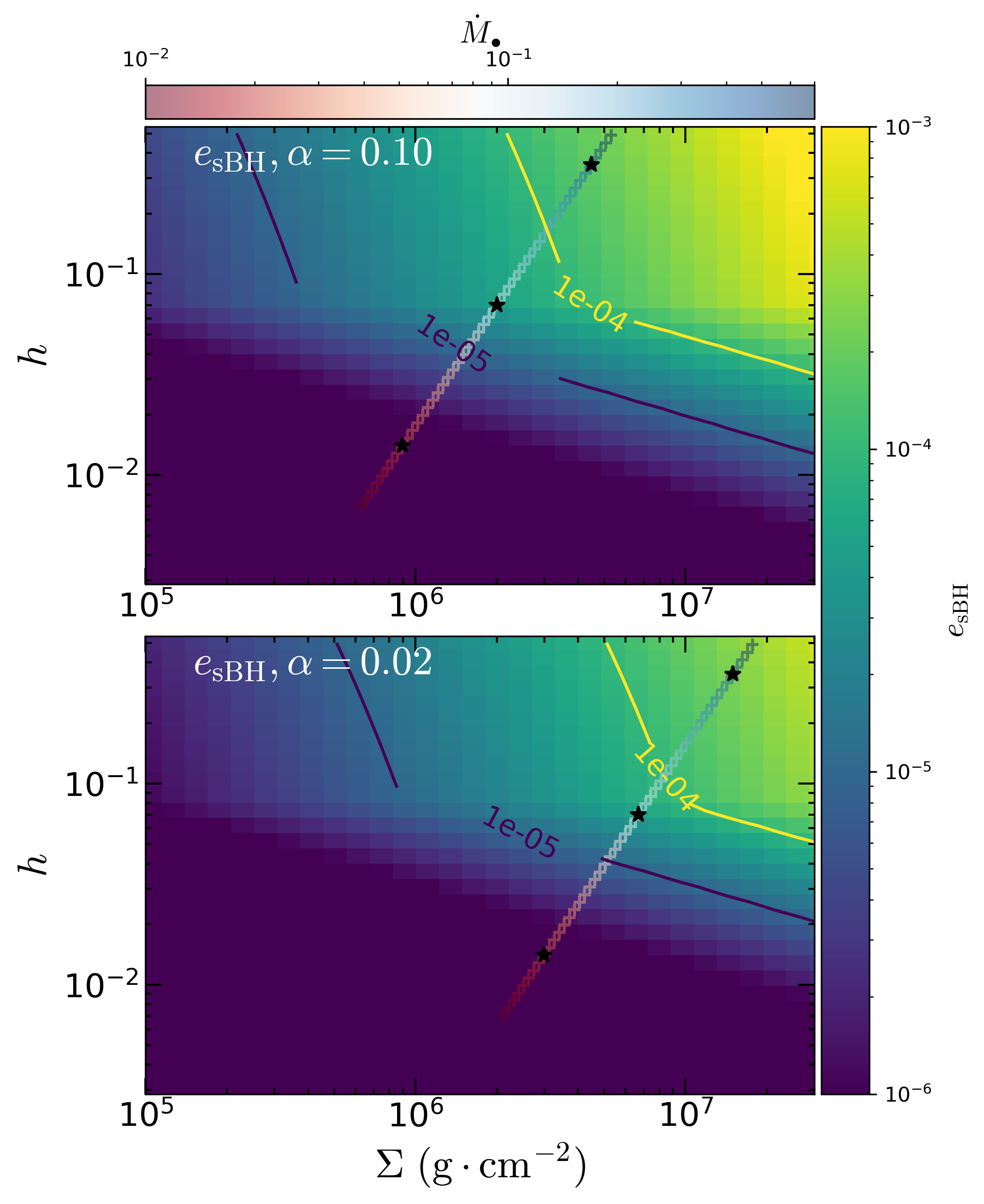}
\caption{The sBH's eccentricity for different disk surface density $\Sigma$ and disk aspect ratio $h$ at $10\ M_{\bullet}$. We have fixed the disk viscosity parameter $\alpha=0.1$ (upper panel) and $\alpha=0.02$ (lower panel) for a $\beta$-disk radial profile. The other parameter are $M_{\bullet}=4\times10^{6}M_{\odot}$, and $q=5\times10^{-6}$. The two lines indicate the detection threshold of EMRI eccentricity $\sim10^{-5}-10^{-4}$. The color line with stars indicates the $\beta$-disk model with $\alpha=0.1$ (upper panel), $\alpha=0.02$ (lower panel) and different $\dot{M}_{\bullet}$. The corresponding value of $\dot{M}_{\bullet}$ is indicated as the colormap on the top. The three stars show the locations of the three disk models with $\dot{M}_{\bullet}=0.02,0.1,0.5\dot{M}_{\rm Edd}$, respectively. }
 \label{fig:ecc_sca}
\end{figure}

To further explore the parameter regime in which the EMRI eccentricity is detectable by future observations, we show the EMRI eccentricity as a function of disk surface density $\Sigma$ and disk aspect ratio $h$ at $10\ M_{\bullet}$ in Fig.~\ref{fig:ecc_sca}. Here we have adopted the typical disk viscosity of $\alpha=0.1$ (upper panel) and $\alpha=0.02$ (lower panel). The radial profile of a $\beta$-disk model is chosen to determine the radius at which $\gamma_{\rm e, gw}\simeq \gamma_{e}$. The eccentricity of the EMRI increases with both the density of the disk surface and the aspect ratio of the disk. For a disk aspect ratio of $h\simeq0.1$ in the inner region of the AGN disk, the eccentricity of the EMRI becomes detectable when the density of the surface of the disk exceeds $\Sigma\gtrsim 5\times10^{5}\ {\rm g\ cm^{-2}}$. Such a high surface density at $10\ M_{\bullet}$ is more likely to occur in a $\beta$-disk model than in a $\alpha$-disk model. A higher disk viscosity $\alpha$ (or stronger turbulence) leads to a greater eccentricity of the EMRI at given $\Sigma$ and $h$. This results from the positive scaling of the equilibrium eccentricity with turbulence strength, as shown in Eq.~(\ref{eq:dr_sca1}-\ref{eq:dr_sca2}). However, it should be noted that for a given $\dot{M}_{\bullet}$, the surface density of a $\beta$-disk decreases with increasing $\alpha$ (see Fig.~\ref{fig:disk_model}), leading to similar EMRI eccentricities in $\beta$-disk models with different viscosities $\alpha$. 

In Fig.~\ref{fig:ecc_sca}, we further map the locations of different $\beta$-disk models in the eccentricity contour. The trajectories for $\alpha=0.1$ (upper panel) and $\alpha=0.02$ (lower panel), corresponding to different values of $\dot{M}_{\bullet}$ in the $\beta$-disk model, are shown as colored lines. The three stars embedded in each line indicate the models with $\dot{M}_{\bullet}=0.02,0.1,0.5\dot{M}_{\rm Edd}$, respectively. As expected, a higher $\dot{M}_{\bullet}$ leads to a higher EMRI eccentricity, and the dependence of the eccentricity on the disk viscosity $\alpha$ at a fixed $\dot{M}_{\bullet}$ is weaker. 
These results suggest that precise measurements of EMRI eccentricities may allow us to constrain the AGN disk parameters and the underlying accretion model. 

We note that \citet{Copparoni:2025jhq} studied the evolution of EMRI using stochastic disk torques based on hydrodynamic simulations of \citep{Derdzinski2021}. However, these hydrodynamic simulations do not implement any MHD turbulence, which is different from the turbulent disk model considered in our work. The stochastic disk forces therein likely originate from the disk fluctuations induced by the perturber and may not share the same properties as the stochastic torques and forces arising from the intrinsic turbulent AGN disks.


\begin{figure}
\centering
\includegraphics[width=0.5\textwidth,clip=true]{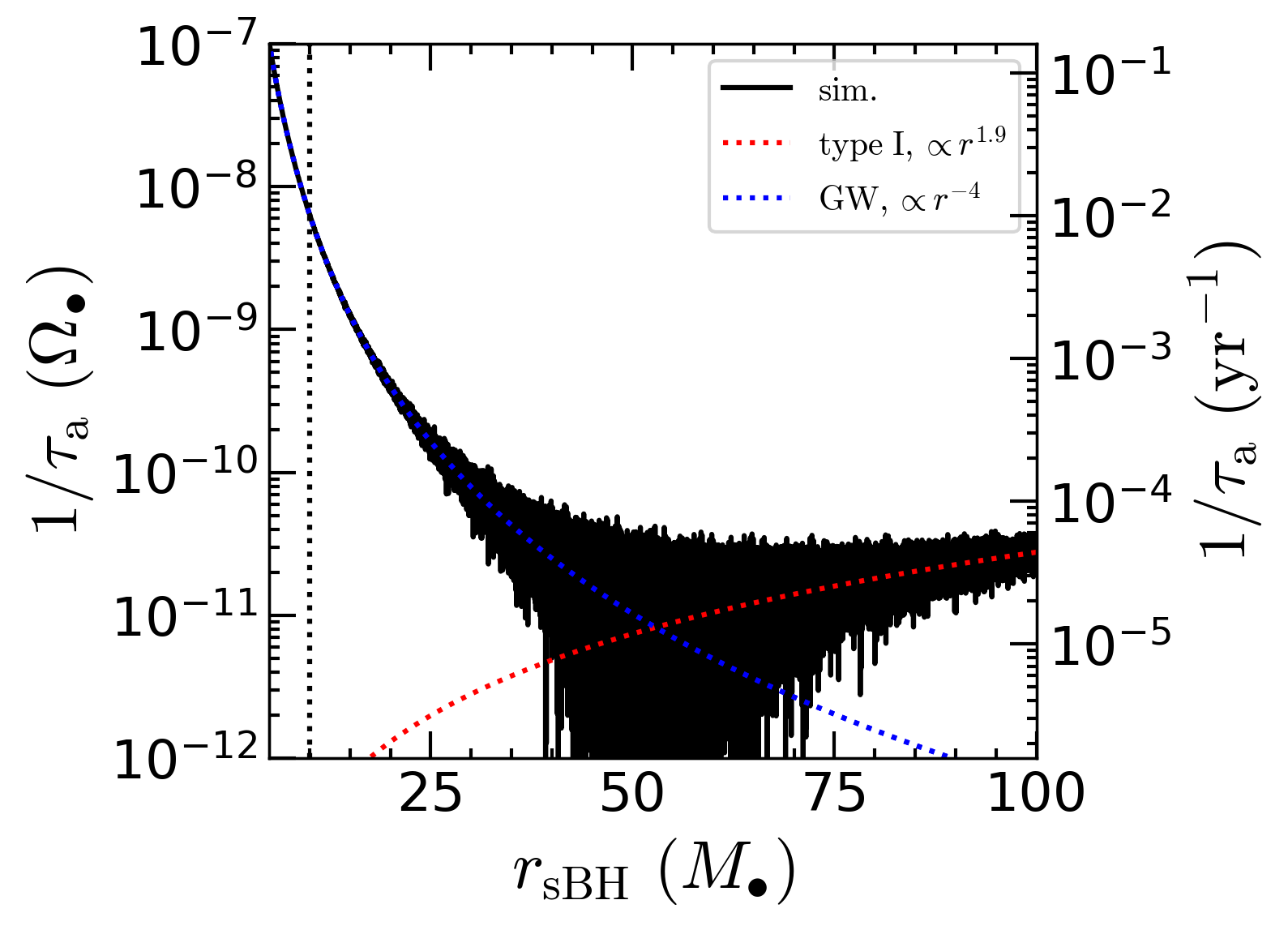}
\caption{The semi-major evolution of the sBH. The black line shows the simulation results for one representative realization in Fig.~\ref{fig:ecc_nbody}. The red and blue dotted lines represent the type I migration rate using the migration torque from the hydrodynamical simulations and the orbital decay due to the gravitational wave radiation, respectively.}
 \label{fig:adot_nbody}
\end{figure}

\begin{figure}
\centering
\includegraphics[width=0.5\textwidth,clip=true]{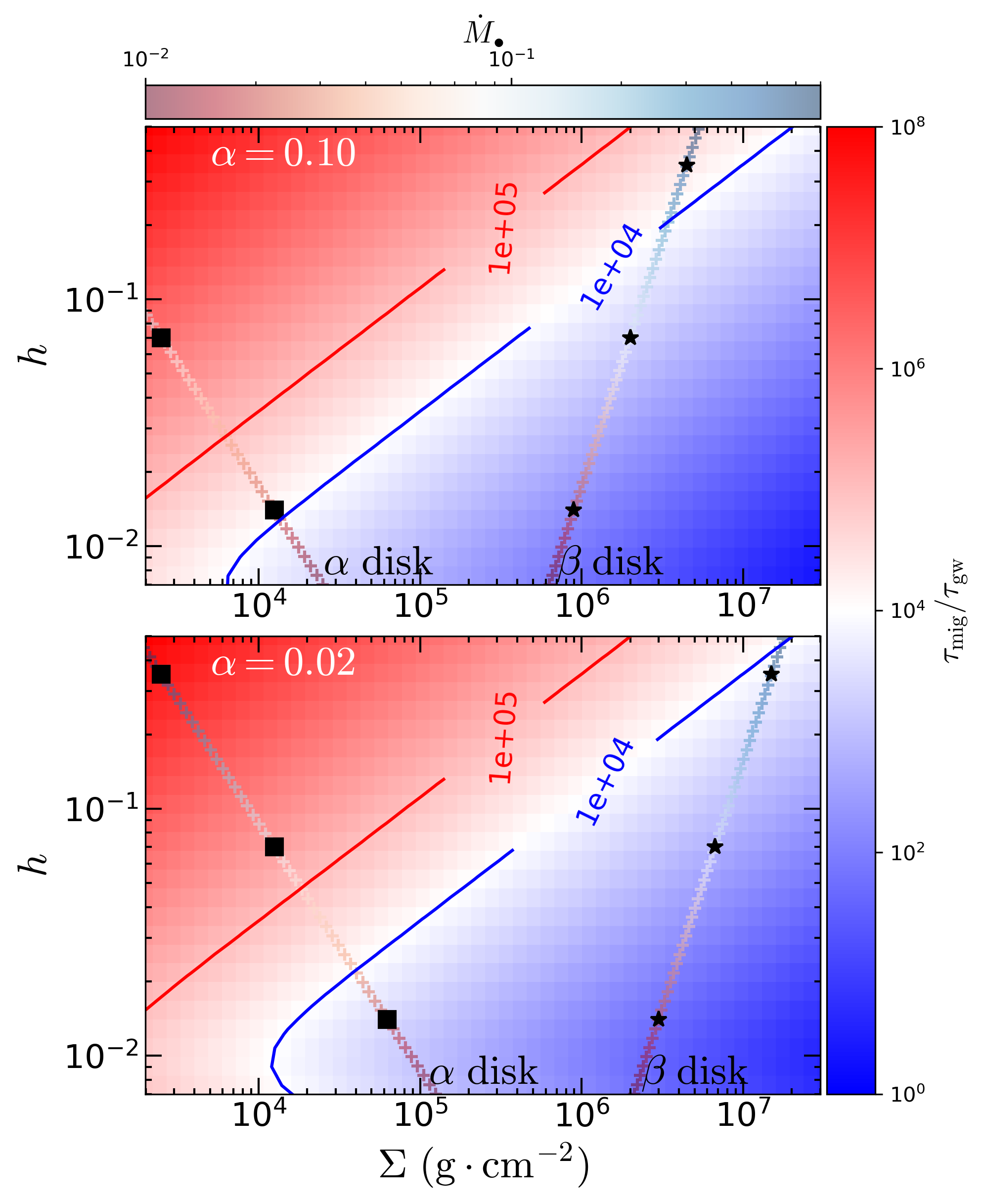}
\caption{The ratio between the disk-driven migration timescale $\tau_{\rm mig}$ and the inspiral timescale due to the gravitational wave radiation $\tau_{\rm gw}$ for different disk surface $\Sigma$ and disk aspect ratio $h$ at $10\ M_{\bullet}$. We have fixed the disk viscosity parameter $\alpha=0.1$ (upper panel) and $\alpha=0.02$ (lower panel). The other parameter are $M_{\bullet}=4\times10^{6}M_{\odot}$, and $q=5\times10^{-6}$. The two lines indicate the detection threshold of EMRI waveform modification due to the disk migration effect if $\tau_{\rm mig}$ is $\simeq 10^{4}-10^{5}$ of $\tau_{\rm gw}$. A smaller ratio (bluer color) suggests a stronger impact on the waveform. The color lines with star symbols indicates the $\alpha$ (the line on the left) and $\beta$ (the line on the right) disk models with $\alpha=0.1$ and different $\dot{M}_{\bullet}$. The corresponding value of $\dot{M}_{\bullet}$ is indicated as the colormap on the top. The three stars show the locations of the three disk models with $\dot{M}_{\bullet}=0.02,0.1,0.5\dot{M}_{\rm Edd}$, respectively.}
 \label{fig:ratio_sca}
\end{figure}

\paragraph{\textbf{Impact on waveform.}}

As sBH migrates inward in the accretion disk, the migration force generally modifies the long-term orbital evolution and, consequently, the gravitational waveform. Since the leading-order phase term of an EMRI scales as $1/q$, it is expected that a phase modulation of order $\mathcal{O}(1)$ radians is detectable. Therefore, a migration force that is at least a fraction $\mathcal{O}(q)$ of the gravitational radiation reaction should, in principle, be observable. This roughly translates to the requirement that the ratio between the gravitational wave damping timescale and the migration timescale be greater than or equal to $10^{-5}$-$10^{-4}$.

For one typical Monte Carlo realization adopted in Fig.~\ref{fig:ecc_nbody}, we show the semi-major axis evolution rate $1/\tau_{\rm a}$ ($\equiv \dot{r}/r$) of EMRI in Fig.~\ref{fig:adot_nbody}. It is clearly seen that disk-driven migration dominates the evolution of the semi-major axis in the outer region, while gravitational wave radiation takes over in the inner region. 
Theoretically, the disk-driven migration rate of the sBH is 
\begin{eqnarray}
     \frac{\dot{r}}{r} &\simeq& 1.7\times 10^{-11} \Omega_{\bullet}\left(\frac{\Sigma_{0}}{6\times10^{5}\ {\rm g\ cm^{-2}}}\right) \\ \nonumber
     &\times& \left(\frac{h_{0}}{0.03}\right)^{-2}\left(\frac{r}{100M_{\bullet}}\right)^{0.5},
     \label{eq:rdot}
\end{eqnarray}
where $\Omega_{\bullet}$ is the Keplerian frequency at $r=M_{\bullet}$. For a $\beta$ disk with $\dot{M}_{\bullet}=0.5\dot{M}_{\rm Edd}$, where $h_{0}\propto r^{-1}$ and $\Sigma_{0}\propto r^{-0.6}$, $\alpha=0.1$, we then have $\dot{r}/r\simeq3\times10^{-11}\Omega_{\bullet}(r/100M_{\bullet})^{1.9}$, which is
fully consistent with the simulation results.
Gravitational wave radiation ($\propto r^{-4}$) dominates at smaller radii. The transition between the two regimes occurs around $r\simeq 50\ M_{\bullet}$.
Although the ratio of the disk-driven migration timescale $\tau_{\rm mig}$ to the gravitational wave inspiral timescale $\tau_{\rm gw}$ at $r\simeq 10\ M_{\bullet}$ is of the order of $\sim 10^{4}$, such a difference can still leave observable imprints on the EMRI waveform.

To explore the parameter space in which the impact of disk-driven migration on the EMRI waveform becomes detectable, we show the ratio $\tau_{\rm mig}/\tau_{\rm gw}$ in Fig.~\ref{fig:ratio_sca} for different values of disk surface density $\Sigma$ and aspect ratio $h$ at $10\ M_{\bullet}$. According to the type I and type II migration formulas, a lower disk aspect ratio $h$ and a higher surface density $\Sigma$ are expected to result in a smaller $\tau_{\rm mig}/\tau_{\rm gw}$, thus producing a stronger impact on the waveform, as illustrated in Fig.~\ref{fig:ratio_sca}. This timescale ratio is generally independent of disk viscosity $\alpha$, as shown in Fig.~\ref{fig:ratio_sca}, except in the very low regime $h$ where the opening of the gap is likely to occur, which is consistent with the type I/II migration rate.
It should be noted that this mapping is independent of the disk’s radial structure or evolution history, since we are only concerned with the parameters of the disk at $10\ M_{\bullet}$. 

Thus, we can further map the different AGN disk models in the contour $\tau_{\rm mig}/\tau_{\rm gw}$. Similarly to Fig.~\ref{fig:ecc_sca}, we show the trajectories of both the $\alpha$ model and the $\beta$ model with different $\dot{M}_{\bullet}$ in Fig.~\ref{fig:ratio_sca}. These trajectories depend on disk viscosity $\alpha$, and therefore occupy different locations in the contour $\tau_{\rm mig}/\tau_{\rm gw}$ for different values of $\alpha$. Unlike eccentricity detection by EMRI, a lower $\dot{M}_{\bullet}$ leads to a smaller $\tau_{\rm mig}/\tau_{\rm gw}$, and hence a stronger impact on the waveform. The three stars embedded in each colored line (representing the $\alpha$ model on the left and the $\beta$ model on the right) correspond to $\dot{M}_{\bullet}=0.02,\ 0.1,\ 0.5\dot{M}_{\rm Edd}$, respectively. For a disk viscosity of $\alpha=0.02$, the gap opening process could occur only for $\dot{M}_{\bullet}\lesssim0.01\dot{M}_{\rm Edd}$ (that is, $h\lesssim0.007$), in which case the migration proceeds at the type II rate \citep{Lin1986,Duffell2013,Kanagawa2015}. The AGN disk with an even lower $\dot{M}_{\bullet}\lesssim0.01\dot{M}_{\rm Edd}$ would accrete via an advection-dominated accretion flow \citep{Yuan2014}, different from the $\alpha/\beta$ disk considered here. A higher $\alpha$ (e.g. $\alpha=0.1$ in the upper panel) would suppress the formation of the gap and, therefore, results in type I migration across the entire range of model parameters considered here. The transition from type I to type II migration can be seen at the lowest $h$ in the lower panel of Fig.~\ref{fig:ratio_sca}. We find that for almost all values of $\dot{M}_{\bullet}$ explored in the $\beta$-disk model, the disk-driven migration has a strong enough effect to produce an observable waveform impact. For the $\alpha$-disk model, the waveform impact remains observable for $\dot{M}_{\bullet} \sim 0.02\dot{M}_{\rm Edd}$. 

In addition to the migration effect, the stochastic force may also influence the secular evolution of the EMRI phase. A subtle issue arises from the fact that the corresponding phase modulation may contain a stochastic component, which may not be compatible with traditional matched filter analyses that rely on deterministic waveforms. It would be interesting to explore how turbulence-induced randomness can be incorporated into waveform modeling, and whether it can be measured with sufficient precision to constrain the properties of the turbulence. A potentially relevant discussion on the inference of stochastic forces can be found in \cite{Seymour:2024kcd}.

\section{Inclination}\label{sec:inc}

Another important observable characterizing various EMRI formation channels is the inclination angle between the EMRI orbital angular momentum vector and the spin axis of the central MBH. Note that in Kerr spacetime, a geodesic generally does not conserve a three-dimensional angular momentum vector. However, the inclination angle $\iota$ can be defined using the relation $\cos \iota := L_z / \sqrt{L_z^2 + \mathcal{Q}}$, where $L_z$ is the component of the angular momentum along the spin axis, and $\mathcal{Q}$ is the Carter constant. This definition reduces to the conventional one in Newtonian mechanics for wide orbits. Moreover, it is straightforward to show that, during a generic inspiral process of an eccentric, inclined EMRI, $\iota$ generally changes by no more than a few degrees \cite{Hughes2000}.
In other words, it is a rather robust observable that is insensitive to the time of measurement within the LISA frequency band.

The distribution of $\iota$ differ significantly between wet and dry formation channels. As discussed in Sec.~\ref{sec:iotawet}, the accretion to the MBH determines the initial orbital inclination of the wet EMRI, and both the Bardeen-Peterson timescale and the lifetime of the disk determine the final distribution. For dry EMRIs, it turns out that the critical factor for the distribution of $\iota$ arises from the critical angular momentum of the last stable orbit. 
This generally leads to an asymmetric distribution, as detailed in Sec.~\ref{sec:iotadry}.

\subsection{Wet EMRI}\label{sec:iotawet}

Two major scenarios are commonly discussed to describe the accretion episodes of MBHs: coherent accretion and chaotic accretion. In coherent accretion, the orbital plane of the incoming gas remains roughly constant across episodes, so the MBH spin tends to align with the angular momentum of the disk. It is also natural to expect that the MBH spin becomes near-extremal after several e-folds of mass growth through coherent accretion.
One subtlety is that MBH mergers, following galaxy mergers, can significantly alter the mass and spin of the remnant black hole. However, since such mergers are infrequent and the typical time between them is much longer than the Bardeen-Petterson (BP) alignment timescale, we do not consider this effect.
In the chaotic scenario, the direction of accretion in each episode is assumed to be random. As a result, the MBH spin is likely misaligned with the normal of the accretion disk at the beginning of each episode, with the inclination gradually evolving due to the BP effect.

As a wet EMRI is generally embedded in the accretion disk during its migration, its initial orbital plane aligns with the disk plane. Therefore, in the coherent accretion scenario, we expect the inclination distribution $P(\cos \iota)$ to peak around $\cos \iota=1$, or $P(\cos \iota) \approx \delta(\cos\iota-1)$.

In the case of chaotic accretion, the angular momentum of the accretion disk is initially misaligned with the spin of the central MBH. This misalignment is gradually reduced due to Lense-Thirring (LT) precession \cite{BardeenPetterson1975}, which arises from the frame-dragging effect in Kerr spacetime. The inner parts of the disk are forced to align with the spin of the MBH, with the characteristic warp radius given by \cite{Natara1998_Alignment, Lodato:2012yr, Lyu2024WetEMRIs}
\begin{align}\label{eq:warp}
R_{\rm warp} \approx 12M_{\bullet} \left(\frac{\alpha}{0.1}\right)^{2/3} \chi^{2/3} \left(\frac{h}{0.1}\right)^{-4/3} \ ,
\end{align}
where $h$ is the aspect ratio of the disk and $\chi$ is the dimensionless spin of the central MBH. 
The warpped disk will exert a torque onto the center MBH and align or anti-align the BH spin with the angular momentum of the outer disk.
The characteristic alignment or anti-alignment timescale over which this process occurs is known as the Bardeen-Petterson timescale $\tau_\text{BP}$\cite{Natara1998_Alignment}:
\begin{align}\label{eq:taubp}
    \tau_\text{BP} \approx\, &5.6\times10^5 \chi^{\frac{11}{16}}\left(\frac{\alpha}{0.03}\right)^{\frac{13}{8}}\left(\frac{L}{0.1L_{\rm Edd}}\right)^{-\frac{7}{8}} \notag \\
    &\times \left(\frac{M_{\bullet}}{10^8M_\odot}\right)^{-\frac{1}{16}}\left(\frac{\epsilon}{0.3}\right)^{\frac{7}{8}}\,,
\end{align}
which depends on the MBH mass $M_{\bullet}$, the dimensionless spin parameter $\chi$,
the viscosity parameter $\alpha$, the MBH luminosity $L$ ($L_{\rm Edd}$ is the Eddington luminosity), 
and the radiative efficiency $\epsilon$. 

Wet EMRIs embedded within the disk remain in the disk during their migration at large radii. The LT torque exerted on the EMRI tends to drive its orbit out of the disk plane, whereas the disk torque acts to realign off-plane orbits back into the plane. The LT torque dominates for $r \le R_{\rm dec}$, where the decoupling radius $R_{\rm dec}$ -which is smaller than the radius where the LT torque overtakes the disk density wave torque - is given by \cite{Lyu2024WetEMRIs}: 
\begin{align}
    R_{\rm dec} = 430 M_{\bullet} \ \chi^{1/8} \left (\frac{\alpha}{0.1}\right )^{1/8} \left (\frac{\dot{M}_{\bullet}}{0.1 \dot{M}_{\rm Edd}}\right )^{1/2}\left (\frac{m}{30 M_\odot}\right )^{-1/8} \ 
\end{align}
According to the analysis in \cite{Hughes2000}, even when a wet EMRI evolves within the decoupling radius, the inclination angle relative to the MBH spin changes by no more than a few degrees. Therefore, the EMRI inclination still closely traces the disk inclination at larger radii.

Assuming that the MBH is completely uncorrelated with the accretion disk at the beginning of each accretion episode, the distribution of the initial inclination angle $\iota_0 \in (0,\pi)$ should be
\begin{align}\label{eq:initial}
    P(\cos\iota_0)=\frac{1}{2}\,.
\end{align}

The evolution of the inclination angle between the center MBH spin and the orbital angular momentum of the outer disk depends on the ratio of the angular momentum $J_{\rm d}/J_\bullet$ \cite{King2005_Alignment}, 
where $J_{\rm d}$ and $J_\bullet$ are the angular momentum of the disk and the angular momentum of the spin of MBH, respectively. If $\cos\iota_0>-J_{\rm d}/2J_\bullet$, the BH will eventually align with the outer disk, and if $\cos\iota_0<-J_{\rm d}/2J_\bullet$, 
the BH will finally anti-align with the outer disk.

In the $\alpha$-disk model, the angular momentum of the outer disk is significantly greater than that of  the MBH, expressed as \cite{Dotti2013BHspinOrientation}
\begin{align}
    \frac{J_{\rm d}}{J_\bullet}\sim&7.3\left(\frac{\alpha}{0.1}\right)^{13/45}\left(\frac{\dot{M}_\bullet}{\dot{M}_{\rm Edd{}}}\times\frac{0.1}{\epsilon}\right)^{-7/45} \notag \\ &\times\left(\frac{M_{\bullet}}{10^6M_\odot}\right)^{-37/45}\chi^{-1}
\end{align}
in which $\epsilon$ is the radiative efficiency, $\dot{M}_\bullet$ is the MBH accretion rate. 

For the typical MBH in this paper ( $M_{\bullet}=10^5-10^7M_\odot$), $J_{\rm d}\gtrsim2J_\bullet$, which means that the BH will finally align with the outer disk for any initial inclination angle $\iota_0$.
The time-dependent evolution of the inclination angle can be described as \cite{King2005_Alignment}

\begin{align}
    \frac{d}{dt}(\cos\iota)=\frac{\sin^2\iota}{\tau}\left(\frac{J_{\rm d}}{J_\bullet}+\cos\iota\right)
\end{align}
where the $\tau$ is a coefficient and should have a relation with the alignment timescale $\tau_{\rm BP}$. To counteract the influence of the angular momentum of the disk and comparing the small angle case, we use $\tau=(1+J_{\rm d}/J_\bullet)\tau_{\rm BP}$. 

The solution of above equation for $J_{\rm d}/J_\bullet\gtrsim2$ can be integrated directly to an implicit form
\begin{align}\label{eq:iotat}
    &-\frac{(J_{\rm d}/J_\bullet-1)\log(1-\cos\iota)-(J_{\rm d}/J_\bullet+1)\log(1+\cos\iota)}{2((J_{\rm d}/J_\bullet)^2-1)} \notag \\
    &-\frac{\log(J_{\rm d}/J_\bullet+\cos\iota)}{(J_{\rm d}/J_\bullet)^2-1}=\frac{t}{\tau} + C
\end{align}
where $C$ is the integral constant and can be obtained by setting $t=0$.

Starting from the initial distribution of the inclination angle described by Eq.~\ref{eq:initial}, the time-dependent evolution of the inclination distribution of the disk can be obtained from
\begin{align}
    P_{\rm disk}(\cos\iota,t) = P(\cos\iota_0)\frac{d\cos\iota_0}{d\cos\iota}
\end{align}
which can be solved numerically.

In order to compute the distribution of $\iota$ of wet EMRIs, we also need to account for the time-dependent formation rate of wet EMRIs. According to the analysis in \cite{Pan202112}, wet-EMRI formation rate $\Gamma_{\rm M_{\bullet}, EMRI}(t)$ typically peaks at the beginning of AGN episodes, and gradually declines due to the insufficient supplies of sBHs from the outer boundary of the nuclear star cluster.
Assuming the disk lifetime to be $T_{\rm disk}$, the distribution of $\iota$ for wet EMRIs is then given by

\begin{align}\label{eq:piotam}
    \langle P(\cos\iota)\rangle_{\rm M_{\bullet}}
    =N_{\rm T}^{-1}\int_0^{T_{\rm disk}} \Gamma_{\rm M_{\bullet}, EMRI}(t) P_{\rm disk}(\cos\iota,t)dt
\end{align}
where the total number of wet EMRIs formed within the lifetime is 
\begin{align}
N_{\rm T} =\int_0^{T_{\rm disk}} \Gamma_{\rm M_{\bullet}, EMRI}(t)dt\,.
\end{align}
Here, $\Gamma_{\rm M_\bullet, EMRI}$ denotes the EMRI formation rate as well, with the subscript ${\rm M_\bullet}$ indicating its dependence on the MBH mass.

Notice that the time-dependent wet-EMRI formation rate based on the above FP calculations explicitly depends on the mass of the MBH, as does the associated $\iota$ distribution in Eq.~(\ref{eq:piotam}). 

\begin{figure}
    \centering
    \includegraphics[width=1.0\linewidth]{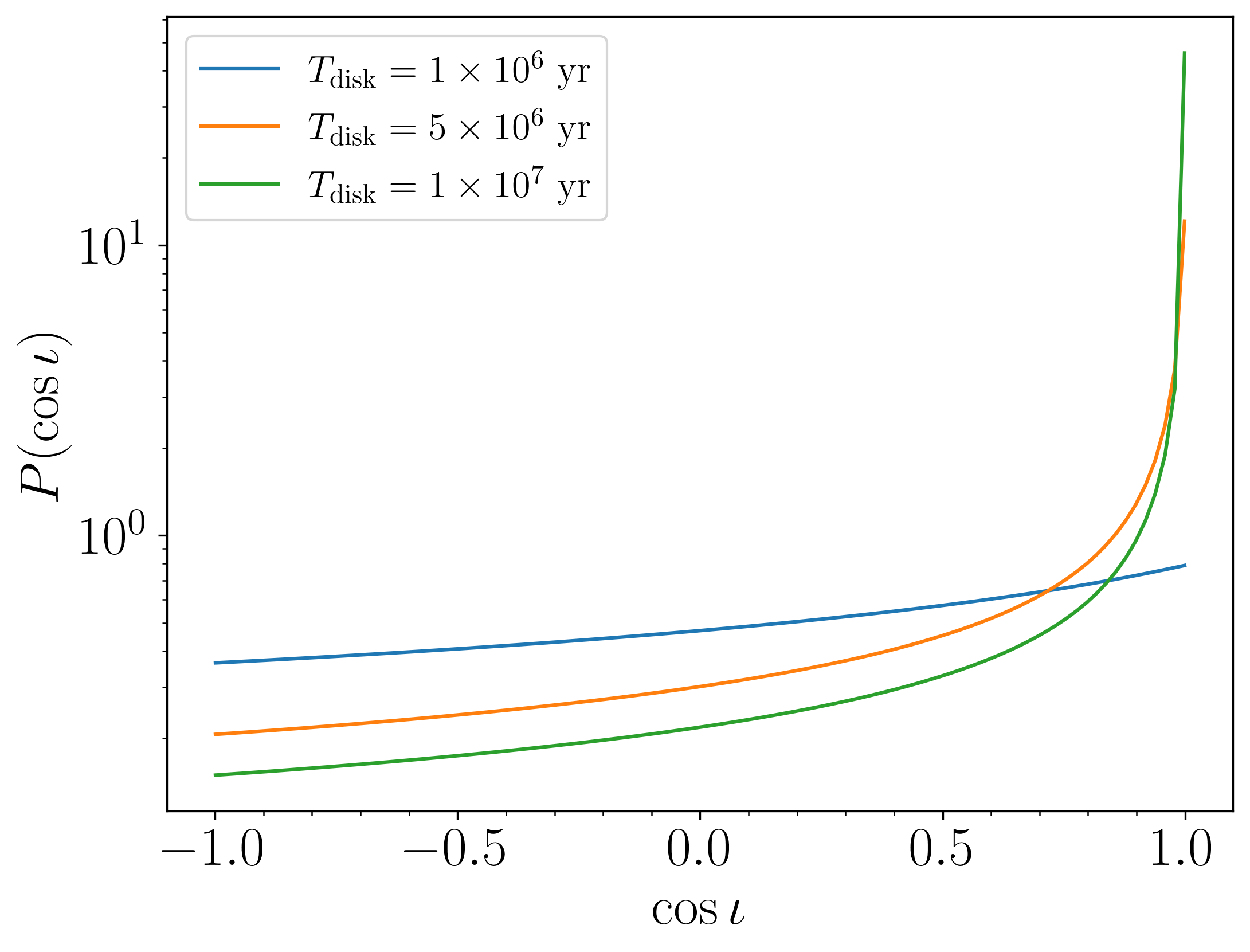}
    \caption{Inclination angle distribution for $M_{\bullet}=1\times10^6M_\odot$ MBH in the context of chotic accretion. We consider three disk lifetime for $T_{\rm disk} = 10^6, 5\times10^6, 10^7 {\rm yr}$. A longer disk lifetime causes the inclination angle distribution to become more concentrated toward $\iota=0$.}
    \label{fig:inc_1e6}
\end{figure}

A sample distribution is illustrated in Fig.~\ref{fig:inc_1e6}, in which case the MBH mass is assumed to be $10^6 M_\odot$, the dimensionless spin is $a=0.8$, and the disk properties are set to $\alpha=0.1, L=0.1L_\text{Edd}, \epsilon=0.1$. We consider three different disk lifetimes: $T_\text{disk}=1\times10^6, \ 5\times10^6, \ 1\times10^7 \text{yr}$. Notice that according to Eq.~(\ref{eq:taubp}), the BP timescale is approximately $1.7\times10^6\text{yr}$. 

For inclined disks, the BP mechanism is generally in operation, and the final state depends on both the disk lifetime and the BP timescale.
In Fig.~\ref{fig:inc_1e6}, we observe that when the disk lifetime is shorter than the BP timescale ($T_\text{disk} < \tau_\text{BP}$), the distribution of $\cos\theta$ remains approximately flat, indicating little alignment. However, when the disk lifetime exceeds the BP timescale ($T_\text{disk} > \tau_\text{BP}$), the inclination angle distribution becomes tilted, with peak corresponding to alignment ($\iota \approx 0$) with the MBH spin. As the disk lifetime increases, the tilted distribution becomes sharper. Therefore, the inclination distribution of wet EMRIs encodes important information about $\tau_{\rm BP}/T_{\rm disk}$.

\begin{figure*}[t]
\centering
\vspace{0.6cm}
\includegraphics[width=0.45\linewidth]{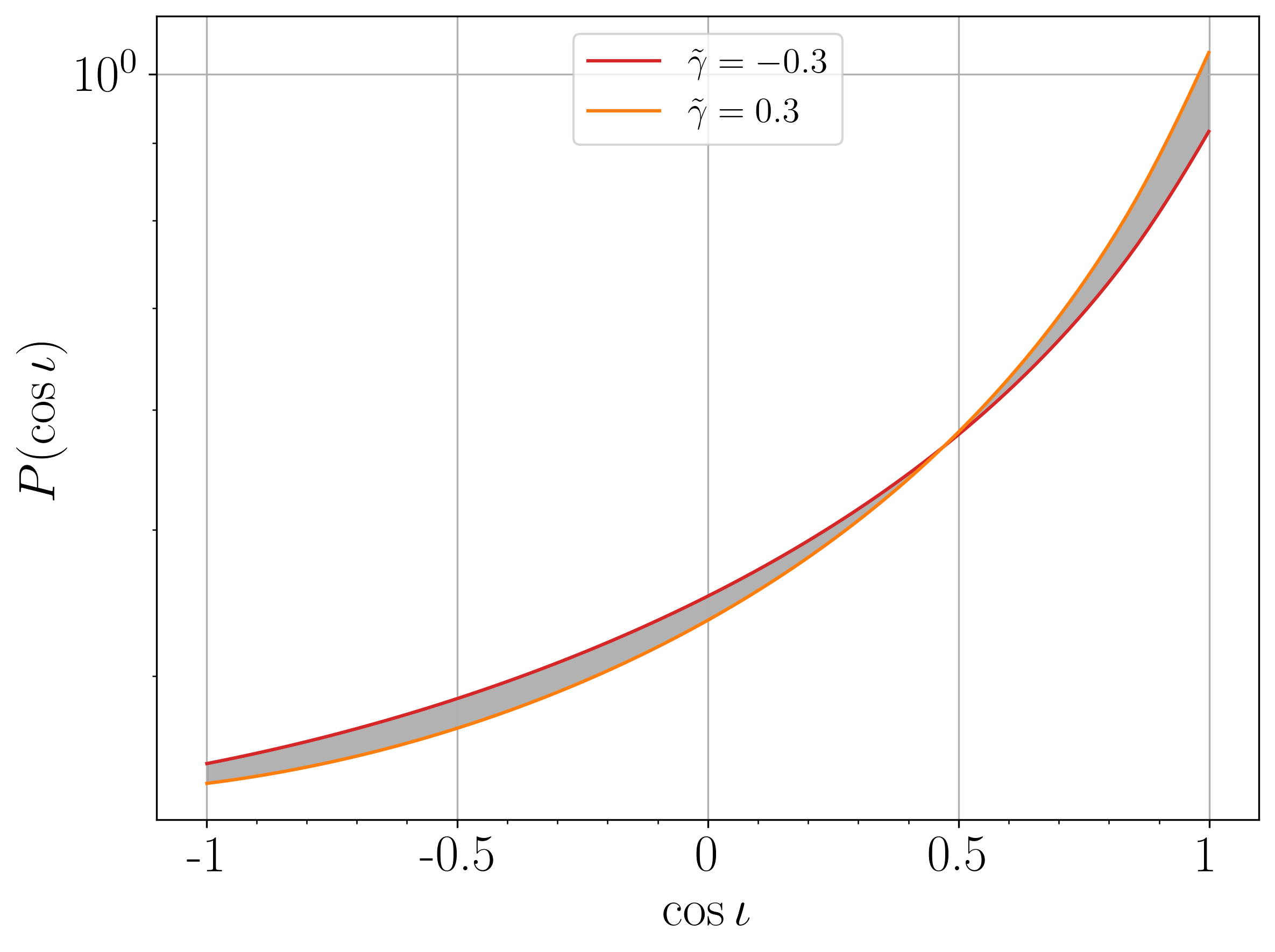}
 \includegraphics[width=0.45\linewidth]{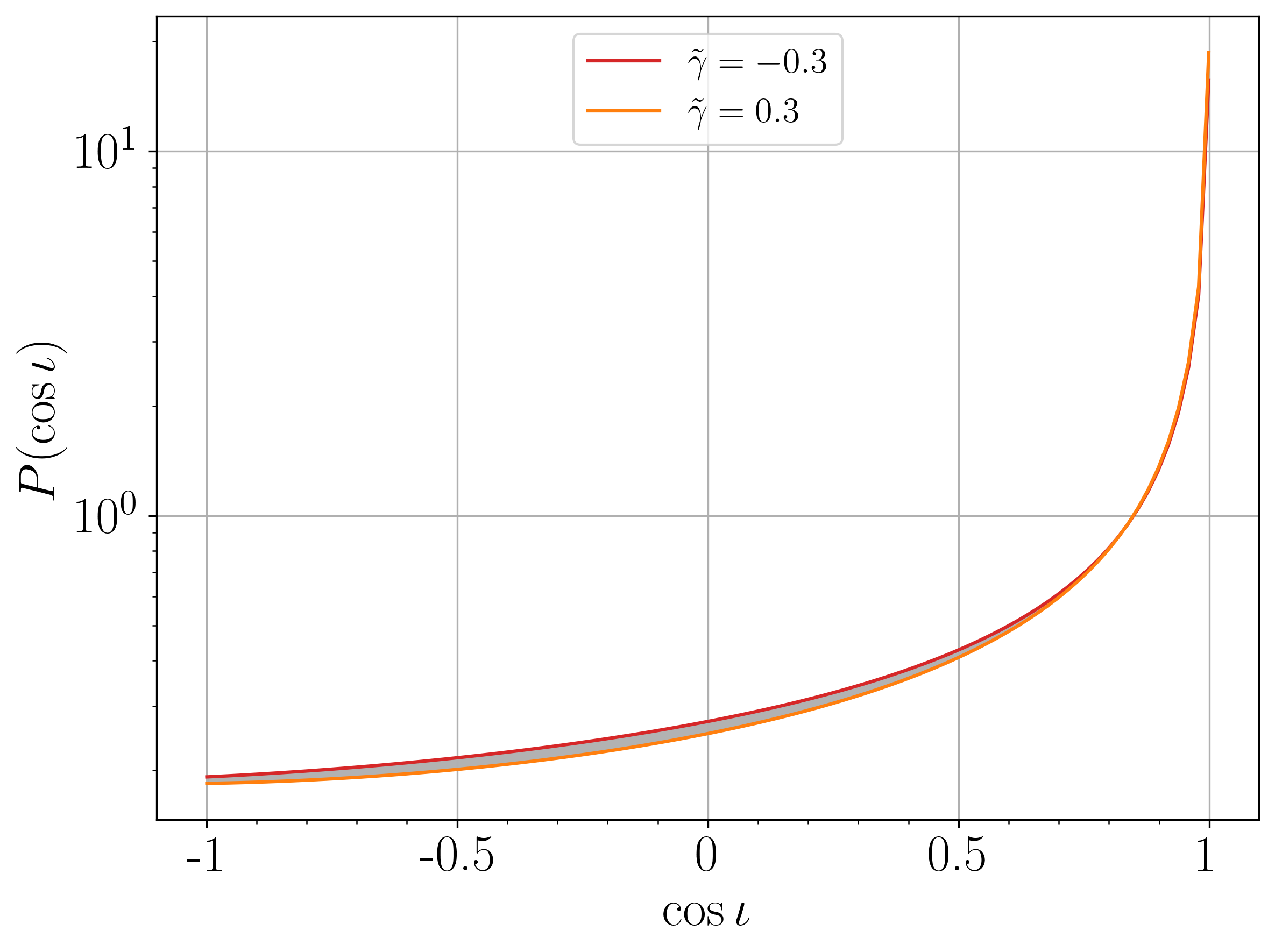}
\vspace{0.3cm}
\caption{Averaged inclination angle distribution functions of wet EMRIs for different MBH mass function models. Results are shown for a disk lifetime of $10^6\, \mathrm{yr}$ (left panel) and $5 \times 10^6\, \mathrm{yr}$ (right panel). When the disk lifetime is sufficiently long, the inclination angle distribution becomes pronounced on $\iota=0$.}
\label{fig:inc_wet}
\end{figure*}

For actual observations, we need to average over the mass function of MBHs to obtain the total inclination distribution of wet EMRIs. Similar to the treatment in Sec.~\ref{sec:drye}, we vary the power-law index of the MBH mass function from $\tilde\gamma=-0.3$ to $\tilde\gamma=0.3$, 
i.e.
\begin{align}
\langle P(\cos\iota)\rangle =N^{-1}\int dM_{\bullet} \frac{d n}{d M_{\bullet}} \langle P(\cos\iota)\rangle_{\rm M_{\bullet}} N_T
\end{align}
where the normalization constant $N$ is 
\begin{align}
    N:=\int dM_{\bullet} \frac{dn}{dM_{\bullet}} N_T\,.
\end{align}
The range of mass integration is from $10^5 M_\odot$ to $10^7 M_\odot$, which is relevant for space-borne gravitational wave observation.

The resulting distribution function is shown in Fig.~\ref{fig:inc_wet} for two disk lifetimes: $T_\text{disk}=10^6 \ \text{yr}$ (left panel) and $T_\text{disk}=5\times10^6 \ \text{yr}$ (right panel).
In the case of a shorter disk lifetime, the deviation in the mass function significantly affects the inclination angle distribution. However, for a longer disk lifetime, the deviation in the mass function only affects the region where $\cos\iota \lesssim 0.5$, due to the alignment effect. 

When the lifetime of the disk is several times longer than the BP timescale, the alignment process is almost complete, and the wet EMRIs are more aligned with the MBH spin, as illustrated in the right panel. 

We should note that the angle of inclination distribution of wet EMRIs in the case of chaotic accretion could be affected by the relationship between $J_{\rm d}$ and $J_\bullet$. If $J_{\rm d}\ll J_\bullet$, the critical angle that distinguishes alignment from anti-alignment of the MBH spin and the disk is $\arccos(0)=\pi/2$, which means that initially prograde cases will align with the MBH spin, while initially retrograde cases will anti-align with it. Consequently, in the chaotic accretion scenario, we may observe a bimodal inclination-angle distribution of wet EMRIs. Therefore, the distribution of the angle of inclination of wet EMRIs can not only reveal the BP effect in AGN disks but also provide information on the relationship between $J_{\rm d}$ and $J_\bullet$.

\subsection{Dry EMRI}\label{sec:iotadry}

Assuming a spherically symmetric nuclear star cluster, the formation rate of dry EMRIs still depends on their inclination with respect to the spin of the MBH \cite{AmaroSeoane2013EMRISpin}. In general, for aligned orbits the angular momentum of the Last Stable Orbits (LSOs) is lower than that of misaligned orbits. 
Fig.~\ref{fig:boundary} shows the loss cone boundary with $J_{\rm lc}=4M_{\bullet}$, but orbits with smaller inclination angles should have a lower loss cone boundary (in the $R$ direction) due to the lower LSO angular momentum.
This implies that the critical energy $E_\text{cr}$ is lower, with a correspondingly larger integrated flux at the boundary $t_{\rm GW} = t_{ J, \rm sca}$. Therefore, higher EMRI rates are expected for prograde orbits than for retrograde orbits.

\begin{figure}[t]
    \centering
    \includegraphics[width=1.0\linewidth]{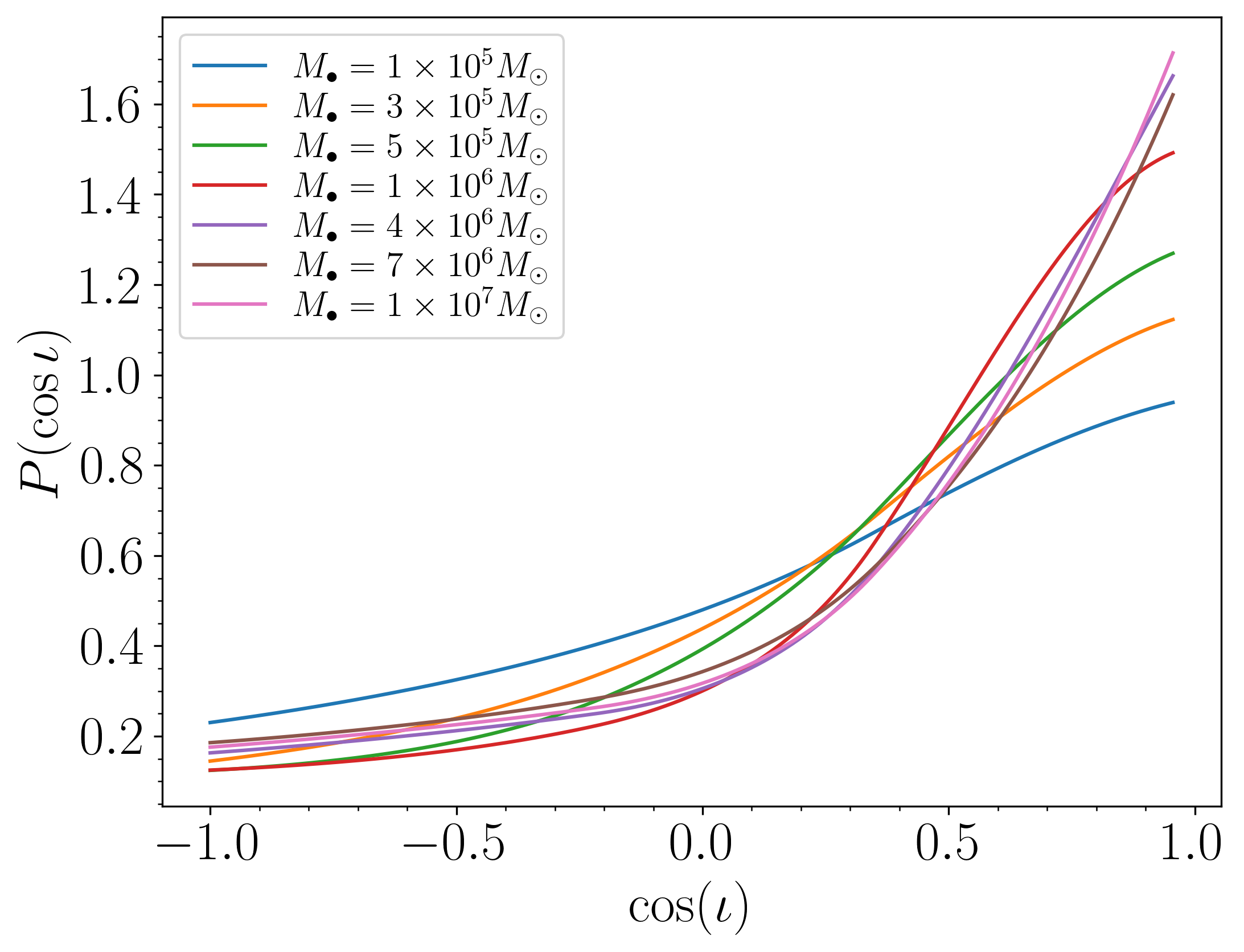}
    \caption{Inclination angle distribution function of dry EMRIs, obtained from Fokker Planck equation for different MBHs with spin parameters $\chi=0.8$. In all cases, the distribution is peaked towards prograde orbits.}
    \label{fig:R_theta}
\end{figure}

The LSO radius and angular momentum for Kerr geodesics can be obtained following the method in \cite{AmaroSeoane2013EMRISpin}, as a function of the black hole spin $a$ and the inclination angle $\iota$. They generally also depend on orbital eccentricity, but in practice, since the EMRIs formed in their early stages are highly eccentric, we essentially adopt the LSO values in the $e=1$ limit.
In Fig.~\ref{fig:R_theta} we present a series of FP calculations following similar procedures in Sec.~\ref{sec:drye}, by choosing $\chi=0.8$ and the $\iota$-dependent LSO angular momentum $J_{\rm lc}(\cos \iota)$. 
We evolve all systems to 5 Gyrs.
For more massive MBHs, the asymmetry of the ``tilted'' distributions becomes significantly larger than for lower-mass MBHs, which is because different MBH systems reach the equilibrium state at different times. During the evolution stage, the sBH component will gradually flow to the inner region from the outer region, reducing the differences in EMRI flux at the loss-cone boundary over time. 
In lower-mass MBH systems, equilibrium is reached relatively quickly, after which the systems evolve in a quasi-equilibrium state. Consequently, the EMRI flux at the loss-cone boundary shows smaller differences compared to a non-equilibrium state. In contrast, higher-mass MBH systems take much longer to reach equilibrium (e.g. $\mathcal{O}(1{\rm Gyrs})$ for heavier MBH and $\mathcal{O}(0.1{\rm Gyrs})$ for lighter MBH). 
As a result, their EMRI flux at the loss-cone boundary exhibits larger variations across different energies, directly affecting the EMRI rate. In fact, orbits with different inclination angles have different critical energies $E_{\rm cr}$, as discussed in the previous section. Therefore, the combined effect of evolution time and inclination angle results in varying degrees of tilt asymmetry in the inclination angle distribution for different MBH systems.

\begin{figure}
    \centering
    \includegraphics[width=1.0\linewidth]{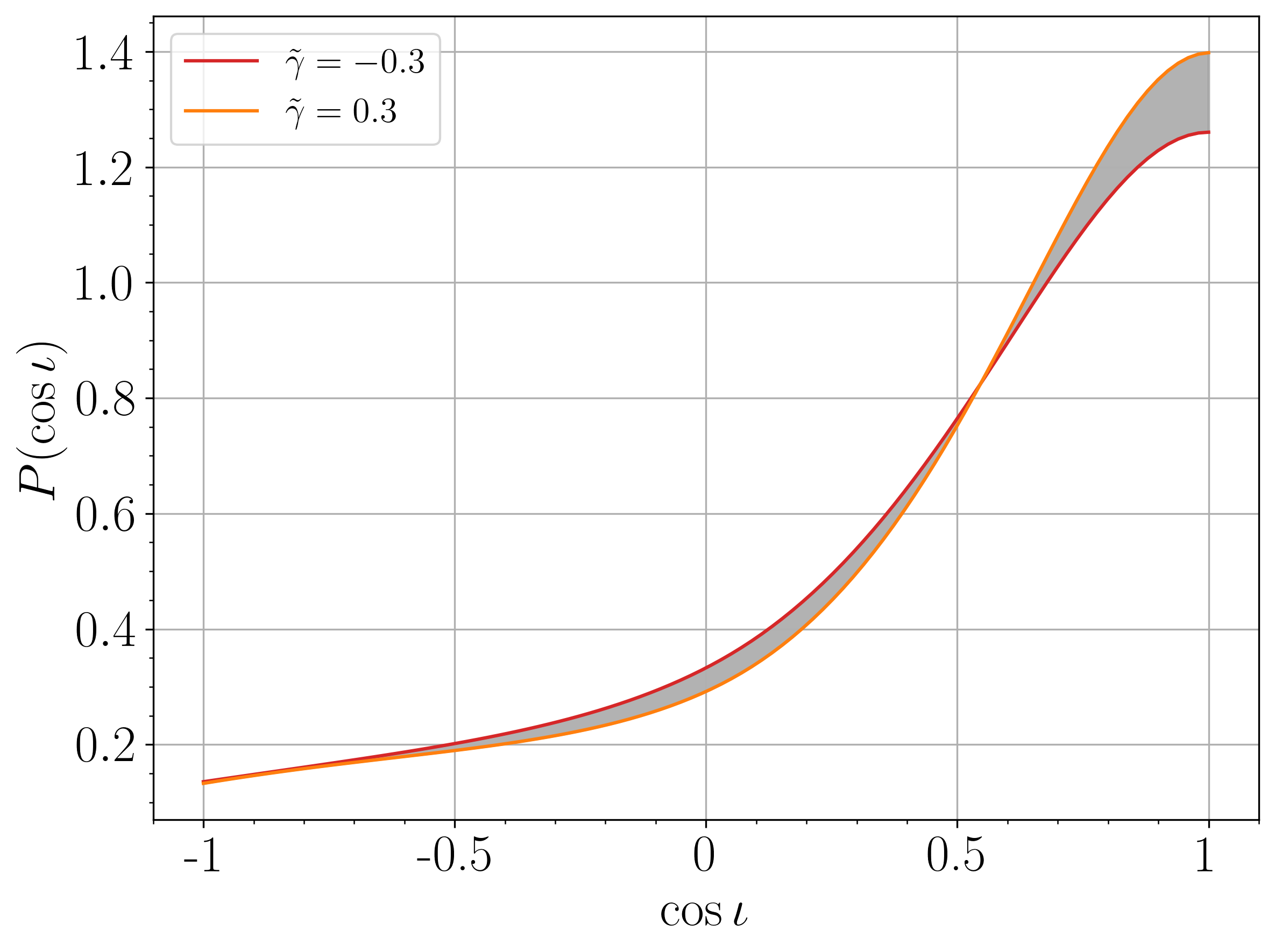}
    \caption{Averaged inclination angle distribution functions of dry EMRIs for different MBH mass function models. }
    \label{fig:inc_dry}
\end{figure}

Similarly to the discussion in Sect.~\ref{sec:iotawet}, we also consider the average distribution of $\iota$ for dry EMRIs, with the power-law index of the MBH mass function ranging from $\tilde\gamma = -0.3$ to $\tilde\gamma = 0.3$. The 
result is shown in Fig.~\ref{fig:inc_dry}. We find a stronger  dependence of $\gamma$ at angles around $\iota \approx \pi/2$ and $\iota \approx 0$, although the asymmetry dependence on the $\cos \iota$ is still significant.

In reality, the spins of MBHs may vary between different systems, but EMRI observations should be able to measure them with great precision, ie $\mathcal{O}(10^{-5})$ \cite{Babak2017_LISA_EMRI}. The above analysis can be modified to account for the spin distribution in order to predict the $\iota$ dependence of dry EMRIs, which can then be compared with the observed population of events from LISA, Taiji, and TianQin \cite{Babak2017_LISA_EMRI, Fan2020TianQinEMRI}.
A possible source of deviation, even when considering uncertainties in $\tilde \gamma$, may arise from intrinsic asymmetry in the nuclear star cluster. For example, if most clusters tend to have stellar disks aligned with the spin of the MBH, the relative ratio between prograde and retrograde EMRIs would be further enhanced.
In this sense, the prediction shown in Fig.~\ref{fig:inc_dry}, along with its spin-modified variants, can serve as baseline benchmarks to investigate the asymmetries of nuclear clusters.

From the analysis of the inclination angle distributions of dry and wet EMRIs, we find that both types exhibit tilted distributions due to different physical origins. For short disk lifetimes, the tilt amplitudes are comparable; however, for long disk lifetime, the asymmetry in wet EMRIs becomes significantly larger than in dry EMRIs.

\section{sBH Mass}\label{sec:lvk}

The population study of BBHs observed by the Advanced LIGO-Virgo-KAGRA (LVK) Collaboration has been one of the central topics in gravitational wave astronomy. The mass function of black holes within these binaries encodes critical information about the underlying formation channels and the physical properties of the formation mechanisms.
Moreover, the mass function of sBHs in EMRI systems should be closely related to the mass spectrum of isolated sBHs within the nuclear cluster.
In this section, we discuss the information that can be extracted from the comparison of the mass functions of stellar BBHs, dry EMRIs, and wet EMRIs.

The mass spectrum of isolated sBHs remains poorly understood. Aside from Galactic X-ray binaries, the only direct observational constraints on the mass distribution of sBHs come from gravitational wave detections by the LVK Collaboration. If the pairing mechanism that forms BBHs can be identified, it may be possible to infer the mass distribution of isolated single black holes from the observed primary mass distribution of BBHs.
In \cite{Roy2025_MidThirtiesCrisis}, the authors investigated a range of possible single black hole formation channels: pulsational pair instability supernovae, chemically homogeneous evolution, Population III stars, globular cluster dynamics, and hierarchical mergers, to determine whether any of them can reproduce the observed BBH mass spectrum through pairing.
They concluded that no single channel can fully account for the population properties currently observed by LVK.

Rather than assuming a specific formation channel, in this paper we adopt a phenomenological approach. We model the mass distribution of single black holes and apply a pairing function to construct binaries. By comparing the resulting primary mass distribution with LVK observations, we infer the underlying single-BH mass spectrum corresponding to the chosen parameterized pairing function.
This inferred mass function of isolated black holes is then used as a benchmark in Sec.\ref{sec:masswet} and Sec.\ref{sec:massdry}, to demonstrate how the sBH mass functions in wet and dry EMRIs can be predicted.
 
We assume a phenomenological ``Power law + Gaussian peak'' model for the distribution of primary masses in LVK BBHs, following \cite{Abbott2021_GWTC2Pop, Abbott2023PopulationGWTC3}
\begin{align}
    P_{\rm LVK}(m_1) &= N S(m_1|\delta_m)[(1-\lambda_{\rm peak})\mathcal{P}(m_1|-\alpha) \notag \\ 
    &+ \lambda_{\rm peak}\mathcal{G}(m_1|\mu_m, \sigma_m)]
\end{align}
in which $\mathcal{P}$ is the normalized power law distribution defined between $m_{\rm min}=4.59M_\odot$  and $m_{\rm max}=86.22M_\odot$ (from GWTC-2 \cite{Abbott2021_GWTC2Pop}) with power-law index $\alpha_{\rm p}$, $\mathcal{G}$ is the normalized Gaussian distribution with mean $\mu_m$ and standard deviation $\sigma_m$, $N$ is the normalized factor, $S$ is the smoothing function that increases from zero to one over the interval $(m_{\rm min}, \ m_{\rm min} + \delta_m)$, 

\begin{equation}
\scalebox{0.9}{$
S(x) =
\begin{cases}
0, & m < m_{\rm min} \\
\bigl[ f(m - m_{\rm min}, \delta_m) + 1 \bigr]^{-1}, & m_{\rm min} \le m < m_{\rm min} + \delta_m \\
1, & m \ge m_{\rm min} + \delta_m
\end{cases}
$}
\end{equation}
with
\begin{align}
    f(m', \delta_m)=\exp\left(\frac{\delta_m}{m'} + \frac{\delta_m}{m'-\delta_m}\right)
\end{align}
where the parameters are $\alpha_{\rm p}=3.5, \mu_m=34, \ \sigma_m=5.69, \ \lambda_{\rm peak}=0.038, \ \delta_m=4.82$ as consistent in \cite{Abbott2021_GWTC2Pop, Abbott2023PopulationGWTC3}.

In addition, although the single/isolated BH mass distribution is also unknown, we apply a similar parameterization for the mass function with the ``Power law + Gaussian peak'' model:
\begin{align}
    F(m) &\propto S(m|\tilde{\delta}_m)[(1-\tilde{\lambda}_{\rm peak})\mathcal{P}(m|-\tilde{\alpha}) \notag \\ 
    &+ \lambda_{\rm peak}\mathcal{G}(m|\tilde{\mu}_m, \tilde{\sigma}_m)]\,,
\end{align}
and with a parametrized pairing function as \cite{Roy2025_MidThirtiesCrisis}
\begin{align}
    g(q) = \left(\frac{1+q}{2}\right)^{\tilde{\beta}}\,.
\end{align}
Here $q$ is the mass ratio $m_2/m_1$ ($q<1$ by definition) and $\tilde{\beta}$ is the power-law index. If $\tilde{\beta}=0$, the distribution is flat and pairing of sBHs with different masses does not have a mass preference. If $\tilde{\beta}>0$, the pairing in the binary system is more likely toward the equal-mass end $q=1$, and if $\tilde{\beta}<0$, the masses are more likely to be asymmetric with a smaller $q$.

The joint mass distribution of the BBH system can be obtained with the single BH mass function and the pairing function:
\begin{align}
    F(m_1, m_2) =N_{\rm b} F(m_1)F(m_2)g(q)
\end{align}
where $N_{\rm b}$ is the normalization constant. The distribution of
the primary mass component $m_1$ is
\begin{align}
    P_{\rm marginal}(m_1) = 2\int_{m_{\rm min}}^{m_1} dm_2 F(m_1, m_2)\,.
\end{align}
The factor of $2$ comes from the symmetry of $m_1$ and $m_2$.
The underlying parameters $\{\tilde{\alpha}_{\rm p}, \ \tilde{\mu}_m, \ \tilde{\sigma}_m, \ \tilde{\lambda}_{\rm peak}, \ \tilde{\delta}_m \}$ are unknown but can be determined (subject to data and fitting uncertainties) by matching $P_{\rm marginal}(m_1)$ with the LVK data.
As a baseline for the following discussion, we shall also adopt the power law index $\tilde\beta=2.6$ as measured by fitting the BBH population data \cite{Roy2025_MidThirtiesCrisis}. Other values within the uncertainty range of \cite{Roy2025_MidThirtiesCrisis} are also possible, although it does not affect the qualitative signatures of the conclusions.

\begin{figure}
    \centering
    \includegraphics[width=1.0\linewidth]{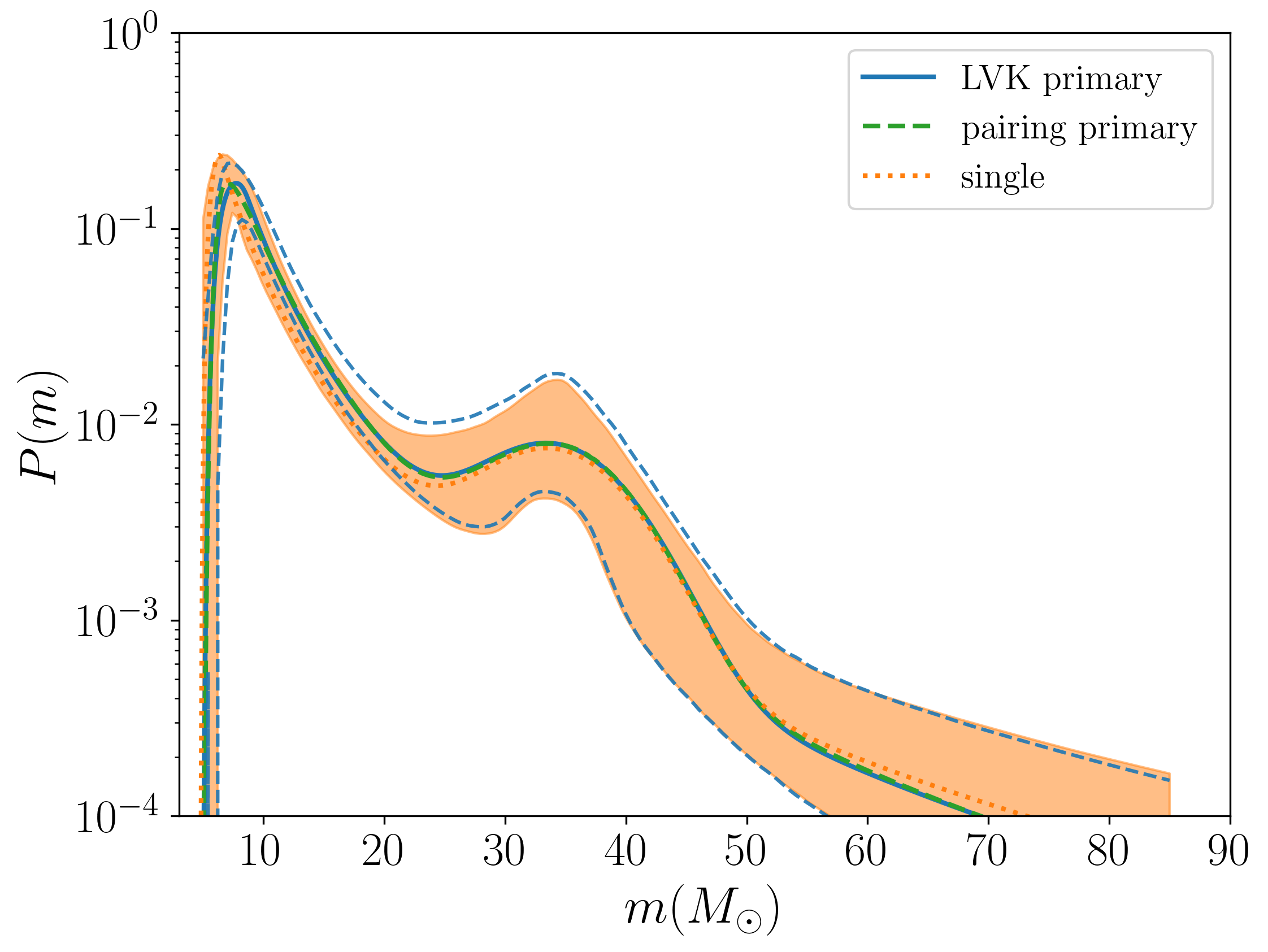}
    
    \caption{This figure shows the agreement between the marginal primary mass distribution and the LVK primary mass distribution for $\tilde\beta = 2.6$ and optimal parameters. The orange dotted line shows the single-BH mass function, peaking at $\sim 10\,M_\odot$ and $30\sim 40M_\odot$. The region between the two blue dashed lines indicates the 90\% confidence interval of the LVK primary mass distribution, while the blue shaded region represents the corresponding interval for the single black hole mass distribution. The latter closely resembles the former, with only a slight downward shift in the Gaussian peak component. Overall, the two regions exhibit a substantial overlap. The orange shaded region is based on a value of $\tilde{\beta} = 2.6$. However, a smaller value of $\tilde{\beta}$ would result in a reduced contribution from the Gaussian component in the single black hole mass distribution. Consequently, the uncertainty in $\tilde{\beta}$ introduces additional uncertainty into the single black hole mass distribution, which is not shown here.}
    \label{fig:mass_fit}
\end{figure}

For example, for a $P_{\rm LVK}(m_1)$ as shown in Fig.~\ref{fig:mass_fit}, and with $\tilde\beta=2.6$, 
we find that the optimal fitting parameters are $\{\tilde{\alpha}_{\rm p}=3.2, 
\tilde{\mu}_m=34, \tilde{\sigma}_m=5.65, \tilde{\lambda}_{\rm peak}=0.058, \tilde{\delta}_m=2.3\}$. 
The solid blue line shows the LVK primary mass distribution, while the green dashed line represents the fitted primary mass distribution obtained by combining the single BH mass distribution with a pairing function, demonstrating promising agreement with the observed data. The orange dotted line corresponds to the inferred isolated BH mass function under the ``Power-law + Gaussian peak'' model with optimal fitting parameters. It is important to note that both the primary mass distribution $P_{\rm LVK}(m_1)$ and the power-law index $\tilde\beta$ have substantial uncertainties \cite{Abbott2021_GWTC2Pop, Abbott2023PopulationGWTC3, Roy2025_MidThirtiesCrisis} due to limited data. As a result, the inferred isolated BH mass function also carries inherent uncertainties (the shaded region in Fig.~\ref{fig:mass_fit}) and should not be interpreted as the  ``true'' distribution. However, it serves as a useful reference point for exploring potential features of the EMRI mass distributions discussed in Sec.\ref{sec:massdry} and Sec.\ref{sec:masswet}.
Similar to $P_{\rm LVK}(m_1)$, the inferred single-BH mass function exhibits a prominent peak around $10\,M_\odot$, as well as a secondary peak near $35\,M_\odot$. Following GWTC-2 \cite{Abbott2021_GWTC2Pop}, we adopt the same upper mass cutoff of $m_{\rm max} = 86.22\,M_\odot$ for the isolated single black hole mass distribution.

We also make one key assumption to facilitate our discussion of EMRI mass functions: we assume that the isolated sBH mass function responsible for producing LVK binaries is the same as that of isolated sBHs within nuclear star clusters. This assumption is not necessarily valid, as the typical environments in which stellar-origin BBHs form may differ significantly from those of nuclear star clusters. However, in the absence of additional observational constraints on the sBH mass function within nuclear clusters, we proceed with the isolated sBH mass function inferred from LVK observations to explore the EMRI mass spectrum in the following two subsections.

\subsection{Dry EMRI}\label{sec:massdry}

The dry EMRI mass spectrum can be obtained by combining the isolated BH mass function with the mass-dependent dry EMRI formation rate. In Sec.~\ref{sec:drye}, the modeling of a nuclear star cluster using the Fokker-Planck equation assumes a stellar mass of $1\ M_\odot$ and an sBH mass of $10\ M_\odot$ (a ``1+10'' model). If the masses of sBHs follow a continuous distribution, a reasonable approximation since stars make up the majority of the population, then different mass components of sBHs evolve independently under an averaged potential. Therefore, we can evaluate the dry EMRI formation rate for a $xM_\odot$ sBH in a ``1+$x$'' model, and then weight this rate by the isolated sBH mass function to obtain the resulting dry EMRI mass spectrum.

\begin{figure}
    \centering
    \includegraphics[width=1.0\linewidth]{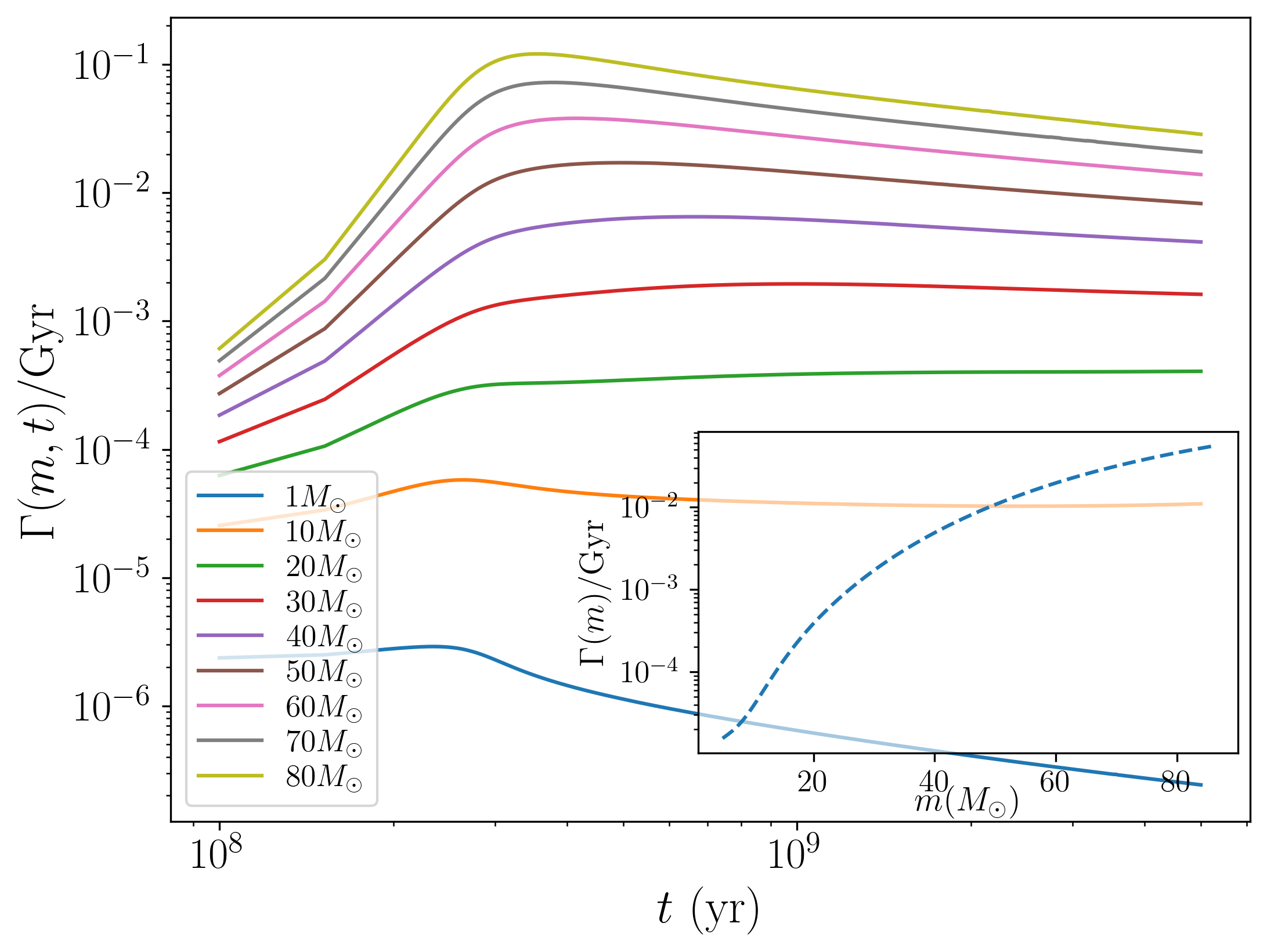}
    \caption{Dry EMRI rates for different mass components. The $y$-axis denotes the EMRI rate per object per Gyr. Heavier components exhibit higher EMRI rates per object, consistent with the effect of mass segregation. The inset shows the time-averaged dry EMRI rate per object as a function of mass within 5 Gyrs.}
    \label{fig:mass_dry_rate}
\end{figure}

The EMRI rates for various mass components are shown in Fig.~\ref{fig:mass_dry_rate}, based on the $1+x$ model described in Sec.~\ref{sec:drye}. Heavier sBHs tend to have higher EMRI rates due to the effect of mass segregation. After the initial ramp-up stage, the rates gradually decline over time as a result of the insufficient supply of sBHs to the nuclear star cluster. 
In the inset of Fig.~\ref{fig:mass_dry_rate}, we show the dry EMRI rate per object averaged over time within a canonical time $5$Gyrs.
This dry EMRI rate acts as a weighting factor in the final dry EMRI mass distribution. By performing a time average, we obtain the final dry EMRI mass spectrum as follows:
\begin{align}
P_{\rm dry}(m) = N_{\rm dry}\int dt P_{\rm single}(m)\Gamma(m,t)\,,
\end{align}
where $\Gamma(m,t)$ is per-object EMRI rate and $N_{\rm dry}$ is the normalization constant. 

\begin{figure}
    \centering
    \includegraphics[width=1.0\linewidth]{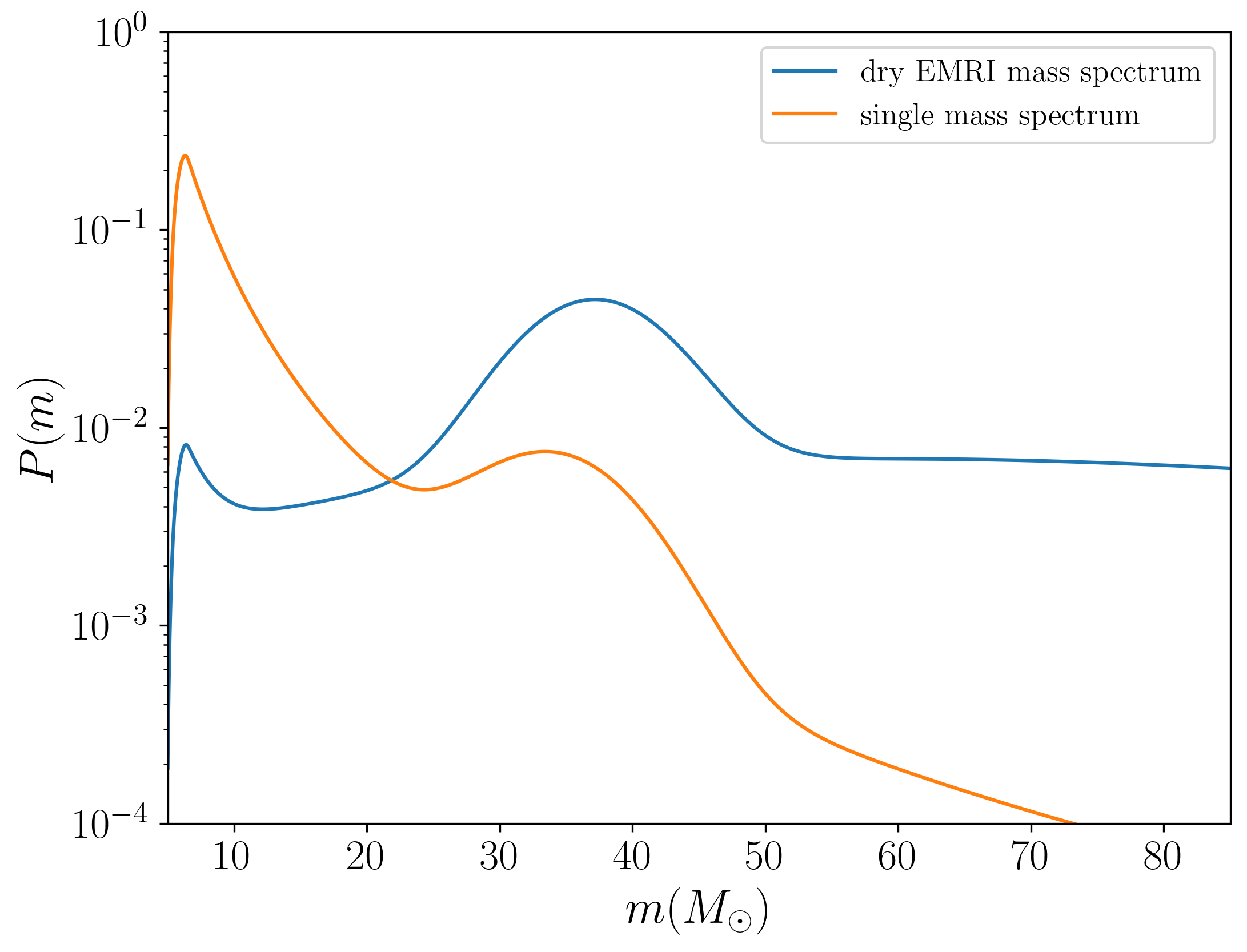}
    \caption{Comparison between the dry EMRI mass spectrum and the isolated single-BH mass spectrum. The dry EMRI spectrum shows two peaks at $m \simeq 10\,M_\odot$ and $m = 30\sim40\,M_\odot$, with the heavier-mass peak being more pronounced than that of the isolated sBH mass function.}
    \label{fig:mass_dry}
\end{figure}

The corresponding dry EMRI mass function is shown in Fig.~\ref{fig:mass_dry}. Compared to the initial isolated mass spectrum, the dry EMRI mass function exhibits a more pronounced peak around $\sim 35M_\odot$, which is higher than the peak in the LVK primary mass function and even higher than the peak at $10M_\odot$. An intriguing implication is that secondary BHs of dry EMRIs are of heavy tailed distribution, therefore lead to more EMRI detections.
With future observations from space-based detectors, the dry EMRI mass function, combined with refined theoretical modeling of the relative rates among different mass components, can be used to infer the mass function of isolated sBHs within nuclear star clusters. This, in turn, can be compared to the mass function derived from LVK observations, serving both as a test of the assumption made in Sec.~\ref{sec:lvk} and as a probe of environmental effects on binary formation.

\subsection{Wet EMRI}\label{sec:masswet}

Isolated sBHs may be captured by the AGN disk, if present, and subsequently migrate toward the central MBH to form wet EMRIs. Therefore, similar to dry EMRIs, the mass function of wet EMRIs also critically depends on the mass function of isolated sBHs within the nuclear star cluster. However, additional factors, namely disk capture efficiency, accretion effects, and merger effects, also influence the final mass distribution. These additional factors will be addressed quantitatively in this section.

\begin{figure}
    \centering
    \includegraphics[width=1.0\linewidth]{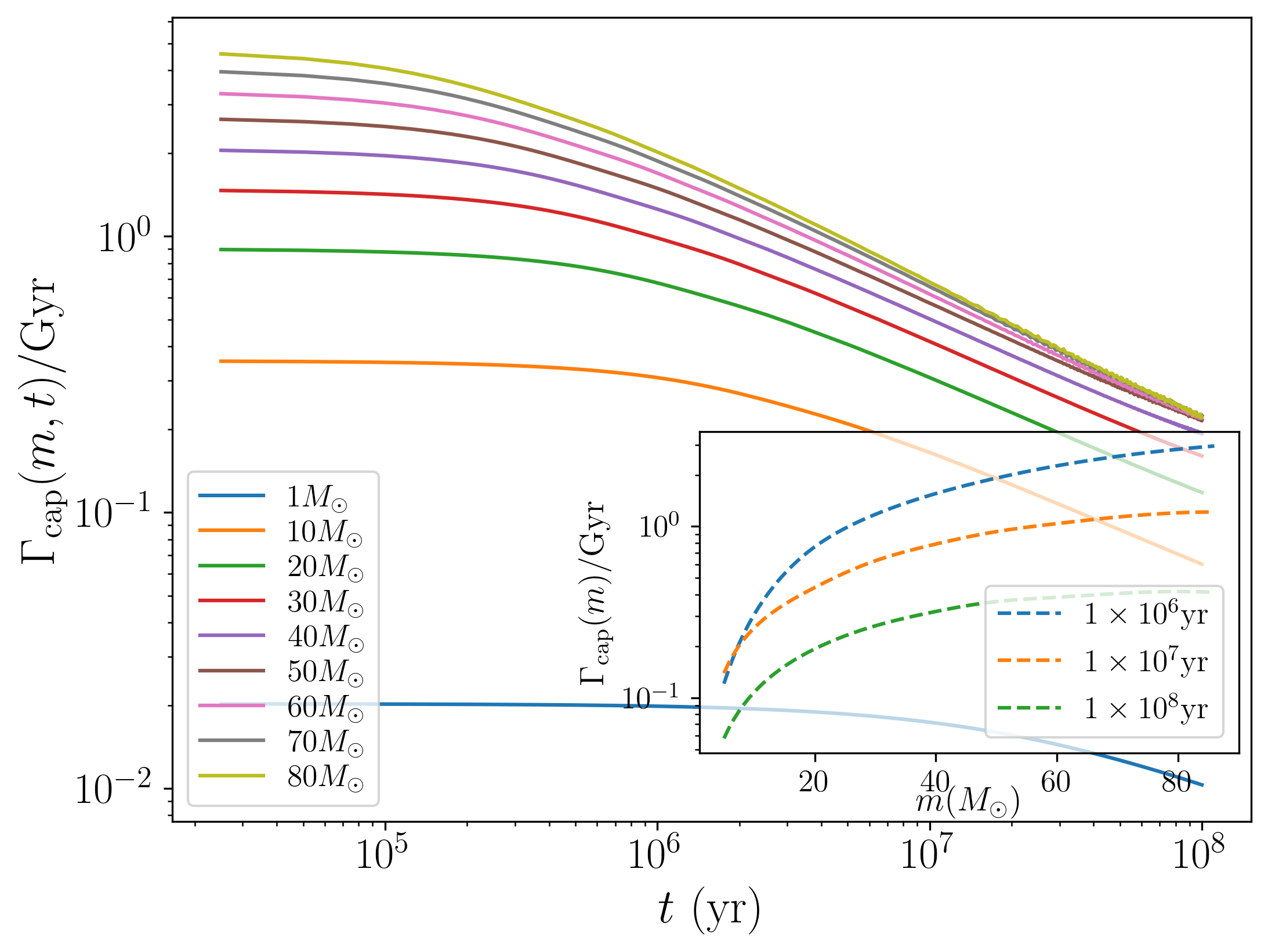}
    \caption{Disk capture rates from the nuclear cluster for different mass components.  The $y$-axis shows the capture rate per object per Gyr. The inset shows the time-averaged (up to the disk lifetime) disk capture rate as a function of sBH mass for different disk lifetimes. }
    \label{fig:mass_rate}
\end{figure}

\begin{figure*}
    \centering
    \includegraphics[width=0.45\linewidth]{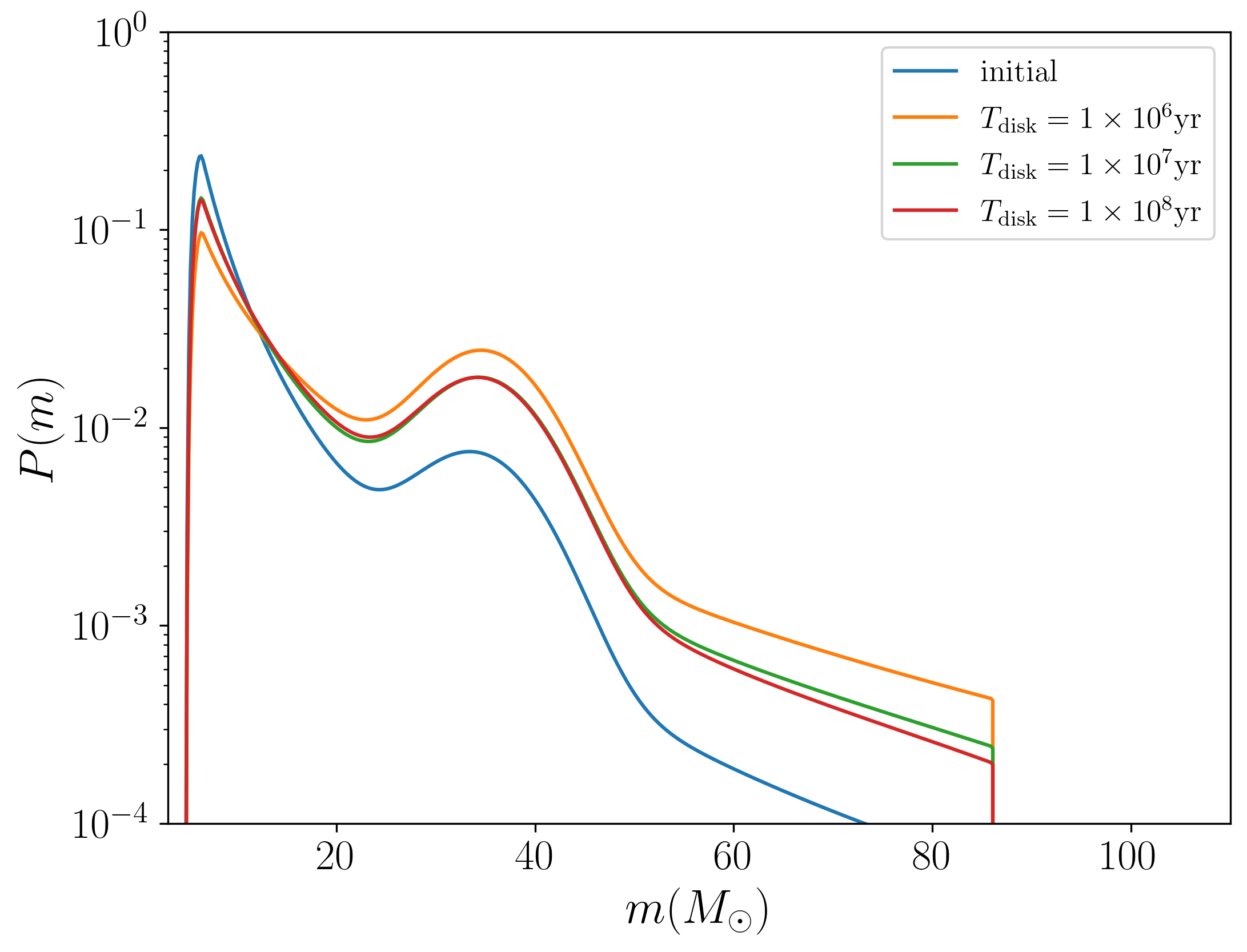}
    \includegraphics[width=0.45\linewidth]{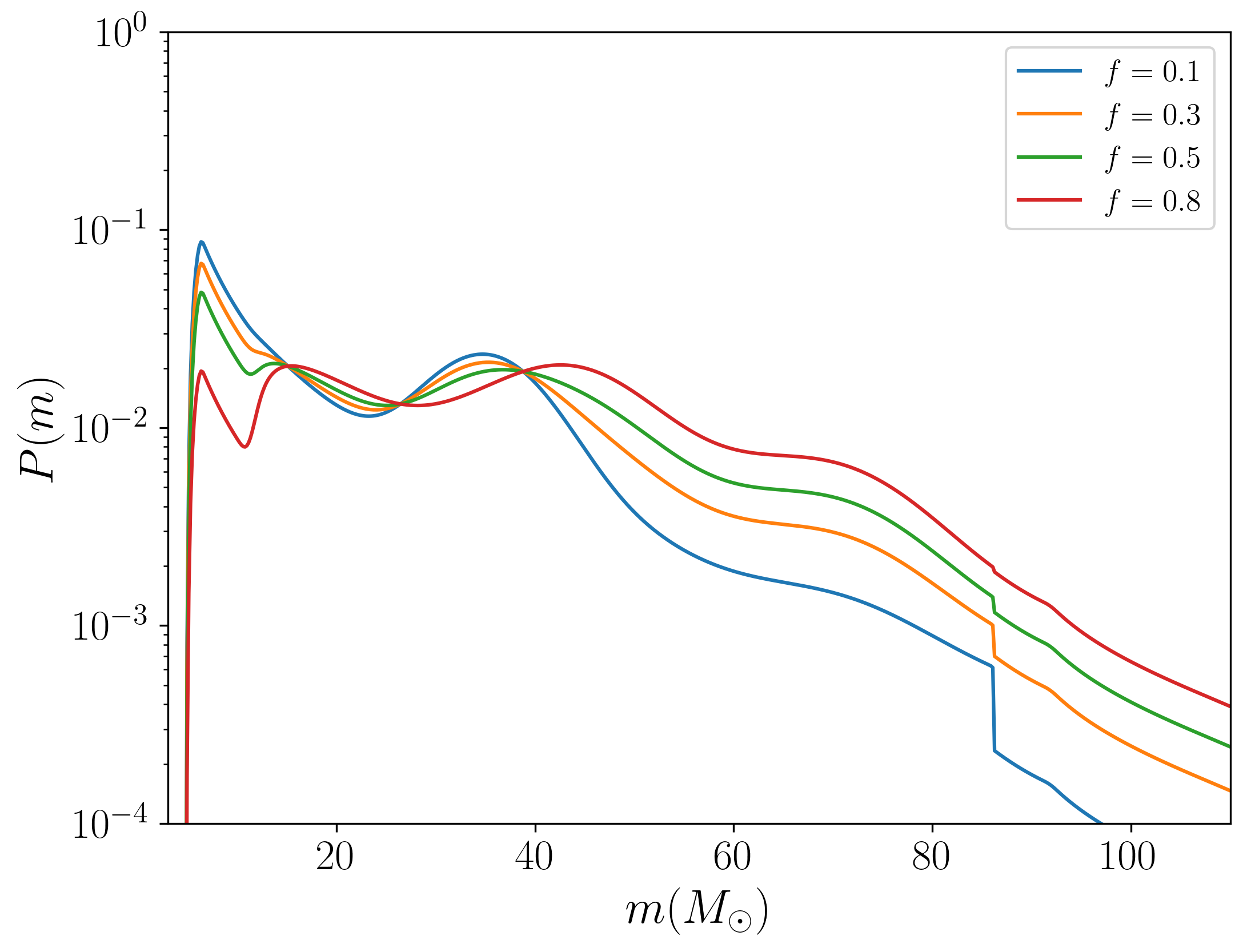}
    \includegraphics[width=0.45\linewidth]{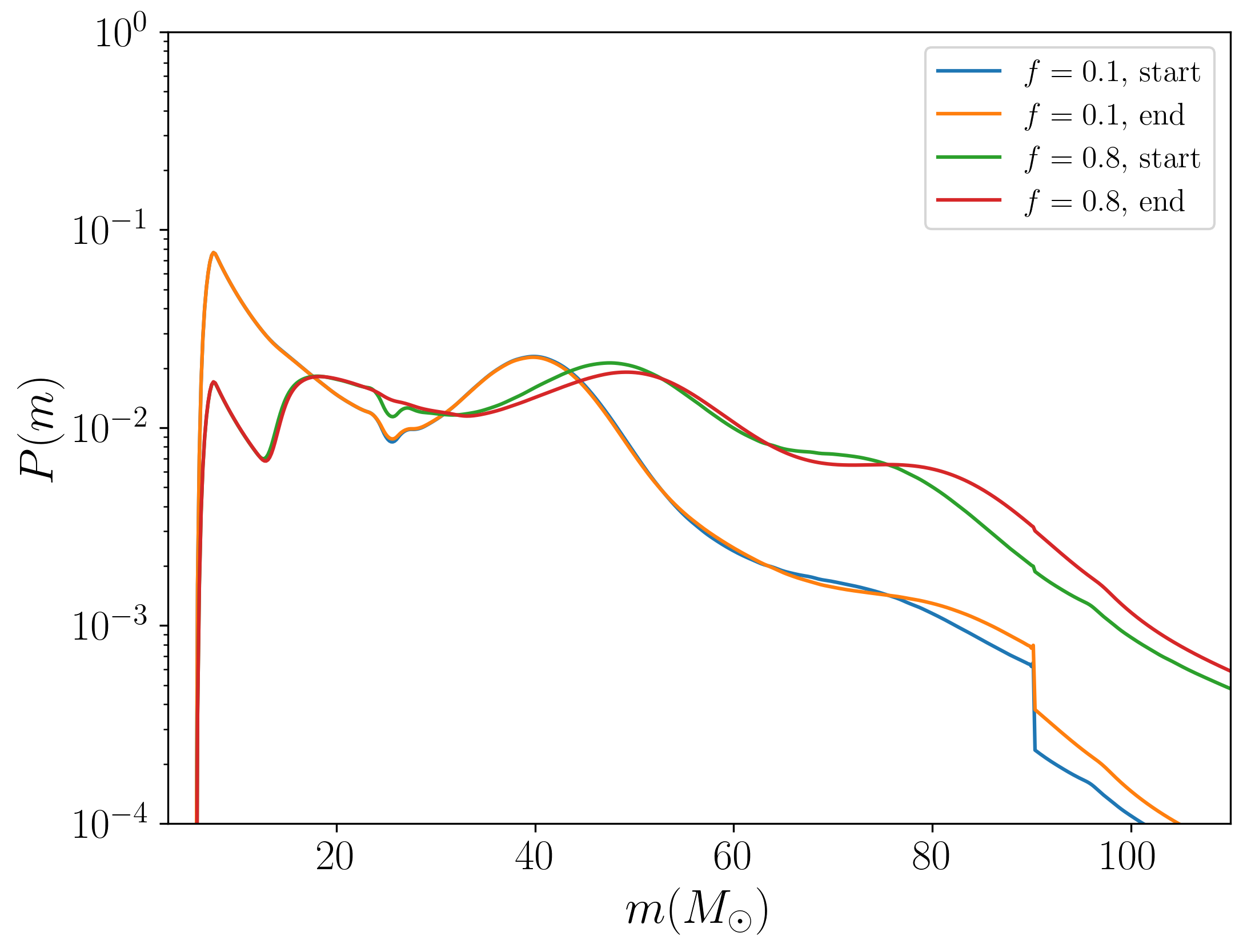}
    \includegraphics[width=0.45\linewidth]{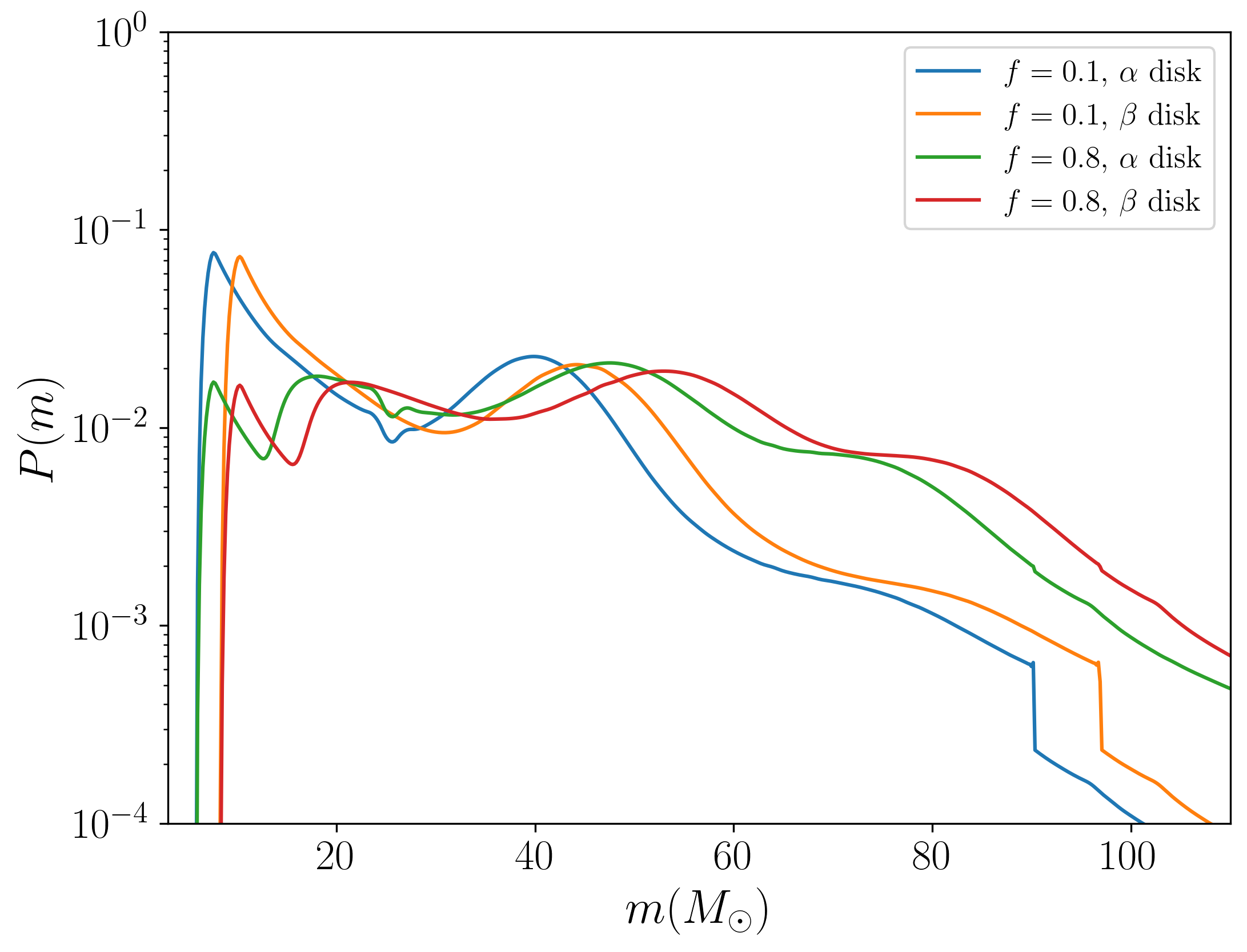}
    \caption{Mass spectra in different processes with various parameters. In our default model, the merger process occurs after the sBH is captured. The merger fraction is $f_{\rm mer} = 0.1$, and the disk is an $\alpha$-disk with $M_{\bullet} = 4\times10^6\,M_\odot, \alpha = 0.1, \dot{M}_\bullet = 0.1\,\dot{M}_{\rm Edd}$, and $T_{\rm disk} = 1\times10^6\,\mathrm{yr}$. The top-left panel shows the capture distribution for different disk lifetimes, as discussed in the paper. The top-right panel shows the merger mass distribution for different merger fractions after the sBH is captured by the AGN disk, indicating that a larger fraction can significantly alter the mass distribution profile. The bottom-left panel shows the final wet EMRI mass distribution for different merger types: ``start'' indicates that the merger occurs after the sBH is captured, and ``end'' indicates that the merger occurs after the sBH has accreted. This panel demonstrates that different merger types lead to negligible differences; therefore, we do not distinguish between them and uniformly treat the merger process as occurring after capture. The bottom-right panel shows the final accretion mass spectrum for different disk types, where the $\beta$-disk exhibits a higher probability at larger masses, due to its higher surface density in the inner region, which allows sBHs to accrete more mass.}
    \label{fig: mass_different_para}
\end{figure*}

Starting from the mass function $P_{\rm single}(m)$ of isolated sBHs in the nuclear cluster, the number of sBHs captured by the accretion disk also depends on the mass of the sBH and the time-dependent per-object capture rate $\Gamma_{\rm cap}(m,t)$, similar to the scenario in dry EMRI.  During the course of an entire AGN episode, the average mass distribution after disk capture depends on the disk lifetime, which is
\begin{align}\label{eq:cap}
    P_{\rm cap}(m) = \frac{\int_{0}^{T_{\rm disk}} dt \ P_{\rm single}(m) \Gamma_{\rm cap}(m,t)}{\int dm \int_{0}^{T_{\rm disk}}dt \ P_{\rm single(m)}\Gamma_{\rm cap}(m,t)}\,.
\end{align}
The per-object captured rate $\Gamma_{\rm cap}(m,t)$ may be determined by applying an FP simulation of the system with both the nuclear star cluster and the incorporated disk, i.e. see the detailed modeling in \cite{Pan202112}. Given similar setups, we compute the capture rates shown in Fig.~\ref{fig:mass_rate}. Its inset displays the time-averaged capture rate as a function of mass for different disk lifetimes.
The resulting mass functions $P_{\rm cap}$ for different disk lifetimes are presented in the upper left panel of Fig.~\ref{fig: mass_different_para}, where a relative shortage of more massive sBHs can be found for scenarios with larger $T_{\rm disk}$. On the other hand, compared to the initial single mass function $P_{\rm single}(m)$, there is already a relative boost in the fraction of more massive sBHs, because inclination-angle diffusion is more efficient for more massive sBHs. 

sBHs captured by the AGN disk generically migrate within the disk due to the back reaction from density-wave emission, absorption, and momentum transfer with the head wind \cite{Kocsis2011DiskEMRI}, and gravitational pull from possible turbulent eddies of the disk.  In particular, the head wind arises from gravitational capture of the surrounding disk material by the sBH, which tends to circularize within the Bondi radius, and form a supercritical accretion flow towards the sBH. In \cite{Pan:2021xhv} we have adopted the ``inflow-outflow'' model \cite{Blandford:1999} to describe the supercritical accretion flow and studied the change in mass and spin of sBH during the migration process. Different simulations have further confirmed the validity of the ``inflow-outflow'' model in the super-Eddington regime \cite{Sadowski2014,Fragile:2025nes}.

\begin{figure}
    \centering
    \includegraphics[width=1.0\linewidth]{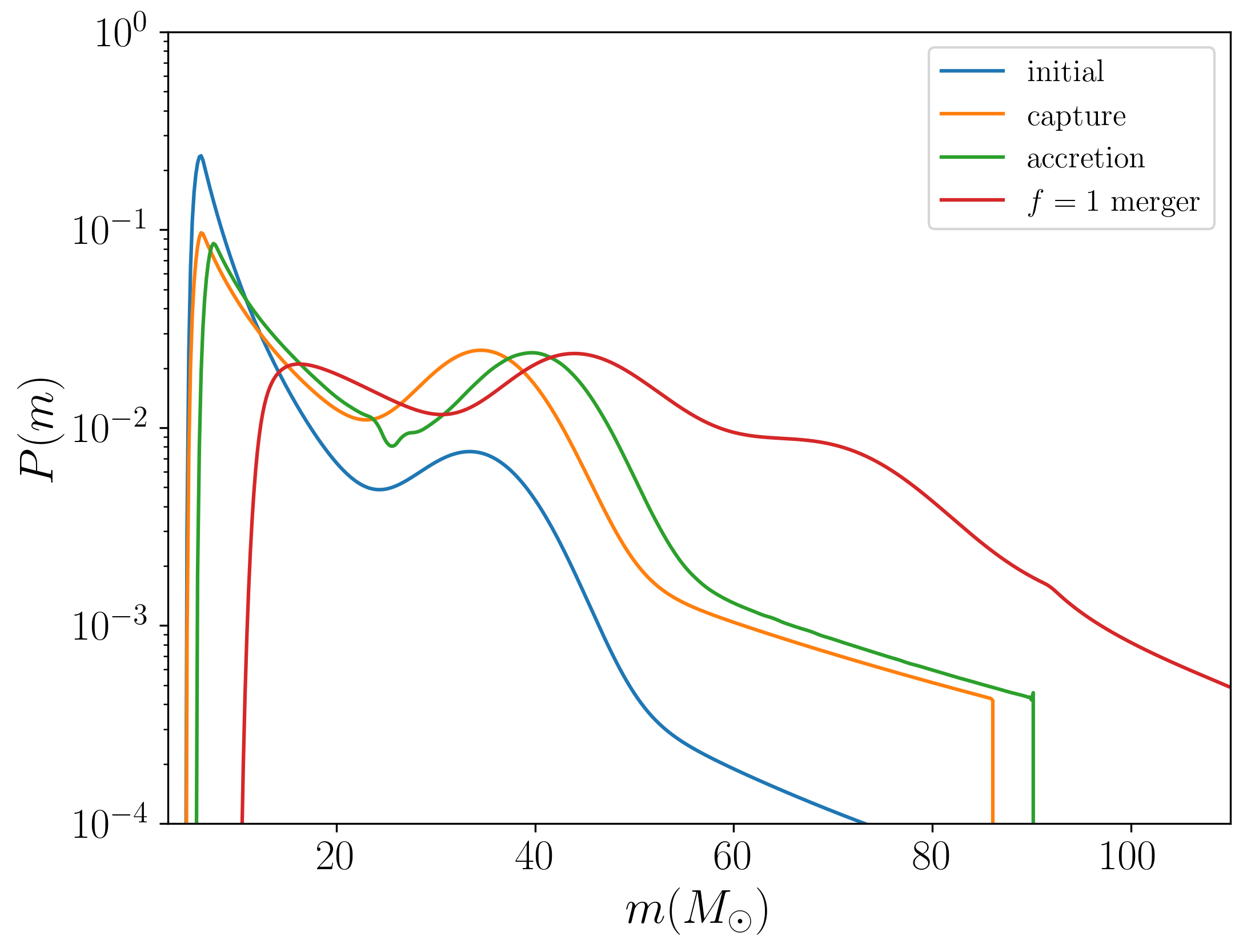}
    \caption{Mass spectra for different processes. The blue line shows the initial isolated single-BH mass distribution in the nuclear cluster. The orange line represents the averaged mass distribution after capture by the disk with a lifetime of $T_{\rm disk} = 1 \times 10^6\,\mathrm{yr}$, showing a boost at higher masses compared to the initial distribution. The green line denotes the accretion mass function after capture, which is shifted toward higher masses. The red line corresponds to the merger mass function with $f_{\rm mer} = 1$ after capture, exhibiting two peaks at about $20\,M_\odot$ and $45\,M_\odot$.}
    \label{fig:mass_toal}
\end{figure}

Assuming the same sBH accretion model used in \cite{Pan:2021xhv}, and incorporating a refined treatment on type II migration (the ``gap-opening'' scenario) as discussed in detail in Appendix.~\ref{sec:type2} \cite{Li2023,Li2024}, we trace the accretion rate of sBHs during their migration processes and obtain their final mass $m_{\rm f}$ as a function of the initial mass $m_{\rm i}$: $m_{\rm f}=m_{\rm f}(m_{\rm i})$, or vice versa $m_{\rm i}=m_{\rm i}(m_{\rm f})$.
It is straightforward to see that the final mass distribution is given by
\begin{align}
P_{\rm acc}(m)= P_{\rm cap}(m_{\rm i}) \frac{dm_{\rm i}(m)}{dm} \,.
\end{align}
As found in \cite{Pan:2021xhv}, accretion generally increases the mass of an sBH by $< 25\%$ for $\alpha$-disk and $< 70\%$ for $\beta$-disk. 
This is also consistent with the mass spectra of wet EMRIs shown in Figs.~\ref{fig: mass_different_para} and \ref{fig:mass_toal}. For reference, we also show the accreted mass in the lower panel of Fig.~\ref{fig:gap}. 

The third critical factor that affects the mass function of wet EMRIs is the pairing and merger of sBHs within the disk. It has been suggested that sBH binaries may form and merge in AGNs, potentially contributing to some of the observed events in the GWTC catalog \cite{Tagawa2020,Li2021,Li2022b,LiR2022,Dittmann2024}. The pairing process arises mainly from the collision of circum-single disks from individual sBHs, which subsequently move sufficiently close to each other \cite{LiJ2023,Mishra2024CircumSingleMHD,Whitehead2024,Wang2025}. However, the formation rate, the subsequent evolution, and the merger rate are still highly uncertain \cite{Tagawa2020}. In this study, we therefore adopt a parametrized approach to (crudely) assume that each sBH has a constant probability $f_{\rm mer}$ of being captured in a binary and eventually merge with its companion. 

Assuming that all isolated sBHs within the disk would form pairs with another sBH that eventually leads to merger, that is, $f_{\rm mer}=1$, and that we only consider 1G(generation)$+$1G scenarios, the resulting wet EMRI mass function is
\begin{align}
    P_{\rm f=1}(m) = \int P_{\rm cap}(x)P_{\rm cap}(m-x) dx
    \label{eq:mer_1}
\end{align}
where we have assumed a flat pairing function for simplicity and have not incorporated accretion in this part of the merger description. 
If the merger fraction is not $100\%$, that is, $f_{\rm mer} \neq 1$, the corresponding mass function becomes a linear combination of $P_{\rm f=1}$ and $P_{\rm cap}$:

\begin{align}
    P_{\rm mer}(m) = (1-f_{\rm mer})P_{\rm cap}(m) + f_{\rm mer}P_{\rm f=1}(m)\ .
\end{align}
As we can find in Fig.~\ref{fig:mass_toal}, $P_{\rm f=1}$ clearly shows displaced peaks at $m=20 M_\odot$ and $m=45 M_\odot$, corresponding to the merger scenarios with $10 M_\odot+10 M_\odot$ and $10 M_\odot+35 M_\odot$. The mass peak corresponding to the $35 M_\odot+35 M_\odot$ merger process is weak due to the large relative abundance of sBHs in the $10 M_\odot$ peak.
In the upper right panel of Fig.~\ref{fig: mass_different_para}, we show results for different fractions of merger $f_{\rm mer}$. Depending on the fraction of merger $f_{\rm mer}$, the $m=20 M_\odot$ peak may manifest itself as a bump around $20 M_\odot$, and another peak may change from $35M_\odot$ to $45M_\odot$.

Notice that the formation of sBH binary may happen at any stage of the migration process, so that the individual sBH in the binary may have acquired a certain amount of mass before reaching the center MBH. In addition, accretion onto these sBHs likely continues during the inspiral evolution of the binary. As a result, it is likely more accurate to apply a mixed distribution between $P_{\rm mer}$ and $P_{\rm acc}$ when analyzing the final wet EMRI mass spectrum. In practice, computing $P_{\rm mer}$ starting from $P_{\rm cap}$ as the initial distribution or replacing $P_{\rm cap}$ in Eq.~\ref{eq:mer_1} by $P_{\rm acc}$ leads to final distributions with only minor differences, as shown in the bottom left panel of Fig.~\ref{fig: mass_different_para}. Therefore, we shall not distinguish the sequence between accretion and merger in our analysis. 

In Fig.~\ref{fig: mass_different_para} we present the mass distribution of wet EMRIs assuming a canonical model with $M_{\bullet}=4 \times 10^6 M_\odot$, $\dot{M}_{\bullet}=0.1 \dot{M}_{\rm Edd}$, $f_{\rm mer}=0.1$, $T_{\rm disk}=10^6 {\rm yr}$ and $\alpha$-disk with $\alpha=0.1$. In order to study various parameter dependences, we also vary $T_{\rm disk}, f_{\rm mer}$ and disk models to compare the resulting mass functions. For example, in the upper left panel, we observe a relative decline in the peak around $35 M_\odot$ for longer disk lifetimes, mainly due to the depletion and insufficient supply of more massive sBHs in the nuclear star cluster. In the upper right panel, the mass functions with respect to different $f_{\rm mer}$ are shown. As $f_{\rm mer}$ increases, the peak around $35 M_\odot$ gradually moves to the right ($\sim 45 M_\odot$), and the peak around $20 M_\odot$ becomes more pronounced. In the bottom left panel, we compute the mass function assuming different sequence of merger and accretion effects. The resulting distributions are largely similar. Finally, in the bottom right panel, we also compare the predictions from two different disk models: the $\alpha$-disk model and the $\beta$-disk mode. In general, the $\beta$-disk model yields a more massive distribution, due to higher AGN disk density, therefore its higher accretion rate onto isolated sBHs in the disk. In general, the mass spectrum of wet EMRI is most sensitive to $f_{\rm mer}$, the merger fraction.

In addition to sBHs that were first formed in the stellar cluster and then captured onto the AGN disk, in situ formation of sBHs within the disk is also possible.
In the outer regions of an AGN disk, where the disk fragments into dense clumps as a result of gravitational instability, stars can form.
The mass function of stars formed in AGN disks may be top-heavy compared to the conventional Salpeter initial mass function \cite{Derdzinski:2022ltb}.
Stars in AGN disks may not undergo chemical evolution because the fresh gas from the disk replenishes stellar cores faster than hydrogen is burned \cite{Cantiello2021,Dittmann2021}. If this is true, stars on AGN disks are effectively ``immortal'' \cite{Jermyn2022} and therefore are not expected to collapse into compact objects and form EMRIs. Otherwise, massive stars in AGN disks may eventually collapse into compact stellar-mass objects, which then migrate toward the central MBH and become wet EMRIs \cite{Levin2007,Derdzinski:2022ltb}.
In this scenario, the mass functions of stars and sBHs embedded in AGN disks are highly uncertain because of the complex stellar environments and interactions between stars \cite{Wang2021,Cantiello2021,Dittmann2021,Jermyn2022}.

\section{Conclusion}\label{sec:conc}

In this work, we have performed a first population study on three important observables of EMRIs: eccentricity, inclination, and sBH mass, all of which exhibit distinct distributions for wet and dry EMRI formation channels. Taking into account their evolutionary histories starting from highly eccentric inspirals, dry EMRIs are likely to retain eccentricities of $e \gtrsim 10^{-2}$ at $r_{\rm p}=10M_{\bullet}$. In contrast, the eccentricities of wet EMRIs are strongly damped by the accretion disk, but can be reexcited through multibody resonances and/or turbulent fluctuations within the disk.
In both cases, the eccentricity is typically below $10^{-2}$ at $r_{\rm p}=10M_{\bullet}$, making it an effective parameter to distinguish between wet and dry formation channels. Furthermore, measurements of wet EMRI eccentricities can provide insight into key disk parameters, as illustrated in Eqs.~(\ref{eq:dr_sca1}, \ref{eq:dr_sca2}).

The distributions of the EMRI inclination angles also exhibit interesting features. For dry EMRIs, since the angular momentum of the LSO depends on the inclination, the EMRI formation rate is intrinsically $\iota$-dependent for spinning MBHs. The ``tilted'' distribution of $\iota$ thus serves as a direct test of the dry EMRI formation model, although it may also be influenced by potential deviations from spherical symmetry in the nuclear star cluster. For wet EMRIs, the $\iota$ distribution is typically aligned or tilted, depending on whether MBH accretion is predominantly coherent or chaotic. In the chaotic accretion scenario, the detailed shape of the resulting tilted distribution can be used to constrain the relative timescales of disk lifetimes and BP alignment.

The mass function of sBHs in wet EMRIs is likely different from that of sBHs in nuclear star clusters and dry EMRIs, primarily due to three key factors: capture efficiency, accretion, and mergers. The disk capture process is more efficient for more massive sBHs, enhancing their relative abundance at the high-mass end. Accretion generally increases the mass of the sBHs, changing the distribution toward larger values. In contrast, sBH binary formation and mergers introduce distinctive features into the mass spectrum. Depending on the fraction of merger $f_{\rm mer}$, new peaks may emerge in the resulting mass function.

Additional EMRI observables are not discussed in this work. For example, the spin of sBHs can be altered by accretion \cite{Pan:2021xhv,Li2022a}, leading to a potentially different spin distribution for wet EMRIs compared to dry EMRIs. However, measuring the spin of sBHs is challenging, as recent studies incorporating sBH spin into EMRI waveforms suggest \cite{Cui:2025bgu, Piovano2021SecondarySpin}, unless the eccentricity is sufficiently large. Therefore, it is unlikely that the spins of wet EMRIs can be accurately measured. In contrast, the spin of MBHs can be determined with extremely high precision, at the level of $\sim 10^{-5}$ \cite{Babak2017_LISA_EMRI}. Since the formation rate of wet EMRIs is largely insensitive to MBH spin, the observed spin distribution of MBHs in wet EMRIs should reflect their intrinsic spin distribution. On the other hand, the formation rate of dry EMRIs depends on the angular momentum of LSO, and therefore on the MBH spin (see the discussion in Sec.~\ref{sec:iotadry}). As a result, the MBH spin distributions inferred from wet and dry EMRIs may differ significantly. Similar argument applies to the mass function of MBH in EMRIs as well, as the formation efficiency of wet EMRIs and dry EMRIs may have a different dependence on $M_{\bullet}$, in addition to being influenced by the intrinsic mass function of MBHs.

In addition to the distributions of individual parameters, correlations between parameters may also exist for different EMRI formation channels. For example, in dry EMRIs, the ``tilt'' in the distribution of $\iota$ is expected to be more pronounced for MBHs with higher spin, implying a positive correlation between $\cos \iota$ and the MBH spin $a$ in their joint distribution. Moreover, if AGN accretion is intrinsically correlated with the mass of the central MBH, that is, if $\dot{M}_{\bullet}$ correlates with $M_{\bullet}$, then the eccentricity of wet EMRIs may also exhibit a non-trivial dependence on $M_{\bullet}$. Such multiparameter correlations and joint distributions remain largely unexplored but may encode valuable information about the underlying EMRI formation channels.

Last but not least, although we have focused only on wet and dry EMRIs, EMRIs formed via the Hill mechanism or the Kozai-Lidov mechanism in SMBH binaries may also constitute a sub-population in the event catalog of space-based detectors. It would be interesting to develop quantitative population models for these alternative formation channels and to investigate the distributions of their associated observables.

\begin{acknowledgments}
        \section{Acknowledgments}
        The authors thank Yanbei Chen and Xinyu Li for interesting discussions. Y.P.L is supported in part by the Natural Science Foundation of China (Grants 12373070, and 12192223), the Natural Science Foundation of Shanghai (grant NO. 23ZR1473700). H.Y. is supported by the Natural Science Foundation of China (Grant 12573048). The calculations have made use of the High Performance Computing Resource in the Core Facility for Advanced Research Computing at Shanghai Astronomical Observatory.
    \end{acknowledgments}

\appendix 
\section{Post-Newtonian Coefficient}\label{app:pn}

In the multibody resonance section, we consider the post-Newtonian force between the sBH and the center MBH. The force depends on their position and velocity in the form of \cite{Pati2002_PNII},  
\begin{align}
    \frac{d^2\vec{r}}{dt^2}=-\frac{M}{r^2}\frac{\vec{r}}{r} &+ \frac{M}{r^2} \left[ \frac{\vec{r}}{r}\left({\rm A}_\text{PN}+{\rm A}_\text{2PN}\right) + v_r\vec{v}\left({\rm B}_\text{PN}+{\rm B}_\text{2PN}\right) \right] \notag \\
    &+ \frac{8}{5}\eta\frac{M}{r^2}\frac{M}{r} \left[ v_r \frac{\vec{r}}{r}{\rm A}_\text{2.5PN} - \vec{v} {\rm B}_\text{2.5PN} \right]
\end{align}
the relative position $\vec{r}=\vec{r}_\text{sBH}-\vec{r}_\text{MBH}$, $r=|\vec{r}|$, the relative velocity $\vec{v}=\vec{v}_\text{sBH}-\vec{v}_\text{MBH}$, the radial velocity $v_r=\vec{v}\cdot \frac{\vec{r}}{r}$, the total mass $M=m+M_{\bullet}$, the symmetric mass ratio $\eta=mM_{\bullet}/M^2$, and the PN coefficients are

\begin{subequations}
\begin{align}
    A_\text{PN}&=-(1+3\eta)v^2 + \frac{3}{2}\eta v_r^2 + 2(2+\eta) \frac{M}{r} \\
    B_\text{PN} &= 2(2-\eta) \\
    A_\text{2PN}&= -\eta (3-4\eta)v^4+\frac{1}{2}\eta(13-4\eta)v^2\frac{M}{r} \notag \\ &+\frac{3}{2}\eta(3-4\eta)v^2v_r^2 + (2+25\eta+2\eta^2)v_r^2\frac{M}{r} \notag \\ &-\frac{15}{8}\eta(1-3\eta)v_r^4 -\frac{3}{4}(12+29\eta)\left(\frac{M}{r}\right)^2 \\
    B_\text{2PN} &= \frac{1}{2}\eta(15+4\eta)v^2-\frac{3}{2}\eta(3+2\eta)v_r^2  \notag \\ 
    &-\frac{1}{2}(4+41\eta+8\eta^2)\frac{M}{r} \\ A_\text{2.5PN}&=3v^2+\frac{17}{3}\frac{M}{r} \\
    B_\text{2.5PN} &= v^2 + 3\frac{M}{r}
\end{align}
\end{subequations}

\section{Equilibrium Eccentricity in Post-Newtonian Analysis}\label{app:equi}

In this paper, we study the resonance between two sBHs orbiting a central MBH, taking into account post-Newtonian corrections. In planetary systems, the case of two planets orbiting a central protostar is very common, and extensive analytical work has been done on such systems, particularly on the equations of motion for the orbital elements under resonance. Here, we include the post-Newtonian terms in equations of motion, which yields good agreement between the equilibrium eccentricity and the actual eccentricity, as shown in Fig.~\ref{fig:res}. The equations of motion for the orbital elements are formulated as \cite{Lin2025ResonanceCapture}
\begin{subequations}
\begin{align}
    \dot{n}_{\rm in} &= -3j\mu_{\rm out}n_{\rm in}^2\mathcal{R}_{a}(e_{\rm in}f_{\rm in}\sin\varphi_{\rm in}+e_{\rm out}f_{\rm out}\sin\varphi_{\rm out}) \notag \\ &+\frac{3n_{\rm in}}{ t_{J,\rm in}}+\frac{3n_{\rm in}e_{\rm in}^2}{ t_{e,\rm in}} \\
    \dot{n}_{\rm out} &= 3(j+1)\mu_{\rm in}n_{\rm out}^2(e_{\rm in}f_{\rm in}\sin\varphi_{\rm in}+e_{\rm out}f_{\rm out}\sin\varphi_{\rm out}) \notag \\ &+ \frac{3n_{\rm out}}{ t_{J,\rm out}} + \frac{3n_{\rm out}e_{\rm out}^2}{ t_{e,\rm out}} \\
    \dot{e}_{\rm in} &= -\mu_{\rm out}n_{\rm in}\mathcal{R}_{a} f_{\rm in}\sin\varphi_{\rm in}-\frac{e_{\rm in}}{ t_{e,\rm in}} \\
    \dot{e}_{\rm out} &= \mu_{\rm in}n_{\rm out}f_{\rm out}\sin\varphi_{\rm out} - \frac{e_{\rm out}}{ t_{e,\rm out}} \\
    \dot{\varpi}_{\rm in} &= \mu_{\rm out}n_{\rm in}\mathcal{R}_{a} f_{\rm in}\frac{\cos\varphi_{\rm in}}{e_{\rm in}} + \frac{3M_{\bullet}^{1.5}}{a_{\rm in}^{2.5}(1-e_{\rm in}^2)} \\
    \dot{\varpi}_{\rm out} &= \mu_{\rm in}n_{\rm out}f_{\rm out}\frac{\cos\varphi_{\rm out}}{e_{\rm out}} + \frac{3M_{\bullet}^{1.5}}{a_{\rm out}^{2.5}(1-e_{\rm out}^2)}
\end{align}
\end{subequations}

We use the labels `in' and `out' to represent the inner and outer objects, respectively.
These equations of motion satisfy only the $j+1:j$ resonance, where ${t_{J}, \ t_{e}}$ are the angular momentum loss timescale and eccentricity damping timescale, due to type I migration forces and gravitational radiation. Here, $a$ is the semi-major axis, $e$ is the eccentricity, $\mathcal{R}_{a} = a_{\rm in}/a_{\rm out}$ is the ratio of the semi-major axis, $n$ is the mean motion, $\varpi$ is the pericenter longitude, $\varphi$ is the resonance angle, $f$ is the resonance coefficient (which can be found in Table 8.1 of \cite{Murray1999}), and $\mu$ is the mass ratio $m/M_{\bullet}$.
We have added only the simple precession term from the 1PN effect to $\dot{\varpi}$.

If the orbital elements reach equilibrium, we can get $\dot{n}_{\rm in}/n_{\rm in}=\dot{n}_{\rm out}/n_{\rm out}$, $\dot{e}_{\rm in}=\dot{e}_{\rm out}=0$, $\dot{\varpi}_{\rm in}=\dot{\varpi}_{\rm out}$, and we can obtain the equilibrium eccentricity equation
\begin{align}
    &\left( j\frac{n_{\rm in}\mathcal{R}_{a}}{q} + (j+1)n_{\rm out}  \right) \left(\frac{qe_{\rm in}^2}{n_{\rm in} \alpha t_{e,\rm in}} + \frac{e_{\rm out}^2}{n_{\rm out} t_{e,\rm out}} \right) \notag \\
    &= \left(\frac{1}{ t_{J,\rm out}}-\frac{1}{ t_{J,\rm in}}\right) + \frac{e_{\rm out}^2}{ t_{e,\rm out}}-\frac{e_{\rm in}^2}{ t_{e,\rm in}}, \\
    &-\sqrt{\frac{\mu_{\rm out}n_{\rm in}^2\mathcal{R}_{a}^2f_{\rm in}^2}{e_{\rm in}^2}-\frac{1}{ t_{e,\rm in}^2}} + \frac{3M_{\bullet}^{1.5}}{a_{\rm in}^{2.5}} \notag \\ &= -\sqrt{\frac{\mu_{\rm in}^2n_{\rm out}^2f_{\rm out}^2}{e_{\rm out}^2}-\frac{1}{ t_{e,\rm out}^2}} + \frac{3M_{\bullet}^{1.5}}{a_{\rm out}^{2.5}},
\end{align}
we have ignored the eccentricity term $(1-e^2)$ in the 1PN precession term for the sake of simplicity in the calculation. Using these two equations, we can obtain the equilibrium eccentricity.

\begin{figure}
    \centering
    \includegraphics[width=1.0\linewidth]{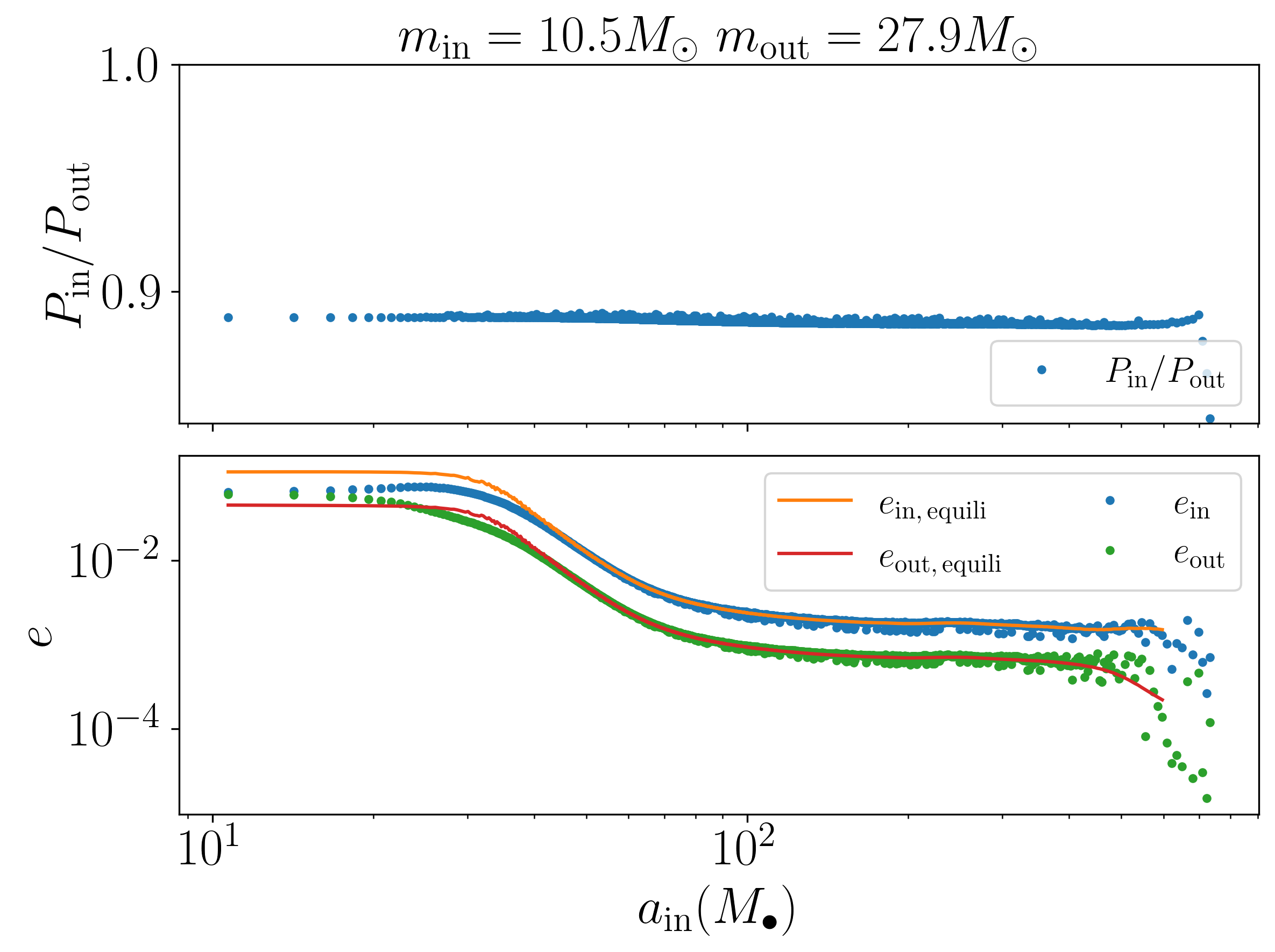}
    \caption{Two sBHs are captured into a 9:8 resonance under the combined effects of disk migration forces and 2.5PN gravitational-wave radiation. The two solid lines in the lower panel represent the equilibrium eccentricities of the inner and outer sBHs, computed using the eccentricity-equilibrium formula described above. Post-Newtonian effects are not included; hence, no 1PN precession term is present in $\dot{\varpi}$.}
    \label{fig:nobreak_new}
\end{figure}

To study the influence of the post-Newtonian effect on the evolution of a resonant pair, we plot the $9:8$ resonance in Fig.~\ref{fig:nobreak_new}, considering only the disk migration force and the 2.5PN gravitational radiation force. The figure clearly shows that the equilibrium eccentricity increases without the 1PN effect, while it decreases when the 1PN effect is included. 

\section{Migration and Accretion}\label{sec:type2}

In the wet EMRI mass spectrum section, we mention that the sBHs captured by the AGN disk will migrate toward the central MBH and accrete gas from the disk. For the accretion model, we adopt the same supercritical accretion method as in \cite{Pan:2021xhv}. A secondary disk forms around a sBH in the AGN disk. The inflow from the outer boundary of the secondary disk will not be completely accreted by the sBH; most of the inflow will escape as an outflow. In \cite{Pan:2021xhv}, the relation between the inflow from the outer boundary, $\dot{m}_{\rm in}$, and the actual accretion rate $\dot{m}_{\rm in, 0}$ was numerically calculated from the “inflow-outflow” model and fitted as 
\begin{equation}
\scalebox{1.0}{$
\dot{m}_{\rm in, 0} =
\begin{cases}
\dot{m}_{\rm in}, & \dot{m}_{\rm in}<\dot{m}_{\rm Edd} \\
\max\{(12.87-8.8\chi)\dot{m}_{\rm in}\frac{m}{r_{\rm obd}}, 1\}, &\dot{m}_{\rm in} \ge \dot{m}_{\rm Edd}
\end{cases}
$}
\end{equation}
in which $\dot{m}_{\rm Edd}$ is the Eddington accretion rate, $\chi$  
is dimensionless spin parameter of the sBH, $r_{\rm obd}$ is the inflow out boundary (see  Eqs.~(26-32) in \cite{Pan:2021xhv} for more details about the out boundary). 

During migration, an sBH will undergo type II or type-I migration, depending on whether it opens a gap. If the sBH is more massive, it may open a gap, which will significantly reduce the surface density around it.

We do not consider type II migration in Section~\ref{sec:mulres} because the post-Newtonian effect operates in the inner region of the AGN disk ($<100M_{\bullet}$), which corresponds to type I migration (without gaps) according to the following criteria. A gap opens if the mass ratio is greater than or equal to \cite{Lin1986,Kocsis2011DiskEMRI,Duffell2013,Kanagawa2015} 
\begin{align}
    q_{\rm crit} = \sqrt{25\alpha\beta^{b}h^5}
\end{align}
in which $q_{\rm crit}$ is the minimum mass ratio that can open a gap, $\alpha$ is the dimensionless disk parameter, $\beta=p_{\rm gas}/p_{\rm tot}$, $b=0$ for $\alpha$-disk and $b=1$ for $\beta$-disk, $h$ is the disk aspect ratio. In our model here, we use $\alpha=0.1$, the center MBH mass $M_{\bullet}=4\times10^6M_\odot$

\begin{figure}
    \centering
    \includegraphics[width=1.0\linewidth]{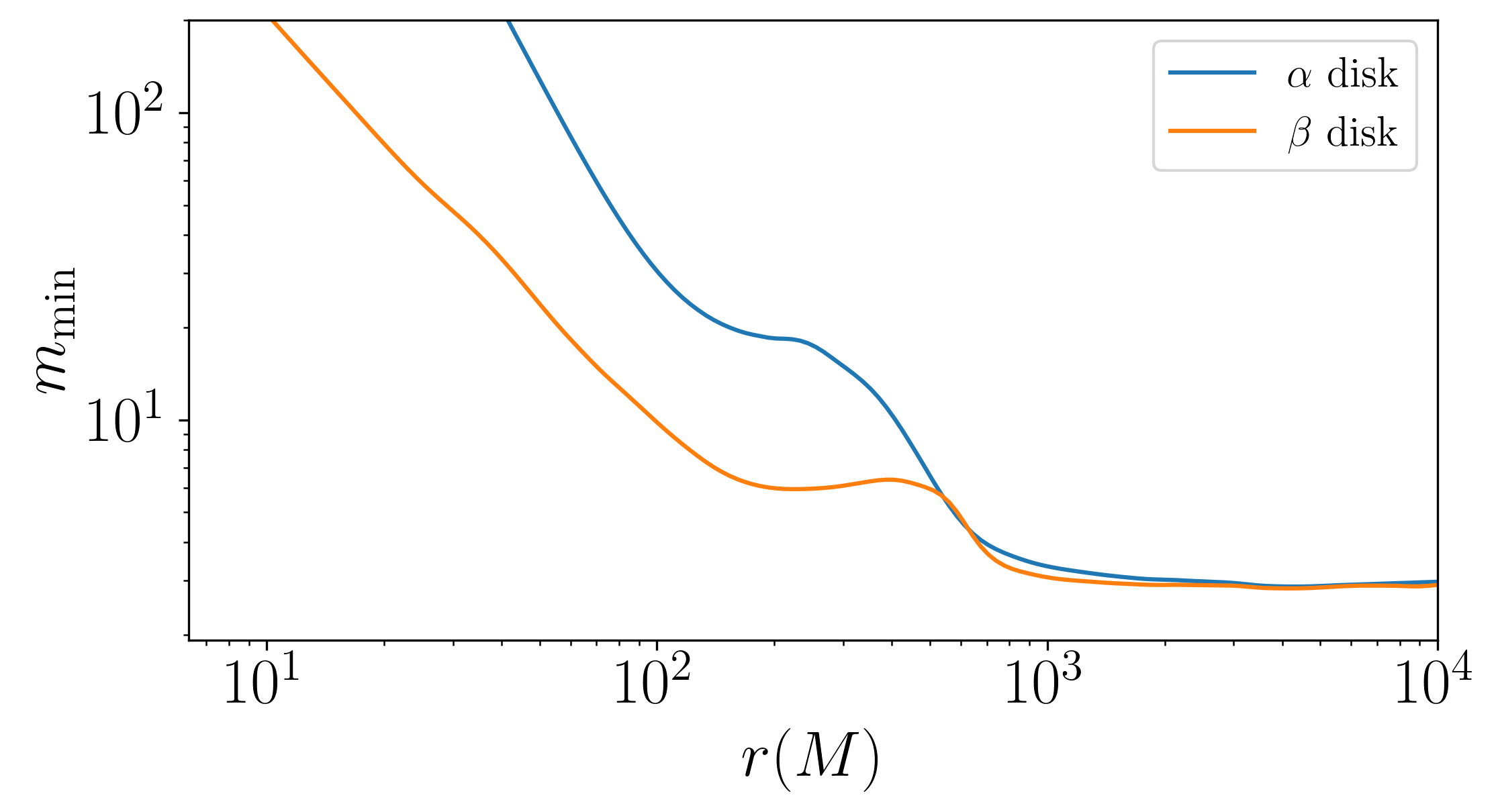}
    \includegraphics[width=1.0\linewidth]{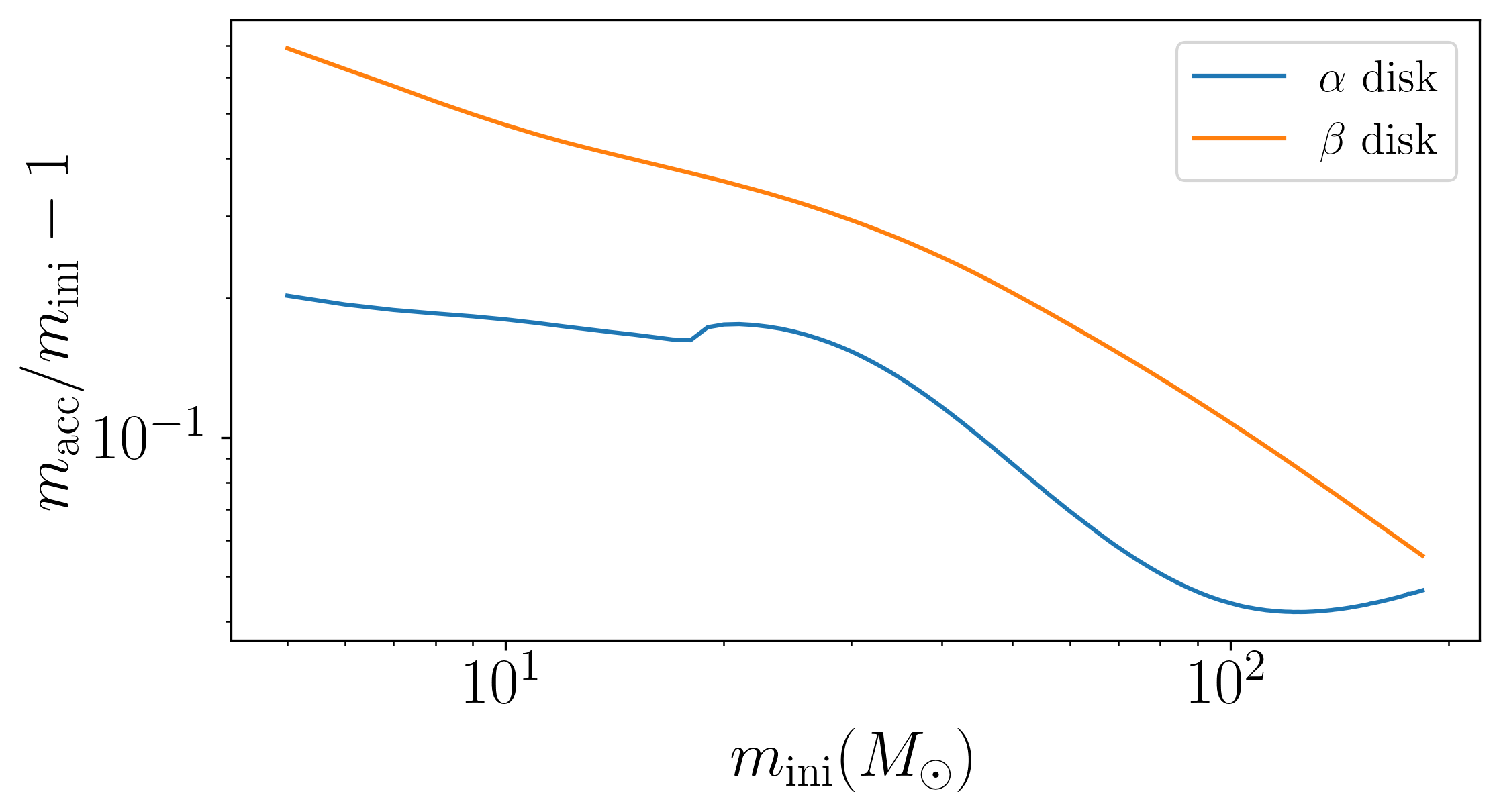}
    \caption{The upper panel shows the minimum sBH mass required to open a gap at different radii for the $\alpha$- and $\beta$-disks. The lower panel shows the ratio of mass increase after accretion relative to the initial mass, with accretion occurring from $1000M_{\bullet}$ to $10M_{\bullet}$ in both the $\alpha$- and $\beta$-disks.}
    \label{fig:gap}
\end{figure}

In the upper panel of Fig.~\ref{fig:gap}, we show the minimum sBH mass that can open a gap at different radii for the $\alpha$- and $\beta$-disk. For simplicity, during the sBH migration, we consider that it first undergoes type II migration until it reaches the radius where it can no longer open a gap and then undergoes type I migration.

For type I migration, the specific torques that act on the sBH have three components: \cite{Kocsis2011DiskEMRI, Pan:2021xhv}\\
(1) Type I migration torque
\begin{align}
    \dot{J}_{\rm mig, I} = C_{\rm I}\frac{m}{M_{\bullet}}\frac{\Sigma}{M_{\bullet}}\frac{r^4\Omega^2}{h^2}
\end{align}
in which $C_{\rm I}$ is the torque coefficient on the order of unity, $m$ is the sBH mass, $M_{\bullet}$ is the center MBH mass, $\Sigma$ is the surface density of the AGN disk, $\Omega$ is the gas Kepler angular velocity, $h$ is the aspect ratio. All disk quantities here are measured at the orbital location of the sBH. \\
(2) Gravitational wave torque
\begin{align}
    \dot{J}_{\rm gw} = -\frac{32}{5}\frac{m}{M_{\bullet}}\left(\frac{M_{\bullet}}{r}\right)^{3.5}
    \label{eq:Jgw}
\end{align}
which is Peter's formula \cite{Peters1964a}, assuming a circular orbit. \\
(3) Wind torque
\begin{align}
    \dot{J}_{\rm wind}=-\frac{r\delta v_\phi \dot{m}_{\rm wind}}{m}
\end{align}
in which $\delta v_\phi$ is the relative bulk velocity in the $\phi$ direction between the local gas and sBH, $\dot{m}_{\rm wind}$ is the head wind strength which determines $\dot{m}_{\rm in}$ at the outside boundary.

The head wind can be obtained from Bondi accretion
\begin{align}
    \dot{m}_{\rm wind}=\dot{m}_{\rm BHL}\times\min\left\{1, \frac{H}{r_{\rm BHL}}\right\}\times\min\left\{1, \frac{r_{\rm Hill}}{r_{\rm BHL}}\right\}
\end{align}
in which $\dot{m}_{\rm BHL}$ is the Bondi accretion rate, $r_{\rm BHL}$ is the Bondi radius, $H$ is the AGN disk height, $r_{\rm Hill}$ is the sBH Hill radius \cite{Pan:2021xhv}.

For type II migration, we only need to consider two components of specific torques acting on the sBH \cite{Kocsis2011DiskEMRI, Pan:2021xhv}: \\
(1) Type II migration torque \citep{Ida2025,Pan2025}  
\begin{align}
    \dot{J}_{\rm mig, II}=\dot{J}_{\rm mig, I}\times\frac{\zeta}{(1+K')(1+\zeta)}
\end{align}
with $K'=0.04q^2h^{-5}\alpha^{-1}$ and $\zeta=3\pi\alpha(1+K') \times \left(\sqrt{2/\pi}(qh^{-3})^{-2}+3^{-1/3}(qh^{-3})^{-2/3}\right)$, in which $q=m/M_{\bullet}$ is the mass ratio. \\
(2) Gravitational wave torque, same formula with Eq.~\ref{eq:Jgw}

The accretion rate of the sBH can be prescribed by combining Bondi accretion and Hill accretion \citep{Li2023,Li2024}
\begin{align}
    \dot{m}_{\rm in}=\frac{1}{1+K'}\left(\frac{\dot{m}_{\rm std}}{\dot{m}_{\rm B0}}+\frac{\dot{m}_{\rm std}}{\dot{m}_{\rm H0}}\right)^{-1} \times \dot{m}_{\rm std}
\end{align}
with 
\begin{align}
    \frac{\dot{m}_{\rm B0}}{\dot{m}_{\rm std}}&=\frac{\sqrt{\pi/2}}{3\pi}(qh^{-3})^2\alpha^{-1} \notag \\
    \frac{\dot{m}_{\rm H0}}{\dot{m}_{\rm std}}&=\frac{3^{1/3}}{3\pi}(qh^{-3})^{2/3}\alpha^{-1}
\end{align}
and $\dot{m}_{\rm std}=3\pi\alpha\Sigma_0H^2\Omega$ is the steady accretion rate of the AGN disk. The outer boundary of the sBH accretion for the sBH that opens the gap (type II migration) is the secondary disk size $\sim r_{\rm Hill}/3$ \citep{Fung2019,Li2023}.

In the lower panel of Fig.~\ref{fig:gap}, we show the ratio of the increase in mass after accretion compared to the initial mass from the outside $r\sim 1000M_{\bullet}$ of the AGN disk to the inside $\sim 10M_{\bullet}$. The sBH in $\beta$-disk always accretes more mass due to the dense gas environment than in $\alpha$-disk.

\section{Comparison with \cite{Mancieri2025_EccentricityEMRI}}\label{app:compare}

In a recent work \cite{Mancieri2025_EccentricityEMRI}, Mancieri \textit{et al.} studied the eccentricity distribution of dry EMRIs at plunge using Monte Carlo simulations and find a distribution peak at $e_{\rm pl} \approx 0.2$ (see Fig.~4 in \cite{Mancieri2025_EccentricityEMRI}). 
We note that their results are broadly consistent with ours. 
In \cite{Mancieri2025_EccentricityEMRI}, the numerical evolution of the nuclear stellar cluster ends when the time reaches ten times the EMRI rate peak time -- where the ``peak time'' refers to when the EMRI rate reach a maximal value -- while in this work we fix the evolution time as 5 Gyr. In addition, a slightly different definition of  GW timescale $t_{\rm GW}$ are used in \cite{Mancieri2025_HangingOnTheCliff} in defining boundaries of computational domain:
``$(1-e)$ decay timescale'' due to gravitational radiation (see Eq.~20 in \cite{Mancieri2025_HangingOnTheCliff}). Another difference lies in the way the semi-major axis $a$ and eccentricity $e$ are calculated from orbital energy $E$ and angular momentum $J$.  In this work, we compute the pericenter $r_p$ and apocenter $r_a$ in the Newtonian potential of the central MBH plus nuclear cluster, then define semi-major axis as $a=(r_a+r_p)/2$ and eccentricity as $e=(r_a-r_p)/(r_a+r_p)$. In Ref.~\cite{Mancieri2025_EccentricityEMRI}, the authors use orbital parameter definitions in Schwarzschild spacetime framework $(E,L)\to (p,e)$ \cite{Cutler1994SchwarzschildRR}, where $p$ is the semi-latus rectum. 

\begin{figure}
    \centering
    \includegraphics[width=0.9\linewidth]{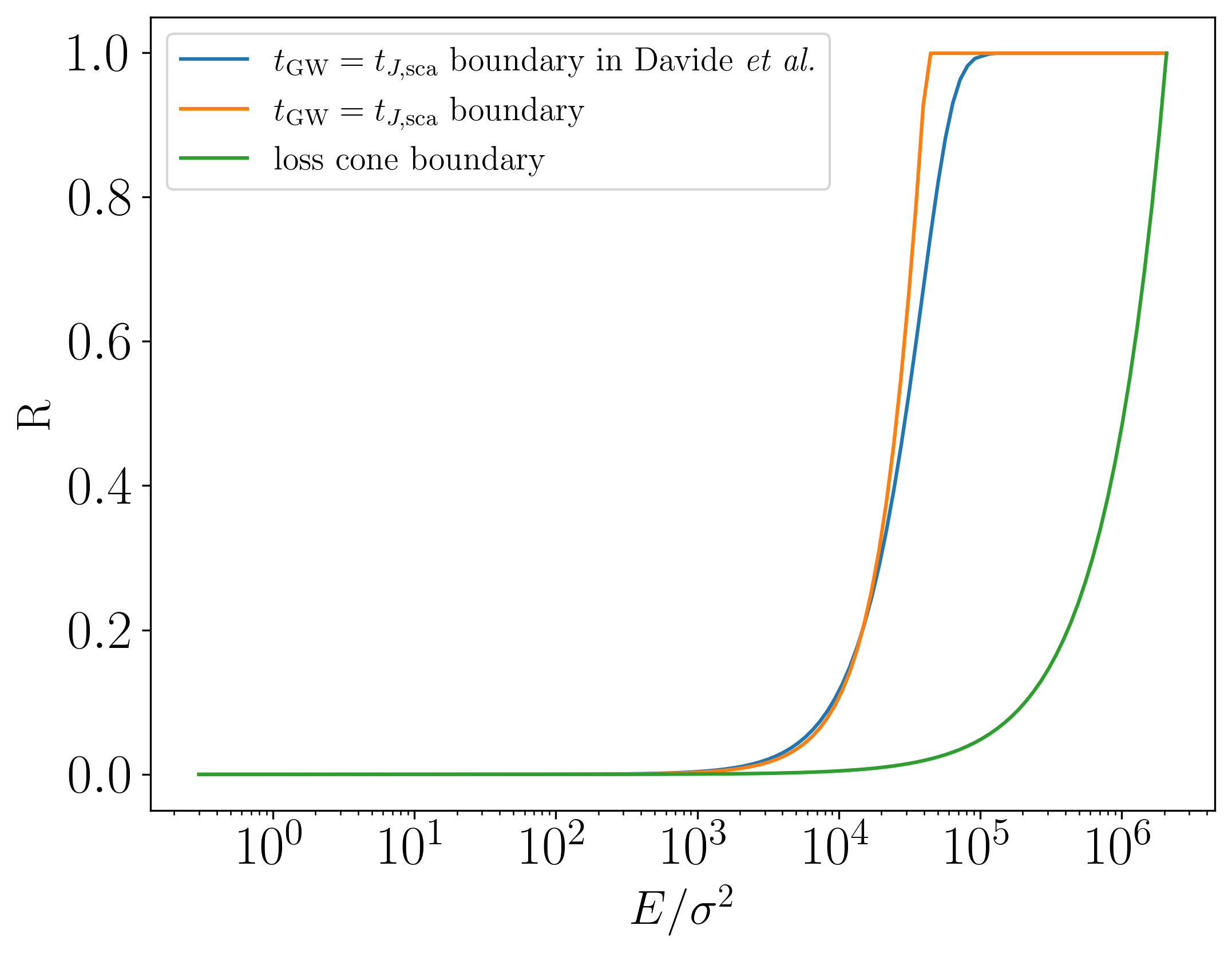}
    \caption{Boundaries adopted in this work. The green solid line marks the loss-cone boundary. The orange line denotes the $t_{\rm GW}=t_{J,\rm sca}$ boundary used in the main text. The blue line shows the $t_{\rm GW}=t_{J,\rm sca}$ boundary as defined in \cite{Mancieri2025_HangingOnTheCliff}, which is employed in this appendix to derive the plunge-eccentricity distribution from the flux along this boundary.}
    \label{fig:bd_compa}
\end{figure}
\begin{figure*}
    \centering
    \includegraphics[width=0.45\textwidth]{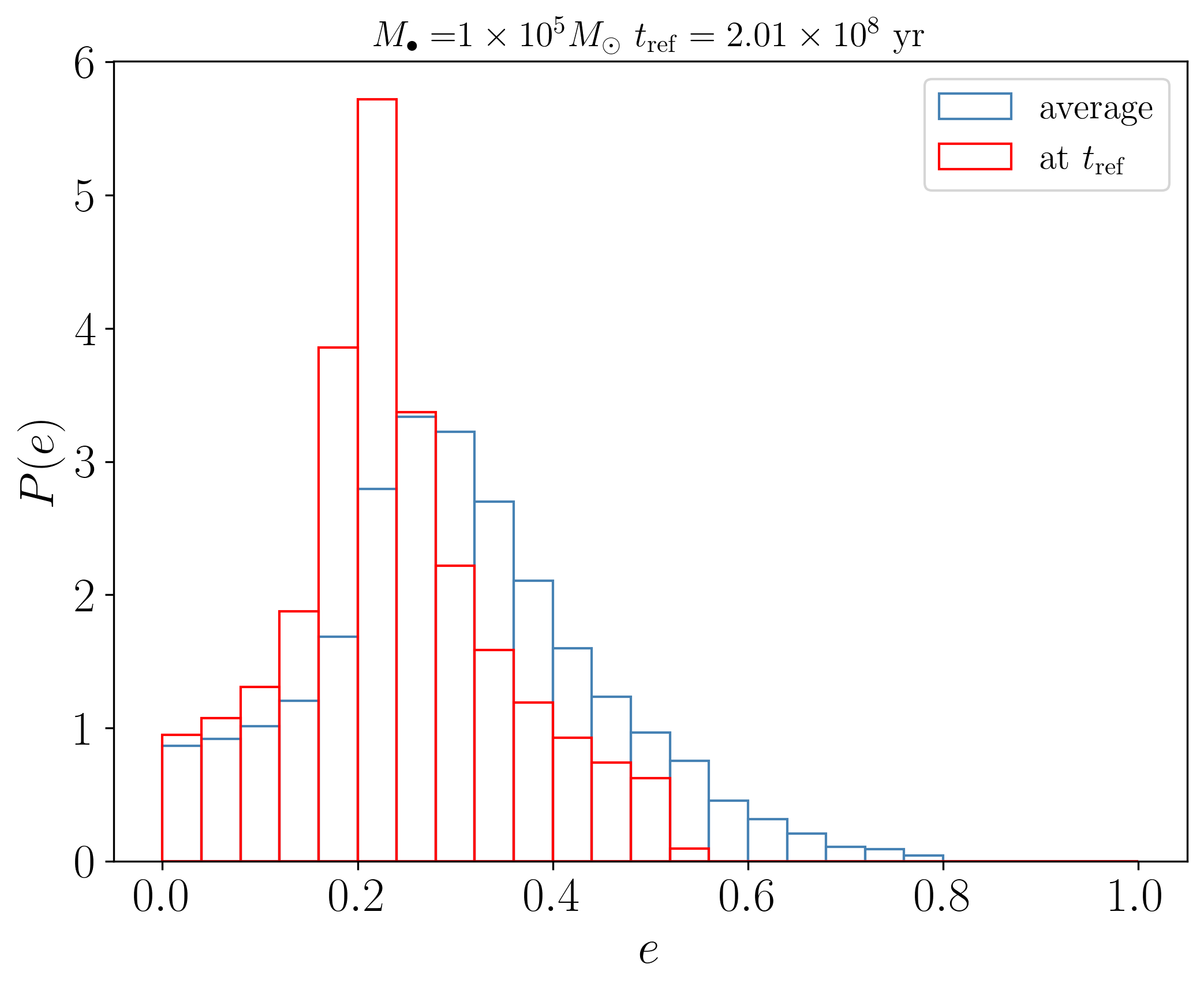}
    \includegraphics[width=0.45\textwidth]{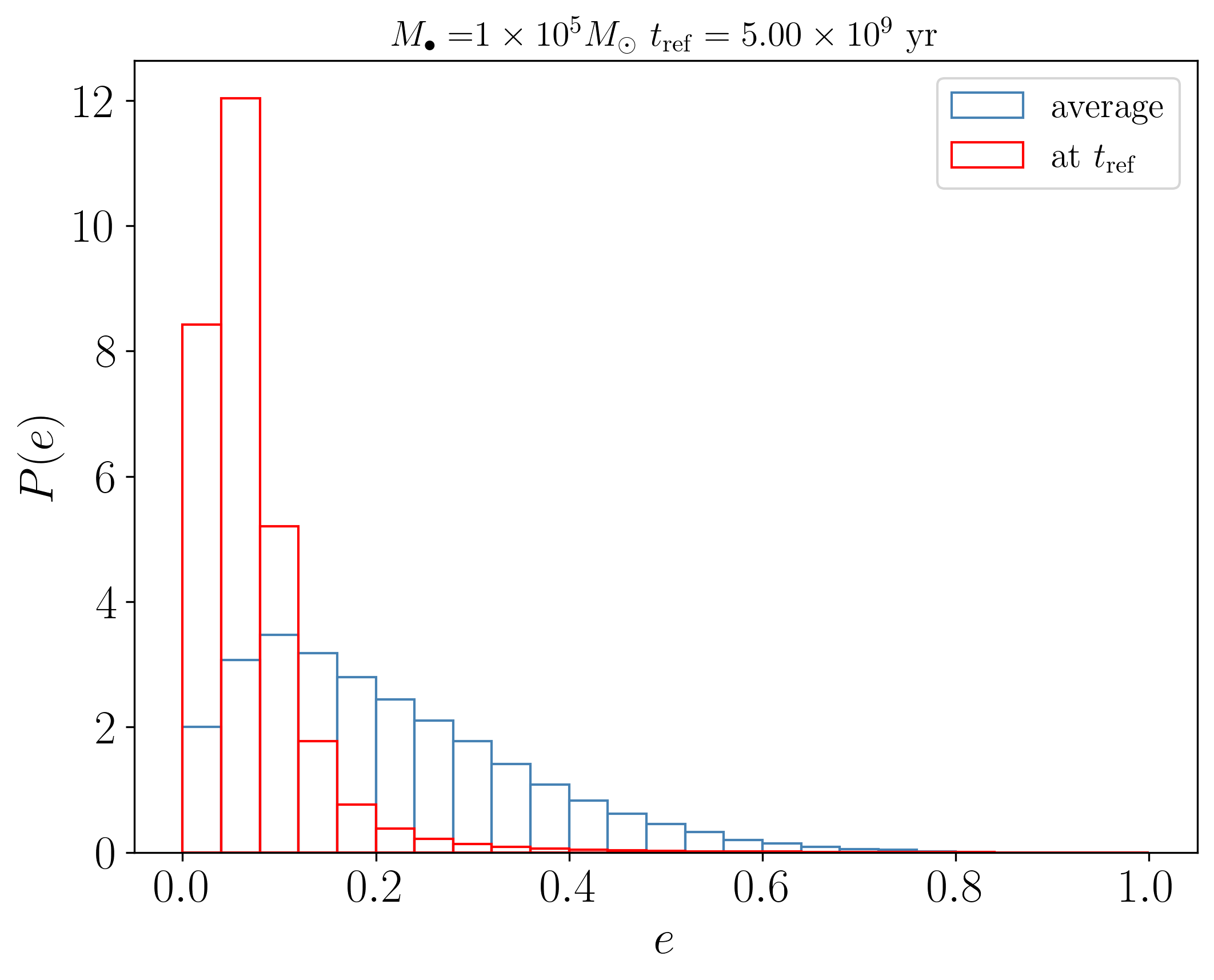}
    \includegraphics[width=0.45\textwidth]{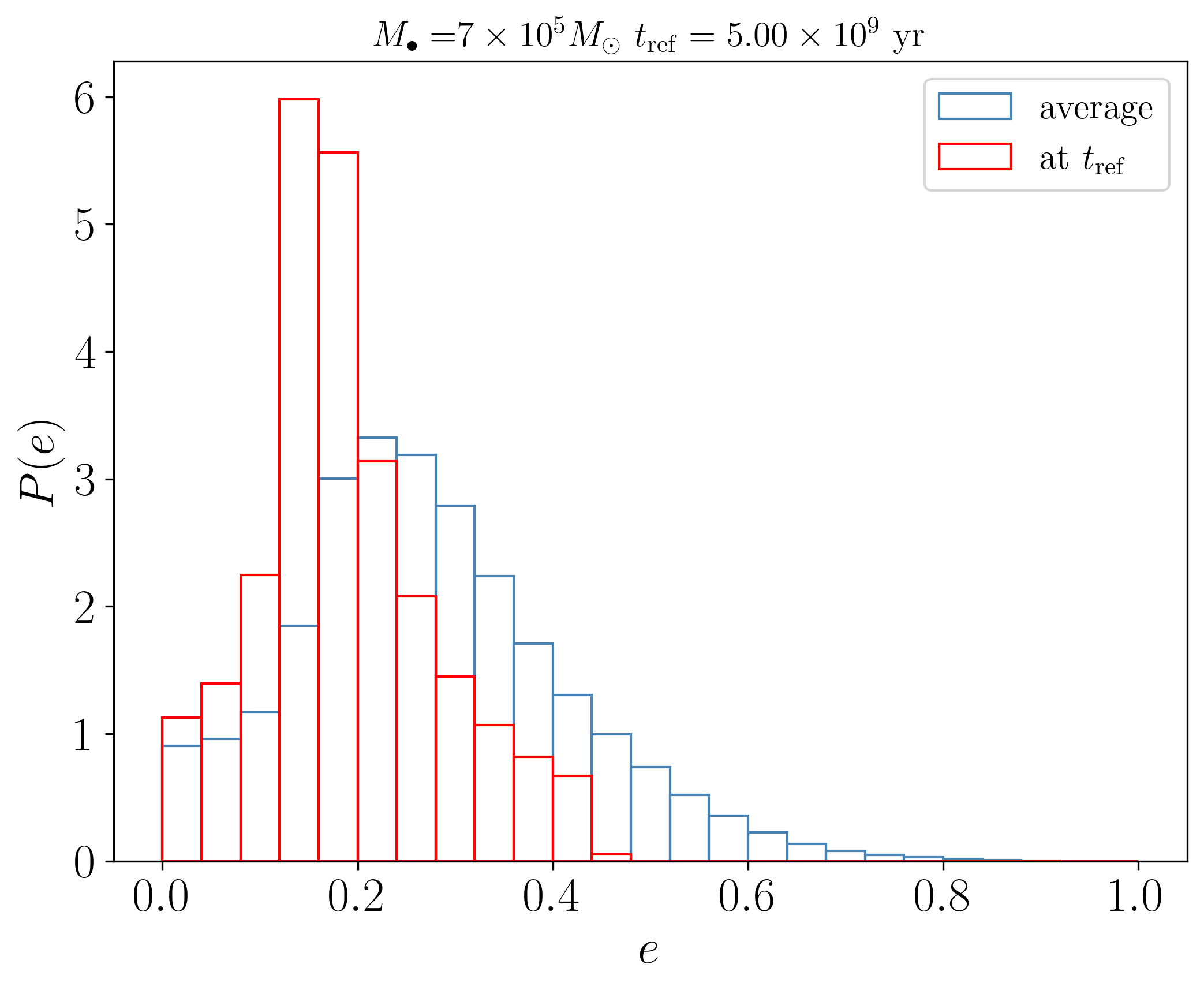}
    \includegraphics[width=0.45\textwidth]{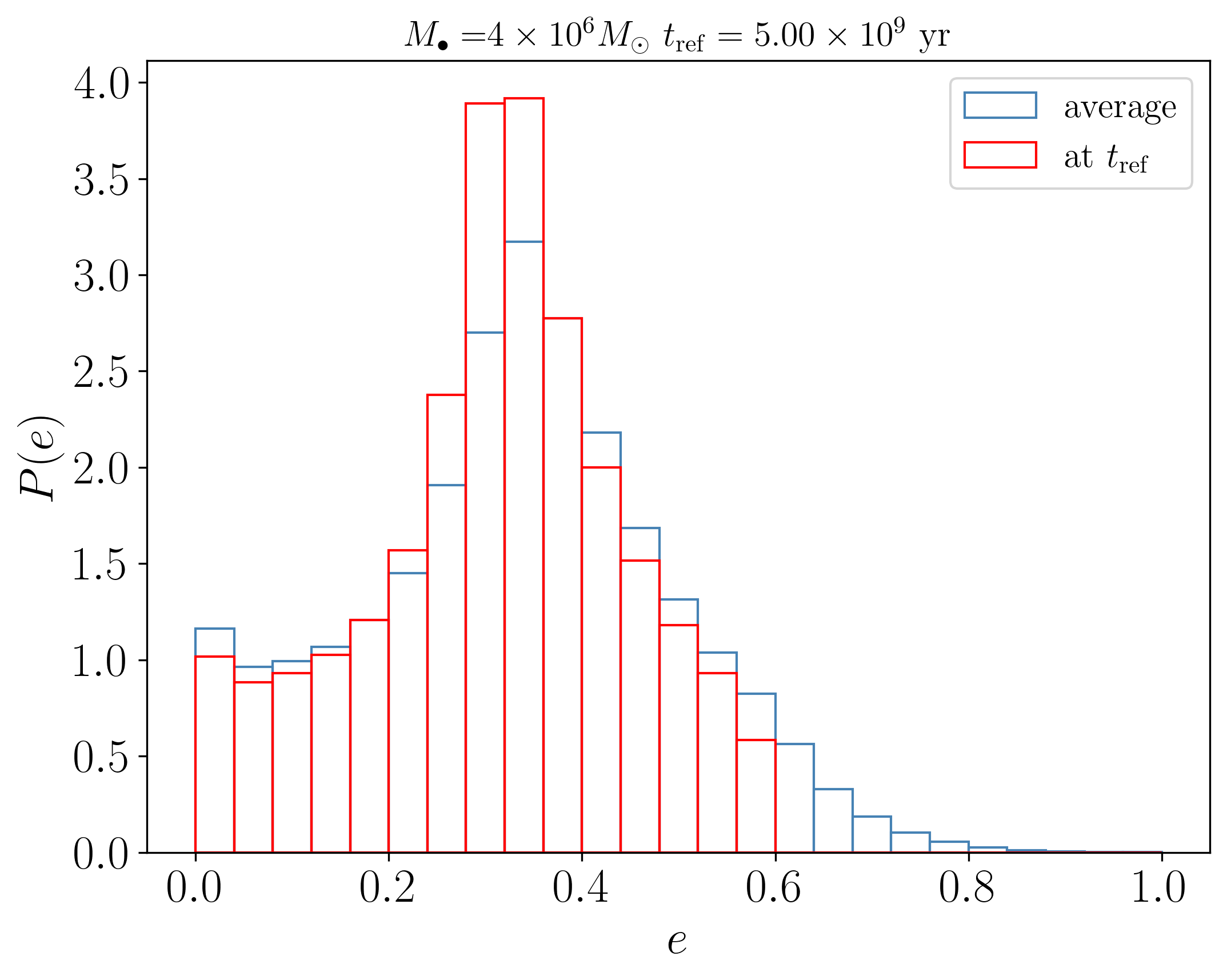}
    \caption{Eccentricity distributions at plunge for different MBH systems. We apply the same treatment as in \cite{Mancieri2025_EccentricityEMRI} and obtain similar distributions. The blue histograms show the time-averaged eccentricity distributions up to the times indicated in the titles, while the red histograms represent the plunge-eccentricity distributions at those times. These figures demonstrate that our FP results are broadly consistent with the Monte Carlo simulations of \cite{Mancieri2025_EccentricityEMRI}.}
    \label{fig:ecc_compa}
\end{figure*}

In Fig.~\ref{fig:bd_compa}, we show the boundaries obtained when adopting different gravitational-radiation timescales. The difference is only visible in the low-eccentricity regime, where a minor fraction of EMRIs are produced. For comparison with \cite{Mancieri2025_EccentricityEMRI}, we now use the same relativistic definitions of $a$, $e$ and calculate the eccentricity distribution of dry EMRIs at plunge, i.e., at $p=6+2e$ \cite{Cutler1994SchwarzschildRR}, and show the results in Fig.~\ref{fig:ecc_compa}. For $M_\bullet=1\times10^5M_\odot$, we display two time snapshots. The snapshot in $t_{\rm ref}\approx2\times10^8{\rm yr}$ corresponds to ten times the EMRI-rate peak time, and the dry EMRI eccentricity profile is similar to that in \cite{Mancieri2025_EccentricityEMRI}. The snapshot at $t_{\rm ref}=5\times10^9 {\rm yr}$ shows the distribution after 5 Gyr of evolution, indicating that the eccentricity distribution also evolves over time, with the peak decreasing. In addition, we also show the figures for two other MBHs. For $M_\bullet=7\times10^5 M_\odot$, the peak EMRI rate time is approximately 5 Gyr, and the figure shows a distribution similar to that for $M_\bullet=1\times10^5 M_\odot$ at 0.2 Gyr. For $M_\bullet=4\times10^6 M_\odot$, 5 Gyr is less than ten times the peak time of the EMRI rate, and the averaged eccentricity distribution shows a higher peak at $e\simeq0.3$.

We thereby confirm that our results are broadly consistent with those in \cite{Mancieri2025_EccentricityEMRI}. We also note that the average distribution evolves with time because of the limit supply. However, the distributions are not fully consistent because the authors include gravitational wave radiation in their Monte Carlo simulations, which can naturally produce the so-called ``cliff EMRI'' \cite{Qunbar2023, Mancieri2025_HangingOnTheCliff}. As a result,  an additional small fraction of EMRIs of high eccentricities are found \cite{Mancieri2025_EccentricityEMRI}.  We plan to update our FP code accordingly in the future.

\section{Physical Quantities}\label{app:phy_q}

In this appendix, we summarize the physical quantities used in this paper and provide their physical meanings in Table~\ref{tab:symbols}. This compilation is intended to serve as a convenient reference for readers and to ensure consistency in notation throughout the main text.

\clearpage

\renewcommand\arraystretch{1.5} 

\begin{longtable*}{C{0.24\textwidth} L{0.24\textwidth}  C{0.24\textwidth} L{0.24\textwidth}}
\caption{\textbf{Symbols}}\label{tab:symbols} \\
\toprule
\hline 
\textbf{Physical Quantity} & \textbf{Physical Meaning} & \textbf{Physical Quantity} & \textbf{Physical Meaning} \\
\midrule
\endfirsthead

\multicolumn{4}{c}%
{{Table \thetable\ \textbf{Continued}}} \\
\toprule
\hline
\textbf{Physical Quantity} & \textbf{Physical Meaning} & \textbf{Physical Quantity} & \textbf{Physical Meaning} \\
\endhead

\endfoot

\bottomrule
\endlastfoot

$M_{\bullet}$ & Mass of center MBH & $t_{\rm GW}$ & Gravitational wave damping timescale for semi-major axis \\[1ex]
\hline
$M$ & Total mass of the MBH and sBH & $t_{J, \rm sca}$ & Gravitational scattering timescale  \\[1ex]
\hline
$\eta$ & Symmetrical mass ratio & $\tilde{\gamma}$ & Power-law index of MBH mass function \\[1ex]
\hline
$m$ & Mass of sBH & $\dot{M}_{\rm Edd}$ &  Eddington accretion rate for the MBH \\[1ex]
\hline
$q$ & Mass ratio between sBH and MBH & $\dot{M}_{\bullet}$ &  Mass accretion rate for the MBH \\[1ex]
\hline
$\alpha$ & Viscosity parameter & $h$ & Disk aspect ratio \\[1ex]
\hline
$E_{\rm cr}$ & Critical energy distinguishing EMRI and Plunge & $H$ & Disk scale height \\[1ex]
\hline
$E_{\rm max}$ & Maximum energy boundary in FP equation & $\Sigma(r), \Sigma_{0}$ & Disk surface density at $r$ and $r_{0}$ \\[1ex]
\hline
$J_{c}(E)$ & Angular momentum on circular orbit at given energy & $\Phi_{\rm turb}$ & Potential for the turbulent disk \\[1ex]
\hline
$J_{\rm lc}$ & Angular momentum on loss cone boundary & $\gamma$ & Dimensionless strength for turbulence \\[1ex]
\hline
$J$ & Angular momentum of sBH & $\omega$ or $\Omega$ & Angular velocity \\[1ex]
\hline
$E$ & Orbital energy of sBH & $\chi$ & Dimensionless spin parameter \\[1ex]
\hline
$R$ & Dimensionless angular momentum variable in FP equation & $n_{k}$ &  Wavenumber of the turbulence mode \\[1ex]
\hline
$N(E,R),f(E,R),C(E,R)$ & Number density in $(E,R)$ phase-space ($N$) and position-velocity space ($f$), weight function ($C$) & $r_{k},\phi_{k}$ & Radial position and azimuthal angle of each turbulence mode \\[1ex]
\hline
$F_{E},F_{R}$ & Phase space flux along $E/R$ direction in FP equation & $\gamma_{e},\tau_{e}$ & Eccentricity damping rate, timescale for the turbulent disk \\[1ex]
\hline
$D_{EE},D_{RR},D_{ER}, D_E, D_R$ & Diffusion and advection coefficients & $S_{\delta r},S_{f_{r}},S_{\tau_\phi}$ & Power spectrum of radial motion, radial force, and azimuthal torque in the turbulent disk \\[1ex]
\hline
$a,a_{\rm in},a_{\rm out}$ & Semi-major axis of the (inner/outer) object & $t_{J},t_{J,\rm in},t_{J,\rm out}$ & Disk-driven damping timescale for angular momentum of (inner/outer) object \\[1ex]
\hline
$e,e_{\rm in},e_{\rm out}$ & Eccentricity of the (inner/outer) object & $t_e,t_{e,\rm in},t_{e,\rm out}$ & Disk-driven damping timescale for eccentricity of (inner/outer) object \\[1ex]
\hline
$\Gamma_{\rm EMRI}$ & EMRI rate & $\iota$ & Inclination angle \\[1ex]
\hline
$\phi_{\rm in},\phi_{\rm out}$ & Resonance angle of inner and outer objects & $f_{r},\tau_{\phi}$ &  Specific gravitational radial force and torque for the turbulent disk \\[1ex]
\hline
$\varpi_{\rm in},\varpi_{\rm out}$ & Longitude of pericenter of inner and outer objects & $f_{r}(\omega),\tau_{\phi}(\omega)$ &  Specific gravitational radial force and torque in the frequency domain for the turbulent disk \\[1ex]
\hline
$r_{a},r_{p}$ & Apocenter, Pericenter of the sBH & $\delta F_{r},\delta \Gamma$ &  Total stochastic radial force and torque for the turbulent disk \\[1ex]
\hline
 $\epsilon$  &  Radiative efficiency of the AGN disk & $F_{r},\Gamma$ & Total radial force and torque for the turbulent disk  \\[1ex]
\hline
$R_{\rm warp}$ & Warped disk radius  & $P_{\rm LVK}$ & LVK primary mass distribution    \\[1ex]
\hline
$R_{\rm dec}$ & Decoupling radius where LT torque dominates over disk torques & $S(x)$ & Smooth function   \\[1ex]
\hline
$\tau_{\rm BP}$ & Bardeen-Petterson timescale & $g(q)$  & Pairing function  \\[1ex]
\hline
$J_{\rm d}$ & Disk angular momentum & $F(m_1, m_2)$ & Joint distribution of binary mass  \\[1ex]
\hline
$J_{\bullet}$ & MBH spin angular momentum & $\tilde\beta$ & Power-law index of pairing function  \\[1ex]
\hline
$\alpha_{p},\mu_{m},\sigma_{m},\delta_{m},\lambda_{\rm peak}$ & Hyper-parameters of LVK primary BH mass distribution & $\tilde\alpha_{p},\tilde\mu_{m},\tilde\sigma_{m},\tilde\delta_{m},\tilde\lambda_{\rm peak}$ & Hyper-parameters of single BH mass distribution  \\[1ex]
\hline
$P_{\rm cap}$ & Wet EMRI mass distribution after being captured & $P_{\rm acc}$ &  Wet EMRI mass distribution after accreting \\[1ex]
\hline
$f_{\rm mer}$ & Merger fraction & $P_{\rm f=1}$ & Wet EMRI mass distribution after merging for $f_{\rm mer}=1$  \\[1ex]
\hline
$q_{\rm crit}$ & Critical mass ratio between type-I and type-II migration & $P_{\rm mer}$ &  Wet EMRI mass distribution after merging \\[1ex]
\hline
$\dot{J}_{\rm mig, I}$ & Type-I migration torque & $\dot{J}_{\rm gw}$ & Gravitational radiation torque \\[1ex]
\hline
$\dot{J}_{\rm wind}$ & Wind torque & $\dot{m}_{\rm wind}$ &  Effective headwind mass flux \\[1ex]
\hline
$\dot{J}_{\rm mig,II}$ & Type-II migration torque & $\dot{m}_{\rm in}$ &  Mass inflow at outer boundary of the sBH \\[1ex]
\hline
$\dot{m}_{\rm std}$ & Steady accretion rate of AGN disk & $T_{\rm disk}$ & Disk lifetime for AGN \\[1ex]
\hline
$\Gamma(m,t)$ & Per-object EMRI rate & $\Gamma_{\rm cap}(m,t)$ & Per-object capture rate \\[1ex]

\end{longtable*}

\clearpage

    \bibliography{reference}

\begin{thebibliography}{109}%
\makeatletter
\providecommand \@ifxundefined [1]{%
 \@ifx{#1\undefined}
}%
\providecommand \@ifnum [1]{%
 \ifnum #1\expandafter \@firstoftwo
 \else \expandafter \@secondoftwo
 \fi
}%
\providecommand \@ifx [1]{%
 \ifx #1\expandafter \@firstoftwo
 \else \expandafter \@secondoftwo
 \fi
}%
\providecommand \natexlab [1]{#1}%
\providecommand \enquote  [1]{``#1''}%
\providecommand \bibnamefont  [1]{#1}%
\providecommand \bibfnamefont [1]{#1}%
\providecommand \citenamefont [1]{#1}%
\providecommand \href@noop [0]{\@secondoftwo}%
\providecommand \href [0]{\begingroup \@sanitize@url \@href}%
\providecommand \@href[1]{\@@startlink{#1}\@@href}%
\providecommand \@@href[1]{\endgroup#1\@@endlink}%
\providecommand \@sanitize@url [0]{\catcode `\\12\catcode `\$12\catcode `\&12\catcode `\#12\catcode `\^12\catcode `\_12\catcode `\%12\relax}%
\providecommand \@@startlink[1]{}%
\providecommand \@@endlink[0]{}%
\providecommand \url  [0]{\begingroup\@sanitize@url \@url }%
\providecommand \@url [1]{\endgroup\@href {#1}{\urlprefix }}%
\providecommand \urlprefix  [0]{URL }%
\providecommand \Eprint [0]{\href }%
\providecommand \doibase [0]{https://doi.org/}%
\providecommand \selectlanguage [0]{\@gobble}%
\providecommand \bibinfo  [0]{\@secondoftwo}%
\providecommand \bibfield  [0]{\@secondoftwo}%
\providecommand \translation [1]{[#1]}%
\providecommand \BibitemOpen [0]{}%
\providecommand \bibitemStop [0]{}%
\providecommand \bibitemNoStop [0]{.\EOS\space}%
\providecommand \EOS [0]{\spacefactor3000\relax}%
\providecommand \BibitemShut  [1]{\csname bibitem#1\endcsname}%
\let\auto@bib@innerbib\@empty
\bibitem [{\citenamefont {{LIGO Scientific Collaboration and Virgo Collaboration}}(2021)}]{Abbott2021_GWTC2Pop}%
  \BibitemOpen
  \bibfield  {author} {\bibinfo {author} {\bibnamefont {{LIGO Scientific Collaboration and Virgo Collaboration}}},\ }\bibfield  {title} {\bibinfo {title} {Population properties of compact objects from the second ligo-virgo gravitational-wave transient catalog},\ }\href {https://doi.org/10.3847/2041-8213/abe949} {\bibfield  {journal} {\bibinfo  {journal} {Astrophysical Journal Letters}\ }\textbf {\bibinfo {volume} {913}},\ \bibinfo {pages} {L7} (\bibinfo {year} {2021})},\ \bibinfo {note} {47 compact binary mergers; false-alarm rate < 1 yr$^{-1}$},\ \Eprint {https://arxiv.org/abs/2010.14533} {arXiv:2010.14533 [astro-ph.HE]} \BibitemShut {NoStop}%
\bibitem [{\citenamefont {{LIGO Scientific Collaboration and Virgo Collaboration and KAGRA Collaboration}}(2023)}]{Abbott2023PopulationGWTC3}%
  \BibitemOpen
  \bibfield  {author} {\bibinfo {author} {\bibnamefont {{LIGO Scientific Collaboration and Virgo Collaboration and KAGRA Collaboration}}},\ }\bibfield  {title} {\bibinfo {title} {The population of merging compact binaries inferred using gravitational waves through gwtc-3},\ }\href {https://doi.org/10.1103/PhysRevX.13.011048} {\bibfield  {journal} {\bibinfo  {journal} {Physical Review X}\ }\textbf {\bibinfo {volume} {13}},\ \bibinfo {pages} {011048} (\bibinfo {year} {2023})},\ \bibinfo {note} {r.Abbott {\it et al.} (LIGO Scientific, Virgo, and KAGRA Collaborations)},\ \Eprint {https://arxiv.org/abs/2111.03634} {arXiv:2111.03634 [astro-ph.HE]} \BibitemShut {NoStop}%
\bibitem [{\citenamefont {{The LIGO Scientific Collaboration, the Virgo Collaboration, and the KAGRA Collaboration}}(2025)}]{LIGOScientific:2025pvj}%
  \BibitemOpen
  \bibfield  {author} {\bibinfo {author} {\bibnamefont {{The LIGO Scientific Collaboration, the Virgo Collaboration, and the KAGRA Collaboration}}} (\bibinfo {collaboration} {LIGO Scientific, VIRGO, KAGRA}),\ }\bibfield  {title} {\bibinfo {title} {{GWTC-4.0: Population Properties of Merging Compact Binaries}},\ }\href@noop {} {\bibfield  {journal} {\bibinfo  {journal} {arXiv preprint}\ } (\bibinfo {year} {2025})},\ \Eprint {https://arxiv.org/abs/2508.18083} {arXiv:2508.18083 [astro-ph.HE]} \BibitemShut {NoStop}%
\bibitem [{\citenamefont {Mapelli}(2020)}]{Mapelli2021BBHformation}%
  \BibitemOpen
  \bibfield  {author} {\bibinfo {author} {\bibfnamefont {M.}~\bibnamefont {Mapelli}},\ }\bibfield  {title} {\bibinfo {title} {Binary black hole mergers: Formation and populations},\ }\href {https://doi.org/10.3389/fspas.2020.00038} {\bibfield  {journal} {\bibinfo  {journal} {Frontiers in Astronomy and Space Sciences}\ }\textbf {\bibinfo {volume} {7}},\ \bibinfo {pages} {38} (\bibinfo {year} {2020})},\ \Eprint {https://arxiv.org/abs/2005.12455} {arXiv:2005.12455 [astro-ph.HE]} \BibitemShut {NoStop}%
\bibitem [{\citenamefont {Mandel}\ and\ \citenamefont {Broekgaarden}(2022)}]{Mandel:2021smh}%
  \BibitemOpen
  \bibfield  {author} {\bibinfo {author} {\bibfnamefont {I.}~\bibnamefont {Mandel}}\ and\ \bibinfo {author} {\bibfnamefont {F.~S.}\ \bibnamefont {Broekgaarden}},\ }\bibfield  {title} {\bibinfo {title} {{Rates of compact object coalescences}},\ }\href {https://doi.org/10.1007/s41114-021-00034-3} {\bibfield  {journal} {\bibinfo  {journal} {Living Rev. Rel.}\ }\textbf {\bibinfo {volume} {25}},\ \bibinfo {pages} {1} (\bibinfo {year} {2022})},\ \Eprint {https://arxiv.org/abs/2107.14239} {arXiv:2107.14239 [astro-ph.HE]} \BibitemShut {NoStop}%
\bibitem [{\citenamefont {Gerosa}\ and\ \citenamefont {Fishbach}(2021)}]{Gerosa:2021mno}%
  \BibitemOpen
  \bibfield  {author} {\bibinfo {author} {\bibfnamefont {D.}~\bibnamefont {Gerosa}}\ and\ \bibinfo {author} {\bibfnamefont {M.}~\bibnamefont {Fishbach}},\ }\bibfield  {title} {\bibinfo {title} {{Hierarchical mergers of stellar-mass black holes and their gravitational-wave signatures}},\ }\href {https://doi.org/10.1038/s41550-021-01398-w} {\bibfield  {journal} {\bibinfo  {journal} {Nature Astron.}\ }\textbf {\bibinfo {volume} {5}},\ \bibinfo {pages} {749} (\bibinfo {year} {2021})},\ \Eprint {https://arxiv.org/abs/2105.03439} {arXiv:2105.03439 [astro-ph.HE]} \BibitemShut {NoStop}%
\bibitem [{\citenamefont {Gr{\"o}bner}\ \emph {et~al.}(2020)\citenamefont {Gr{\"o}bner}, \citenamefont {Ishibashi}, \citenamefont {Tiwari}, \citenamefont {Haney},\ and\ \citenamefont {Jetzer}}]{Grobner:2020drr}%
  \BibitemOpen
  \bibfield  {author} {\bibinfo {author} {\bibfnamefont {M.}~\bibnamefont {Gr{\"o}bner}}, \bibinfo {author} {\bibfnamefont {W.}~\bibnamefont {Ishibashi}}, \bibinfo {author} {\bibfnamefont {S.}~\bibnamefont {Tiwari}}, \bibinfo {author} {\bibfnamefont {M.}~\bibnamefont {Haney}},\ and\ \bibinfo {author} {\bibfnamefont {P.}~\bibnamefont {Jetzer}},\ }\bibfield  {title} {\bibinfo {title} {{Binary black hole mergers in AGN accretion discs: gravitational wave rate density estimates}},\ }\href {https://doi.org/10.1051/0004-6361/202037681} {\bibfield  {journal} {\bibinfo  {journal} {Astron. Astrophys.}\ }\textbf {\bibinfo {volume} {638}},\ \bibinfo {pages} {A119} (\bibinfo {year} {2020})},\ \Eprint {https://arxiv.org/abs/2005.03571} {arXiv:2005.03571 [astro-ph.GA]} \BibitemShut {NoStop}%
\bibitem [{\citenamefont {Antonini}\ and\ \citenamefont {Rasio}(2016)}]{Antonini:2016gqe}%
  \BibitemOpen
  \bibfield  {author} {\bibinfo {author} {\bibfnamefont {F.}~\bibnamefont {Antonini}}\ and\ \bibinfo {author} {\bibfnamefont {F.~A.}\ \bibnamefont {Rasio}},\ }\bibfield  {title} {\bibinfo {title} {{Merging black hole binaries in galactic nuclei: implications for advanced-LIGO detections}},\ }\href {https://doi.org/10.3847/0004-637X/831/2/187} {\bibfield  {journal} {\bibinfo  {journal} {Astrophys. J.}\ }\textbf {\bibinfo {volume} {831}},\ \bibinfo {pages} {187} (\bibinfo {year} {2016})},\ \Eprint {https://arxiv.org/abs/1606.04889} {arXiv:1606.04889 [astro-ph.HE]} \BibitemShut {NoStop}%
\bibitem [{\citenamefont {Babak}\ \emph {et~al.}(2017)\citenamefont {Babak}, \citenamefont {Gair}, \citenamefont {Sesana}, \citenamefont {Barausse}, \citenamefont {Sopuerta}, \citenamefont {Berry}, \citenamefont {Berti}, \citenamefont {Amaro‑Seoane}, \citenamefont {Petiteau},\ and\ \citenamefont {Klein}}]{Babak2017_LISA_EMRI}%
  \BibitemOpen
  \bibfield  {author} {\bibinfo {author} {\bibfnamefont {S.}~\bibnamefont {Babak}}, \bibinfo {author} {\bibfnamefont {J.~R.}\ \bibnamefont {Gair}}, \bibinfo {author} {\bibfnamefont {A.}~\bibnamefont {Sesana}}, \bibinfo {author} {\bibfnamefont {E.}~\bibnamefont {Barausse}}, \bibinfo {author} {\bibfnamefont {C.~F.}\ \bibnamefont {Sopuerta}}, \bibinfo {author} {\bibfnamefont {C.~P.~L.}\ \bibnamefont {Berry}}, \bibinfo {author} {\bibfnamefont {E.}~\bibnamefont {Berti}}, \bibinfo {author} {\bibfnamefont {P.}~\bibnamefont {Amaro‑Seoane}}, \bibinfo {author} {\bibfnamefont {A.}~\bibnamefont {Petiteau}},\ and\ \bibinfo {author} {\bibfnamefont {A.}~\bibnamefont {Klein}},\ }\bibfield  {title} {\bibinfo {title} {Science with the space‑based interferometer {LISA}. v: Extreme mass‑ratio inspirals},\ }\href {https://doi.org/10.1103/PhysRevD.95.103012} {\bibfield  {journal} {\bibinfo  {journal} {Physical Review D}\ }\textbf {\bibinfo {volume} {95}},\ \bibinfo {pages} {103012} (\bibinfo {year} {2017})},\ \bibinfo {note} {published 31 May 2017},\ \Eprint {https://arxiv.org/abs/1703.09722} {arXiv:1703.09722 [gr-qc]} \BibitemShut {NoStop}%
\bibitem [{\citenamefont {Fan}\ \emph {et~al.}(2020)\citenamefont {Fan}, \citenamefont {Hu}, \citenamefont {Barausse}, \citenamefont {Sesana}, \citenamefont {Zhang}, \citenamefont {Zhang}, \citenamefont {Zi},\ and\ \citenamefont {Mei}}]{Fan2020TianQinEMRI}%
  \BibitemOpen
  \bibfield  {author} {\bibinfo {author} {\bibfnamefont {H.-M.}\ \bibnamefont {Fan}}, \bibinfo {author} {\bibfnamefont {Y.-M.}\ \bibnamefont {Hu}}, \bibinfo {author} {\bibfnamefont {E.}~\bibnamefont {Barausse}}, \bibinfo {author} {\bibfnamefont {A.}~\bibnamefont {Sesana}}, \bibinfo {author} {\bibfnamefont {J.-d.}\ \bibnamefont {Zhang}}, \bibinfo {author} {\bibfnamefont {X.}~\bibnamefont {Zhang}}, \bibinfo {author} {\bibfnamefont {T.-G.}\ \bibnamefont {Zi}},\ and\ \bibinfo {author} {\bibfnamefont {J.}~\bibnamefont {Mei}},\ }\bibfield  {title} {\bibinfo {title} {Science with the tianqin observatory: Preliminary result on extreme-mass-ratio inspirals},\ }\href {https://doi.org/10.1103/PhysRevD.102.063016} {\bibfield  {journal} {\bibinfo  {journal} {Phys. Rev. D}\ }\textbf {\bibinfo {volume} {102}},\ \bibinfo {pages} {063016} (\bibinfo {year} {2020})},\ \Eprint {https://arxiv.org/abs/2005.08212} {arXiv:2005.08212 [astro-ph.HE]} \BibitemShut {NoStop}%
\bibitem [{\citenamefont {Cutler}\ \emph {et~al.}(1994)\citenamefont {Cutler}, \citenamefont {Kennefick},\ and\ \citenamefont {Poisson}}]{Cutler1994SchwarzschildRR}%
  \BibitemOpen
  \bibfield  {author} {\bibinfo {author} {\bibfnamefont {C.}~\bibnamefont {Cutler}}, \bibinfo {author} {\bibfnamefont {D.}~\bibnamefont {Kennefick}},\ and\ \bibinfo {author} {\bibfnamefont {E.}~\bibnamefont {Poisson}},\ }\bibfield  {title} {\bibinfo {title} {Gravitational radiation reaction for bound motion around a schwarzschild black hole},\ }\href@noop {} {\bibfield  {journal} {\bibinfo  {journal} {Physical Review D}\ }\textbf {\bibinfo {volume} {50}},\ \bibinfo {pages} {3816} (\bibinfo {year} {1994})},\ \bibinfo {note} {detailed calculation of orbital evolution under gravitational radiation reaction in Schwarzschild spacetime}\BibitemShut {NoStop}%
\bibitem [{\citenamefont {Miller}\ \emph {et~al.}(2003)\citenamefont {Miller}, \citenamefont {Nichol}, \citenamefont {Gomez}, \citenamefont {Hopkins},\ and\ \citenamefont {Bernardi}}]{MillerAGNfrac}%
  \BibitemOpen
  \bibfield  {author} {\bibinfo {author} {\bibfnamefont {C.~J.}\ \bibnamefont {Miller}}, \bibinfo {author} {\bibfnamefont {R.~C.}\ \bibnamefont {Nichol}}, \bibinfo {author} {\bibfnamefont {P.}~\bibnamefont {Gomez}}, \bibinfo {author} {\bibfnamefont {A.}~\bibnamefont {Hopkins}},\ and\ \bibinfo {author} {\bibfnamefont {M.}~\bibnamefont {Bernardi}},\ }\bibfield  {title} {\bibinfo {title} {The environment of agns in the sloan digital sky survey},\ }\href {https://doi.org/10.1086/377358} {\bibfield  {journal} {\bibinfo  {journal} {Astrophysical Journal Letters}\ }\textbf {\bibinfo {volume} {596}},\ \bibinfo {pages} {L79} (\bibinfo {year} {2003})},\ \Eprint {https://arxiv.org/abs/astro-ph/0307124} {arXiv:astro-ph/0307124 [astro-ph]} \BibitemShut {NoStop}%
\bibitem [{\citenamefont {{Levin}}(2007)}]{Levin2007}%
  \BibitemOpen
  \bibfield  {author} {\bibinfo {author} {\bibfnamefont {Y.}~\bibnamefont {{Levin}}},\ }\bibfield  {title} {\bibinfo {title} {{Starbursts near supermassive black holes: young stars in the Galactic Centre, and gravitational waves in LISA band}},\ }\href {https://doi.org/10.1111/j.1365-2966.2006.11155.x} {\bibfield  {journal} {\bibinfo  {journal} {\mnras}\ }\textbf {\bibinfo {volume} {374}},\ \bibinfo {pages} {515} (\bibinfo {year} {2007})},\ \Eprint {https://arxiv.org/abs/astro-ph/0603583} {arXiv:astro-ph/0603583 [astro-ph]} \BibitemShut {NoStop}%
\bibitem [{\citenamefont {Pan}\ and\ \citenamefont {Yang}(2021{\natexlab{a}})}]{Pan202101}%
  \BibitemOpen
  \bibfield  {author} {\bibinfo {author} {\bibfnamefont {Z.}~\bibnamefont {Pan}}\ and\ \bibinfo {author} {\bibfnamefont {H.}~\bibnamefont {Yang}},\ }\bibfield  {title} {\bibinfo {title} {Formation rate of extreme mass ratio inspirals in active galactic nuclei},\ }\href {https://doi.org/10.1103/PhysRevD.103.103018} {\bibfield  {journal} {\bibinfo  {journal} {Phys.\ Rev.\ D}\ }\textbf {\bibinfo {volume} {103}},\ \bibinfo {pages} {103018} (\bibinfo {year} {2021}{\natexlab{a}})},\ \bibinfo {note} {arXiv:2101.09146},\ \Eprint {https://arxiv.org/abs/2101.09146} {arXiv:2101.09146 [astro-ph.HE]} \BibitemShut {NoStop}%
\bibitem [{\citenamefont {Pan}\ \emph {et~al.}(2021)\citenamefont {Pan}, \citenamefont {Lyu},\ and\ \citenamefont {Yang}}]{Pan:2021oob}%
  \BibitemOpen
  \bibfield  {author} {\bibinfo {author} {\bibfnamefont {Z.}~\bibnamefont {Pan}}, \bibinfo {author} {\bibfnamefont {Z.}~\bibnamefont {Lyu}},\ and\ \bibinfo {author} {\bibfnamefont {H.}~\bibnamefont {Yang}},\ }\bibfield  {title} {\bibinfo {title} {{Wet extreme mass ratio inspirals may be more common for spaceborne gravitational wave detection}},\ }\href {https://doi.org/10.1103/PhysRevD.104.063007} {\bibfield  {journal} {\bibinfo  {journal} {Phys. Rev. D}\ }\textbf {\bibinfo {volume} {104}},\ \bibinfo {pages} {063007} (\bibinfo {year} {2021})},\ \Eprint {https://arxiv.org/abs/2104.01208} {arXiv:2104.01208 [astro-ph.HE]} \BibitemShut {NoStop}%
\bibitem [{\citenamefont {Derdzinski}\ and\ \citenamefont {Mayer}(2023)}]{Derdzinski:2022ltb}%
  \BibitemOpen
  \bibfield  {author} {\bibinfo {author} {\bibfnamefont {A.}~\bibnamefont {Derdzinski}}\ and\ \bibinfo {author} {\bibfnamefont {L.}~\bibnamefont {Mayer}},\ }\bibfield  {title} {\bibinfo {title} {{In situ extreme mass ratio inspirals via subparsec formation and migration of stars in thin, gravitationally unstable AGN discs}},\ }\href {https://doi.org/10.1093/mnras/stad749} {\bibfield  {journal} {\bibinfo  {journal} {Mon. Not. Roy. Astron. Soc.}\ }\textbf {\bibinfo {volume} {521}},\ \bibinfo {pages} {4522} (\bibinfo {year} {2023})},\ \Eprint {https://arxiv.org/abs/2205.10382} {arXiv:2205.10382 [astro-ph.GA]} \BibitemShut {NoStop}%
\bibitem [{\citenamefont {Hills}(1988)}]{Hills1988TidalDisruption}%
  \BibitemOpen
  \bibfield  {author} {\bibinfo {author} {\bibfnamefont {J.~G.}\ \bibnamefont {Hills}},\ }\bibfield  {title} {\bibinfo {title} {Hyper-velocity and tidal stars from binaries disrupted by a massive galactic black hole},\ }\href@noop {} {\bibfield  {journal} {\bibinfo  {journal} {Nature}\ } (\bibinfo {year} {1988})},\ \bibinfo {note} {proposed the tidal separation (Hills mechanism) channel for EMRI formation via binary disruption}\BibitemShut {NoStop}%
\bibitem [{\citenamefont {Naoz}\ \emph {et~al.}(2022)\citenamefont {Naoz}, \citenamefont {Rose}, \citenamefont {Michaely}, \citenamefont {Melchor}, \citenamefont {Ramirez-Ruiz}, \citenamefont {Mockler},\ and\ \citenamefont {Schnittman}}]{Naoz2022EKL_EMRI}%
  \BibitemOpen
  \bibfield  {author} {\bibinfo {author} {\bibfnamefont {S.}~\bibnamefont {Naoz}}, \bibinfo {author} {\bibfnamefont {S.~C.}\ \bibnamefont {Rose}}, \bibinfo {author} {\bibfnamefont {E.}~\bibnamefont {Michaely}}, \bibinfo {author} {\bibfnamefont {D.}~\bibnamefont {Melchor}}, \bibinfo {author} {\bibfnamefont {E.}~\bibnamefont {Ramirez-Ruiz}}, \bibinfo {author} {\bibfnamefont {B.}~\bibnamefont {Mockler}},\ and\ \bibinfo {author} {\bibfnamefont {J.~D.}\ \bibnamefont {Schnittman}},\ }\bibfield  {title} {\bibinfo {title} {The combined effects of two-body relaxation processes and the eccentric kozai-lidov mechanism on the emri rate},\ }\href {https://doi.org/10.3847/2041-8213/ac574b} {\bibfield  {journal} {\bibinfo  {journal} {Astrophysical Journal Letters}\ }\textbf {\bibinfo {volume} {927}},\ \bibinfo {pages} {L18} (\bibinfo {year} {2022})},\ \Eprint {https://arxiv.org/abs/2202.12303} {arXiv:2202.12303 [astro-ph.HE]} \BibitemShut {NoStop}%
\bibitem [{\citenamefont {{Raveh}}\ and\ \citenamefont {{Perets}}(2021)}]{Raveh2021}%
  \BibitemOpen
  \bibfield  {author} {\bibinfo {author} {\bibfnamefont {Y.}~\bibnamefont {{Raveh}}}\ and\ \bibinfo {author} {\bibfnamefont {H.~B.}\ \bibnamefont {{Perets}}},\ }\bibfield  {title} {\bibinfo {title} {{Extreme mass-ratio gravitational-wave sources: mass segregation and post binary tidal-disruption captures}},\ }\href {https://doi.org/10.1093/mnras/staa4001} {\bibfield  {journal} {\bibinfo  {journal} {\mnras}\ }\textbf {\bibinfo {volume} {501}},\ \bibinfo {pages} {5012} (\bibinfo {year} {2021})},\ \Eprint {https://arxiv.org/abs/2011.13952} {arXiv:2011.13952 [astro-ph.GA]} \BibitemShut {NoStop}%
\bibitem [{\citenamefont {Mazzolari}\ \emph {et~al.}(2022)\citenamefont {Mazzolari}, \citenamefont {Bonetti}, \citenamefont {Sesana}, \citenamefont {Colombo}, \citenamefont {Dotti}, \citenamefont {Lodato},\ and\ \citenamefont {Izquierdo-Villalba}}]{Mazzolari2022MBHB_EMRI}%
  \BibitemOpen
  \bibfield  {author} {\bibinfo {author} {\bibfnamefont {G.}~\bibnamefont {Mazzolari}}, \bibinfo {author} {\bibfnamefont {M.}~\bibnamefont {Bonetti}}, \bibinfo {author} {\bibfnamefont {A.}~\bibnamefont {Sesana}}, \bibinfo {author} {\bibfnamefont {R.~M.}\ \bibnamefont {Colombo}}, \bibinfo {author} {\bibfnamefont {M.}~\bibnamefont {Dotti}}, \bibinfo {author} {\bibfnamefont {G.}~\bibnamefont {Lodato}},\ and\ \bibinfo {author} {\bibfnamefont {D.}~\bibnamefont {Izquierdo-Villalba}},\ }\bibfield  {title} {\bibinfo {title} {Extreme mass ratio inspirals triggered by massive black hole binaries: from relativistic dynamics to cosmological rates},\ }\href {https://doi.org/10.1093/mnras/stac2295} {\bibfield  {journal} {\bibinfo  {journal} {Monthly Notices of the Royal Astronomical Society}\ }\textbf {\bibinfo {volume} {516}},\ \bibinfo {pages} {1959} (\bibinfo {year} {2022})},\ \Eprint {https://arxiv.org/abs/2204.05343} {arXiv:2204.05343 [astro-ph.HE]} \BibitemShut {NoStop}%
\bibitem [{\citenamefont {Bode}\ and\ \citenamefont {Wegg}(2014)}]{BodeWegg2014EMRIinMBHB}%
  \BibitemOpen
  \bibfield  {author} {\bibinfo {author} {\bibfnamefont {J.~N.}\ \bibnamefont {Bode}}\ and\ \bibinfo {author} {\bibfnamefont {C.}~\bibnamefont {Wegg}},\ }\bibfield  {title} {\bibinfo {title} {Production of extreme mass-ratio inspirals in supermassive black hole binaries},\ }\href {https://doi.org/10.1093/mnras/stt2227} {\bibfield  {journal} {\bibinfo  {journal} {Monthly Notices of the Royal Astronomical Society}\ }\textbf {\bibinfo {volume} {438}},\ \bibinfo {pages} {573} (\bibinfo {year} {2014})}\BibitemShut {NoStop}%
\bibitem [{\citenamefont {Rodriguez}\ \emph {et~al.}(2016)\citenamefont {Rodriguez}, \citenamefont {Zevin}, \citenamefont {Pankow}, \citenamefont {Kalogera},\ and\ \citenamefont {Rasio}}]{Rodriguez2016SpinFormation}%
  \BibitemOpen
  \bibfield  {author} {\bibinfo {author} {\bibfnamefont {C.~L.}\ \bibnamefont {Rodriguez}}, \bibinfo {author} {\bibfnamefont {M.}~\bibnamefont {Zevin}}, \bibinfo {author} {\bibfnamefont {C.}~\bibnamefont {Pankow}}, \bibinfo {author} {\bibfnamefont {V.}~\bibnamefont {Kalogera}},\ and\ \bibinfo {author} {\bibfnamefont {F.~A.}\ \bibnamefont {Rasio}},\ }\bibfield  {title} {\bibinfo {title} {Illuminating black hole binary formation channels with spins in advanced ligo},\ }\href {https://doi.org/10.3847/2041-8205/832/1/L2} {\bibfield  {journal} {\bibinfo  {journal} {Astrophysical Journal Letters}\ }\textbf {\bibinfo {volume} {832}},\ \bibinfo {pages} {L2} (\bibinfo {year} {2016})},\ \Eprint {https://arxiv.org/abs/1609.05916} {arXiv:1609.05916 [astro-ph.HE]} \BibitemShut {NoStop}%
\bibitem [{\citenamefont {Kocsis}\ \emph {et~al.}(2011)\citenamefont {Kocsis}, \citenamefont {Yunes},\ and\ \citenamefont {Loeb}}]{Kocsis2011DiskEMRI}%
  \BibitemOpen
  \bibfield  {author} {\bibinfo {author} {\bibfnamefont {B.}~\bibnamefont {Kocsis}}, \bibinfo {author} {\bibfnamefont {N.}~\bibnamefont {Yunes}},\ and\ \bibinfo {author} {\bibfnamefont {A.}~\bibnamefont {Loeb}},\ }\bibfield  {title} {\bibinfo {title} {Observable signatures of emri black hole binaries embedded in thin accretion disks},\ }\href {https://doi.org/10.1103/PhysRevD.84.024032} {\bibfield  {journal} {\bibinfo  {journal} {Physical Review D}\ }\textbf {\bibinfo {volume} {84}},\ \bibinfo {pages} {024032} (\bibinfo {year} {2011})},\ \Eprint {https://arxiv.org/abs/1104.2322} {arXiv:1104.2322 [astro-ph.HE]} \BibitemShut {NoStop}%
\bibitem [{\citenamefont {Derdzinski}\ \emph {et~al.}(2021)\citenamefont {Derdzinski}, \citenamefont {D'Orazio}, \citenamefont {Duffell}, \citenamefont {Haiman},\ and\ \citenamefont {MacFadyen}}]{Derdzinski2020GasDiskEMRI}%
  \BibitemOpen
  \bibfield  {author} {\bibinfo {author} {\bibfnamefont {A.}~\bibnamefont {Derdzinski}}, \bibinfo {author} {\bibfnamefont {D.}~\bibnamefont {D'Orazio}}, \bibinfo {author} {\bibfnamefont {P.}~\bibnamefont {Duffell}}, \bibinfo {author} {\bibfnamefont {Z.}~\bibnamefont {Haiman}},\ and\ \bibinfo {author} {\bibfnamefont {A.}~\bibnamefont {MacFadyen}},\ }\bibfield  {title} {\bibinfo {title} {Evolution of gas disc-embedded intermediate mass ratio inspirals in the lisa band},\ }\href {https://doi.org/10.1093/mnras/staa3585} {\bibfield  {journal} {\bibinfo  {journal} {Monthly Notices of the Royal Astronomical Society}\ }\textbf {\bibinfo {volume} {501}},\ \bibinfo {pages} {3540} (\bibinfo {year} {2021})},\ \Eprint {https://arxiv.org/abs/2005.11333} {arXiv:2005.11333 [astro-ph.HE]} \BibitemShut {NoStop}%
\bibitem [{\citenamefont {Sakimoto}\ and\ \citenamefont {Coroniti}(1981)}]{Sakimoto1981MagneticViscosity}%
  \BibitemOpen
  \bibfield  {author} {\bibinfo {author} {\bibfnamefont {P.~J.}\ \bibnamefont {Sakimoto}}\ and\ \bibinfo {author} {\bibfnamefont {F.~V.}\ \bibnamefont {Coroniti}},\ }\bibfield  {title} {\bibinfo {title} {Accretion disk models for qsos and active galactic nuclei: The role of magnetic viscosity},\ }\href {https://doi.org/10.1086/159005} {\bibfield  {journal} {\bibinfo  {journal} {Astrophysical Journal}\ }\textbf {\bibinfo {volume} {247}},\ \bibinfo {pages} {19} (\bibinfo {year} {1981})},\ \bibinfo {note} {proposes magnetic viscosity in QSO/AGN accretion disk models, leading to thermally stable, optically thick inner regions}\BibitemShut {NoStop}%
\bibitem [{\citenamefont {Dotti}\ \emph {et~al.}(2013)\citenamefont {Dotti}, \citenamefont {Colpi}, \citenamefont {Pallini}, \citenamefont {Perego},\ and\ \citenamefont {Volonteri}}]{Dotti2013BHspinOrientation}%
  \BibitemOpen
  \bibfield  {author} {\bibinfo {author} {\bibfnamefont {M.}~\bibnamefont {Dotti}}, \bibinfo {author} {\bibfnamefont {M.}~\bibnamefont {Colpi}}, \bibinfo {author} {\bibfnamefont {S.}~\bibnamefont {Pallini}}, \bibinfo {author} {\bibfnamefont {A.}~\bibnamefont {Perego}},\ and\ \bibinfo {author} {\bibfnamefont {M.}~\bibnamefont {Volonteri}},\ }\bibfield  {title} {\bibinfo {title} {On the orientation and magnitude of the black hole spin in galactic nuclei},\ }\href {https://doi.org/10.1088/0004-637X/762/2/68} {\bibfield  {journal} {\bibinfo  {journal} {Astrophysical Journal}\ }\textbf {\bibinfo {volume} {762}},\ \bibinfo {pages} {68} (\bibinfo {year} {2013})},\ \Eprint {https://arxiv.org/abs/1211.4871} {arXiv:1211.4871 [astro-ph.CO]} \BibitemShut {NoStop}%
\bibitem [{\citenamefont {Natarajan}\ and\ \citenamefont {Pringle}(1998)}]{Natara1998_Alignment}%
  \BibitemOpen
  \bibfield  {author} {\bibinfo {author} {\bibfnamefont {P.}~\bibnamefont {Natarajan}}\ and\ \bibinfo {author} {\bibfnamefont {J.~E.}\ \bibnamefont {Pringle}},\ }\bibfield  {title} {\bibinfo {title} {The alignment of disk and black hole spins in active galactic nuclei},\ }\href {https://doi.org/10.1086/311658} {\bibfield  {journal} {\bibinfo  {journal} {The Astrophysical Journal}\ }\textbf {\bibinfo {volume} {506}},\ \bibinfo {pages} {L97} (\bibinfo {year} {1998})},\ \bibinfo {note} {arXiv:astro-ph/9808187},\ \Eprint {https://arxiv.org/abs/astro-ph/9808187} {arXiv:astro-ph/9808187 [astro-ph]} \BibitemShut {NoStop}%
\bibitem [{\citenamefont {Cui}\ \emph {et~al.}(2025)\citenamefont {Cui}, \citenamefont {Han},\ and\ \citenamefont {Pan}}]{Cui:2025bgu}%
  \BibitemOpen
  \bibfield  {author} {\bibinfo {author} {\bibfnamefont {Q.}~\bibnamefont {Cui}}, \bibinfo {author} {\bibfnamefont {W.-B.}\ \bibnamefont {Han}},\ and\ \bibinfo {author} {\bibfnamefont {Z.}~\bibnamefont {Pan}},\ }\bibfield  {title} {\bibinfo {title} {{Secondary spins of extreme mass-ratio inspirals: A probe to the formation channels}},\ }\href {https://doi.org/10.1103/PhysRevD.111.103044} {\bibfield  {journal} {\bibinfo  {journal} {Phys. Rev. D}\ }\textbf {\bibinfo {volume} {111}},\ \bibinfo {pages} {103044} (\bibinfo {year} {2025})},\ \Eprint {https://arxiv.org/abs/2502.00856} {arXiv:2502.00856 [astro-ph.HE]} \BibitemShut {NoStop}%
\bibitem [{\citenamefont {Hopman}\ and\ \citenamefont {Alexander}(2005)}]{Hopman2005_OrbitalStatistics}%
  \BibitemOpen
  \bibfield  {author} {\bibinfo {author} {\bibfnamefont {C.}~\bibnamefont {Hopman}}\ and\ \bibinfo {author} {\bibfnamefont {T.}~\bibnamefont {Alexander}},\ }\bibfield  {title} {\bibinfo {title} {The orbital statistics of stellar inspiral and relaxation near a massive black hole: Characterizing gravitational wave sources},\ }\href {https://doi.org/10.1086/431475} {\bibfield  {journal} {\bibinfo  {journal} {The Astrophysical Journal}\ }\textbf {\bibinfo {volume} {629}},\ \bibinfo {pages} {362} (\bibinfo {year} {2005})},\ \bibinfo {note} {arXiv:astro-ph/0503672},\ \Eprint {https://arxiv.org/abs/astro-ph/0503672} {arXiv:astro-ph/0503672 [astro-ph]} \BibitemShut {NoStop}%
\bibitem [{\citenamefont {Amaro-Seoane}\ and\ \citenamefont {Preto}(2011)}]{Amaro-Seoane:2010dzj}%
  \BibitemOpen
  \bibfield  {author} {\bibinfo {author} {\bibfnamefont {P.}~\bibnamefont {Amaro-Seoane}}\ and\ \bibinfo {author} {\bibfnamefont {M.}~\bibnamefont {Preto}},\ }\bibfield  {title} {\bibinfo {title} {{The impact of realistic models of mass segregation on the event rate of extreme-mass ratio inspirals and cusp re-growth}},\ }\href {https://doi.org/10.1088/0264-9381/28/9/094017} {\bibfield  {journal} {\bibinfo  {journal} {Class. Quant. Grav.}\ }\textbf {\bibinfo {volume} {28}},\ \bibinfo {pages} {094017} (\bibinfo {year} {2011})},\ \Eprint {https://arxiv.org/abs/1010.5781} {arXiv:1010.5781 [astro-ph.CO]} \BibitemShut {NoStop}%
\bibitem [{\citenamefont {Amaro-Seoane}(2018)}]{Amaro-Seoane:2012lgq}%
  \BibitemOpen
  \bibfield  {author} {\bibinfo {author} {\bibfnamefont {P.}~\bibnamefont {Amaro-Seoane}},\ }\bibfield  {title} {\bibinfo {title} {{Relativistic dynamics and extreme mass ratio inspirals}},\ }\href {https://doi.org/10.1007/s41114-018-0013-8} {\bibfield  {journal} {\bibinfo  {journal} {Living Rev. Rel.}\ }\textbf {\bibinfo {volume} {21}},\ \bibinfo {pages} {4} (\bibinfo {year} {2018})},\ \Eprint {https://arxiv.org/abs/1205.5240} {arXiv:1205.5240 [astro-ph.CO]} \BibitemShut {NoStop}%
\bibitem [{\citenamefont {{Bar-Or}}\ and\ \citenamefont {{Alexander}}(2016)}]{Bar-Or2016}%
  \BibitemOpen
  \bibfield  {author} {\bibinfo {author} {\bibfnamefont {B.}~\bibnamefont {{Bar-Or}}}\ and\ \bibinfo {author} {\bibfnamefont {T.}~\bibnamefont {{Alexander}}},\ }\bibfield  {title} {\bibinfo {title} {{Steady-state Relativistic Stellar Dynamics Around a Massive Black hole}},\ }\href {https://doi.org/10.3847/0004-637X/820/2/129} {\bibfield  {journal} {\bibinfo  {journal} {\apj}\ }\textbf {\bibinfo {volume} {820}},\ \bibinfo {eid} {129} (\bibinfo {year} {2016})},\ \Eprint {https://arxiv.org/abs/1508.01390} {arXiv:1508.01390 [astro-ph.GA]} \BibitemShut {NoStop}%
\bibitem [{\citenamefont {G{\"u}ltekin}\ \emph {et~al.}(2009)\citenamefont {G{\"u}ltekin}, \citenamefont {Richstone}, \citenamefont {Gebhardt}, \citenamefont {Lauer}, \citenamefont {Tremaine}, \citenamefont {Aller}, \citenamefont {Bender}, \citenamefont {Dressler}, \citenamefont {Faber}, \citenamefont {Filippenko}, \citenamefont {Green}, \citenamefont {Ho}, \citenamefont {Kormendy}, \citenamefont {Magorrian}, \citenamefont {Pinkney},\ and\ \citenamefont {Siopis}}]{Gultekin2009MSigmaML}%
  \BibitemOpen
  \bibfield  {author} {\bibinfo {author} {\bibfnamefont {K.}~\bibnamefont {G{\"u}ltekin}}, \bibinfo {author} {\bibfnamefont {D.~O.}\ \bibnamefont {Richstone}}, \bibinfo {author} {\bibfnamefont {K.}~\bibnamefont {Gebhardt}}, \bibinfo {author} {\bibfnamefont {T.~R.}\ \bibnamefont {Lauer}}, \bibinfo {author} {\bibfnamefont {S.}~\bibnamefont {Tremaine}}, \bibinfo {author} {\bibfnamefont {M.~C.}\ \bibnamefont {Aller}}, \bibinfo {author} {\bibfnamefont {R.}~\bibnamefont {Bender}}, \bibinfo {author} {\bibfnamefont {A.}~\bibnamefont {Dressler}}, \bibinfo {author} {\bibfnamefont {S.~M.}\ \bibnamefont {Faber}}, \bibinfo {author} {\bibfnamefont {A.~V.}\ \bibnamefont {Filippenko}}, \bibinfo {author} {\bibfnamefont {R.~F.}\ \bibnamefont {Green}}, \bibinfo {author} {\bibfnamefont {L.~C.}\ \bibnamefont {Ho}}, \bibinfo {author} {\bibfnamefont {J.}~\bibnamefont {Kormendy}}, \bibinfo {author} {\bibfnamefont {J.}~\bibnamefont {Magorrian}}, \bibinfo {author} {\bibfnamefont {J.}~\bibnamefont {Pinkney}},\ and\ \bibinfo {author} {\bibfnamefont {C.}~\bibnamefont {Siopis}},\ }\bibfield  {title} {\bibinfo {title} {The m-\ensuremath{\sigma} and m-l relations in galactic bulges and determinations of their intrinsic scatter},\ }\href {https://doi.org/10.1088/0004-637X/698/1/198} {\bibfield  {journal} {\bibinfo  {journal} {The Astrophysical Journal}\ }\textbf {\bibinfo {volume} {698}},\ \bibinfo {pages} {198} (\bibinfo {year} {2009})},\ \Eprint {https://arxiv.org/abs/0903.4897} {arXiv:0903.4897 [astro-ph.GA]} \BibitemShut {NoStop}%
\bibitem [{\citenamefont {Cohn}\ and\ \citenamefont {Kulsrud}(1978)}]{Cohn1978_StellarDistBH}%
  \BibitemOpen
  \bibfield  {author} {\bibinfo {author} {\bibfnamefont {H.}~\bibnamefont {Cohn}}\ and\ \bibinfo {author} {\bibfnamefont {R.~M.}\ \bibnamefont {Kulsrud}},\ }\bibfield  {title} {\bibinfo {title} {The stellar distribution around a black hole: Numerical integration of the fokker-planck equation},\ }\href {https://doi.org/10.1086/156685} {\bibfield  {journal} {\bibinfo  {journal} {The Astrophysical Journal}\ }\textbf {\bibinfo {volume} {226}},\ \bibinfo {pages} {1087} (\bibinfo {year} {1978})}\BibitemShut {NoStop}%
\bibitem [{\citenamefont {Cohn}(1979)}]{Cohn1979}%
  \BibitemOpen
  \bibfield  {author} {\bibinfo {author} {\bibfnamefont {H.}~\bibnamefont {Cohn}},\ }\bibfield  {title} {\bibinfo {title} {Numerical integration of the fokker-planck equation and the evolution of star clusters},\ }\href {https://doi.org/10.1086/157587} {\bibfield  {journal} {\bibinfo  {journal} {The Astrophysical Journal}\ }\textbf {\bibinfo {volume} {234}},\ \bibinfo {pages} {1036} (\bibinfo {year} {1979})}\BibitemShut {NoStop}%
\bibitem [{\citenamefont {Pan}\ \emph {et~al.}(2022)\citenamefont {Pan}, \citenamefont {Lyu},\ and\ \citenamefont {Yang}}]{Pan202112}%
  \BibitemOpen
  \bibfield  {author} {\bibinfo {author} {\bibfnamefont {Z.}~\bibnamefont {Pan}}, \bibinfo {author} {\bibfnamefont {Z.}~\bibnamefont {Lyu}},\ and\ \bibinfo {author} {\bibfnamefont {H.}~\bibnamefont {Yang}},\ }\bibfield  {title} {\bibinfo {title} {Mass-gap extreme mass ratio inspirals},\ }\href {https://doi.org/10.1103/PhysRevD.105.083005} {\bibfield  {journal} {\bibinfo  {journal} {Phys. Rev. D}\ }\textbf {\bibinfo {volume} {105}},\ \bibinfo {pages} {083005} (\bibinfo {year} {2022})},\ \bibinfo {note} {arXiv:2112.10237},\ \Eprint {https://arxiv.org/abs/2112.10237} {arXiv:2112.10237 [astro-ph.HE]} \BibitemShut {NoStop}%
\bibitem [{\citenamefont {Zwick}\ \emph {et~al.}(2020)\citenamefont {Zwick}, \citenamefont {Capelo}, \citenamefont {Bortolas}, \citenamefont {Mayer},\ and\ \citenamefont {Amaro-Seoane}}]{Zwick2020}%
  \BibitemOpen
  \bibfield  {author} {\bibinfo {author} {\bibfnamefont {L.}~\bibnamefont {Zwick}}, \bibinfo {author} {\bibfnamefont {P.~R.}\ \bibnamefont {Capelo}}, \bibinfo {author} {\bibfnamefont {E.}~\bibnamefont {Bortolas}}, \bibinfo {author} {\bibfnamefont {L.}~\bibnamefont {Mayer}},\ and\ \bibinfo {author} {\bibfnamefont {P.}~\bibnamefont {Amaro-Seoane}},\ }\bibfield  {title} {\bibinfo {title} {Improved gravitational radiation time-scales: significance for {LISA} and {LIGO}-{Virgo} sources},\ }\href {https://doi.org/10.1093/mnras/staa1314} {\bibfield  {journal} {\bibinfo  {journal} {Monthly Notices of the Royal Astronomical Society}\ }\textbf {\bibinfo {volume} {495}},\ \bibinfo {pages} {2321} (\bibinfo {year} {2020})},\ \Eprint {https://arxiv.org/abs/1911.06024} {arXiv:1911.06024 [astro-ph.GA]} \BibitemShut {NoStop}%
\bibitem [{\citenamefont {Zwick}\ \emph {et~al.}(2021)\citenamefont {Zwick}, \citenamefont {Capelo}, \citenamefont {Bortolas}, \citenamefont {Vázquez-Aceves}, \citenamefont {Mayer},\ and\ \citenamefont {Amaro-Seoane}}]{Zwick2021a}%
  \BibitemOpen
  \bibfield  {author} {\bibinfo {author} {\bibfnamefont {L.}~\bibnamefont {Zwick}}, \bibinfo {author} {\bibfnamefont {P.~R.}\ \bibnamefont {Capelo}}, \bibinfo {author} {\bibfnamefont {E.}~\bibnamefont {Bortolas}}, \bibinfo {author} {\bibfnamefont {V.}~\bibnamefont {Vázquez-Aceves}}, \bibinfo {author} {\bibfnamefont {L.}~\bibnamefont {Mayer}},\ and\ \bibinfo {author} {\bibfnamefont {P.}~\bibnamefont {Amaro-Seoane}},\ }\bibfield  {title} {\bibinfo {title} {Improved gravitational radiation time-scales ii: spin-orbit contributions and environmental perturbations},\ }\href {https://doi.org/10.1093/mnras/stab1602} {\bibfield  {journal} {\bibinfo  {journal} {Monthly Notices of the Royal Astronomical Society}\ }\textbf {\bibinfo {volume} {506}},\ \bibinfo {pages} {1007} (\bibinfo {year} {2021})},\ \bibinfo {note} {arXiv:2102.00015v2 [astro-ph.GA] (published 24 Jun 2021)},\ \Eprint {https://arxiv.org/abs/2102.00015} {arXiv:2102.00015 [astro-ph.GA]} \BibitemShut {NoStop}%
\bibitem [{\citenamefont {V{\'a}zquez-Aceves}\ \emph {et~al.}(2022)\citenamefont {V{\'a}zquez-Aceves}, \citenamefont {Zwick}, \citenamefont {Bortolas}, \citenamefont {Capelo}, \citenamefont {Amaro-Seoane}, \citenamefont {Mayer},\ and\ \citenamefont {Chen}}]{VazquezAceves2022}%
  \BibitemOpen
  \bibfield  {author} {\bibinfo {author} {\bibfnamefont {V.}~\bibnamefont {V{\'a}zquez-Aceves}}, \bibinfo {author} {\bibfnamefont {L.}~\bibnamefont {Zwick}}, \bibinfo {author} {\bibfnamefont {E.}~\bibnamefont {Bortolas}}, \bibinfo {author} {\bibfnamefont {P.~R.}\ \bibnamefont {Capelo}}, \bibinfo {author} {\bibfnamefont {P.}~\bibnamefont {Amaro-Seoane}}, \bibinfo {author} {\bibfnamefont {L.}~\bibnamefont {Mayer}},\ and\ \bibinfo {author} {\bibfnamefont {X.}~\bibnamefont {Chen}},\ }\bibfield  {title} {\bibinfo {title} {Revised event rates for extreme and extremely large mass-ratio inspirals},\ }\href {https://doi.org/10.1093/mnras/stab3485} {\bibfield  {journal} {\bibinfo  {journal} {Monthly Notices of the Royal Astronomical Society}\ }\textbf {\bibinfo {volume} {510}},\ \bibinfo {pages} {2379} (\bibinfo {year} {2022})},\ \bibinfo {note} {arXiv:2108.00135 [astro-ph.GA] (submitted 31 July 2021)},\ \Eprint {https://arxiv.org/abs/2108.00135} {arXiv:2108.00135 [astro-ph.GA]} \BibitemShut {NoStop}%
\bibitem [{\citenamefont {Katz}\ \emph {et~al.}(2021)\citenamefont {Katz}, \citenamefont {Chua}, \citenamefont {Speri}, \citenamefont {Warburton},\ and\ \citenamefont {Hughes}}]{Katz2021FastEMRIWaveforms}%
  \BibitemOpen
  \bibfield  {author} {\bibinfo {author} {\bibfnamefont {M.~L.}\ \bibnamefont {Katz}}, \bibinfo {author} {\bibfnamefont {A.~J.~K.}\ \bibnamefont {Chua}}, \bibinfo {author} {\bibfnamefont {L.}~\bibnamefont {Speri}}, \bibinfo {author} {\bibfnamefont {N.}~\bibnamefont {Warburton}},\ and\ \bibinfo {author} {\bibfnamefont {S.~A.}\ \bibnamefont {Hughes}},\ }\bibfield  {title} {\bibinfo {title} {Fastemriwaveforms: New tools for millihertz gravitational-wave data analysis},\ }\href {https://doi.org/10.1103/PhysRevD.104.064047} {\bibfield  {journal} {\bibinfo  {journal} {Physical Review D}\ }\textbf {\bibinfo {volume} {104}},\ \bibinfo {pages} {064047} (\bibinfo {year} {2021})},\ \Eprint {https://arxiv.org/abs/2104.04582} {arXiv:2104.04582 [gr-qc]} \BibitemShut {NoStop}%
\bibitem [{\citenamefont {Peters}(1964)}]{Peters1964a}%
  \BibitemOpen
  \bibfield  {author} {\bibinfo {author} {\bibfnamefont {P.~C.}\ \bibnamefont {Peters}},\ }\bibfield  {title} {\bibinfo {title} {Gravitational radiation and the motion of two point masses},\ }\href {https://doi.org/10.1103/PhysRev.136.B1224} {\bibfield  {journal} {\bibinfo  {journal} {Physical Review}\ }\textbf {\bibinfo {volume} {136}},\ \bibinfo {pages} {B1224} (\bibinfo {year} {1964})}\BibitemShut {NoStop}%
\bibitem [{\citenamefont {Mancieri}\ \emph {et~al.}(2025{\natexlab{a}})\citenamefont {Mancieri}, \citenamefont {Broggi}, \citenamefont {Vinciguerra}, \citenamefont {Sesana},\ and\ \citenamefont {Bonetti}}]{Mancieri2025_EccentricityEMRI}%
  \BibitemOpen
  \bibfield  {author} {\bibinfo {author} {\bibfnamefont {D.}~\bibnamefont {Mancieri}}, \bibinfo {author} {\bibfnamefont {L.}~\bibnamefont {Broggi}}, \bibinfo {author} {\bibfnamefont {M.}~\bibnamefont {Vinciguerra}}, \bibinfo {author} {\bibfnamefont {A.}~\bibnamefont {Sesana}},\ and\ \bibinfo {author} {\bibfnamefont {M.}~\bibnamefont {Bonetti}},\ }\bibfield  {title} {\bibinfo {title} {Eccentricity distribution of extreme mass ratio inspirals},\ }\href@noop {} {\bibfield  {journal} {\bibinfo  {journal} {arXiv preprint}\ } (\bibinfo {year} {2025}{\natexlab{a}})},\ \bibinfo {note} {submitted},\ \Eprint {https://arxiv.org/abs/2509.02394} {arXiv:2509.02394 [astro-ph.HE]} \BibitemShut {NoStop}%
\bibitem [{\citenamefont {Cresswell}\ \emph {et~al.}(2007)\citenamefont {Cresswell}, \citenamefont {Dirksen}, \citenamefont {Kley},\ and\ \citenamefont {Nelson}}]{Cresswell2007EccIncEvol}%
  \BibitemOpen
  \bibfield  {author} {\bibinfo {author} {\bibfnamefont {P.}~\bibnamefont {Cresswell}}, \bibinfo {author} {\bibfnamefont {G.}~\bibnamefont {Dirksen}}, \bibinfo {author} {\bibfnamefont {W.}~\bibnamefont {Kley}},\ and\ \bibinfo {author} {\bibfnamefont {R.~P.}\ \bibnamefont {Nelson}},\ }\bibfield  {title} {\bibinfo {title} {On the evolution of eccentric and inclined protoplanets embedded in protoplanetary disks},\ }\href {https://doi.org/10.1051/0004-6361:20077015} {\bibfield  {journal} {\bibinfo  {journal} {Astronomy \& Astrophysics}\ }\textbf {\bibinfo {volume} {473}},\ \bibinfo {pages} {329} (\bibinfo {year} {2007})},\ \bibinfo {note} {hydrodynamical simulations demonstrating rapid exponential damping of eccentricity and inclination}\BibitemShut {NoStop}%
\bibitem [{\citenamefont {{Li}}\ \emph {et~al.}(2019)\citenamefont {{Li}}, \citenamefont {{Li}}, \citenamefont {{Li}},\ and\ \citenamefont {{Lin}}}]{Li2019}%
  \BibitemOpen
  \bibfield  {author} {\bibinfo {author} {\bibfnamefont {Y.-P.}\ \bibnamefont {{Li}}}, \bibinfo {author} {\bibfnamefont {H.}~\bibnamefont {{Li}}}, \bibinfo {author} {\bibfnamefont {S.}~\bibnamefont {{Li}}},\ and\ \bibinfo {author} {\bibfnamefont {D.~N.~C.}\ \bibnamefont {{Lin}}},\ }\bibfield  {title} {\bibinfo {title} {{On the Dust Signatures Induced by Eccentric Super-Earths in Protoplanetary Disks}},\ }\href {https://doi.org/10.3847/1538-4357/ab4bc8} {\bibfield  {journal} {\bibinfo  {journal} {\apj}\ }\textbf {\bibinfo {volume} {886}},\ \bibinfo {eid} {62} (\bibinfo {year} {2019})},\ \Eprint {https://arxiv.org/abs/1910.03130} {arXiv:1910.03130 [astro-ph.EP]} \BibitemShut {NoStop}%
\bibitem [{\citenamefont {Bellovary}\ \emph {et~al.}(2016)\citenamefont {Bellovary}, \citenamefont {Low}, \citenamefont {McKernan},\ and\ \citenamefont {Ford}}]{Bell2016}%
  \BibitemOpen
  \bibfield  {author} {\bibinfo {author} {\bibfnamefont {J.~M.}\ \bibnamefont {Bellovary}}, \bibinfo {author} {\bibfnamefont {M.-M.~M.}\ \bibnamefont {Low}}, \bibinfo {author} {\bibfnamefont {B.}~\bibnamefont {McKernan}},\ and\ \bibinfo {author} {\bibfnamefont {K.}~\bibnamefont {Ford}},\ }\bibfield  {title} {\bibinfo {title} {Migration traps in disks around supermassive black holes},\ }\href {https://doi.org/10.3847/2041-8205/819/2/L17} {\bibfield  {journal} {\bibinfo  {journal} {The Astrophysical Journal Letters}\ }\textbf {\bibinfo {volume} {819}},\ \bibinfo {pages} {L17} (\bibinfo {year} {2016})},\ \Eprint {https://arxiv.org/abs/1511.00005} {arXiv:1511.00005 [astro-ph.GA]} \BibitemShut {NoStop}%
\bibitem [{\citenamefont {Rein}\ and\ \citenamefont {Liu}(2012)}]{ReinLiu2012}%
  \BibitemOpen
  \bibfield  {author} {\bibinfo {author} {\bibfnamefont {H.}~\bibnamefont {Rein}}\ and\ \bibinfo {author} {\bibfnamefont {S.-F.}\ \bibnamefont {Liu}},\ }\bibfield  {title} {\bibinfo {title} {Rebound: an open-source multi-purpose n-body code for collisional dynamics},\ }\href {https://doi.org/10.1051/0004-6361/201118085} {\bibfield  {journal} {\bibinfo  {journal} {Astronomy \& Astrophysics}\ }\textbf {\bibinfo {volume} {537}},\ \bibinfo {pages} {A128} (\bibinfo {year} {2012})},\ \Eprint {https://arxiv.org/abs/1110.4876} {arXiv:1110.4876 [astro-ph.EP]} \BibitemShut {NoStop}%
\bibitem [{\citenamefont {Pati}\ and\ \citenamefont {Will}(2001)}]{Pati2002_PNII}%
  \BibitemOpen
  \bibfield  {author} {\bibinfo {author} {\bibfnamefont {M.~E.}\ \bibnamefont {Pati}}\ and\ \bibinfo {author} {\bibfnamefont {C.~M.}\ \bibnamefont {Will}},\ }\href@noop {} {\bibinfo {title} {Post-newtonian gravitational radiation and equations of motion via direct integration of the relaxed einstein equations. {II}. two-body equations of motion to second post-newtonian order, and radiation-reaction to 3.5 post-newtonian order}} (\bibinfo {year} {2001}),\ \bibinfo {note} {arXiv:gr-qc/0201001},\ \Eprint {https://arxiv.org/abs/gr-qc/0201001} {arXiv:gr-qc/0201001 [gr-qc]} \BibitemShut {NoStop}%
\bibitem [{\citenamefont {Mishra}\ and\ \citenamefont {Calcino}(2024)}]{Mishra2024CircumSingleMHD}%
  \BibitemOpen
  \bibfield  {author} {\bibinfo {author} {\bibfnamefont {B.}~\bibnamefont {Mishra}}\ and\ \bibinfo {author} {\bibfnamefont {J.}~\bibnamefont {Calcino}},\ }\bibfield  {title} {\bibinfo {title} {Nature vs nurture: Three dimensional mhd simulations of misaligned embedded circum-single disks within an agn disk},\ }\href {https://arxiv.org/abs/2409.05614} {\bibfield  {journal} {\bibinfo  {journal} {arXiv e-prints}\ } (\bibinfo {year} {2024})},\ \bibinfo {note} {3D MHD simulations showing that black hole close encounters in AGN disks often involve circum-single disks, requiring their modeling},\ \Eprint {https://arxiv.org/abs/2409.05614} {2409.05614} \BibitemShut {NoStop}%
\bibitem [{\citenamefont {{Li}}\ \emph {et~al.}(2023{\natexlab{a}})\citenamefont {{Li}}, \citenamefont {{Dempsey}}, \citenamefont {{Li}}, \citenamefont {{Lai}},\ and\ \citenamefont {{Li}}}]{LiJ2023}%
  \BibitemOpen
  \bibfield  {author} {\bibinfo {author} {\bibfnamefont {J.}~\bibnamefont {{Li}}}, \bibinfo {author} {\bibfnamefont {A.~M.}\ \bibnamefont {{Dempsey}}}, \bibinfo {author} {\bibfnamefont {H.}~\bibnamefont {{Li}}}, \bibinfo {author} {\bibfnamefont {D.}~\bibnamefont {{Lai}}},\ and\ \bibinfo {author} {\bibfnamefont {S.}~\bibnamefont {{Li}}},\ }\bibfield  {title} {\bibinfo {title} {{Hydrodynamical Simulations of Black Hole Binary Formation in AGN Disks}},\ }\href {https://doi.org/10.3847/2041-8213/acb934} {\bibfield  {journal} {\bibinfo  {journal} {\apjl}\ }\textbf {\bibinfo {volume} {944}},\ \bibinfo {eid} {L42} (\bibinfo {year} {2023}{\natexlab{a}})},\ \Eprint {https://arxiv.org/abs/2211.10357} {arXiv:2211.10357 [astro-ph.HE]} \BibitemShut {NoStop}%
\bibitem [{\citenamefont {{Wang}}\ \emph {et~al.}(2025)\citenamefont {{Wang}}, \citenamefont {{Ma}}, \citenamefont {{Li}}, \citenamefont {{Wu}}, \citenamefont {{Li}}, \citenamefont {{Lei}},\ and\ \citenamefont {{Wu}}}]{Wang2025}%
  \BibitemOpen
  \bibfield  {author} {\bibinfo {author} {\bibfnamefont {M.}~\bibnamefont {{Wang}}}, \bibinfo {author} {\bibfnamefont {Y.}~\bibnamefont {{Ma}}}, \bibinfo {author} {\bibfnamefont {H.}~\bibnamefont {{Li}}}, \bibinfo {author} {\bibfnamefont {Q.}~\bibnamefont {{Wu}}}, \bibinfo {author} {\bibfnamefont {Y.-P.}\ \bibnamefont {{Li}}}, \bibinfo {author} {\bibfnamefont {X.}~\bibnamefont {{Lei}}},\ and\ \bibinfo {author} {\bibfnamefont {J.}~\bibnamefont {{Wu}}},\ }\bibfield  {title} {\bibinfo {title} {{Simulation of Binary-single Interactions in AGN Disk. I. Gas-enhanced Binary Orbital Hardening}},\ }\href {https://doi.org/10.3847/1538-4357/adbf8e} {\bibfield  {journal} {\bibinfo  {journal} {\apj}\ }\textbf {\bibinfo {volume} {983}},\ \bibinfo {eid} {114} (\bibinfo {year} {2025})},\ \Eprint {https://arxiv.org/abs/2501.10703} {arXiv:2501.10703 [astro-ph.HE]} \BibitemShut {NoStop}%
\bibitem [{\citenamefont {Cresswell}\ and\ \citenamefont {Nelson}(2006)}]{Cresswell2006MultiProtoplanets}%
  \BibitemOpen
  \bibfield  {author} {\bibinfo {author} {\bibfnamefont {P.}~\bibnamefont {Cresswell}}\ and\ \bibinfo {author} {\bibfnamefont {R.~P.}\ \bibnamefont {Nelson}},\ }\bibfield  {title} {\bibinfo {title} {On the evolution of multiple protoplanets embedded in a protostellar disc},\ }\href {https://doi.org/10.1051/0004-6361:20054551} {\bibfield  {journal} {\bibinfo  {journal} {Astronomy \& Astrophysics}\ }\textbf {\bibinfo {volume} {450}},\ \bibinfo {pages} {833} (\bibinfo {year} {2006})}\BibitemShut {NoStop}%
\bibitem [{\citenamefont {Yang}\ \emph {et~al.}(2019)\citenamefont {Yang}, \citenamefont {Bonga}, \citenamefont {Peng},\ and\ \citenamefont {Li}}]{Yang2019RelativisticMMR}%
  \BibitemOpen
  \bibfield  {author} {\bibinfo {author} {\bibfnamefont {H.}~\bibnamefont {Yang}}, \bibinfo {author} {\bibfnamefont {B.}~\bibnamefont {Bonga}}, \bibinfo {author} {\bibfnamefont {Z.}~\bibnamefont {Peng}},\ and\ \bibinfo {author} {\bibfnamefont {G.}~\bibnamefont {Li}},\ }\bibfield  {title} {\bibinfo {title} {Relativistic mean motion resonance},\ }\href {https://doi.org/10.1103/PhysRevD.100.124056} {\bibfield  {journal} {\bibinfo  {journal} {Physical Review D}\ }\textbf {\bibinfo {volume} {100}},\ \bibinfo {pages} {124056} (\bibinfo {year} {2019})},\ \bibinfo {note} {studies relativistic mean motion resonances in systems with stellar mass black holes orbiting a supermassive black hole, with implications for gravitational wave observations.},\ \Eprint {https://arxiv.org/abs/1910.07337} {arXiv:1910.07337 [gr-qc]} \BibitemShut {NoStop}%
\bibitem [{\citenamefont {{Balbus}}\ and\ \citenamefont {{Hawley}}(1991)}]{Balbus1991}%
  \BibitemOpen
  \bibfield  {author} {\bibinfo {author} {\bibfnamefont {S.~A.}\ \bibnamefont {{Balbus}}}\ and\ \bibinfo {author} {\bibfnamefont {J.~F.}\ \bibnamefont {{Hawley}}},\ }\bibfield  {title} {\bibinfo {title} {{A Powerful Local Shear Instability in Weakly Magnetized Disks. I. Linear Analysis}},\ }\href {https://doi.org/10.1086/170270} {\bibfield  {journal} {\bibinfo  {journal} {\apj}\ }\textbf {\bibinfo {volume} {376}},\ \bibinfo {pages} {214} (\bibinfo {year} {1991})}\BibitemShut {NoStop}%
\bibitem [{\citenamefont {{Ben{\'\i}tez-Llambay}}\ and\ \citenamefont {{Masset}}(2016)}]{FARGO3D}%
  \BibitemOpen
  \bibfield  {author} {\bibinfo {author} {\bibfnamefont {P.}~\bibnamefont {{Ben{\'\i}tez-Llambay}}}\ and\ \bibinfo {author} {\bibfnamefont {F.~S.}\ \bibnamefont {{Masset}}},\ }\bibfield  {title} {\bibinfo {title} {{FARGO3D: A New GPU-oriented MHD Code}},\ }\href {https://doi.org/10.3847/0067-0049/223/1/11} {\bibfield  {journal} {\bibinfo  {journal} {\apjs}\ }\textbf {\bibinfo {volume} {223}},\ \bibinfo {eid} {11} (\bibinfo {year} {2016})},\ \Eprint {https://arxiv.org/abs/1602.02359} {arXiv:1602.02359 [astro-ph.IM]} \BibitemShut {NoStop}%
\bibitem [{\citenamefont {{Shakura}}\ and\ \citenamefont {{Sunyaev}}(1973)}]{shakura1973}%
  \BibitemOpen
  \bibfield  {author} {\bibinfo {author} {\bibfnamefont {N.~I.}\ \bibnamefont {{Shakura}}}\ and\ \bibinfo {author} {\bibfnamefont {R.~A.}\ \bibnamefont {{Sunyaev}}},\ }\bibfield  {title} {\bibinfo {title} {{Black holes in binary systems. Observational appearance.}},\ }\href@noop {} {\bibfield  {journal} {\bibinfo  {journal} {\aap}\ }\textbf {\bibinfo {volume} {24}},\ \bibinfo {pages} {337} (\bibinfo {year} {1973})}\BibitemShut {NoStop}%
\bibitem [{\citenamefont {{Ogihara}}\ \emph {et~al.}(2007)\citenamefont {{Ogihara}}, \citenamefont {{Ida}},\ and\ \citenamefont {{Morbidelli}}}]{Ogihara2007}%
  \BibitemOpen
  \bibfield  {author} {\bibinfo {author} {\bibfnamefont {M.}~\bibnamefont {{Ogihara}}}, \bibinfo {author} {\bibfnamefont {S.}~\bibnamefont {{Ida}}},\ and\ \bibinfo {author} {\bibfnamefont {A.}~\bibnamefont {{Morbidelli}}},\ }\bibfield  {title} {\bibinfo {title} {{Accretion of terrestrial planets from oligarchs in a turbulent disk}},\ }\href {https://doi.org/10.1016/j.icarus.2006.12.006} {\bibfield  {journal} {\bibinfo  {journal} {\icarus}\ }\textbf {\bibinfo {volume} {188}},\ \bibinfo {pages} {522} (\bibinfo {year} {2007})},\ \Eprint {https://arxiv.org/abs/astro-ph/0612619} {arXiv:astro-ph/0612619 [astro-ph]} \BibitemShut {NoStop}%
\bibitem [{\citenamefont {{Baruteau}}\ and\ \citenamefont {{Lin}}(2010)}]{BaruteauLin2010}%
  \BibitemOpen
  \bibfield  {author} {\bibinfo {author} {\bibfnamefont {C.}~\bibnamefont {{Baruteau}}}\ and\ \bibinfo {author} {\bibfnamefont {D.~N.~C.}\ \bibnamefont {{Lin}}},\ }\bibfield  {title} {\bibinfo {title} {{Protoplanetary Migration in Turbulent Isothermal Disks}},\ }\href {https://doi.org/10.1088/0004-637X/709/2/759} {\bibfield  {journal} {\bibinfo  {journal} {\apj}\ }\textbf {\bibinfo {volume} {709}},\ \bibinfo {pages} {759} (\bibinfo {year} {2010})},\ \Eprint {https://arxiv.org/abs/0912.0964} {arXiv:0912.0964 [astro-ph.EP]} \BibitemShut {NoStop}%
\bibitem [{\citenamefont {{Pierens}}\ \emph {et~al.}(2011)\citenamefont {{Pierens}}, \citenamefont {{Baruteau}},\ and\ \citenamefont {{Hersant}}}]{Pierens2011}%
  \BibitemOpen
  \bibfield  {author} {\bibinfo {author} {\bibfnamefont {A.}~\bibnamefont {{Pierens}}}, \bibinfo {author} {\bibfnamefont {C.}~\bibnamefont {{Baruteau}}},\ and\ \bibinfo {author} {\bibfnamefont {F.}~\bibnamefont {{Hersant}}},\ }\bibfield  {title} {\bibinfo {title} {{On the dynamics of resonant super-Earths in disks with turbulence driven by stochastic forcing}},\ }\href {https://doi.org/10.1051/0004-6361/201116611} {\bibfield  {journal} {\bibinfo  {journal} {\aap}\ }\textbf {\bibinfo {volume} {531}},\ \bibinfo {eid} {A5} (\bibinfo {year} {2011})},\ \Eprint {https://arxiv.org/abs/1103.4923} {arXiv:1103.4923 [astro-ph.EP]} \BibitemShut {NoStop}%
\bibitem [{\citenamefont {{Chen}}\ and\ \citenamefont {{Lin}}(2023)}]{ChenLin2023}%
  \BibitemOpen
  \bibfield  {author} {\bibinfo {author} {\bibfnamefont {Y.-X.}\ \bibnamefont {{Chen}}}\ and\ \bibinfo {author} {\bibfnamefont {D.~N.~C.}\ \bibnamefont {{Lin}}},\ }\bibfield  {title} {\bibinfo {title} {{Chaotic gas accretion by black holes embedded in AGN discs as cause of low-spin signatures in gravitational wave events}},\ }\href {https://doi.org/10.1093/mnras/stad992} {\bibfield  {journal} {\bibinfo  {journal} {\mnras}\ }\textbf {\bibinfo {volume} {522}},\ \bibinfo {pages} {319} (\bibinfo {year} {2023})},\ \Eprint {https://arxiv.org/abs/2303.17097} {arXiv:2303.17097 [astro-ph.HE]} \BibitemShut {NoStop}%
\bibitem [{\citenamefont {{Wu}}\ \emph {et~al.}(2024)\citenamefont {{Wu}}, \citenamefont {{Chen}},\ and\ \citenamefont {{Lin}}}]{Wu2024chaotic}%
  \BibitemOpen
  \bibfield  {author} {\bibinfo {author} {\bibfnamefont {Y.}~\bibnamefont {{Wu}}}, \bibinfo {author} {\bibfnamefont {Y.-X.}\ \bibnamefont {{Chen}}},\ and\ \bibinfo {author} {\bibfnamefont {D.~N.~C.}\ \bibnamefont {{Lin}}},\ }\bibfield  {title} {\bibinfo {title} {{Chaotic Type I migration in turbulent discs}},\ }\href {https://doi.org/10.1093/mnrasl/slad183} {\bibfield  {journal} {\bibinfo  {journal} {\mnras}\ }\textbf {\bibinfo {volume} {528}},\ \bibinfo {pages} {L127} (\bibinfo {year} {2024})},\ \Eprint {https://arxiv.org/abs/2311.15747} {arXiv:2311.15747 [astro-ph.EP]} \BibitemShut {NoStop}%
\bibitem [{\citenamefont {{Chen}}\ \emph {et~al.}(2025)\citenamefont {{Chen}}, \citenamefont {{Wu}}, \citenamefont {{Li}}, \citenamefont {{Lin}}, \citenamefont {{Alexander}}, \citenamefont {{Nayakshin}},\ and\ \citenamefont {{Dai}}}]{Chen2025}%
  \BibitemOpen
  \bibfield  {author} {\bibinfo {author} {\bibfnamefont {Y.-X.}\ \bibnamefont {{Chen}}}, \bibinfo {author} {\bibfnamefont {Y.}~\bibnamefont {{Wu}}}, \bibinfo {author} {\bibfnamefont {Y.-P.}\ \bibnamefont {{Li}}}, \bibinfo {author} {\bibfnamefont {D.~N.~C.}\ \bibnamefont {{Lin}}}, \bibinfo {author} {\bibfnamefont {R.}~\bibnamefont {{Alexander}}}, \bibinfo {author} {\bibfnamefont {S.}~\bibnamefont {{Nayakshin}}},\ and\ \bibinfo {author} {\bibfnamefont {F.}~\bibnamefont {{Dai}}},\ }\bibfield  {title} {\bibinfo {title} {{Capture and escape of planetary mean-motion resonances in turbulent discs}},\ }\href {https://doi.org/10.1093/mnras/staf867} {\bibfield  {journal} {\bibinfo  {journal} {\mnras}\ }\textbf {\bibinfo {volume} {540}},\ \bibinfo {pages} {1998} (\bibinfo {year} {2025})},\ \Eprint {https://arxiv.org/abs/2505.13952} {arXiv:2505.13952 [astro-ph.EP]} \BibitemShut {NoStop}%
\bibitem [{\citenamefont {{Laughlin}}\ \emph {et~al.}(2004)\citenamefont {{Laughlin}}, \citenamefont {{Steinacker}},\ and\ \citenamefont {{Adams}}}]{Laughlin2004}%
  \BibitemOpen
  \bibfield  {author} {\bibinfo {author} {\bibfnamefont {G.}~\bibnamefont {{Laughlin}}}, \bibinfo {author} {\bibfnamefont {A.}~\bibnamefont {{Steinacker}}},\ and\ \bibinfo {author} {\bibfnamefont {F.~C.}\ \bibnamefont {{Adams}}},\ }\bibfield  {title} {\bibinfo {title} {{Type I Planetary Migration with MHD Turbulence}},\ }\href {https://doi.org/10.1086/386316} {\bibfield  {journal} {\bibinfo  {journal} {\apj}\ }\textbf {\bibinfo {volume} {608}},\ \bibinfo {pages} {489} (\bibinfo {year} {2004})},\ \Eprint {https://arxiv.org/abs/astro-ph/0308406} {arXiv:astro-ph/0308406 [astro-ph]} \BibitemShut {NoStop}%
\bibitem [{\citenamefont {{de Val-Borro}}\ \emph {et~al.}(2006)\citenamefont {{de Val-Borro}}, \citenamefont {{Edgar}}, \citenamefont {{Artymowicz}}, \citenamefont {{Ciecielag}}, \citenamefont {{Cresswell}}, \citenamefont {{D'Angelo}}, \citenamefont {{Delgado-Donate}}, \citenamefont {{Dirksen}}, \citenamefont {{Fromang}}, \citenamefont {{Gawryszczak}}, \citenamefont {{Klahr}}, \citenamefont {{Kley}}, \citenamefont {{Lyra}}, \citenamefont {{Masset}}, \citenamefont {{Mellema}}, \citenamefont {{Nelson}}, \citenamefont {{Paardekooper}}, \citenamefont {{Peplinski}}, \citenamefont {{Pierens}}, \citenamefont {{Plewa}}, \citenamefont {{Rice}}, \citenamefont {{Sch{\"a}fer}},\ and\ \citenamefont {{Speith}}}]{deValBorro2006}%
  \BibitemOpen
  \bibfield  {author} {\bibinfo {author} {\bibfnamefont {M.}~\bibnamefont {{de Val-Borro}}}, \bibinfo {author} {\bibfnamefont {R.~G.}\ \bibnamefont {{Edgar}}}, \bibinfo {author} {\bibfnamefont {P.}~\bibnamefont {{Artymowicz}}}, \bibinfo {author} {\bibfnamefont {P.}~\bibnamefont {{Ciecielag}}}, \bibinfo {author} {\bibfnamefont {P.}~\bibnamefont {{Cresswell}}}, \bibinfo {author} {\bibfnamefont {G.}~\bibnamefont {{D'Angelo}}}, \bibinfo {author} {\bibfnamefont {E.~J.}\ \bibnamefont {{Delgado-Donate}}}, \bibinfo {author} {\bibfnamefont {G.}~\bibnamefont {{Dirksen}}}, \bibinfo {author} {\bibfnamefont {S.}~\bibnamefont {{Fromang}}}, \bibinfo {author} {\bibfnamefont {A.}~\bibnamefont {{Gawryszczak}}}, \bibinfo {author} {\bibfnamefont {H.}~\bibnamefont {{Klahr}}}, \bibinfo {author} {\bibfnamefont {W.}~\bibnamefont {{Kley}}}, \bibinfo {author} {\bibfnamefont {W.}~\bibnamefont {{Lyra}}}, \bibinfo {author} {\bibfnamefont {F.}~\bibnamefont {{Masset}}}, \bibinfo {author} {\bibfnamefont {G.}~\bibnamefont {{Mellema}}}, \bibinfo {author} {\bibfnamefont {R.~P.}\ \bibnamefont {{Nelson}}}, \bibinfo {author} {\bibfnamefont {S.~J.}\ \bibnamefont {{Paardekooper}}}, \bibinfo {author} {\bibfnamefont {A.}~\bibnamefont {{Peplinski}}}, \bibinfo {author} {\bibfnamefont {A.}~\bibnamefont {{Pierens}}}, \bibinfo {author} {\bibfnamefont {T.}~\bibnamefont {{Plewa}}}, \bibinfo {author} {\bibfnamefont {K.}~\bibnamefont {{Rice}}}, \bibinfo {author} {\bibfnamefont {C.}~\bibnamefont {{Sch{\"a}fer}}},\ and\ \bibinfo {author} {\bibfnamefont {R.}~\bibnamefont {{Speith}}},\ }\bibfield  {title} {\bibinfo {title} {{A comparative study of disc-planet interaction}},\ }\href {https://doi.org/10.1111/j.1365-2966.2006.10488.x} {\bibfield  {journal} {\bibinfo  {journal} {\mnras}\ }\textbf {\bibinfo {volume} {370}},\ \bibinfo {pages} {529} (\bibinfo {year} {2006})},\ \Eprint {https://arxiv.org/abs/astro-ph/0605237} {arXiv:astro-ph/0605237 [astro-ph]} \BibitemShut {NoStop}%
\bibitem [{\citenamefont {{Li}}\ \emph {et~al.}(2025)\citenamefont {{Li}}, \citenamefont {{Yang}},\ and\ \citenamefont {{Pan}}}]{Li2025}%
  \BibitemOpen
  \bibfield  {author} {\bibinfo {author} {\bibfnamefont {Y.-P.}\ \bibnamefont {{Li}}}, \bibinfo {author} {\bibfnamefont {H.}~\bibnamefont {{Yang}}},\ and\ \bibinfo {author} {\bibfnamefont {Z.}~\bibnamefont {{Pan}}},\ }\bibfield  {title} {\bibinfo {title} {{Extreme mass-ratio inspirals in active galactic nucleus disks: The role of circumsingle disks}},\ }\href {https://doi.org/10.1103/PhysRevD.111.063074} {\bibfield  {journal} {\bibinfo  {journal} {\prd}\ }\textbf {\bibinfo {volume} {111}},\ \bibinfo {eid} {063074} (\bibinfo {year} {2025})},\ \Eprint {https://arxiv.org/abs/2503.04042} {arXiv:2503.04042 [astro-ph.HE]} \BibitemShut {NoStop}%
\bibitem [{\citenamefont {{Papaloizou}}\ and\ \citenamefont {{Larwood}}(2000)}]{Papaloizou2000}%
  \BibitemOpen
  \bibfield  {author} {\bibinfo {author} {\bibfnamefont {J.~C.~B.}\ \bibnamefont {{Papaloizou}}}\ and\ \bibinfo {author} {\bibfnamefont {J.~D.}\ \bibnamefont {{Larwood}}},\ }\bibfield  {title} {\bibinfo {title} {{On the orbital evolution and growth of protoplanets embedded in a gaseous disc}},\ }\href {https://doi.org/10.1046/j.1365-8711.2000.03466.x} {\bibfield  {journal} {\bibinfo  {journal} {\mnras}\ }\textbf {\bibinfo {volume} {315}},\ \bibinfo {pages} {823} (\bibinfo {year} {2000})},\ \Eprint {https://arxiv.org/abs/astro-ph/9911431} {arXiv:astro-ph/9911431 [astro-ph]} \BibitemShut {NoStop}%
\bibitem [{\citenamefont {{Ida}}\ \emph {et~al.}(2020)\citenamefont {{Ida}}, \citenamefont {{Muto}}, \citenamefont {{Matsumura}},\ and\ \citenamefont {{Brasser}}}]{Ida2020}%
  \BibitemOpen
  \bibfield  {author} {\bibinfo {author} {\bibfnamefont {S.}~\bibnamefont {{Ida}}}, \bibinfo {author} {\bibfnamefont {T.}~\bibnamefont {{Muto}}}, \bibinfo {author} {\bibfnamefont {S.}~\bibnamefont {{Matsumura}}},\ and\ \bibinfo {author} {\bibfnamefont {R.}~\bibnamefont {{Brasser}}},\ }\bibfield  {title} {\bibinfo {title} {{A new and simple prescription for planet orbital migration and eccentricity damping by planet-disc interactions based on dynamical friction}},\ }\href {https://doi.org/10.1093/mnras/staa1073} {\bibfield  {journal} {\bibinfo  {journal} {\mnras}\ }\textbf {\bibinfo {volume} {494}},\ \bibinfo {pages} {5666} (\bibinfo {year} {2020})},\ \Eprint {https://arxiv.org/abs/2004.07481} {arXiv:2004.07481 [astro-ph.EP]} \BibitemShut {NoStop}%
\bibitem [{\citenamefont {{Tanaka}}\ \emph {et~al.}(2002)\citenamefont {{Tanaka}}, \citenamefont {{Takeuchi}},\ and\ \citenamefont {{Ward}}}]{Tanaka2002}%
  \BibitemOpen
  \bibfield  {author} {\bibinfo {author} {\bibfnamefont {H.}~\bibnamefont {{Tanaka}}}, \bibinfo {author} {\bibfnamefont {T.}~\bibnamefont {{Takeuchi}}},\ and\ \bibinfo {author} {\bibfnamefont {W.~R.}\ \bibnamefont {{Ward}}},\ }\bibfield  {title} {\bibinfo {title} {{Three-Dimensional Interaction between a Planet and an Isothermal Gaseous Disk. I. Corotation and Lindblad Torques and Planet Migration}},\ }\href {https://doi.org/10.1086/324713} {\bibfield  {journal} {\bibinfo  {journal} {\apj}\ }\textbf {\bibinfo {volume} {565}},\ \bibinfo {pages} {1257} (\bibinfo {year} {2002})}\BibitemShut {NoStop}%
\bibitem [{\citenamefont {{Paardekooper}}\ \emph {et~al.}(2023)\citenamefont {{Paardekooper}}, \citenamefont {{Dong}}, \citenamefont {{Duffell}}, \citenamefont {{Fung}}, \citenamefont {{Masset}}, \citenamefont {{Ogilvie}},\ and\ \citenamefont {{Tanaka}}}]{Paardekooper2023}%
  \BibitemOpen
  \bibfield  {author} {\bibinfo {author} {\bibfnamefont {S.}~\bibnamefont {{Paardekooper}}}, \bibinfo {author} {\bibfnamefont {R.}~\bibnamefont {{Dong}}}, \bibinfo {author} {\bibfnamefont {P.}~\bibnamefont {{Duffell}}}, \bibinfo {author} {\bibfnamefont {J.}~\bibnamefont {{Fung}}}, \bibinfo {author} {\bibfnamefont {F.~S.}\ \bibnamefont {{Masset}}}, \bibinfo {author} {\bibfnamefont {G.}~\bibnamefont {{Ogilvie}}},\ and\ \bibinfo {author} {\bibfnamefont {H.}~\bibnamefont {{Tanaka}}},\ }\bibfield  {title} {\bibinfo {title} {{Planet-Disk Interactions and Orbital Evolution}},\ }in\ \href {https://doi.org/10.48550/arXiv.2203.09595} {\emph {\bibinfo {booktitle} {Protostars and Planets VII}}},\ \bibinfo {series} {Astronomical Society of the Pacific Conference Series}, Vol.\ \bibinfo {volume} {534},\ \bibinfo {editor} {edited by\ \bibinfo {editor} {\bibfnamefont {S.}~\bibnamefont {{Inutsuka}}}, \bibinfo {editor} {\bibfnamefont {Y.}~\bibnamefont {{Aikawa}}}, \bibinfo {editor} {\bibfnamefont {T.}~\bibnamefont {{Muto}}}, \bibinfo {editor} {\bibfnamefont {K.}~\bibnamefont {{Tomida}}},\ and\ \bibinfo {editor} {\bibfnamefont {M.}~\bibnamefont {{Tamura}}}}\ (\bibinfo {year} {2023})\ p.\ \bibinfo {pages} {685},\ \Eprint {https://arxiv.org/abs/2203.09595} {arXiv:2203.09595 [astro-ph.EP]} \BibitemShut {NoStop}%
\bibitem [{\citenamefont {{Rein}}\ and\ \citenamefont {{Tamayo}}(2015)}]{ReinTamayo2015}%
  \BibitemOpen
  \bibfield  {author} {\bibinfo {author} {\bibfnamefont {H.}~\bibnamefont {{Rein}}}\ and\ \bibinfo {author} {\bibfnamefont {D.}~\bibnamefont {{Tamayo}}},\ }\bibfield  {title} {\bibinfo {title} {{WHFAST: a fast and unbiased implementation of a symplectic Wisdom-Holman integrator for long-term gravitational simulations}},\ }\href {https://doi.org/10.1093/mnras/stv1257} {\bibfield  {journal} {\bibinfo  {journal} {\mnras}\ }\textbf {\bibinfo {volume} {452}},\ \bibinfo {pages} {376} (\bibinfo {year} {2015})},\ \Eprint {https://arxiv.org/abs/1506.01084} {arXiv:1506.01084 [astro-ph.EP]} \BibitemShut {NoStop}%
\bibitem [{\citenamefont {{Tamayo}}\ \emph {et~al.}(2020)\citenamefont {{Tamayo}}, \citenamefont {{Rein}}, \citenamefont {{Shi}},\ and\ \citenamefont {{Hernand ez}}}]{Tamayo2020}%
  \BibitemOpen
  \bibfield  {author} {\bibinfo {author} {\bibfnamefont {D.}~\bibnamefont {{Tamayo}}}, \bibinfo {author} {\bibfnamefont {H.}~\bibnamefont {{Rein}}}, \bibinfo {author} {\bibfnamefont {P.}~\bibnamefont {{Shi}}},\ and\ \bibinfo {author} {\bibfnamefont {D.~M.}\ \bibnamefont {{Hernand ez}}},\ }\bibfield  {title} {\bibinfo {title} {{REBOUNDx: a library for adding conservative and dissipative forces to otherwise symplectic N-body integrations}},\ }\href {https://doi.org/10.1093/mnras/stz2870} {\bibfield  {journal} {\bibinfo  {journal} {\mnras}\ }\textbf {\bibinfo {volume} {491}},\ \bibinfo {pages} {2885} (\bibinfo {year} {2020})},\ \Eprint {https://arxiv.org/abs/1908.05634} {arXiv:1908.05634 [astro-ph.EP]} \BibitemShut {NoStop}%
\bibitem [{\citenamefont {Copparoni}\ \emph {et~al.}(2025)\citenamefont {Copparoni}, \citenamefont {Barausse}, \citenamefont {Speri}, \citenamefont {Sberna},\ and\ \citenamefont {Derdzinski}}]{Copparoni:2025jhq}%
  \BibitemOpen
  \bibfield  {author} {\bibinfo {author} {\bibfnamefont {L.}~\bibnamefont {Copparoni}}, \bibinfo {author} {\bibfnamefont {E.}~\bibnamefont {Barausse}}, \bibinfo {author} {\bibfnamefont {L.}~\bibnamefont {Speri}}, \bibinfo {author} {\bibfnamefont {L.}~\bibnamefont {Sberna}},\ and\ \bibinfo {author} {\bibfnamefont {A.}~\bibnamefont {Derdzinski}},\ }\bibfield  {title} {\bibinfo {title} {{Implications of stochastic gas torques for asymmetric binaries in the LISA band}},\ }\href {https://doi.org/10.1103/PhysRevD.111.104079} {\bibfield  {journal} {\bibinfo  {journal} {Phys. Rev. D}\ }\textbf {\bibinfo {volume} {111}},\ \bibinfo {pages} {104079} (\bibinfo {year} {2025})},\ \Eprint {https://arxiv.org/abs/2502.10087} {arXiv:2502.10087 [gr-qc]} \BibitemShut {NoStop}%
\bibitem [{\citenamefont {{Derdzinski}}\ \emph {et~al.}(2021)\citenamefont {{Derdzinski}}, \citenamefont {{D'Orazio}}, \citenamefont {{Duffell}}, \citenamefont {{Haiman}},\ and\ \citenamefont {{MacFadyen}}}]{Derdzinski2021}%
  \BibitemOpen
  \bibfield  {author} {\bibinfo {author} {\bibfnamefont {A.}~\bibnamefont {{Derdzinski}}}, \bibinfo {author} {\bibfnamefont {D.}~\bibnamefont {{D'Orazio}}}, \bibinfo {author} {\bibfnamefont {P.}~\bibnamefont {{Duffell}}}, \bibinfo {author} {\bibfnamefont {Z.}~\bibnamefont {{Haiman}}},\ and\ \bibinfo {author} {\bibfnamefont {A.}~\bibnamefont {{MacFadyen}}},\ }\bibfield  {title} {\bibinfo {title} {{Evolution of gas disc-embedded intermediate mass ratio inspirals in the LISA band}},\ }\href {https://doi.org/10.1093/mnras/staa3976} {\bibfield  {journal} {\bibinfo  {journal} {\mnras}\ }\textbf {\bibinfo {volume} {501}},\ \bibinfo {pages} {3540} (\bibinfo {year} {2021})},\ \Eprint {https://arxiv.org/abs/2005.11333} {arXiv:2005.11333 [astro-ph.HE]} \BibitemShut {NoStop}%
\bibitem [{\citenamefont {{Lin}}\ and\ \citenamefont {{Papaloizou}}(1986)}]{Lin1986}%
  \BibitemOpen
  \bibfield  {author} {\bibinfo {author} {\bibfnamefont {D.~N.~C.}\ \bibnamefont {{Lin}}}\ and\ \bibinfo {author} {\bibfnamefont {J.}~\bibnamefont {{Papaloizou}}},\ }\bibfield  {title} {\bibinfo {title} {{On the Tidal Interaction between Protoplanets and the Protoplanetary Disk. III. Orbital Migration of Protoplanets}},\ }\href {https://doi.org/10.1086/164653} {\bibfield  {journal} {\bibinfo  {journal} {\apj}\ }\textbf {\bibinfo {volume} {309}},\ \bibinfo {pages} {846} (\bibinfo {year} {1986})}\BibitemShut {NoStop}%
\bibitem [{\citenamefont {{Duffell}}\ and\ \citenamefont {{MacFadyen}}(2013)}]{Duffell2013}%
  \BibitemOpen
  \bibfield  {author} {\bibinfo {author} {\bibfnamefont {P.~C.}\ \bibnamefont {{Duffell}}}\ and\ \bibinfo {author} {\bibfnamefont {A.~I.}\ \bibnamefont {{MacFadyen}}},\ }\bibfield  {title} {\bibinfo {title} {{Gap Opening by Extremely Low-mass Planets in a Viscous Disk}},\ }\href {https://doi.org/10.1088/0004-637X/769/1/41} {\bibfield  {journal} {\bibinfo  {journal} {\apj}\ }\textbf {\bibinfo {volume} {769}},\ \bibinfo {eid} {41} (\bibinfo {year} {2013})},\ \Eprint {https://arxiv.org/abs/1302.1934} {arXiv:1302.1934 [astro-ph.EP]} \BibitemShut {NoStop}%
\bibitem [{\citenamefont {{Kanagawa}}\ \emph {et~al.}(2015)\citenamefont {{Kanagawa}}, \citenamefont {{Muto}}, \citenamefont {{Tanaka}}, \citenamefont {{Tanigawa}}, \citenamefont {{Takeuchi}}, \citenamefont {{Tsukagoshi}},\ and\ \citenamefont {{Momose}}}]{Kanagawa2015}%
  \BibitemOpen
  \bibfield  {author} {\bibinfo {author} {\bibfnamefont {K.~D.}\ \bibnamefont {{Kanagawa}}}, \bibinfo {author} {\bibfnamefont {T.}~\bibnamefont {{Muto}}}, \bibinfo {author} {\bibfnamefont {H.}~\bibnamefont {{Tanaka}}}, \bibinfo {author} {\bibfnamefont {T.}~\bibnamefont {{Tanigawa}}}, \bibinfo {author} {\bibfnamefont {T.}~\bibnamefont {{Takeuchi}}}, \bibinfo {author} {\bibfnamefont {T.}~\bibnamefont {{Tsukagoshi}}},\ and\ \bibinfo {author} {\bibfnamefont {M.}~\bibnamefont {{Momose}}},\ }\bibfield  {title} {\bibinfo {title} {{Mass Estimates of a Giant Planet in a Protoplanetary Disk from the Gap Structures}},\ }\href {https://doi.org/10.1088/2041-8205/806/1/L15} {\bibfield  {journal} {\bibinfo  {journal} {\apjl}\ }\textbf {\bibinfo {volume} {806}},\ \bibinfo {eid} {L15} (\bibinfo {year} {2015})},\ \Eprint {https://arxiv.org/abs/1505.04482} {arXiv:1505.04482 [astro-ph.EP]} \BibitemShut {NoStop}%
\bibitem [{\citenamefont {{Yuan}}\ and\ \citenamefont {{Narayan}}(2014)}]{Yuan2014}%
  \BibitemOpen
  \bibfield  {author} {\bibinfo {author} {\bibfnamefont {F.}~\bibnamefont {{Yuan}}}\ and\ \bibinfo {author} {\bibfnamefont {R.}~\bibnamefont {{Narayan}}},\ }\bibfield  {title} {\bibinfo {title} {{Hot Accretion Flows Around Black Holes}},\ }\href {https://doi.org/10.1146/annurev-astro-082812-141003} {\bibfield  {journal} {\bibinfo  {journal} {\araa}\ }\textbf {\bibinfo {volume} {52}},\ \bibinfo {pages} {529} (\bibinfo {year} {2014})},\ \Eprint {https://arxiv.org/abs/1401.0586} {arXiv:1401.0586 [astro-ph.HE]} \BibitemShut {NoStop}%
\bibitem [{\citenamefont {{Seymour}}\ and\ \citenamefont {{Chen}}(2024)}]{Seymour:2024kcd}%
  \BibitemOpen
  \bibfield  {author} {\bibinfo {author} {\bibfnamefont {B.~C.}\ \bibnamefont {{Seymour}}}\ and\ \bibinfo {author} {\bibfnamefont {Y.}~\bibnamefont {{Chen}}},\ }\bibfield  {title} {\bibinfo {title} {{Gravitational-wave signatures of non-violent non-locality}},\ }\href {https://doi.org/10.48550/arXiv.2411.13714} {\bibfield  {journal} {\bibinfo  {journal} {arXiv e-prints}\ ,\ \bibinfo {eid} {arXiv:2411.13714}} (\bibinfo {year} {2024})},\ \Eprint {https://arxiv.org/abs/2411.13714} {arXiv:2411.13714 [gr-qc]} \BibitemShut {NoStop}%
\bibitem [{\citenamefont {Hughes}(2000)}]{Hughes2000}%
  \BibitemOpen
  \bibfield  {author} {\bibinfo {author} {\bibfnamefont {S.~A.}\ \bibnamefont {Hughes}},\ }\bibfield  {title} {\bibinfo {title} {Evolution of circular, nonequatorial orbits of kerr black holes due to gravitational-wave emission},\ }\href {https://doi.org/10.1103/PhysRevD.61.084004} {\bibfield  {journal} {\bibinfo  {journal} {Phys. Rev. D}\ }\textbf {\bibinfo {volume} {61}},\ \bibinfo {pages} {084004} (\bibinfo {year} {2000})},\ \Eprint {https://arxiv.org/abs/gr-qc/9910091} {arXiv:gr-qc/9910091 [gr-qc]} \BibitemShut {NoStop}%
\bibitem [{\citenamefont {Bardeen}\ and\ \citenamefont {Petterson}(1975)}]{BardeenPetterson1975}%
  \BibitemOpen
  \bibfield  {author} {\bibinfo {author} {\bibfnamefont {J.~M.}\ \bibnamefont {Bardeen}}\ and\ \bibinfo {author} {\bibfnamefont {J.~A.}\ \bibnamefont {Petterson}},\ }\bibfield  {title} {\bibinfo {title} {The lense--thirring effect and accretion disks around kerr black holes},\ }\href {https://doi.org/10.1086/181711} {\bibfield  {journal} {\bibinfo  {journal} {The Astrophysical Journal Letters}\ }\textbf {\bibinfo {volume} {195}},\ \bibinfo {pages} {L65} (\bibinfo {year} {1975})}\BibitemShut {NoStop}%
\bibitem [{\citenamefont {Lodato}\ and\ \citenamefont {Gerosa}(2013)}]{Lodato:2012yr}%
  \BibitemOpen
  \bibfield  {author} {\bibinfo {author} {\bibfnamefont {G.}~\bibnamefont {Lodato}}\ and\ \bibinfo {author} {\bibfnamefont {D.}~\bibnamefont {Gerosa}},\ }\bibfield  {title} {\bibinfo {title} {{Black hole mergers: do gas discs lead to spin alignment?}},\ }\href {https://doi.org/10.1093/mnrasl/sls018} {\bibfield  {journal} {\bibinfo  {journal} {Mon. Not. Roy. Astron. Soc.}\ }\textbf {\bibinfo {volume} {429}},\ \bibinfo {pages} {L30} (\bibinfo {year} {2013})},\ \Eprint {https://arxiv.org/abs/1211.0284} {arXiv:1211.0284 [astro-ph.CO]} \BibitemShut {NoStop}%
\bibitem [{\citenamefont {Lyu}\ \emph {et~al.}(2024)\citenamefont {Lyu}, \citenamefont {Pan}, \citenamefont {Mao}, \citenamefont {Jiang},\ and\ \citenamefont {Yang}}]{Lyu2024WetEMRIs}%
  \BibitemOpen
  \bibfield  {author} {\bibinfo {author} {\bibfnamefont {Z.}~\bibnamefont {Lyu}}, \bibinfo {author} {\bibfnamefont {Z.}~\bibnamefont {Pan}}, \bibinfo {author} {\bibfnamefont {J.}~\bibnamefont {Mao}}, \bibinfo {author} {\bibfnamefont {N.}~\bibnamefont {Jiang}},\ and\ \bibinfo {author} {\bibfnamefont {H.}~\bibnamefont {Yang}},\ }\bibfield  {title} {\bibinfo {title} {Science opportunities of wet extreme mass-ratio inspirals},\ }\href {https://doi.org/10.48550/arXiv.2501.03252} {\bibfield  {journal} {\bibinfo  {journal} {arXiv e-prints}\ } (\bibinfo {year} {2024})},\ \bibinfo {note} {submitted on 27 Dec 2024},\ \Eprint {https://arxiv.org/abs/2501.03252} {2501.03252} \BibitemShut {NoStop}%
\bibitem [{\citenamefont {King}\ \emph {et~al.}(2005)\citenamefont {King}, \citenamefont {Lubow}, \citenamefont {Ogilvie},\ and\ \citenamefont {Pringle}}]{King2005_Alignment}%
  \BibitemOpen
  \bibfield  {author} {\bibinfo {author} {\bibfnamefont {A.~R.}\ \bibnamefont {King}}, \bibinfo {author} {\bibfnamefont {S.~H.}\ \bibnamefont {Lubow}}, \bibinfo {author} {\bibfnamefont {G.~I.}\ \bibnamefont {Ogilvie}},\ and\ \bibinfo {author} {\bibfnamefont {J.~E.}\ \bibnamefont {Pringle}},\ }\bibfield  {title} {\bibinfo {title} {Aligning spinning black holes and accretion discs},\ }\href {https://doi.org/10.1111/j.1365-2966.2005.09378.x} {\bibfield  {journal} {\bibinfo  {journal} {Mon. Not. Roy. Astron. Soc.}\ }\textbf {\bibinfo {volume} {363}},\ \bibinfo {pages} {49} (\bibinfo {year} {2005})},\ \bibinfo {note} {arXiv:astro-ph/0507098},\ \Eprint {https://arxiv.org/abs/astro-ph/0507098} {arXiv:astro-ph/0507098 [astro-ph]} \BibitemShut {NoStop}%
\bibitem [{\citenamefont {Amaro-Seoane}\ \emph {et~al.}(2013)\citenamefont {Amaro-Seoane}, \citenamefont {Sopuerta},\ and\ \citenamefont {Freitag}}]{AmaroSeoane2013EMRISpin}%
  \BibitemOpen
  \bibfield  {author} {\bibinfo {author} {\bibfnamefont {P.}~\bibnamefont {Amaro-Seoane}}, \bibinfo {author} {\bibfnamefont {C.~F.}\ \bibnamefont {Sopuerta}},\ and\ \bibinfo {author} {\bibfnamefont {M.~D.}\ \bibnamefont {Freitag}},\ }\bibfield  {title} {\bibinfo {title} {The role of the supermassive black hole spin in the estimation of the emri event rate},\ }\href {https://doi.org/10.1093/mnras/sts572} {\bibfield  {journal} {\bibinfo  {journal} {Monthly Notices of the Royal Astronomical Society}\ }\textbf {\bibinfo {volume} {429}},\ \bibinfo {pages} {3155} (\bibinfo {year} {2013})},\ \bibinfo {note} {accepted 21 May 2012; published in MNRAS 429, 3155-3165 (2013)},\ \Eprint {https://arxiv.org/abs/arXiv:1205.4713} {arXiv:arXiv:1205.4713 [astro-ph.CO]} \BibitemShut {NoStop}%
\bibitem [{\citenamefont {Roy}\ \emph {et~al.}(2025)\citenamefont {Roy}, \citenamefont {van Son},\ and\ \citenamefont {Farr}}]{Roy2025_MidThirtiesCrisis}%
  \BibitemOpen
  \bibfield  {author} {\bibinfo {author} {\bibfnamefont {S.~K.}\ \bibnamefont {Roy}}, \bibinfo {author} {\bibfnamefont {L.~A.~C.}\ \bibnamefont {van Son}},\ and\ \bibinfo {author} {\bibfnamefont {W.~M.}\ \bibnamefont {Farr}},\ }\bibfield  {title} {\bibinfo {title} {A mid-thirties crisis: Dissecting the properties of gravitational wave sources near the 35 solar mass peak},\ }\href {https://doi.org/10.48550/arXiv.2507.01086} {\bibfield  {journal} {\bibinfo  {journal} {arXiv}\ } (\bibinfo {year} {2025})},\ \bibinfo {note} {preprint, arXiv:2507.01086},\ \Eprint {https://arxiv.org/abs/2507.01086} {2507.01086 [astro-ph.HE]} \BibitemShut {NoStop}%
\bibitem [{\citenamefont {Pan}\ and\ \citenamefont {Yang}(2021{\natexlab{b}})}]{Pan:2021xhv}%
  \BibitemOpen
  \bibfield  {author} {\bibinfo {author} {\bibfnamefont {Z.}~\bibnamefont {Pan}}\ and\ \bibinfo {author} {\bibfnamefont {H.}~\bibnamefont {Yang}},\ }\bibfield  {title} {\bibinfo {title} {{Supercritical Accretion of Stellar-mass Compact Objects in Active Galactic Nuclei}},\ }\href {https://doi.org/10.3847/1538-4357/ac249c} {\bibfield  {journal} {\bibinfo  {journal} {Astrophys. J.}\ }\textbf {\bibinfo {volume} {923}},\ \bibinfo {pages} {173} (\bibinfo {year} {2021}{\natexlab{b}})},\ \Eprint {https://arxiv.org/abs/2108.00267} {arXiv:2108.00267 [astro-ph.HE]} \BibitemShut {NoStop}%
\bibitem [{\citenamefont {{Blandford}}\ and\ \citenamefont {{Begelman}}(1999)}]{Blandford:1999}%
  \BibitemOpen
  \bibfield  {author} {\bibinfo {author} {\bibfnamefont {R.~D.}\ \bibnamefont {{Blandford}}}\ and\ \bibinfo {author} {\bibfnamefont {M.~C.}\ \bibnamefont {{Begelman}}},\ }\bibfield  {title} {\bibinfo {title} {{On the fate of gas accreting at a low rate on to a black hole}},\ }\href {https://doi.org/10.1046/j.1365-8711.1999.02358.x} {\bibfield  {journal} {\bibinfo  {journal} {\mnras}\ }\textbf {\bibinfo {volume} {303}},\ \bibinfo {pages} {L1} (\bibinfo {year} {1999})},\ \Eprint {https://arxiv.org/abs/astro-ph/9809083} {arXiv:astro-ph/9809083 [astro-ph]} \BibitemShut {NoStop}%
\bibitem [{\citenamefont {Sądowski}\ \emph {et~al.}(2014)\citenamefont {Sądowski}, \citenamefont {Narayan}, \citenamefont {McKinney},\ and\ \citenamefont {Tchekhovskoy}}]{Sadowski2014}%
  \BibitemOpen
  \bibfield  {author} {\bibinfo {author} {\bibfnamefont {A.}~\bibnamefont {Sądowski}}, \bibinfo {author} {\bibfnamefont {R.}~\bibnamefont {Narayan}}, \bibinfo {author} {\bibfnamefont {J.~C.}\ \bibnamefont {McKinney}},\ and\ \bibinfo {author} {\bibfnamefont {A.}~\bibnamefont {Tchekhovskoy}},\ }\bibfield  {title} {\bibinfo {title} {Numerical simulations of super‑critical black hole accretion flows in general relativity},\ }\href {https://doi.org/10.1093/mnras/stt2479} {\bibfield  {journal} {\bibinfo  {journal} {Monthly Notices of the Royal Astronomical Society}\ }\textbf {\bibinfo {volume} {439}},\ \bibinfo {pages} {503} (\bibinfo {year} {2014})},\ \Eprint {https://arxiv.org/abs/1311.5900} {arXiv:1311.5900 [astro-ph.HE]} \BibitemShut {NoStop}%
\bibitem [{\citenamefont {{Fragile}}\ \emph {et~al.}(2025)\citenamefont {{Fragile}}, \citenamefont {{Middleton}}, \citenamefont {{Bollimpalli}},\ and\ \citenamefont {{Smith}}}]{Fragile:2025nes}%
  \BibitemOpen
  \bibfield  {author} {\bibinfo {author} {\bibfnamefont {P.~C.}\ \bibnamefont {{Fragile}}}, \bibinfo {author} {\bibfnamefont {M.~J.}\ \bibnamefont {{Middleton}}}, \bibinfo {author} {\bibfnamefont {D.~A.}\ \bibnamefont {{Bollimpalli}}},\ and\ \bibinfo {author} {\bibfnamefont {Z.}~\bibnamefont {{Smith}}},\ }\bibfield  {title} {\bibinfo {title} {{Long time-scale numerical simulations of large supercritical accretion discs}},\ }\href {https://doi.org/10.1093/mnras/staf890} {\bibfield  {journal} {\bibinfo  {journal} {\mnras}\ }\textbf {\bibinfo {volume} {540}},\ \bibinfo {pages} {2820} (\bibinfo {year} {2025})},\ \Eprint {https://arxiv.org/abs/2505.08859} {arXiv:2505.08859 [astro-ph.HE]} \BibitemShut {NoStop}%
\bibitem [{\citenamefont {{Li}}\ \emph {et~al.}(2023{\natexlab{b}})\citenamefont {{Li}}, \citenamefont {{Chen}},\ and\ \citenamefont {{Lin}}}]{Li2023}%
  \BibitemOpen
  \bibfield  {author} {\bibinfo {author} {\bibfnamefont {Y.-P.}\ \bibnamefont {{Li}}}, \bibinfo {author} {\bibfnamefont {Y.-X.}\ \bibnamefont {{Chen}}},\ and\ \bibinfo {author} {\bibfnamefont {D.~N.~C.}\ \bibnamefont {{Lin}}},\ }\bibfield  {title} {\bibinfo {title} {{3D global simulations of accretion onto gap-opening planets: implications for circumplanetary disc structures and accretion rates}},\ }\href {https://doi.org/10.1093/mnras/stad3049} {\bibfield  {journal} {\bibinfo  {journal} {\mnras}\ }\textbf {\bibinfo {volume} {526}},\ \bibinfo {pages} {5346} (\bibinfo {year} {2023}{\natexlab{b}})},\ \Eprint {https://arxiv.org/abs/2310.02822} {arXiv:2310.02822 [astro-ph.EP]} \BibitemShut {NoStop}%
\bibitem [{\citenamefont {{Li}}\ \emph {et~al.}(2024)\citenamefont {{Li}}, \citenamefont {{Chen}},\ and\ \citenamefont {{Lin}}}]{Li2024}%
  \BibitemOpen
  \bibfield  {author} {\bibinfo {author} {\bibfnamefont {Y.-P.}\ \bibnamefont {{Li}}}, \bibinfo {author} {\bibfnamefont {Y.-X.}\ \bibnamefont {{Chen}}},\ and\ \bibinfo {author} {\bibfnamefont {D.~N.~C.}\ \bibnamefont {{Lin}}},\ }\bibfield  {title} {\bibinfo {title} {{Concurrent Accretion and Migration of Giant Planets in Their Natal Disks with Consistent Accretion Torque}},\ }\href {https://doi.org/10.3847/1538-4357/ad5a06} {\bibfield  {journal} {\bibinfo  {journal} {\apj}\ }\textbf {\bibinfo {volume} {971}},\ \bibinfo {eid} {130} (\bibinfo {year} {2024})},\ \Eprint {https://arxiv.org/abs/2406.12716} {arXiv:2406.12716 [astro-ph.EP]} \BibitemShut {NoStop}%
\bibitem [{\citenamefont {Tagawa}\ \emph {et~al.}(2020)\citenamefont {Tagawa}, \citenamefont {Haiman},\ and\ \citenamefont {Kocsis}}]{Tagawa2020}%
  \BibitemOpen
  \bibfield  {author} {\bibinfo {author} {\bibfnamefont {H.}~\bibnamefont {Tagawa}}, \bibinfo {author} {\bibfnamefont {Z.}~\bibnamefont {Haiman}},\ and\ \bibinfo {author} {\bibfnamefont {B.}~\bibnamefont {Kocsis}},\ }\bibfield  {title} {\bibinfo {title} {Formation and evolution of compact‑object binaries in agn disks},\ }\href {https://doi.org/10.3847/1538-4357/ab9b8c} {\bibfield  {journal} {\bibinfo  {journal} {The Astrophysical Journal}\ }\textbf {\bibinfo {volume} {898}},\ \bibinfo {pages} {25} (\bibinfo {year} {2020})},\ \Eprint {https://arxiv.org/abs/1912.08218} {arXiv:1912.08218 [astro-ph.GA]} \BibitemShut {NoStop}%
\bibitem [{\citenamefont {{Li}}\ \emph {et~al.}(2021)\citenamefont {{Li}}, \citenamefont {{Dempsey}}, \citenamefont {{Li}}, \citenamefont {{Li}},\ and\ \citenamefont {{Li}}}]{Li2021}%
  \BibitemOpen
  \bibfield  {author} {\bibinfo {author} {\bibfnamefont {Y.-P.}\ \bibnamefont {{Li}}}, \bibinfo {author} {\bibfnamefont {A.~M.}\ \bibnamefont {{Dempsey}}}, \bibinfo {author} {\bibfnamefont {S.}~\bibnamefont {{Li}}}, \bibinfo {author} {\bibfnamefont {H.}~\bibnamefont {{Li}}},\ and\ \bibinfo {author} {\bibfnamefont {J.}~\bibnamefont {{Li}}},\ }\bibfield  {title} {\bibinfo {title} {{Orbital Evolution of Binary Black Holes in Active Galactic Nucleus Disks: A Disk Channel for Binary Black Hole Mergers?}},\ }\href {https://doi.org/10.3847/1538-4357/abed48} {\bibfield  {journal} {\bibinfo  {journal} {\apj}\ }\textbf {\bibinfo {volume} {911}},\ \bibinfo {eid} {124} (\bibinfo {year} {2021})},\ \Eprint {https://arxiv.org/abs/2101.09406} {arXiv:2101.09406 [astro-ph.HE]} \BibitemShut {NoStop}%
\bibitem [{\citenamefont {{Li}}\ \emph {et~al.}(2022{\natexlab{a}})\citenamefont {{Li}}, \citenamefont {{Dempsey}}, \citenamefont {{Li}}, \citenamefont {{Li}},\ and\ \citenamefont {{Li}}}]{Li2022b}%
  \BibitemOpen
  \bibfield  {author} {\bibinfo {author} {\bibfnamefont {Y.-P.}\ \bibnamefont {{Li}}}, \bibinfo {author} {\bibfnamefont {A.~M.}\ \bibnamefont {{Dempsey}}}, \bibinfo {author} {\bibfnamefont {H.}~\bibnamefont {{Li}}}, \bibinfo {author} {\bibfnamefont {S.}~\bibnamefont {{Li}}},\ and\ \bibinfo {author} {\bibfnamefont {J.}~\bibnamefont {{Li}}},\ }\bibfield  {title} {\bibinfo {title} {{Hot Circumsingle Disks Drive Binary Black Hole Mergers in Active Galactic Nucleus Disks}},\ }\href {https://doi.org/10.3847/2041-8213/ac60fd} {\bibfield  {journal} {\bibinfo  {journal} {\apjl}\ }\textbf {\bibinfo {volume} {928}},\ \bibinfo {eid} {L19} (\bibinfo {year} {2022}{\natexlab{a}})},\ \Eprint {https://arxiv.org/abs/2112.11057} {arXiv:2112.11057 [astro-ph.HE]} \BibitemShut {NoStop}%
\bibitem [{\citenamefont {{Li}}\ and\ \citenamefont {{Lai}}(2022)}]{LiR2022}%
  \BibitemOpen
  \bibfield  {author} {\bibinfo {author} {\bibfnamefont {R.}~\bibnamefont {{Li}}}\ and\ \bibinfo {author} {\bibfnamefont {D.}~\bibnamefont {{Lai}}},\ }\bibfield  {title} {\bibinfo {title} {{Hydrodynamical evolution of black-hole binaries embedded in AGN discs}},\ }\href {https://doi.org/10.1093/mnras/stac2577} {\bibfield  {journal} {\bibinfo  {journal} {\mnras}\ }\textbf {\bibinfo {volume} {517}},\ \bibinfo {pages} {1602} (\bibinfo {year} {2022})},\ \Eprint {https://arxiv.org/abs/2202.07633} {arXiv:2202.07633 [astro-ph.HE]} \BibitemShut {NoStop}%
\bibitem [{\citenamefont {{Dittmann}}\ \emph {et~al.}(2024)\citenamefont {{Dittmann}}, \citenamefont {{Dempsey}},\ and\ \citenamefont {{Li}}}]{Dittmann2024}%
  \BibitemOpen
  \bibfield  {author} {\bibinfo {author} {\bibfnamefont {A.~J.}\ \bibnamefont {{Dittmann}}}, \bibinfo {author} {\bibfnamefont {A.~M.}\ \bibnamefont {{Dempsey}}},\ and\ \bibinfo {author} {\bibfnamefont {H.}~\bibnamefont {{Li}}},\ }\bibfield  {title} {\bibinfo {title} {{The Evolution of Inclined Binary Black Holes in the Disks of Active Galactic Nuclei}},\ }\href {https://doi.org/10.3847/1538-4357/ad23ce} {\bibfield  {journal} {\bibinfo  {journal} {\apj}\ }\textbf {\bibinfo {volume} {964}},\ \bibinfo {eid} {61} (\bibinfo {year} {2024})},\ \Eprint {https://arxiv.org/abs/2310.03832} {arXiv:2310.03832 [astro-ph.HE]} \BibitemShut {NoStop}%
\bibitem [{\citenamefont {{Whitehead}}\ \emph {et~al.}(2024)\citenamefont {{Whitehead}}, \citenamefont {{Rowan}}, \citenamefont {{Boekholt}},\ and\ \citenamefont {{Kocsis}}}]{Whitehead2024}%
  \BibitemOpen
  \bibfield  {author} {\bibinfo {author} {\bibfnamefont {H.}~\bibnamefont {{Whitehead}}}, \bibinfo {author} {\bibfnamefont {C.}~\bibnamefont {{Rowan}}}, \bibinfo {author} {\bibfnamefont {T.}~\bibnamefont {{Boekholt}}},\ and\ \bibinfo {author} {\bibfnamefont {B.}~\bibnamefont {{Kocsis}}},\ }\bibfield  {title} {\bibinfo {title} {{Gas assisted binary black hole formation in AGN discs}},\ }\href {https://doi.org/10.1093/mnras/stae1430} {\bibfield  {journal} {\bibinfo  {journal} {\mnras}\ }\textbf {\bibinfo {volume} {531}},\ \bibinfo {pages} {4656} (\bibinfo {year} {2024})},\ \Eprint {https://arxiv.org/abs/2309.11561} {arXiv:2309.11561 [astro-ph.GA]} \BibitemShut {NoStop}%
\bibitem [{\citenamefont {{Cantiello}}\ \emph {et~al.}(2021)\citenamefont {{Cantiello}}, \citenamefont {{Jermyn}},\ and\ \citenamefont {{Lin}}}]{Cantiello2021}%
  \BibitemOpen
  \bibfield  {author} {\bibinfo {author} {\bibfnamefont {M.}~\bibnamefont {{Cantiello}}}, \bibinfo {author} {\bibfnamefont {A.~S.}\ \bibnamefont {{Jermyn}}},\ and\ \bibinfo {author} {\bibfnamefont {D.~N.~C.}\ \bibnamefont {{Lin}}},\ }\bibfield  {title} {\bibinfo {title} {{Stellar Evolution in AGN Disks}},\ }\href {https://doi.org/10.3847/1538-4357/abdf4f} {\bibfield  {journal} {\bibinfo  {journal} {\apj}\ }\textbf {\bibinfo {volume} {910}},\ \bibinfo {eid} {94} (\bibinfo {year} {2021})},\ \Eprint {https://arxiv.org/abs/2009.03936} {arXiv:2009.03936 [astro-ph.SR]} \BibitemShut {NoStop}%
\bibitem [{\citenamefont {{Dittmann}}\ \emph {et~al.}(2021)\citenamefont {{Dittmann}}, \citenamefont {{Cantiello}},\ and\ \citenamefont {{Jermyn}}}]{Dittmann2021}%
  \BibitemOpen
  \bibfield  {author} {\bibinfo {author} {\bibfnamefont {A.~J.}\ \bibnamefont {{Dittmann}}}, \bibinfo {author} {\bibfnamefont {M.}~\bibnamefont {{Cantiello}}},\ and\ \bibinfo {author} {\bibfnamefont {A.~S.}\ \bibnamefont {{Jermyn}}},\ }\bibfield  {title} {\bibinfo {title} {{Accretion onto Stars in the Disks of Active Galactic Nuclei}},\ }\href {https://doi.org/10.3847/1538-4357/ac042c} {\bibfield  {journal} {\bibinfo  {journal} {\apj}\ }\textbf {\bibinfo {volume} {916}},\ \bibinfo {eid} {48} (\bibinfo {year} {2021})},\ \Eprint {https://arxiv.org/abs/2102.12484} {arXiv:2102.12484 [astro-ph.GA]} \BibitemShut {NoStop}%
\bibitem [{\citenamefont {{Jermyn}}\ \emph {et~al.}(2022)\citenamefont {{Jermyn}}, \citenamefont {{Dittmann}}, \citenamefont {{McKernan}}, \citenamefont {{Ford}},\ and\ \citenamefont {{Cantiello}}}]{Jermyn2022}%
  \BibitemOpen
  \bibfield  {author} {\bibinfo {author} {\bibfnamefont {A.~S.}\ \bibnamefont {{Jermyn}}}, \bibinfo {author} {\bibfnamefont {A.~J.}\ \bibnamefont {{Dittmann}}}, \bibinfo {author} {\bibfnamefont {B.}~\bibnamefont {{McKernan}}}, \bibinfo {author} {\bibfnamefont {K.~E.~S.}\ \bibnamefont {{Ford}}},\ and\ \bibinfo {author} {\bibfnamefont {M.}~\bibnamefont {{Cantiello}}},\ }\bibfield  {title} {\bibinfo {title} {{Effects of an Immortal Stellar Population in AGN Disks}},\ }\href {https://doi.org/10.3847/1538-4357/ac5d40} {\bibfield  {journal} {\bibinfo  {journal} {\apj}\ }\textbf {\bibinfo {volume} {929}},\ \bibinfo {eid} {133} (\bibinfo {year} {2022})},\ \Eprint {https://arxiv.org/abs/2203.06187} {arXiv:2203.06187 [astro-ph.GA]} \BibitemShut {NoStop}%
\bibitem [{\citenamefont {{Wang}}\ \emph {et~al.}(2021)\citenamefont {{Wang}}, \citenamefont {{Liu}}, \citenamefont {{Ho}},\ and\ \citenamefont {{Du}}}]{Wang2021}%
  \BibitemOpen
  \bibfield  {author} {\bibinfo {author} {\bibfnamefont {J.-M.}\ \bibnamefont {{Wang}}}, \bibinfo {author} {\bibfnamefont {J.-R.}\ \bibnamefont {{Liu}}}, \bibinfo {author} {\bibfnamefont {L.~C.}\ \bibnamefont {{Ho}}},\ and\ \bibinfo {author} {\bibfnamefont {P.}~\bibnamefont {{Du}}},\ }\bibfield  {title} {\bibinfo {title} {{Accretion-modified Stars in Accretion Disks of Active Galactic Nuclei: Slowly Transient Appearance}},\ }\href {https://doi.org/10.3847/2041-8213/abee81} {\bibfield  {journal} {\bibinfo  {journal} {\apjl}\ }\textbf {\bibinfo {volume} {911}},\ \bibinfo {eid} {L14} (\bibinfo {year} {2021})},\ \Eprint {https://arxiv.org/abs/2103.07708} {arXiv:2103.07708 [astro-ph.HE]} \BibitemShut {NoStop}%
\bibitem [{\citenamefont {{Li}}\ \emph {et~al.}(2022{\natexlab{b}})\citenamefont {{Li}}, \citenamefont {{Chen}}, \citenamefont {{Lin}},\ and\ \citenamefont {{Wang}}}]{Li2022a}%
  \BibitemOpen
  \bibfield  {author} {\bibinfo {author} {\bibfnamefont {Y.-P.}\ \bibnamefont {{Li}}}, \bibinfo {author} {\bibfnamefont {Y.-X.}\ \bibnamefont {{Chen}}}, \bibinfo {author} {\bibfnamefont {D.~N.~C.}\ \bibnamefont {{Lin}}},\ and\ \bibinfo {author} {\bibfnamefont {Z.}~\bibnamefont {{Wang}}},\ }\bibfield  {title} {\bibinfo {title} {{Spin Evolution of Stellar-mass Black Holes Embedded in AGN Disks: Orbital Eccentricity Produces Retrograde Circumstellar Flows}},\ }\href {https://doi.org/10.3847/2041-8213/ac5b61} {\bibfield  {journal} {\bibinfo  {journal} {\apjl}\ }\textbf {\bibinfo {volume} {928}},\ \bibinfo {eid} {L1} (\bibinfo {year} {2022}{\natexlab{b}})},\ \Eprint {https://arxiv.org/abs/2203.05539} {arXiv:2203.05539 [astro-ph.HE]} \BibitemShut {NoStop}%
\bibitem [{\citenamefont {Piovano}\ \emph {et~al.}(2021)\citenamefont {Piovano}, \citenamefont {Brito}, \citenamefont {Maselli},\ and\ \citenamefont {Pani}}]{Piovano2021SecondarySpin}%
  \BibitemOpen
  \bibfield  {author} {\bibinfo {author} {\bibfnamefont {G.~A.}\ \bibnamefont {Piovano}}, \bibinfo {author} {\bibfnamefont {R.}~\bibnamefont {Brito}}, \bibinfo {author} {\bibfnamefont {A.}~\bibnamefont {Maselli}},\ and\ \bibinfo {author} {\bibfnamefont {P.}~\bibnamefont {Pani}},\ }\bibfield  {title} {\bibinfo {title} {Assessing the detectability of the secondary spin in extreme mass-ratio inspirals with fully-relativistic numerical waveforms},\ }\href {https://doi.org/10.1103/PhysRevD.104.124019} {\bibfield  {journal} {\bibinfo  {journal} {Phys. Rev. D}\ }\textbf {\bibinfo {volume} {104}},\ \bibinfo {pages} {124019} (\bibinfo {year} {2021})},\ \Eprint {https://arxiv.org/abs/2105.07083} {arXiv:2105.07083 [gr-qc]} \BibitemShut {NoStop}%
\bibitem [{\citenamefont {Lin}\ \emph {et~al.}(2025)\citenamefont {Lin}, \citenamefont {Liu},\ and\ \citenamefont {Zheng}}]{Lin2025ResonanceCapture}%
  \BibitemOpen
  \bibfield  {author} {\bibinfo {author} {\bibfnamefont {L.}~\bibnamefont {Lin}}, \bibinfo {author} {\bibfnamefont {B.}~\bibnamefont {Liu}},\ and\ \bibinfo {author} {\bibfnamefont {Z.}~\bibnamefont {Zheng}},\ }\href {https://arxiv.org/abs/2501.12650} {\bibinfo {title} {Resonance capture and stability analysis for planet pairs under type i disk migration}} (\bibinfo {year} {2025}),\ \bibinfo {note} {arXiv:2501.12650},\ \Eprint {https://arxiv.org/abs/2501.12650} {arXiv:2501.12650 [astro-ph.EP]} \BibitemShut {NoStop}%
\bibitem [{\citenamefont {Murray}\ and\ \citenamefont {Dermott}(1999)}]{Murray1999}%
  \BibitemOpen
  \bibfield  {author} {\bibinfo {author} {\bibfnamefont {C.~D.}\ \bibnamefont {Murray}}\ and\ \bibinfo {author} {\bibfnamefont {S.~F.}\ \bibnamefont {Dermott}},\ }\href {https://doi.org/10.1017/CBO9781139174817} {\emph {\bibinfo {title} {Solar System Dynamics}}}\ (\bibinfo  {publisher} {Cambridge University Press},\ \bibinfo {address} {Cambridge, UK},\ \bibinfo {year} {1999})\BibitemShut {NoStop}%
\bibitem [{\citenamefont {{Ida}}\ \emph {et~al.}(2025)\citenamefont {{Ida}}, \citenamefont {{Li}}, \citenamefont {{Pan}}, \citenamefont {{Chen}},\ and\ \citenamefont {{Lin}}}]{Ida2025}%
  \BibitemOpen
  \bibfield  {author} {\bibinfo {author} {\bibfnamefont {S.}~\bibnamefont {{Ida}}}, \bibinfo {author} {\bibfnamefont {Y.-P.}\ \bibnamefont {{Li}}}, \bibinfo {author} {\bibfnamefont {J.-P.}\ \bibnamefont {{Pan}}}, \bibinfo {author} {\bibfnamefont {Y.-X.}\ \bibnamefont {{Chen}}},\ and\ \bibinfo {author} {\bibfnamefont {D.}~\bibnamefont {{Lin}}},\ }\bibfield  {title} {\bibinfo {title} {{}},\ }\href@noop {} {\bibfield  {journal} {\bibinfo  {journal} {submitted}\ } (\bibinfo {year} {2025})}\BibitemShut {NoStop}%
\bibitem [{\citenamefont {{Pan}}\ \emph {et~al.}(2025)\citenamefont {{Pan}}, \citenamefont {{Li}}, \citenamefont {{Chen}}, \citenamefont {{Ida}},\ and\ \citenamefont {{Lin}}}]{Pan2025}%
  \BibitemOpen
  \bibfield  {author} {\bibinfo {author} {\bibfnamefont {J.-P.}\ \bibnamefont {{Pan}}}, \bibinfo {author} {\bibfnamefont {Y.-P.}\ \bibnamefont {{Li}}}, \bibinfo {author} {\bibfnamefont {Y.-X.}\ \bibnamefont {{Chen}}}, \bibinfo {author} {\bibfnamefont {S.}~\bibnamefont {{Ida}}},\ and\ \bibinfo {author} {\bibfnamefont {D.~N.~C.}\ \bibnamefont {{Lin}}},\ }\bibfield  {title} {\bibinfo {title} {{ }},\ }\href@noop {} {\bibfield  {journal} {\bibinfo  {journal} {submitted}\ } (\bibinfo {year} {2025})}\BibitemShut {NoStop}%
\bibitem [{\citenamefont {{Fung}}\ \emph {et~al.}(2019)\citenamefont {{Fung}}, \citenamefont {{Zhu}},\ and\ \citenamefont {{Chiang}}}]{Fung2019}%
  \BibitemOpen
  \bibfield  {author} {\bibinfo {author} {\bibfnamefont {J.}~\bibnamefont {{Fung}}}, \bibinfo {author} {\bibfnamefont {Z.}~\bibnamefont {{Zhu}}},\ and\ \bibinfo {author} {\bibfnamefont {E.}~\bibnamefont {{Chiang}}},\ }\bibfield  {title} {\bibinfo {title} {{Circumplanetary Disk Dynamics in the Isothermal and Adiabatic Limits}},\ }\href {https://doi.org/10.3847/1538-4357/ab53da} {\bibfield  {journal} {\bibinfo  {journal} {\apj}\ }\textbf {\bibinfo {volume} {887}},\ \bibinfo {eid} {152} (\bibinfo {year} {2019})},\ \Eprint {https://arxiv.org/abs/1909.09655} {arXiv:1909.09655 [astro-ph.EP]} \BibitemShut {NoStop}%
\bibitem [{\citenamefont {Mancieri}\ \emph {et~al.}(2025{\natexlab{b}})\citenamefont {Mancieri}, \citenamefont {Broggi}, \citenamefont {Bonetti},\ and\ \citenamefont {Sesana}}]{Mancieri2025_HangingOnTheCliff}%
  \BibitemOpen
  \bibfield  {author} {\bibinfo {author} {\bibfnamefont {D.}~\bibnamefont {Mancieri}}, \bibinfo {author} {\bibfnamefont {L.}~\bibnamefont {Broggi}}, \bibinfo {author} {\bibfnamefont {M.}~\bibnamefont {Bonetti}},\ and\ \bibinfo {author} {\bibfnamefont {A.}~\bibnamefont {Sesana}},\ }\bibfield  {title} {\bibinfo {title} {Hanging on the cliff: Extreme mass ratio inspiral formation with local two-body relaxation and post-newtonian dynamics},\ }\href {https://doi.org/10.1051/0004-6361/202452306} {\bibfield  {journal} {\bibinfo  {journal} {Astronomy \& Astrophysics}\ }\textbf {\bibinfo {volume} {694}},\ \bibinfo {pages} {A272} (\bibinfo {year} {2025}{\natexlab{b}})},\ \Eprint {https://arxiv.org/abs/2409.09122} {arXiv:2409.09122 [astro-ph.HE]} \BibitemShut {NoStop}%
\bibitem [{\citenamefont {Qunbar}\ and\ \citenamefont {Stone}(2023)}]{Qunbar2023}%
  \BibitemOpen
  \bibfield  {author} {\bibinfo {author} {\bibfnamefont {I.}~\bibnamefont {Qunbar}}\ and\ \bibinfo {author} {\bibfnamefont {N.~C.}\ \bibnamefont {Stone}},\ }\bibfield  {title} {\bibinfo {title} {Enhanced extreme mass ratio inspiral rates into intermediate mass black holes},\ }\href {https://arxiv.org/abs/2304.13062} {\bibfield  {journal} {\bibinfo  {journal} {arXiv e-prints}\ }\textbf {\bibinfo {volume} {2023}},\ \bibinfo {pages} {arXiv:2304.13062} (\bibinfo {year} {2023})},\ \Eprint {https://arxiv.org/abs/2304.13062} {arXiv:2304.13062 [astro-ph.GA]} \BibitemShut {NoStop}%
\end{thebibliography}%

\end{document}